\documentclass{lmcs}
\pdfoutput=1

\usepackage{lastpage}
\lmcsdoi{20}{4}{6}
\lmcsheading{}{\pageref{LastPage}}{}{}%
{Oct.~07,~2022}{Nov.~13,~2024}{}

\usepackage[T1]{fontenc}
\usepackage[utf8]{inputenc}
\usepackage{style/notation}
\usepackage{colortbl}
\usepackage{booktabs}

\theoremstyle{plain}

\begin{document}

\title[Asynchronous Sessions: Deadlock Freedom in Cyclic Process Networks]
    {Asynchronous Session-Based Concurrency: \texorpdfstring{\\}{} Deadlock Freedom in Cyclic Process Networks}

\author[B.~van~den~Heuvel]{Bas van den Heuvel\lmcsorcid{0000-0002-8264-7371}}[a]
\author[J.A.~P\'erez]{Jorge A.\ P\'erez\lmcsorcid{0000-0002-1452-6180}}[b]
\address{HKA Karlsruhe and University of Freiburg, Germany}
\address{University of Groningen, The Netherlands}

\begin{abstract}
    \noindent
    We tackle the challenge of ensuring the deadlock-freedom property for message-passing processes that communicate asynchronously in cyclic process networks.
    Our contributions are twofold.
    First, we present Asynchronous Priority-based Classical Processes (\APCP), a session-typed process framework that supports asynchronous communication, delegation, and recursion in cyclic process networks.
    Building upon the Curry-Howard correspondences between linear logic and session types, we establish essential meta-theoretical results for \APCP, most notably deadlock freedom.
    Second, we present a new concurrent $\lambda$-calculus with asynchronous session types, dubbed \FIRST.
    We illustrate \FIRST by example and establish its meta-theoretical results;
    in particular, we show how to soundly transfer the deadlock-freedom guarantee from \APCP. To this end, we develop a translation of terms in \FIRST into  processes in \APCP that satisfies  a strong formulation of operational correspondence.
\end{abstract}

\maketitle

\section{Introduction}
\label{s:intro}

Modern software systems often comprise independent components that coordinate by exchanging messages.
The $\pi$-calculus~\cite{journal/ic/MilnerPW92,book/Milner89} is a mature formalism for specifying and reasoning about message-passing processes; in particular, it offers a rigorous foundation for designing \emph{type systems} that statically enforce communication correctness.
A well-known approach in this line is \emph{session types}~\cite{conf/concur/Honda93,conf/esop/HondaVK98,conf/secret/YoshidaV07}, which specify the structure of the two-party communication protocols implemented by the channels of a process.
In this paper, we are interested in session types as a governing discipline in concurrent and functional paradigms, in conjunction with two important aspects of message-passing concurrency, namely the \emph{network topologies} formed by interacting processes and the underlying discipline of \emph{asynchronous communication}.

The study of session types has gained considerable attention after the discovery by Caires and Pfenning~\cite{conf/concur/CairesP10} and Wadler~\cite{conf/icfp/Wadler12} of Curry-Howard correspondences between session types and linear logic~\cite{journal/tcs/Girard87}.
The present work is motivated by (and develops further) two salient features of type systems derived from these correspondences, namely (i)~their remarkably effective approach to establishing the \emph{deadlock-freedom} property for processes, and (ii)~their clear connections with \emph{functional calculi} with concurrency. These two aspects form the central themes of this paper, and we discuss them in order.

\paragraph{Deadlock Freedom}
Curry-Howard approaches to session types   induce a very precise form of interaction between parallel processes: they interpret the cut rule of linear logic as the interaction of two processes on \emph{exactly one channel}.
While this design elegantly rules out the insidious circular dependencies that lead to deadlocks, there is a catch: typable processes cannot be connected to form \emph{cyclic} network topologies---only tree-shaped networks are allowed.
Hence, type systems based upon Curry-Howard foundations reject whole classes of  process networks that are cyclic but deadlock-free~\cite{conf/express/DardhaP15,journal/jlamp/DardhaP22}.
This includes important concurrency patterns, such as those exemplified by Milner's cyclic scheduler~\cite{book/Milner89}.

The problem of designing type systems that rule out circular dependencies and deadlocks while allowing for cyclic topologies has received considerable attention.
Works by Kobayashi and others  have developed advanced solutions; see, e.g.,~\cite{conf/concur/Kobayashi06,conf/lics/Padovani14,conf/fossacs/DardhaG18}.
In a nutshell, these works exploit orderings based on {priority annotations} on types to detect and avoid circular dependencies.
Dardha and Gay~\cite{conf/fossacs/DardhaG18} have shown how to incorporate this priority-based approach in the realm of session type systems based on linear logic; it boils down to replacing the cut rule with a cycle rule and adding priority checks in other selected typing rules.
Their work thus extends the class of typable processes to cover cyclic network topologies, while retaining strong ties with linear logic.

Unfortunately, none of the methods proposed until now consider \emph{session types with asynchronous communication}.
Addressing asynchronous communication is of clear practical relevance: it is the standard in most distributed systems and web-based applications nowadays.
In a process calculi setting, asynchronous communication means that output prefixes are non-blocking~\cite{conf/ecoop/HondaT91,conf/occ/HondaT91,report/Boudol92}, and that exchanged messages implicitly or explicitly reside in an auxiliary structure, such as a buffer or a queue~\cite{chapter/BeauxisPV08}.
In the context of session types, asynchrony moreover means that the ordering of messages \emph{within a session} should be respected, but messages from \emph{different sessions} need not be ordered~\cite{conf/forte/KouzapasYH11}.

To address this gap,
in the first part of the paper
we define a new session-typed asynchronous $\pi$-calculus, \APCP \emph{(Asynchronous Priority-based Classical Processes)}, for which we develop its fundamental meta-theoretical results.
The design of \APCP builds upon insights developed in several prior works:
\begin{itemize}
    \item
        Advanced type systems that exploit annotations on types to enforce deadlock freedom of cyclic process networks, already mentioned;

    \item
        Dardha and Gay's \PCP (Priority-based Classical Processes)~\cite{conf/fossacs/DardhaG18}, also already mentioned, which incorporates into Wadler's \CP~\cite{conf/icfp/Wadler12} (Classical Processes; derived from classical linear logic) Padovani's simplification of Kobayashi's type annotations~\cite{conf/lics/Padovani14}.

    \item
        DeYoung \etal's asynchronous semantics for session communication, defined in context of the correspondence between intuitionistic linear logic and session types~\cite{conf/csl/DeYoungCPT12}.
\end{itemize}
Our calculus
\APCP combines these semantics for asynchronous communication with \PCP's  priority-based type system.
The design of \APCP uncovers fundamental properties of type systems for asynchronous communication.
A particular insight is the following: because outputs are non-blocking, \APCP simplifies priority management while preserving deadlock freedom.
Additionally, as an orthogonal feature, \APCP increases expressivity by supporting tail recursion without compromising deadlock freedom.
We motivate these features of \APCP by discussing Milner's cyclic scheduler in \Cref{s:APCP:milner}.

\paragraph{Functional Calculi with Concurrency}
Session types are paradigm-independent, in the sense that they can be accommodated on top of programming models and languages in different paradigms---concurrent, object-oriented, and functional.
In the functional setting, a milestone is the asynchronous concurrent $\lambda$-calculus with sessions by Gay and Vasconcelos~\cite{journal/jfp/GayV10}, which in the following we shall call \LAST.\footnote{\LAST stands for `Linear Asynchronous Session Types'.}
\LAST is a call-by-value calculus
in which collections of threads (configurations) communicate following session protocols by relying on buffered channels.
The type system of \LAST ensures that well-typed configurations respect their ascribed protocols (protocol fidelity) but
does not guarantee deadlock freedom, i.e., it allows the typing of cyclic  configurations with circular dependencies.

\LAST has been brought back to the spotlight through Wadler's work on \GV (Good Variation), a synchronous variant of \LAST without cyclic configurations.
Terms in \GV are guaranteed to be deadlock-free via a (typed) translation into \CP~\cite{conf/icfp/Wadler12}.
Subsequently, Kokke and Dardha formulated \PGV (Priority-based \GV), an extension of \GV that strictly augments the class of deadlock-free computations with cyclic configurations by leveraging priorities~\cite{conf/forte/KokkeD21,report/KokkeD21}, following Padovani and Novara~\cite{conf/forte/PadovaniN15}.

In the second part of the paper, we study asynchronous, deadlock-free communication with support for cyclic topologies in the  setting of a prototypical functional programming language.
We present \FIRST, a new call-by-name variant of \LAST.
Notably, session communication in \FIRST is asynchronous---a feature not accounted for by \GV and \PGV (see also \Cref{s:APCP:rw} for extended comparisons).

We equip \FIRST with a deliberately simple type system, with functional and session types, which ensures type preservation/protocol fidelity but not deadlock freedom (just like the type system for \LAST).
To address this gap, we develop a way of soundly transferring the deadlock-freedom property from \APCP to \FIRST.
This transference of results hinges on a translation of \FIRST programs into \APCP processes, in the style for Milner's seminal work~\cite{conf/icalp/Milner90,journal/mscs/Milner92}. The translation clarifies the role of \APCP as an abstract model for asynchronous, functional concurrency; it  satisfies  in particular a tight form of \emph{operational correspondence} that follows the well-known formulation by Gorla~\cite{journal/ic/Gorla10}.
This way, we can ensure that a (class of) well-typed \FIRST  programs with cyclic configurations satisfies deadlock freedom.
While the development of \FIRST  is of interest in itself (it improves over \GV and \PGV, as just discussed), it is also a significant test for \APCP, its expressiveness and meta-theoretical results.

\paragraph{Contributions}
In summary, in this paper we make the following contributions:
\begin{enumerate}
    \item
        The process calculus \APCP, its associated type system, and its essential meta-theoretical properties: type preservation (\Cref{t:APCP:sr}) and deadlock freedom (\Cref{t:APCP:df}).

    \item
        The functional calculus \FIRST, its associated type system, and its meta-theoretical property of type preservation (\Cref{as:FIRST:tp}).
        \item A translation of \FIRST into \APCP that enjoys operational correspondence properties (\Cref{t:FIRST:transCompl,t:FIRST:transSound}) and unlocks the transfer of deadlock freedom from \APCP to \FIRST (\Cref{t:FIRST:df}).
\end{enumerate}

\paragraph{Organization}
In \Cref{s:examples}, we motivate \APCP  and  \FIRST    by example.
\Cref{s:APCP} defines the language of \APCP processes and its type system, and establishes its meta-theoretical properties.
\Cref{s:LAST} recalls \LAST and its type system, as proposed in \cite{journal/jfp/GayV10}, and briefly discusses the issue of devising an operationally correct translation into \APCP.
Building upon this background, \Cref{s:FIRST} presents \FIRST as a call-by-name variant of \LAST, develops its meta-theoretical results, and gives a correct translation of \FIRST into \APCP.
\Cref{s:APCP:rw} discusses related work and \Cref{s:APCP:concl} draws conclusions.
\Cref{as:FIRST} collects omitted definitions and proofs for \FIRST.

\paragraph{Origin of the results}
The current paper combines, revises, and extends our papers~\cite{conf/ice/vdHeuvelP21,conf/express/vdHeuvelP22}. Here we provide fully detailed proofs, expanded examples, and consolidated comparisons of related works.
While the paper~\cite{conf/ice/vdHeuvelP21} offered an abridged introduction to \APCP, the paper
\cite{conf/express/vdHeuvelP22} developed \CGV, a functional calculus with concurrency (also in the style of Gay and Vasconcelos' \LAST), and gave a correct translation into \APCP.
Novelties with respect to~\cite{conf/ice/vdHeuvelP21} include a revised treatment of type preservation and recursion in \APCP.
With respect to \cite{conf/express/vdHeuvelP22}, in this presentation we have revised \CGV into \FIRST, in order to showcase a more clear connection with \LAST and a well-known reduction strategy (call by name).
As a result, \Cref{s:FIRST,s:LAST} are entirely new to this presentation.

\section{Motivating Examples}
\label{s:examples}

In this section, we informally describe \APCP and \FIRST, the two calculi presented in this paper, using motivating examples that illustrate their distinctive features and expressiveness.

\subsection{Milner's Cyclic Scheduler in \APCP}
\label{s:APCP:milner}

We motivate \APCP by  considering Milner's cyclic scheduler~\cite{book/Milner89}, a recursive process that   relies on a cyclic network to perform asynchronous communications.
This example is inspired by Dardha and Gay~\cite{conf/fossacs/DardhaG18}, who use \PCP to type a synchronous, non-recursive version of the scheduler.

\begin{figure}[t]
    \begin{minipage}[b]{.5\textwidth}
        \begin{tikzpicture}[
            proc/.style={fill=none, draw=none, shape=circle, inner sep=-.5},
            thick/.style={-, line width=1pt},
            thicker/.style={-, line width=1.2pt},
            chan/.style={-, thick, inner sep=3mm},
            chanb/.style={chan, inner sep=2mm},
            chant/.style={chanb, thicker},
            aChan/.style={color=aChanClr},
            bChan/.style={color=bChanClr},
            cChan/.style={color=cChanClr},
            dChan/.style={color=dChanClr},
        ]

            \node [proc] (A1) at (-4, 5.25) {$A_1$};
            \node [proc] (A2) at (-2, 6.5) {$A_2$};
            \node [proc] (A3) at (0, 5.25) {$A_3$};
            \node [proc] (A4) at (0, 3.25) {$A_4$};
            \node [proc] (A5) at (-2, 2) {$A_5$};
            \node [proc] (A6) at (-4, 3.25) {$A_6$};
            \node [proc] (P1) at (-5.5, 6) {$P_1$};
            \node [proc] (P2) at (-2, 8) {$P_2$};
            \node [proc] (P3) at (1.5, 6) {$P_3$};
            \node [proc] (P4) at (1.5, 2.5) {$P_4$};
            \node [proc] (P5) at (-2, .5) {$P_5$};
            \node [proc] (P6) at (-5.5, 2.5) {$P_6$};

            \draw [chan, bend left, looseness=0.75] (A1) to node [dChan, below, pos=.3] {$d_1$} (A2);
            \draw [chanb, bend left, looseness=0.75] (A1) to node [cChan, below, pos=.9] {$c_1$} (A2);
            \draw [chanb, bend left, looseness=0.75] (A2) to node [dChan, below, pos=.15] {$d_2$} (A3);
            \draw [chan, bend left, looseness=0.75] (A2) to node [cChan, below, pos=.7] {$c_2$} (A3);
            \draw [chanb, bend left=15, looseness=0.75] (A3) to node [dChan, left, pos=.2] {$d_3$} (A4);
            \draw [chanb, bend left=15, looseness=0.75] (A3) to node [cChan, left, pos=.8] {$c_3$} (A4);
            \draw [chan, bend left, looseness=0.75] (A4) to node [dChan, above, pos=.3] {$d_4$} (A5);
            \draw [chanb, bend left, looseness=0.75] (A4) to node [cChan, above, pos=.9] {$c_4$} (A5);
            \draw [chanb, bend right=330, looseness=0.75] (A5) to node [dChan, above, pos=.1] {$d_5$} (A6);
            \draw [chan, bend right=330, looseness=0.75] (A5) to node [cChan, above, pos=.7] {$c_5$} (A6);
            \draw [chanb, bend left=15, looseness=0.75] (A6) to node [dChan, right, pos=.15] {$d_6$} (A1);
            \draw [chanb, bend left=15, looseness=0.75] (A6) to node [cChan, right, pos=.8] {$c_6$} (A1);
            \draw [chan] (A1) to node [aChan, below, pos=.3] {$a_1$} (P1);
            \draw [chan] (A1) to node [bChan, below, pos=1] {$b_1$} (P1);
            \draw [chant] (A2) to node [aChan, left, pos=.1] {$a_2$} (P2);
            \draw [chant] (A2) to node [bChan, left, pos=.9] {$b_2$} (P2);
            \draw [chan] (A3) to node [aChan, above, pos=0] {$a_3$} (P3);
            \draw [chan] (A3) to node [bChan, above, pos=.7] {$b_3$} (P3);
            \draw [chan] (A4) to node [aChan, above, pos=.3] {$a_4$} (P4);
            \draw [chan] (A4) to node [bChan, above, pos=1] {$b_4$} (P4);
            \draw [chant] (A5) to node [aChan, right, pos=.1] {$a_5$} (P5);
            \draw [chant] (A5) to node [bChan, right, pos=.9] {$b_5$} (P5);
            \draw [chan] (A6) to node [aChan, below, pos=0] {$a_6$} (P6);
            \draw [chan] (A6) to node [bChan, below, pos=.7] {$b_6$} (P6);
        \end{tikzpicture}
    \end{minipage}%
    \hfill%
    \begin{minipage}[b]{.45\textwidth}
        \caption{
            \label{f:APCP:milner}
            Milner's cyclic sche\-duler with 6 workers.
            Lines denote channels connecting processes on the indicated names.
        }
    \end{minipage}
\end{figure}

\paragraph{The Scheduler, Informally}
The scheduler consists of $n \geq 1$ worker processes $P_i$ (the workers, for short), each attached to a partial scheduler $A_i$.
The partial schedulers connect to each other in a ring structure, together forming the cyclic scheduler.
Connections consist of pairs of endpoints; we further refer to these endpoints by the names that represent them.

The scheduler then lets the workers perform their tasks in rounds, each new round triggered by the leading partial scheduler $A_1$ (the \emph{leader}) once each worker finishes their previous task.
We refer to the non-leading partial schedulers $A_{i+1}$ for $1 \leq i < n$ as the \emph{followers}.

\Cref{f:APCP:milner} illustrates the process network of Milner's cyclic scheduler with 6 workers ($n=6$).
Each partial scheduler $A_i$ has a name \aChan{$a_i$} to connect with the worker $P_i$'s name \bChan{$b_i$}.
The leader $A_1$ has a name \cChan{$c_n$} to connect with $A_n$ and a name \dChan{$d_1$} to connect with $A_2$ (or with $A_1$ if $n=1$; we further elide this case for brevity).
Each follower $A_{i+1}$ has a name \cChan{$c_i$} to connect with $A_i$ and a name \dChan{$d_{i+1}$} to connect with $A_{i+2}$ (or with $A_1$ if $i+1=n$; we also elide this case).

In each round, each follower $A_{i+1}$ awaits a start signal from $A_i$, and then asynchronously signals $P_{i+1}$ and $A_{i+2}$ to start.
After awaiting acknowledgment from $P_{i+1}$ and a next round signal from $A_i$, the follower then signals next round to $A_{i+2}$.
The leader $A_1$, which starts each round of tasks, signals $A_2$ and $P_1$ to start, and, after awaiting acknowledgment from $P_1$, signals a next round to $A_2$.
Then, the leader awaits $A_n$'s start and next round signals.
It is crucial that $A_1$ does not await $A_n$'s start signal before starting $P_1$, as the leader would otherwise not be able to initiate rounds of tasks.

\paragraph{Syntax of \APCP}
Before formally specifying the scheduler, we briefly introduce the syntax of \APCP, and how asynchronous communication works; \Cref{s:APCP} gives formal definitions.

Let us write $a,b,c,\ldots,x, y, z, \ldots$ to denote names.
We write $\pOut x[a,b]$ and $\pIn y(w,z) ; P$ to denote processes for sending and receiving, respectively.
The names $a$ and $w$ stand for the payloads, whereas $b$ and $z$ stand for \emph{continuations}, i.e., the names on which the rest of the session should be performed.
This \emph{continuation-passing style} for asynchronous communication is required to ensure the correct ordering of messages within a session.
A communication redex in \APCP is thus of the form
$\pRes{xy} ( \pOut x[a,b] \| \pIn y(w,z) ; P )$:
the restriction $\pRes{xy}$ serves to declare that $x$ and $y$ are dual names of the same channel, and $\cdot \|\cdot$ denotes parallel composition.
This process reduces to $P \{a/w, b/z\}$: $P$'s names $w$ and $z$ are substituted for by $a$ and $b$, respectively.
Notice how, since the send is a standalone  process, it cannot block any other communications in the process.

The communication of labels $\ell, \ell', \ldots$ follows the same principle: the process $\pSel x[b] < \ell$ denotes the output of a label $\ell$ on $x$, and the process $\pBra y(z) > \ell ; P$ blocks until a label $\ell$ is received on $y$ before continuing as $P$; here again, $b$ and $z$ are continuation names that are sent and received together with $\ell$.
Finally, process $\pRec X(\tilde{x}) ; P$ denotes a recursive definition.
Here, $P$ has access to the names in $\tilde{x}$ and may contain recursive calls $\pCall X<\tilde{y}>$ to indicate a repetition of $P$.
Upon such a recursive call $\pCall X<\tilde{y}>$, the names in $\tilde{y}$ are assigned to $\tilde{x}$ in the next round of $P$.

\paragraph{The  Scheduler in \APCP}
We now formally specify the partial schedulers.
Because each sent label requires a restriction to bind the selection's continuation name to the rest of the session, these processes may look rather complicated.
For example, (part of) the leader is specified as follows (where $\dChan{d'_1},\aChan{a''_1}$ are used in the omitted remainder):
\[
    A_1 \deq \pRec X(\aChan{a_1},\cChan{c_n},\dChan{d_1}) ; \pRes{a\dChan{d'_1}} ( \pSel \dChan{d_1}[a] < \sff{start} \| \pRes{b\aChan{a'_1}} ( \pSel \aChan{a_1}[b] < \sff{start} \| \pBra \aChan{a'_1}(\aChan{a''_1}) > \sff{ack} ; \ldots ) )
\]
To improve readability, we rely on the notation $\pSel* x < \ell \cdot P$, which is syntactic sugar that elides the continuation name involved (cf.\ \Cref{n:APCP:sugar}); our use of the floating dot `\,$\cdot$\,' is intended to stress that the communication of $\ell$ along $x$ does not block any further prefixes in $P$, and the overline `\,$\ol{\phantom{x}}$\,' to indicate that there is a hidden restriction.
Similarly, the notation $\pBra* x > \ell ; P$ elides the continuation name involved in receiving labels, though it remains blocking hence the `\,$;$\,'.

The leader and followers are then specified as follows:
\def\cyclicAsyncMilnerScheds{%
    \begin{align*}
        A_1 &\deq \pRec X(\aChan{a_1},\cChan{c_n},\dChan{d_1}) ; \begin{array}[t]{@{}l@{}}
            \pSel* \dChan{d_1} < \sff{start} \cdot \pSel* \aChan{a_1} < \sff{start} \cdot \pBra* \aChan{a_1} > \sff{ack} ; \pSel* \dChan{d_1} < \sff{next} \cdot {}
            \\
            \pBra* \cChan{c_n} > \sff{start} ; \pBra* \cChan{c_n} > \sff{next} ; \pCall X<\aChan{a_1},\cChan{c_n},\dChan{d_1}>
        \end{array}
        \\
        A_{i+1} &\deq \pRec X(\aChan{a_{i+1}},\cChan{c_i},\dChan{d_{i+1}}) ; \begin{array}[t]{@{}l@{}}
            \pBra* \cChan{c_i} > \sff{start} ; \pSel* \aChan{a_{i+1}} < \sff{start} \cdot \pSel*\dChan{d_{i+1}} < \sff{start} \cdot \pBra* \aChan{a_{i+1}} > \sff{ack} ; {}
            \\
            \pBra* \cChan{c_i} > \sff{next} ; \pSel* \dChan{d_{i+1}} < \sff{next} \cdot \pCall X<\aChan{a_{i+1}},\cChan{c_i},\dChan{d_{i+1}}>
        \end{array}
        \qquad \forall 1 \leq i < n
    \end{align*}
}
\cyclicAsyncMilnerScheds

Assuming that each worker $P_i$ is specified such that it behaves as expected on the name \bChan{$b_i$}, we formally specify the complete scheduler as a ring of partial schedulers connected to workers:
\def\cyclicAsyncMilnerSched{%
    \[
        Sched_n \deq \pRes{\cChan{c_1}\dChan{d_1}} \ldots \pRes{\cChan{c_n}\dChan{d_n}} \big( \pRes{\aChan{a_1}\bChan{b_1}} ( A_1 \| P_1 ) \| \ldots \| \pRes{\aChan{a_n}\bChan{b_n}} ( A_n \| P_n ) \big)
    \]
}
\cyclicAsyncMilnerSched
We return to this example in \Cref{s:APCP:APCP:milner}, where we type check the scheduler using \APCP to show that it is deadlock-free (which is not obvious from its process definition).

\subsection{A Bookshop Scenario in \FIRST}
\label{s:FIRST:example}

Our new calculus \FIRST is a call-by-name variant of Gay and Vasconcelos' \LAST~\cite{journal/jfp/GayV10} with linear resources.
We briefly motivate the design of \FIRST by adapting the running example of~\cite{journal/jfp/GayV10}, which involves  a mother interacting with a  bookshop to buy a book for her son.

First, we define a term representing the shop.
The shop has an endpoint $\term{s}$ on which it communicates with a client.
First, the shop receives a book title and then offers a choice between buying the book or only accessing its blurb (the text on the book's back cover).
If the client decides to buy, the shop receives credit card information and sends the book to the client.
Otherwise, if the client requests the blurb, the shop sends its text.
In \FIRST, we can define this shop as follows:
\[
    \term{\sff{Shop}(s)} \deq \term{
        \begin{array}[t]{@{}l@{}}
            \tLet (\textit{title},s_1) = \tRecv s \tIn
            \\
            \tCase s_1 \tOf \{
                \begin{array}[t]{@{}l@{}}
                    \sff{buy}: \lam s_2. \begin{array}[t]{@{}l@{}}
                        \tLet (\textit{card},s_3) = \tRecv s_2 \tIn
                        \\
                        \tLet s_4 = \tSend \text{book}(\textit{title}) \, s_3 \tIn
                        \\
                        \tClose s_4 ; ()
                        ,
                    \end{array}
                    \\
                    \sff{blurb}: \lam s_2. \begin{array}[t]{@{}l@{}}
                        \tLet s_3 = \tSend \text{blurb}(\textit{title}) \, s_2 \tIn
                        \\
                        \tClose s_3 ; ()
                    \}
                \end{array}
            \end{array}
        \end{array}
    }
\]
where notations $\term{\text{book}(\textit{title})}$ and $\term{\text{blurb}(\textit{title})}$ are syntactic sugar for lookup functions, implemented, e.g., as labeled selections.

The functional behavior of \FIRST is standard, so we only explain the message-passing components in the above term:
\begin{itemize}

    \item
        `$\term{\tRecv s}$' waits for a message to be ready on $\term{s}$, and returns a pair containing the message and an endpoint on which to continue the session.

    \item
        `$\term{\tCase s \tOf \{ \ldots \}}$' waits for a label to be ready on $\term{s}$, determining a continuation.
        Continuations are defined as abstractions, which will be applied to the session's continuation once the label has been received.

    \item
        `$\term{\tSend M \, s}$' buffers the message $M$ on endpoint $\term{s}$, which can be received asynchronously on a connected endpoint.

    \item
        `$\term{\tClose s ; M}$' waits until the session on the endpoint $\term{s}$ can be closed before running $\term{M}$, which acts as a continuation.

\end{itemize}
\noindent 
Next, we define a term abstracting the son's behavior.
Term $\term{\sff{Son}(s',m')}$, given below, has an endpoint~$\term{s'}$ on which he communicates with the shop, and an endpoint $\term{m'}$ for communication with his mother.
The son sends a book title to the shop, and then selects to buy.
Notation `$\term{\tSel \ell \, \sff{s}}$' denotes the buffering of the label $\term{\ell}$ on endpoint $\term{s}$, to be received asynchronously on a connected endpoint.
After this selection, to let his mother pay for him, the son proceeds to send the shop's endpoint to his mother.
Finally, he receives the book from his mother, and returns it as a result of the computations.
\[
    \term{\sff{Son}(s',m')} \deq \term{
        \begin{array}[t]{@{}l@{}}
            \tLet s'_1 = \tSend \text{``Dune''} \, s' \tIn
            \\
            \tLet s'_2 = \tSel \sff{buy} \, s'_1 \tIn
            \\
            \tLet m'_1 = \tSend s'_2 \, m' \tIn
            \\
            \tLet (\textit{book},m'_2) = \tRecv m'_1 \tIn
            \\
            \tClose m'_2 ; \textit{book}
        \end{array}
    }
\]

Finally, we define the mother, who has an endpoint $\term{m}$ to communicate with her son.
She receives the shop endpoint from her son, and then sends her credit card information to the shop.
She then receives the book from the shop, and sends it to her son.
\[
    \term{\sff{Mother}(m)} \deq \term{
        \begin{array}[t]{@{}l@{}}
            \tLet (x,m_1) = \tRecv m \tIn
            \\
            \tLet x_1 = \tSend \text{visa} \, x \tIn
            \\
            \tLet (\textit{book},x_2) = \tRecv x_1 \tIn
            \\
            \tLet m_2 = \tSend \textit{book} \, m_1 \tIn
            \\
            \tClose m_2 ; \tClose x_2 ; ()
        \end{array}
    }
\]
Now, we only have to compose these terms together, connecting all the endpoints appropriately.
This is achieved by term $\term{\sff{Sys}}$ below, which relies on two additional constructs.
\begin{itemize}

    \item
        `$\term{\tNew}$' creates a new channel, and returns a pair containing the channel's two endpoints; asynchronous communication is achieved by connecting the endpoints through an ordered buffer.

    \item
        `$\term{\tFork M ; N}$' denotes splitting of $\term{M}$ as a \emph{thread}, which runs concurrently to the immediate continuation $\term{N}$.

\end{itemize}
Concretely, we have:
\[
    \term{\sff{Sys}} \deq \term{
        \begin{array}[t]{@{}l@{}}
            \tLet (s,s') = \tNew \tIn \tFork \sff{Shop}(s) ;
            \\
            \tLet (m,m') = \tNew \tIn \tFork \sff{Mother}(m) ;
            \\
            \sff{Son}(s',m')
        \end{array}
    }
\]

Note that the son cannot be forked, as he returns the result of the computation (the book).
In \Cref{s:LAST,s:FIRST} we will discuss the behavior and typing of these terms.

\smallskip
After presenting \APCP next, in \Cref{s:FIRST}
we formally define \FIRST and  illustrate the language and its runtime semantics: how new channels are created, new threads are forked, and messages are written to and read from buffers.

\section{\APCP: Asynchronous Priority-based Classical Processes}
\label{s:APCP}

In this section, we define \APCP, a session-typed $\pi$-calculus in which processes communicate asynchronously  on connected channel endpoints.
We further refer to endpoints by the names that represent them.
As already discussed, the output of messages (names and labels) is non-blocking, and explicit continuations ensure the ordering of messages within a session.
In our type system, names are assigned types that specify two-party protocols, in the style of binary session types~\cite{conf/concur/Honda93}, following the Curry-Howard correspondences between linear logic and session types~\cite{conf/concur/CairesP10,conf/icfp/Wadler12}.

\APCP combines the salient features of Dardha and Gay's \PCP~\cite{conf/fossacs/DardhaG18} with DeYoung \etal's semantics for asynchronous communication~\cite{conf/csl/DeYoungCPT12}.
Recursion---not present in the works by Dardha and Gay and DeYoung \etal---is an orthogonal feature, with syntax inspired by the work of Toninho \etal~\cite{conf/tgc/ToninhoCP14}.

As in \PCP, types in \APCP rely on \emph{priority} annotations, which enable cyclic connections while ruling out circular dependencies between sessions.
A key insight of our work is that asynchrony induces different priority management than synchrony: while \PCP's blocking outputs are compared to their continuation processes, \APCP's non-blocking outputs are only compared to their continuation \emph{sessions} (see \Cref{r:APCP:simplerPrios}).

Properties of well-typed \APCP processes are \emph{type preservation} (\Cref{t:APCP:sr}) and \emph{deadlock freedom} (\Cref{t:APCP:df}).
This includes cyclically connected processes, which priority-annotated types guarantee free from circular dependencies that may cause deadlock.

\subsection{The Process Language}
\label{s:APCP:APCP:proclang}

\begin{figure}[t!]
    Process syntax:
    \begin{align*}
        P, Q &
        \begin{array}[t]{@{}l@{}lr@{\kern2em}l@{\kern1em}lr@{}}
            {} ::= {} &
            \pOut x[a,b] & \text{send}
            & \sepr &
            \pIn x(y,z) ; P & \text{receive}
            \\ \sepr* &
            \pSel x[b] < \ell & \text{selection}
            & \sepr &
            \pBra x(z) > {\{ i : P \}_{i \in I}} & \text{branch}
            \\ \sepr* &
            \pRes{xy} P & \text{restriction}
            & \sepr &
            P \| Q & \text{parallel}
            \\ \sepr* &
            \0 & \text{inaction}
            & \sepr &
            \pFwd [x<>y] & \text{forwarder}
            \\ \sepr* &
            \pRec X(\tilde{z}) ; P & \text{recursive definition}
            & \sepr &
            \pCall X<\tilde{z}> & \text{recursive call}
        \end{array}
    \end{align*}

    \phantom{.}\dashes 

    Structural congruence:
    \begin{mathpar}
        \begin{bussproof}[cong-alpha]
            \bussAssume{
                P \equiv_\alpha Q
            }
            \bussUn{
                P \equiv Q
            }
        \end{bussproof}
        \and
        \begin{bussproof}[cong-par-unit]
            \bussAx{
                P \| \0 \equiv P
            }
        \end{bussproof}
        \and
        \begin{bussproof}[cong-par-comm]
            \bussAx{
                P \| Q \equiv Q \| P
            }
        \end{bussproof}
        \and
        \begin{bussproof}[cong-par-assoc]
            \bussAx{
                P \| (Q \| R) \equiv (P \| Q) \| R
            }
        \end{bussproof}
        \and
        \begin{bussproof}[cong-scope]
            \bussAssume{
                x,y \notin \fn(P)
            }
            \bussUn{
                P \| \pRes{xy} Q \equiv \pRes{xy} ( P \| Q )
            }
        \end{bussproof}
        \and
        \begin{bussproof}[cong-res-comm]
            \bussAx{
                \pRes{xy} \pRes{zw} P \equiv \pRes{zw} \pRes{xy} P
            }
        \end{bussproof}
        \and
        \begin{bussproof}[cong-res-symm]
            \bussAx{
                \pRes{xy} P \equiv \pRes{yx} P
            }
        \end{bussproof}
        \and
        \begin{bussproof}[cong-res-inact]
            \bussAx{
                \pRes{xy} \0 \equiv \0
            }
        \end{bussproof}
        \and
        \begin{bussproof}[cong-fwd-symm]
            \bussAx{
                \pFwd [x<>y] \equiv \pFwd [y<>x]
            }
        \end{bussproof}
        \and
        \begin{bussproof}[cong-res-fwd]
            \bussAx{
                \pRes{xy} \pFwd [x<>y] \equiv \0
            }
        \end{bussproof}
        \and
        \begin{bussproof}[cong-unfold]
            \bussAx{
                \pRec X(x_1,\ldots,x_n) ; P \equiv P \big\{ \big( \pRec X(y_1,\ldots,y_n) ; P \{ y_1/x_1,\ldots,y_n/x_n \} \big) / \pCall X<y_1,\ldots,y_n> \big\}
            }
        \end{bussproof}
    \end{mathpar}

    \phantom{.}\dashes 

    Reduction:
    \begin{mathpar}
        \begin{bussproof}[red-send-recv]
            \bussAx{
                \pRes{xy} ( \pOut x[a,b] \| \pIn y(z,y') ; Q )
                \redd
                Q \{a/z,b/y'\}
            }
        \end{bussproof}
        \and
        \begin{bussproof}[red-sel-bra]
            \bussAssume{
                j \in I
            }
            \bussUn{
                \pRes{xy} ( \pSel x[b] < j \| \pBra y(y') > {\{ i : Q_i \}_{i \in I}} )
                \redd
                Q_j \{b/y'\}
            }
        \end{bussproof}
        \and
        \begin{bussproof}[red-fwd]
            \bussAssume{
                y \neq z
            }
            \bussUn{
                \pRes{xy}( \pFwd [x<>z] \| P )
                \redd
                P \{z/y\}
            }
        \end{bussproof}
        \and
        \begin{bussproof}[red-cong]
            \bussAssume{
                P \equiv P'
            }
            \bussAssume{
                P' \redd Q'
            }
            \bussAssume{
                Q' \equiv Q
            }
            \bussTern{
                P \redd Q
            }
        \end{bussproof}
        \and
        \begin{bussproof}[red-res]
            \bussAssume{
                P \redd Q
            }
            \bussUn{
                \pRes{xy} P \redd \pRes{xy} Q
            }
        \end{bussproof}
        \and
        \begin{bussproof}[red-par]
            \bussAssume{
                P \redd Q
            }
            \bussUn{
                P \| R \redd Q \| R
            }
        \end{bussproof}
    \end{mathpar}

    \caption{Definition of \APCP's process language.}
    \label{f:procdef}
\end{figure}

We write $a,b,c,\ldots,x, y, z, \ldots$ to denote (channel) \emph{names} (also known as \emph{names}); by convention we use the early letters of the alphabet for the objects of output-like prefixes.
Also, we write $\tilde{x}, \tilde{y}, \tilde{z}, \ldots$ to denote sequences of names.
In \APCP,  communication is
\emph{asynchronous} (cf.~\cite{conf/ecoop/HondaT91,conf/occ/HondaT91,report/Boudol92}) and
\emph{dyadic}: each communication involves the transmission of a pair of names, a message name and a continuation name.
With a slight abuse of notation, we sometimes write $x_i \in  \tilde{x}$ to refer to a specific element in the sequence $\tilde{x}$.
Also, we write $i, j, k, \ldots$ to denote \emph{labels} for choices and $I, J, K, \ldots$ to denote sets of labels.
We write $X, Y, \ldots$ to denote \emph{recursion variables}, and $P,Q, \ldots$ to denote processes.

\begin{defi}[\APCP Syntax]
    \label{d:APCP:syntax}
    The syntax of \APCP processes is as follows:
    \begin{align*}
        P, Q &
        \begin{array}[t]{@{}l@{}lr@{\kern2em}l@{\kern1em}lr@{}}
            {} ::= {} &
            \pOut x[a,b] & \text{send}
            & \sepr &
            \pIn x(y,z) ; P & \text{receive}
            \\ \sepr* &
            \pSel x[b] < \ell & \text{selection}
            & \sepr &
            \pBra x(z) > {\{ i : P \}_{i \in I}} & \text{branch}
            \\ \sepr* &
            \pRes{xy} P & \text{restriction}
            & \sepr &
            P \| Q & \text{parallel}
            \\ \sepr* &
            \0 & \text{inaction}
            & \sepr &
            \pFwd [x<>y] & \text{forwarder}
            \\ \sepr* &
            \pRec X(\tilde{z}) ; P & \text{recursive definition}
            & \sepr &
            \pCall X<\tilde{z}> & \text{recursive call}
        \end{array}
    \end{align*}
\end{defi}
\noindent 
\Cref{f:procdef} (top) gives the syntax of processes.
The send $\pOut x[a,b]$ sends along $x$ a message name $a$ and a continuation name $b$.
The receive~$\pIn x(y,z) ; P$ blocks until on $x$ a message and continuation name are received (referred to in~$P$ as the placeholders $y$ and $z$, respectively), binding $y$ and $z$ in $P$.
The selection~\mbox{$\pSel x[b] < i$} sends along $x$ a label $i$ and a continuation name $b$.
The branch $\pBra x(z) > {\{ i : P_i \}_{i \in I}}$ blocks until it receives on $x$ a label $i \in I$ and a continuation name (referred to in $P_i$ as the placeholder~$z$), binding $z$ in each $P_i$.
In the rest of this paper, we refer to sends, receives, selections, and branches---including their continuations, if any---as \emph{prefixes}.
We refer to sends and selections collectively as \emph{outputs}, and to receives and branches as \emph{inputs}.

\noindent 
Restriction $\pRes{xy} P$ binds $x$ and $y$ in $P$, thus declaring them as the two names of a channel and enabling communication, as in~\cite{journal/ic/Vasconcelos12}.
The process $P \| Q$ denotes the parallel composition of $P$ and $Q$.
The process $\0$ denotes inaction.
The forwarder $\pFwd [x<>y]$ is a primitive copycat process that links together $x$ and $y$.
We say a forwarder $\pFwd [x<>y]$ in $P$ is \emph{independent} if $P$ does not bind $x$ and $y$ together through restriction (and \emph{dependent} if it does).
The process $\pRec X(\tilde{z}) ; P$ denotes a recursive definition, binding occurrences of $X$ in $P$; the names $\tilde{z}$ form a context for $P$.
Then $P$ may contain recursive calls $\pCall X<\tilde{z}>$ that indicate a repetition of $P$, providing the names $\tilde{z}$ as context.
We only consider contractive recursion, disallowing processes with subexpressions of the form $\pRec X_1(\tilde{z}) ; \ldots ; \pRec X_n(\tilde{z}) ; \pCall X_1<\tilde{z}>$.

Names and recursion variables
are free unless otherwise stated (i.e., unless they are bound somehow).
We write $\fn(P)$
and $\frv(P)$
for the sets of free names
and free recursion variables
of $P$, respectively,
and $\bn(P)$ for the set of bound names of $P$.
Also, we write $P \{ x/y \}$ to denote the capture-avoiding substitution of the free occurrences of $y$ in $P$ for $x$.
Notation $P \big\{ \big( \pRec X(y_1,\ldots,y_n) ; P' \big) / \pCall X<y_1,\ldots,y_n> \big\}$ denotes the substitution of occurrences of recursive calls $\pCall X<y_1,\ldots,y_n>$ in $P$ with the recursive definition $\pRec X(y_1,\ldots,y_n) ; P'$, which we call \emph{unfolding} recursion.
We write sequences of substitutions $P \{ x_1/y_1 \} \ldots \{ x_n/y_n \}$ as $P \{ x_1/y_1, \ldots, x_n/y_n \}$.

Except for asynchrony and recursion, there are minor differences with respect to the languages of Dardha and Gay~\cite{conf/fossacs/DardhaG18} and DeYoung \etal~\cite{conf/csl/DeYoungCPT12}.
Unlike Dardha and Gay's, our syntax does not include empty send and receive prefixes that explicitly close channels; this simplifies the type system.
We also do not include the operator for replicated servers, denoted $\pSrv x(y) ; P$, which is present in~\cite{conf/fossacs/DardhaG18,conf/csl/DeYoungCPT12}.
Although replication can be handled without difficulties, we omit it here; we prefer focusing on recursion, because it fits well with the examples we consider.
See \Cref{s:APCP:APCP:exts} for further discussion.

\paragraph{A Convenient Notation}

In the send $\pOut x[a,b]$ and the selection
$\pSel x[b] < \ell$, names $a$ and $b$ are \emph{free}. They can be bound to a continuation process using parallel composition and restriction, as in $\pRes{ay} \pRes{bz} ( \pOut x[a,b] \| P_{y,z} )$
and
$\pRes{bz} ( \pSel x[b] < \ell] \| Q_{z} )$,
respectively.
We introduce useful notations that abstract away from elide these constructs and continuation names:

\begin{notation}[Derivable Bound Communication]
    \label{n:APCP:sugar}
    We use the following syntactic sugar:
    \begin{align*}
        \pOut* x[y] \cdot P
        & \deq \pRes{ya} \pRes{zb} ( \pOut x[a,b] \| P \{ z/x \} )
        &
        \pSel* x < \ell \cdot P
        & \deq \pRes{zb} ( \pSel x[b] < \ell \| P \{ z/x \} )
        \\
        \pIn x(y) ; P
        & \deq \pIn x(y,z) ; P \{ z/x \}
        &
        \pBra* x > {\{ i : P_i \}_{i \in I}}
        & \deq \pBra x(z) > {\{ i : P_i \{ z/x \} \}_{i \in I}}
    \end{align*}
\end{notation}

\noindent
Note our use of `$\,\cdot\,$' instead of `$\,;\,$' in sending and selection to stress that they are non-blocking operators.
As we will see, these derived constructs are typable (\Cref{t:APCP:derivable})
\paragraph{Operational Semantics}

We define a reduction relation for processes ($P \redd Q$) that formalizes how complementary outputs/inputs on connected names may synchronize.
As usual for $\pi$-calculi, reduction relies on \emph{structural congruence} ($P \equiv Q$), which relates processes with minor syntactic differences.
Structural congruence is the smallest congruence on the syntax of processes (\Cref{f:procdef} (top)) satisfying the axioms in \Cref{f:procdef} (center).

\begin{defi}[Structural Congruence ($\equiv$) for \APCP]
    \label{d:APCP:APCP:strcong}
    Structural congruence for \APCP, denoted $P \equiv Q$, is the smallest congruence on the syntax of processes (\Cref{d:APCP:syntax}) satisfying the axioms in \Cref{f:APCP:strcong}.
    \begin{figure}[t]
        \begin{mathpar}
            \begin{bussproof}[cong-alpha]
                \bussAssume{
                    P \equiv_\alpha Q
                }
                \bussUn{
                    P \equiv Q
                }
            \end{bussproof}
            \and
            \begin{bussproof}[cong-par-unit]
                \bussAx{
                    P \| \0 \equiv P
                }
            \end{bussproof}
            \and
            \begin{bussproof}[cong-par-comm]
                \bussAx{
                    P \| Q \equiv Q \| P
                }
            \end{bussproof}
            \and
            \begin{bussproof}[cong-par-assoc]
                \bussAx{
                    P \| (Q \| R) \equiv (P \| Q) \| R
                }
            \end{bussproof}
            \and
            \begin{bussproof}[cong-scope]
                \bussAssume{
                    x,y \notin \fn(P)
                }
                \bussUn{
                    P \| \pRes{xy} Q \equiv \pRes{xy} ( P \| Q )
                }
            \end{bussproof}
            \and
            \begin{bussproof}[cong-res-comm]
                \bussAx{
                    \pRes{xy} \pRes{zw} P \equiv \pRes{zw} \pRes{xy} P
                }
            \end{bussproof}
            \and
            \begin{bussproof}[cong-res-symm]
                \bussAx{
                    \pRes{xy} P \equiv \pRes{yx} P
                }
            \end{bussproof}
            \and
            \begin{bussproof}[cong-res-inact]
                \bussAx{
                    \pRes{xy} \0 \equiv \0
                }
            \end{bussproof}
            \and
            \begin{bussproof}[cong-fwd-symm]
                \bussAx{
                    \pFwd [x<>y] \equiv \pFwd [y<>x]
                }
            \end{bussproof}
            \and
            \begin{bussproof}[cong-res-fwd]
                \bussAx{
                    \pRes{xy} \pFwd [x<>y] \equiv \0
                }
            \end{bussproof}
            \and
            \begin{bussproof}[cong-unfold]
                \bussAx{
                    \pRec X(x_1,\ldots,x_n) ; P \equiv P \big\{ \big( \pRec X(y_1,\ldots,y_n) ; P \{ y_1/x_1,\ldots,y_n/x_n \} \big) / \pCall X<y_1,\ldots,y_n> \big\}
                }
            \end{bussproof}
        \end{mathpar}
        \caption{Structural congruence for \APCP: axioms.}\label{f:APCP:strcong}
    \end{figure}
\end{defi}
\noindent 
Structural congruence defines the following properties for processes.
Processes are equivalent up to $\alpha$ equivalence (Rule~\ruleLabel{cong-alpha}).
Parallel composition is associative (Rule~\ruleLabel{cong-par-assoc}) and commutative (Rule~\ruleLabel{cong-par-comm}), with unit $\0$ (Rule~\ruleLabel{cong-par-unit}).
A parallel process may be moved into or out of a restriction as long as the bound channels do not occur free in the moved process (Rule~\ruleLabel{cong-scope}): this is \emph{scope inclusion} and \emph{scope extrusion}, respectively.
Restrictions on inactive processes may be dropped (Rule~\ruleLabel{cong-res-inact}), and the order of names in restrictions and of consecutive restrictions does not matter (Rules~\ruleLabel{cong-res-symm} and~\ruleLabel{cong-res-comm}, respectively).
Forwarders are symmetric (Rule~\ruleLabel{cong-fwd-symm}), and equivalent to inaction if both names are bound together through restriction (Rule~\ruleLabel{cong-res-fwd}).
Finally, a recursive definition is equivalent to its unfolding (Rule~\ruleLabel{cong-unfold}), replacing any recursive calls with copies of the recursive definition, where the recursive definition's contextual names are pairwise substituted for by the call's names.

As we will see next, the semantics of \APCP is closed under structural congruence.
This means that processes are \emph{equi-recursive}; however, \APCP's typing discipline (described in \Cref{s:APCP:APCP:types}) treats recursive types as \emph{iso-recursive} (see, e.g., Pierce~\cite{book/Pierce02}).

We define the reduction relation $P \redd Q$ by the axioms and closure rules in \Cref{f:procdef} (bottom).
We write $\redd^\ast$ for the reflexive, transitive closure of $\redd$.
Rule~$\ruleLabel{red-send-recv}$ synchronizes a send and a receive on connected names and substitutes the message and continuation names.
Rule~$\ruleLabel{red-sel-bra}$ synchronizes a selection and a branch:
the received label determines the continuation process, substituting the continuation name appropriately.
Rule~$\ruleLabel{red-fwd}$ implements the forwarder as a substitution.
Rules~$\ruleLabel{red-cong}$, $\ruleLabel{red-res}$, and~$\ruleLabel{red-par}$ close reduction under structural congruence, restriction, and parallel composition, respectively.

Having communication of free names in sends and selections is different from communication in the works by Dardha and Gay~\cite{conf/fossacs/DardhaG18} and DeYoung \etal~\cite{conf/csl/DeYoungCPT12}, where, following an internal mobility discipline~\cite{journal/tcs/Boreale98}, communication involves bound names only.
Still, notice that free output is expressible in those works by combining bound output and forwarding.

\NewDocumentCommand{\cyclicAsyncXReddP}{s}{%
    \[
        P \IfBooleanTF{#1}{=}{\deq} \pRes{zu} \begin{array}[t]{@{}l@{}}
            \big(
                \pRes{xy} \begin{array}[t]{@{}l@{}}
                    \big(
                        \pRes{ax'} ( \pOut x[v_1,a] \| \pOut x'[v_2,b] )
                        \\
                        {} \| \pRes{cz'} ( \pOut z[v_3,c] \| \pIn y(w_1,y') ; \pIn y'(w_2,y'') ; Q )
                    \big)
                \end{array}
                \\
                {} \| \pIn u(w_3,u') ; R
            \big)
        \end{array}
    \]
}
\begin{exa}
    \label{x:APCP:redd}
    To illustrate the preservation of order \emph{within} a session and the asynchrony \emph{between} different sessions, we consider the following process:
    \cyclicAsyncXReddP
    The process $P$ defines two consecutive sends on a session from $x$ to $y$, and an asynchronous send on a session from $z$ to $u$.
    Two reductions are possible from $P$:
    \begin{align}
        P &\redd \pRes{zu} \big(
            \pRes{ax'} \big(
                \pOut x'[v_2,b] \| \pRes{cz'} ( \pOut z[v_3,c] \| \pIn a(w_2,y'') ; Q \{ v_1/w_1 \} )
            \big) \| \pIn u(w_3,u') ; R
        \big)
        \label{eq:APCP:xRed1}
        \\
        P &\redd \pRes{cz'} \big(
            \pRes{xy} \big(
                \pRes{ax'} ( \pOut x[v_1,a] \| \pOut x'[v_2,b] ) \| \pIn y(w_1,y') ; \pIn y'(w_2,y'') ; Q
            \big) \| R \{ v_3/w_3 , c/u' \}
        \big)
        \label{eq:APCP:xRed2}
    \end{align}
    The reduction~\eqref{eq:APCP:xRed1} entails the synchronization of the send on $x$ and the receive on $y$; afterwards, the send on $x'$ is connected to the receive on $a$ that prefixes $Q$.
    The reduction~\eqref{eq:APCP:xRed2} entails the synchronization of the send on $z$ and the receive on $u$, connecting $Q$ and $R$ on a new session between $z'$ and $c$.
    Note that from $P$ there is no reduction involving the send on $x'$, since $x'$ is connected to the continuation name of the send on $x$ and is thus not (yet) paired with a dual receive.

    Using the sugared syntax from \Cref{n:APCP:sugar}, we can write
    \[
        P = \pRes{zu} ( \pRes{xy} ( \pOut* x[v_1] \cdot \pOut* x[v_2] \cdot \0 \| \pOut* z[v_3] \cdot \pIn y(w_1) ; \pIn y(w_2) ; Q' ) \| \pIn u(w_3) ; R' )
    \]
    where $Q' \deq Q \{ y/y'' \}$ and $R' \deq R \{ u/u' \}$.

    The following diagram illustrates all the possible reduction paths from $P$; horizontal reductions concern the session between $x$ and $y$, and diagonal reductions concern the session between $z$ and $u$:
    \[
        \begin{array}{@{}l@{\kern1pt}l@{}l@{\kern1pt}l@{}l@{\kern1pt}l@{}l@{}}
            P & \redd & P_1 & \redd & P_3
            \\[-5pt]
            & \rotatebox[origin=l]{-30}{$\redd$}
            &
            & \rotatebox[origin=l]{-30}{$\redd$}
            &
            & \rotatebox[origin=l]{-30}{$\redd$}
            \\[-5pt]
            & & P_2 & \redd & P_4 & \redd & P_5
        \end{array}
    \]
    where processes $P_1$, $P_2$, $P_3$, $P_4$, and $P_5$ are as follows:
    \begin{align*}
        P_1 &\deq \pRes{zu}( \pRes{xy}( \pOut* x[v_2] \cdot \0 \| \pOut* z[v_3] \cdot \pIn y(w_2) ; Q' \{ v_1/w_1 \} ) \| \pIn u(w_3) ; R' )
        \\
        P_2 &\deq \pRes{xy}( \pOut* x[v_1] \cdot \pOut* x[v_2] \cdot \0 \| \pIn y(w_1) ; \pIn y(w_2) ; Q' ) \| R' \{ v_3/w_3 \}
        \\
        P_3 &\deq \pRes{zu}( \pOut* z[v_3] \cdot Q' \{ v_1/w_1 , v_2/w_2 \} \| \pIn u(w_3) ; R' )
        \\
        P_4 &\deq \pRes{xy}( \pOut* x[v_2] \cdot \0 \| \pIn y(w_2) ; Q' \{ v_1/w_1 \} ) \| R' \{ v_3/w_3 \}
        \\
        P_5 &\deq Q' \{ v_1/w_1 , v_2/w_2 \} \| R' \{ v_3/w_3 \}
        \tag*{\qedhere}
    \end{align*}
\end{exa}
\noindent 
Presentations of reduction for session-typed $\pi$-calculi derived from the Curry-Howard interpretations of linear logic often include rules that correspond to \emph{commuting conversions} in linear logic (cf.~\cite{conf/concur/CairesP10,conf/icfp/Wadler12,conf/fossacs/DardhaG18,conf/csl/DeYoungCPT12}), which allow rewriting processes in such a way that blocking prefixes on free names are ``pulled out'' of restrictions.
Commuting conversions can be easily included for \APCP (cf.\ \cite{conf/ice/vdHeuvelP21}), but we do not consider them here.
While Dardha and Gay~\cite{conf/fossacs/DardhaG18} rely on commuting conversions to prove deadlock freedom, the proof of deadlock freedom for \APCP takes a different approach and does not require commuting conversions---see \Cref{s:APCP:APCP:results} for a detailed discussion.

\subsection{The Type System}
\label{s:APCP:APCP:types}

\APCP types processes by assigning binary session types to names.
Following Curry-Howard interpretations, we present session types as linear logic propositions (cf., e.g., Caires \etal~\cite{journal/mscs/CairesPT16}, Wadler~\cite{conf/icfp/Wadler12}, Caires and P\'erez~\cite{conf/esop/CairesP17}, and Dardha and Gay~\cite{conf/fossacs/DardhaG18}).
We extend these propositions with recursion and \emph{priority} annotations on connectives.
Intuitively, prefixes typed with lower priority should not be blocked by those with higher priority.

We write $\pri, \pi, \rho, \ldots$ to denote priorities, and $\omega$ to denote the ultimate priority that is greater than all other priorities  and cannot be increased further.
That is, $\forall \pri \in \mbb{N}.~\omega > \pri$ and $\forall \pri \in \mbb{N}.~\omega + \pri = \omega$.

\begin{defi}[Session Types for \APCP]
    \label{d:APCP:types}
    The following grammar defines the syntax of \emph{session types} $A,B$.
    Let $\pri \in \mbb{N}$.
    \begin{align*}
        A,B
        &::= A \tensor^\pri B \sepr A \parr^\pri B \sepr \oplus^\pri \{ i : A \}_{i \in I} \sepr \&^\pri \{ i : A \}_{i \in I} \sepr \bullet \sepr \mu X . A \sepr X
    \end{align*}
\end{defi}

\noindent
A name of type $A \tensor^\pri B$ (resp.\ $A \parr^\pri B$) first sends (resp.\ receives) a name of type $A$ and then behaves as $B$.
A name of type $\oplus^\pri \{ i : A_i \}_{i \in I}$ selects a label $i \in I$ and then behaves as $A_i$.
A name of type $\&^\pri \{ i : A_i \}_{i \in I}$ offers a choice: after receiving a label $i \in I$, the name behaves as $A_i$.
A name of type $\bullet$ is closed; it does not require a priority, as closed names do not exhibit behavior and thus are non-blocking.

Unlike Caires and Pfenning~\cite{conf/concur/CairesP10} and Dardha and Gay \cite{conf/fossacs/DardhaG18}, \APCP does not associate any behavior with closed sessions (i.e., no closing synchronizations).
Moreover, as we will see, \APCP's type system allows arbitrary parallel composition.
Atkey \etal~\cite{chapter/AtkeyLM16} have shown that, in presence of arbitrary parallel composition (i.e., linear logic's Rule~\ruleLabel{mix} that combines sequents arbitrarily, usually interpreted as session typing rule for arbitrary parallel composition), the dual propositions $\1$ and $\bot$ (usually associated with complementary prefixes for closing sessions) are equivalent.
Hence, since in \APCP session closing is silent, we follow Caires~\cite{report/Caires14} in conflating the types $\1$ and $\bot$ to the single, self-dual type $\bullet$ for closed sessions.

Type $\mu X . A$ denotes a recursive type, in which $A$ may contain occurrences of the recursion variable $X$.
As customary, $\mu$ is a binder: it induces the standard notions of $\alpha$ equivalence, substitution (denoted $A \{ B/X \}$), and free recursion variables (denoted $\frv(A)$).
We work with tail-recursive, contractive types, disallowing types of the form $\mu X_1 \ldots \mu X_n . X_1$ and $\mu X . X \tensor^\pri A$.
Recursive types are treated iso-recursively: there will be an explicit typing rule that unfolds recursive types, and recursive types are not equal to their unfolding.
We postpone formalizing the unfolding of recursive types, as it requires additional definitions to ensure consistency of priorities upon unfolding.

\emph{Duality}, the cornerstone notion of session types and linear logic, ensures that the two names of a channel have
complementary behaviors.
Furthermore, dual types must have matching priority annotations.
The following inductive definition of duality suffices for our tail-recursive types (cf.\ Gay \etal~\cite{conf/places/GayTV20}).

\begin{defi}[Duality]
    \label{d:APCP:duality}
    The \emph{dual} of session type $A$, denoted $\ol{A}$, is defined inductively as follows:
    \begin{align*}
        \ol{A \tensor^\pri B}
        &\deq
        \ol{A} \parr^\pri \ol{B}
        &
        \ol{\oplus^\pri \{ i : A_i \}_{i \in I}}
        &\deq
        \&^\pri \{ i : \ol{A_i} \}_{i \in I}
        &
        \ol{\bullet}
        &\deq
        \bullet
        &
        \ol{\mu X . A}
        &\deq
        \mu X . \ol{A}
        \\
        \ol{A \parr^\pri B}
        &\deq
        \ol{A} \tensor^\pri \ol{B}
        &
        \ol{\&^\pri \{ i : A_i \}_{i \in I}}
        &\deq
        \oplus^\pri \{ i : \ol{A_i} \}_{i \in I}
        &
        &
        &
        \ol{X}
        &\deq
        X
    \end{align*}
\end{defi}

The priority of a type is determined by the priority of the type's outermost connective:

\begin{defi}[Priorities]
    \label{d:APCP:priority}
    For session type $A$, $\pr(A)$ denotes its \emph{priority}:
    \begin{align*}
        \pr(A \tensor^\pri B)
        \deq
        \pr(A \parr^\pri B)
        &\deq
        \pri
        &
        \pr(\mu X . A)
        &\deq
        \pr(A)
        \\
        \pr(\oplus^\pri \{ i : A_i \}_{i \in I})
        \deq
        \pr(\&^\pri \{ i : A_i \}_{i \in I})
        &\deq
        \pri
        &
        \pr(\bullet)
        \deq
        \pr(X)
        &\deq
        \omega
    \end{align*}
\end{defi}

\noindent
The priority of $\bullet$ and $X$ is the constant $\omega$: they denote the ``final'', non-blocking part of protocols.
Although $\tensor$ and $\oplus$ also denote non-blocking prefixes, they do block their continuation until they are received.
Hence, their priority is not constant.

We now turn to formalizing the unfolding of recursive types.
Recall the intuition that prefixes typed with lower priority should not be blocked by those with higher priority.
Based on this rationale, we observe that the unfolding of the recursive type $\mu X . A$ should not result in  $A \{ ( \mu X . A )/X \}$, as usual; rather, the priorities of the unfolded type should be \emph{increased} to ensure a global ordering between actions.

\begin{exa}
    \label{x:APCP:equiRec}
    Consider the recursive type $\mu X . A \parr^0 X$.
    By unfolding this type without increasing the priority, we obtain $A \parr^0 (\mu X . A \parr^0 X)$, a type in which the priorities do not ensure a global ordering between the two receives.
    In contrast, increasing the priority in the unfolded type as in, e.g., $A \parr^0 (\mu X . A \parr^1 X)$, does ensure a global ordering.
\end{exa}

We make this intuition precise by defining the \emph{lift} of priorities in types:

\begin{defi}[Lift]
    \label{d:APCP:lift}
    For proposition $A$ and $t \in \mbb{N}$, we define $\lift{t}A$ as the \emph{lift} operation:
    \begin{align*}
        \lift{t}(A \tensor^\pri B)
        &\deq
        (\lift{t}A) \tensor^{\pri+t} (\lift{t}B)
        &
        \lift{t}(\oplus^\pri \{ i : A_i \}_{i \in I})
        &\deq
        \oplus^{\pri+t} \{ i : \lift{t}A_i \}_{i \in I}
        &
        \lift{t}\bullet
        &\deq
        \bullet
        \\
        \lift{t}(A \parr^\pri B)
        &\deq
        (\lift{t}A) \parr^{\pri+t} (\lift{t}B)
        &
        \lift{t}(\&^\pri \{ i : A_i \}_{i \in I})
        &\deq
        \&^{\pri+t} \{ i : \lift{t}A_i \}_{i \in I}
        \\
        \lift{t}(\mu X . A)
        &\deq
        \mu X . (\lift{t}A)
        &
        \lift{t}X
        &\deq
        X
    \end{align*}
\end{defi}

\noindent
Henceforth, the unfolding of  $\mu X . A$ is $A \{ \big(\mu X . (\lift{t} A) \big)/X \}$, denoted $\unfold^t(\mu X . A)$, where $t \in \mbb{N}$ depends on the highest priority of the types occurring in a typing context.
We recall that we do not consider types to be equi-recursive: recursive types are not equal to their unfolding.
Recursive types can only be unfolded by typing rules, discussed next.

We now define the top priority of a type, i.e., the highest priority appearing in a type:

\begin{defi}[Top Priority]
    \label{d:APCP:top}
    For session type $A$, $\prtop(A)$ denotes its \emph{top priority}:
    \begin{align*}
        \prtop(A \tensor^\pri B)
        \deq
        \prtop(A \parr^\pri B)
        &\deq
        \max(\prtop(A), \prtop(B), \pri)
        \\
        \prtop(\oplus^\pri \{ i : A_i \}_{i \in I})
        \deq
        \prtop(\&^\pri \{ i : A_i \}_{i \in I})
        &\deq
        \max(\max_{i \in I}(\prtop(A_i)), \pri)
        \\
        \prtop(\mu X . A)
        &\deq
        \prtop(A)
        &
        \prtop(\bullet)
        \deq
        \prtop(X)
        &\deq
        0
    \end{align*}
\end{defi}

\noindent
Notice how the top priority of $\bullet$ and $X$ is 0, in contrast to their priority (as given by \Cref{d:APCP:priority}): as we will see next, they do not contribute to the increase in priority needed for unfolding recursive types.

\paragraph{Typing Rules}

The typing rules of \APCP ensure that prefixes with lower priority are not blocked by those with higher priority (cf.\ Dardha and Gay~\cite{conf/fossacs/DardhaG18}).
To this end, they enforce the following laws:
\begin{enumerate}

    \item\label{i:law:sendSel}
        Sends and selections with priority $\pri$ must have continuations/payloads with priority strictly larger than $\pri$;

    \item\label{i:law:recvBra}
        A prefix with priority $\pri$ must be prefixed only by receives and branches with priority strictly smaller than $\pri$;

    \item\label{i:law:resFwd}
        Dual prefixes leading to a synchronization must have equal priorities (cf.\ \Cref{d:APCP:duality}).

\end{enumerate}
Judgments are of the form $\Omega \vdash P :: \Gamma$, where:
\begin{itemize}

    \item
        $P$ is a process;

    \item
        $\Gamma$ is a context that assigns types to channels ($x:A$);

    \item
        $\Omega$ is a context that assigns tuples of types to recursion variables ($X:(A, B, \ldots)$).

\end{itemize}
A judgment $\Omega \vdash P :: \Gamma$ then means that $P$ can be typed in accordance with the type assignments for names recorded in $\Gamma$ and the recursion variables in $\Omega$.
Intuitively, the latter context ensures that types concur between the context names of recursive definitions and calls.
Both contexts $\Gamma$ and $\Omega$ obey \emph{exchange}: assignments may be silently reordered.
$\Gamma$ is \emph{linear}, disallowing \emph{weakening} (i.e., all assignments must be used, except names typed $\bullet$) and \emph{contraction} (i.e., assignments may not be duplicated).
$\Omega$ allows weakening and contraction, because a recursive definition may be called \emph{zero or more} times.

The empty context is written $\emptyset$.
In writing $\Gamma, x:A$ we assume that $x \notin \dom(\Gamma)$ (and similarly for $\Omega$).
We write $\lift{t} \Gamma$ to denote the component-wise extension of lift (\Cref{d:APCP:lift}) to typing contexts.
Also, we write $\pr(\Gamma)$ to denote the least of the priorities of all types in $\Gamma$ (\Cref{d:APCP:priority}).
An assignment $\tilde{z}:\tilde{A}$ means $z_1:A_1, \ldots, z_k:A_k$.
We define the top priority of a sequence of types $\prtop(\tilde{A})$ as $\max_{A_i \in \tilde{A}}(\prtop(A_i))$.

\begin{figure}[t!]
    \begin{mathpar}
        \begin{bussproof}[typ-send]
            \bussAssume{
                \pri < \pr(A),\pr(B)
            }
            \bussUn{
                \Omega
                \vdash \pOut x[y,z] :: x:A \tensor^\pri B, y:\ol{A}, z:\ol{B}
            }
        \end{bussproof}
        \and
        \begin{bussproof}[typ-recv]
            \bussAssume{
                \Omega
                \vdash P :: \Gamma, y:A, z:B
            }
            \bussAssume{
                \pri < \pr(\Gamma)
            }
            \bussBin{
                \Omega
                \vdash \pIn x(y, z) ; P :: \Gamma, x:A \parr^\pri B
            }
        \end{bussproof}
        \and
        \begin{bussproof}[typ-sel]
            \bussAssume{
                j \in I
            }
            \bussAssume{
                \pri < \pr(A_j)
            }
            \bussBin{
                \Omega
                \vdash \pSel x[z] < j :: x:\oplus^\pri \{ i : A_i \}_{i \in I}, z:\ol{A_j}
            }
        \end{bussproof}
        \and
        \begin{bussproof}[typ-bra]
            \bussAssume{
                \forall i \in I.~ \Omega \vdash P_i :: \Gamma, z:A_i
            }
            \bussAssume{
                \pri < \pr(\Gamma)
            }
            \bussBin{
                \Omega
                \vdash \pBra x(z) > {\{ i : P_i \}_{i \in I}} :: \Gamma, x:\&^\pri \{ i : A_i \}_{i \in I}
            }
        \end{bussproof}
        \and
        \begin{bussproof}[typ-end]
            \bussAssume{
                \Omega
                \vdash P :: \Gamma
            }
            \bussUn{
                \Omega
                \vdash P :: \Gamma, x:\bullet
            }
        \end{bussproof}
        \and
        \begin{bussproof}[typ-par]
            \bussAssume{
                \Omega
                \vdash P :: \Gamma
            }
            \bussAssume{
                \Omega
                \vdash Q :: \Delta
            }
            \bussBin{
                \Omega
                \vdash P \| Q :: \Gamma, \Delta
            }
        \end{bussproof}
        \and
        \begin{bussproof}[typ-res]
            \bussAssume{
                \Omega
                \vdash P :: \Gamma, x:A, y:\ol{A}
            }
            \bussUn{
                \Omega
                \vdash \pRes{xy} P :: \Gamma
            }
        \end{bussproof}
        \and
        \begin{bussproof}[typ-inact]
            \bussAx{
                \Omega
                \vdash \0 :: \emptyset
            }
        \end{bussproof}
        \and
        \begin{bussproof}[typ-fwd]
            \bussAx{
                \Omega
                \vdash \pFwd [x<>y] :: x:\ol{A}, y:A
            }
        \end{bussproof}
        \and
        \begin{bussproof}[typ-rec]
            \bussAssume{
                \Omega, X:{\tilde{A}} \vdash P :: \tilde{z}:\tilde{U}
            }
            \bussAssume{
                t \in \mbb{N} > \prtop(\tilde{A})
            }
            \bussAssume{
                \forall U_i \in \tilde{U}.~ U_i = \unfold^t(\mu X . A_i)
            }
            \bussTern{
                \Omega \vdash \pRec X(\tilde{z}) ; P :: \tilde{z}:\widetilde{\mu X . A}
            }
        \end{bussproof}
        \and
        \begin{bussproof}[typ-var]
            \bussAssume{
                t \in \mbb{N}
            }
            \bussAssume{
                \forall U_i \in \tilde{U}.~ U_i = \mu X . \lift{t} A_i
            }
            \bussBin{
                \Omega, X:\tilde{A} \vdash \pCall X<\tilde{z}> :: \tilde{z}:\tilde{U}
            }
        \end{bussproof}
        \\ \dashes \\
        \begin{bussproof}[typ-send$\star$]
            \bussAssume{
                \Omega
                \vdash P :: \Gamma, y:A, x:B
            }
            \bussAssume{
                \pri < \pr(A),\pr(B)
            }
            \bussBin{
                \Omega
                \vdash \pOut* x[y] \cdot P :: \Gamma, x:A \tensor^\pri B
            }
        \end{bussproof}
        \and
        \begin{bussproof}[typ-sel$\star$]
            \bussAssume{
                \Omega
                \vdash P :: \Gamma, x:A_j
            }
            \bussAssume{
                j \in I
            }
            \bussAssume{
                \pri < \pr(A_j)
            }
            \bussTern{
                \Omega
                \vdash \pSel* x < j \cdot P :: \Gamma, x:\oplus^\pri \{ i : A_i \}_{i \in I}
            }
        \end{bussproof}
        \and
        \begin{bussproof}[typ-lift]
            \bussAssume{
                \Omega
                \vdash P :: \Gamma
            }
            \bussAssume{
                t \in \mbb{N}
            }
            \bussBin{
                \Omega
                \vdash P :: \lift{t} \Gamma
            }
        \end{bussproof}
    \end{mathpar}

    \caption{The typing rules of \APCP (top) and derivable rules (bottom).}
    \label{f:APCP:typingRules}
\end{figure}

\Cref{f:APCP:typingRules}~(top) gives the typing rules.
We describe the typing rules from a \emph{bottom-up} perspective.
Rule~\ruleLabel{typ-send} types a send; this rule does not have premises to provide a continuation process, leaving the free names to be bound to a continuation process using Rules~\ruleLabel{typ-par} and~\ruleLabel{typ-res} (discussed hereafter).
Similarly, Rule~\ruleLabel{typ-sel} types an unbound selection.
Both rules require that the priority of the subject is lower than the priorities of both objects (continuation and payload)---this enforces Law~\labelcref{i:law:sendSel}.
Rules~\ruleLabel{typ-recv} and~\ruleLabel{typ-bra} type receives and branches, respectively.
In both cases, the used name's priority must be lower than the priorities of the other types in the continuation's typing context---this enforces Law~\labelcref{i:law:recvBra}.

Rule~\ruleLabel{typ-par} types the parallel composition of two processes that do not share assignments on the same names.
Rule~\ruleLabel{typ-res} types a restriction, where the two restricted names must be of dual type and thus have matching priority---this enforces Law~\labelcref{i:law:resFwd}.
Rule~\ruleLabel{typ-end} silently removes a closed name from the typing context.
Rule~\ruleLabel{typ-inact} types an inactive process with no names.
Rule~\ruleLabel{typ-fwd} types forwarding between names of dual type---this also enforces Law~\labelcref{i:law:resFwd}.
\begin{exa}
    To illustrate the typing rules discussed so far, we recall process $P$ from \Cref{x:APCP:redd}:
    \cyclicAsyncXReddP*

    We give the typing of the two consecutive sends on $x$ (omitting the context $\Omega$):
    \[
        \begin{bussproof}
            \bussAssume{
                \pri < \pr(A_1),\pi
            }
            \bussUn[\ruleLabel{typ-send}]{
                \vdash \pOut x[v_1,a] :: \begin{array}[t]{@{}l@{}}
                    x:A_1 \tensor^\pri A_2 \tensor^\pi B,
                    \\
                    v_1:\ol{A_1}, a:\ol{A_2 \tensor^\pi B}
                \end{array}
            }
            \bussAssume{
                \pi < \pr(A_2),\pr(B)
            }
            \bussUn[\ruleLabel{typ-send}]{
                \vdash \pOut x'[v_2,b] :: \begin{array}[t]{@{}l@{}}
                    x':A_2 \tensor^\pi B,
                    \\
                    v_2:\ol{A_2}, b:\ol{B}
                \end{array}
            }
            \bussBin[\ruleLabel{typ-par}]{
                \vdash \pOut x[v_1,a] \| \pOut x'[v_2,b] :: \begin{array}[t]{@{}l@{}}
                    v_1:\ol{A_1}, v_2:\ol{A_2}, b:\ol{B}, x:A_1 \tensor^\pri A_2 \tensor^\pi B,
                    \\
                    a:\ol{A_2 \tensor^\pi B}, x':A_2 \tensor^\pi B
                \end{array}
            }
            \bussUn[\ruleLabel{typ-res}]{
                \vdash \pRes{ax'} ( \pOut x[v_1,a] \| \pOut x'[v_2,b] ) :: v_1:\ol{A_1}, v_2:\ol{A_2}, b:\ol{B}, x:A_1 \tensor^\pri A_2 \tensor^\pi B
            }
        \end{bussproof}
    \]
    As discussed before, this typing leaves the (free) names $v_1$, $v_2$, and $b$ to be accounted for by the context.

    Now let us derive the typing of the consecutive receives on $y$, i.e., of the subprocess $\pIn y(w_1,y') ; \pIn y'(w_2,y'') ; Q$.
    Because $x$ and  $y$ are dual names in $P$, the type of $y$ should be dual to the type of $x$ above:
    \[
        \begin{bussproof}
            \bussAssume{
                \vdash Q :: \Gamma, w_1:\ol{A_1}, w_2:\ol{A_2}, y'':\ol{B}
            }
            \bussAssume{
                \pi < \pr(\Gamma, w_1:\ol{A_1})
            }
            \bussBin[\ruleLabel{typ-recv}]{
                \vdash \pIn y'(w_2,y'') ; Q :: \Gamma, w_1:\ol{A_1}, y':\ol{A_2} \parr^\pi \ol{B}
            }
            \bussAssume{
                \pri < \pr(\Gamma)
            }
            \bussBin[\ruleLabel{typ-recv}]{
                \vdash \pIn y(w_1,y') ; \pIn y'(w_2,y'') ; Q :: \Gamma, y:\ol{A_1} \parr^\pri \ol{A_2} \parr^\pi \ol{B}
            }
        \end{bussproof}
    \]

    These two derivations tell us that $\pri < \pi < \pr(A_1),\pr(A_2),\pr(B),\pr(\Gamma)$.
    This way, the type system ensures that none of the sessions in $Q$ can be connected to sessions that block the sends on $x,x'$, which may leave the door open for a deadlock otherwise.
    In \Cref{s:FIRST:df}, \Cref{x:FIRST:deadlocked} illustrates such a situation in the context of \FIRST and how the type system catches it.
\end{exa}
Consider the usual Rule~\ruleLabel{typ-cut} in type systems based on linear logic~\cite{conf/concur/CairesP10,conf/icfp/Wadler12}:
\begin{mathpar}
        \begin{bussproof}[typ-cut]
            \bussAssume{
                \Omega
                \vdash P :: \Gamma, x:A
            }
            \bussAssume{
                \Omega
                \vdash Q :: \Delta, y:\ol{A}
            }
            \bussBin{
                \Omega
                \vdash \pRes{xy}(P \| Q) :: \Gamma, \Delta
            }
        \end{bussproof}
\end{mathpar}
Note that a single application of Rule~\ruleLabel{typ-par} followed by Rule~\ruleLabel{typ-res} coincides with Rule~\ruleLabel{typ-cut}.
Without  annotations and conditions related to priority, Rules~\ruleLabel{typ-par} and~\ruleLabel{typ-res} give rise to deadlocks, as the following example shows.

\begin{exa}
    \label{x:APCP:noPrios}
    Consider the following process, arguably the paradigmatic example of a deadlock:
    \[
        Q \deq \pRes{xy} \pRes{zw} ( \pIn x(u) ; \pOut* z[u'] \cdot \0 \| \pIn w(v) ; \pOut* y[v'] \cdot \0 )
    \]
    Without priorities (and priority checks), this process can be typed using Rules~\ruleLabel{typ-par} and~\ruleLabel{typ-res} (omitting ``\textsc{typ-}'' from rule labels):
    \[
        \begin{bussproof}
            \def\defaultHypSeparation{\hskip1ex}
            \bussAx[\ruleLabel{inact}]{
                \vdash \0 :: \emptyset
            }
            \doubleLine
            \bussUn[$\ruleLabel{end}^4$]{
                \vdash \0 :: z:\bullet , x:\bullet , u:\bullet , u':\bullet
            }
            \bussUn[\ruleLabel{send$\star$}]{
                \vdash \pOut* z[u'] \cdot \0 :: z:\bullet \tensor \bullet , x:\bullet , u:\bullet
            }
            \bussUn[\ruleLabel{recv}]{
                \vdash \pIn x(u) ; \pOut* z[u'] \cdot \0 :: z:\bullet \tensor \bullet , x:\bullet \parr \bullet
            }
            \bussAx[\ruleLabel{inact}]{
                \vdash \0 :: \emptyset
            }
            \doubleLine
            \bussUn[$\ruleLabel{end}^4$]{
                \vdash \0 :: w:\bullet , y:\bullet , v:\bullet , v':\bullet
            }
            \bussUn[\ruleLabel{send$\star$}]{
                \vdash \pOut* y[v'] \cdot \0 :: w:\bullet , y:\bullet \tensor \bullet , v:\bullet
            }
            \bussUn[\ruleLabel{recv}]{
                \vdash \pIn w(v) ; \pOut* y[v'] \cdot \0 :: w:\bullet \tensor \bullet , y:\bullet \parr \bullet
            }
            \doubleLine
            \bussBin[\ruleLabel{par}+$\ruleLabel{res}^2$]{
                \vdash Q :: \emptyset
            }
        \end{bussproof}
    \]
    On the other hand, were we to restrict parallel composition and restriction using Rule~\ruleLabel{typ-cut}, $Q$ would not be typable: Rule~\ruleLabel{typ-cut} can only type one of the restrictions, not both.
    With priorities, $Q$ would not be typable either, due to the requirements induced by Rule~\ruleLabel{typ-recv}: (i)~the priority $\pri$ of the input on $x$ is smaller than the priority $\pi$ of the send on $z$ (left-hand side above), and (ii)~the priority $\pi$ of the input on $w$ is smaller than the priority $\pri$ of the send on $y$ (right-hand side above).
    Clearly, these requirements combined are unsatisfiable.
\end{exa}

Rules~\ruleLabel{typ-rec} and~\ruleLabel{typ-var} type recursive definitions and  recursive calls, respectively.
To justify their formulation, let us consider naive formulations for each of them:
\begin{mathpar}
    \begin{bussproof}[typ-rec-naive]
        \bussAssume{
            \Omega \vdash P :: \tilde{z}:\tilde{A}
        }
        \bussUn{
            \Omega , X:|\tilde{z}| \vdash \pRec X(\tilde{z}) ; P :: \tilde{z}:\widetilde{\mu X . A}
        }
    \end{bussproof}
    \and
    \begin{bussproof}[typ-var-naive]
        \bussAx{
            \Omega , X:|\tilde{z}| \vdash \pCall X<\tilde{z}> :: \tilde{z}:\tilde{X}
        }
    \end{bussproof}
\end{mathpar}
Rule~\ruleLabel{typ-rec-naive} requires a name typed $\mu X . A$ at the recursive definition to be typed simply $A$ in the recursive body.
The associated Rule~\ruleLabel{typ-var-naive} then requires all names to be typed $X$, using the recursive context to make sure that the number of names concurs between recursive definition and call.
However, as the following example shows, such a combination of naive rules discards priority annotations to a point where it is possible to type processes that deadlock:

\begin{exa}
    \label{x:APCP:naiveRec}
    Consider the processes
    \begin{align*}
        P &\deq \pRec X(x,y) ; \pOut* x[a] \cdot \pIn x(b) ; \pOut* y[c] \cdot \pIn y(d) ; \pCall X<y,x>,
        \\
        Q &\deq \pRec X(u,v) ; \pIn u(a) ; \pOut* u[b] \cdot \pIn v(c) ; \pOut* v[d] \cdot \pCall X<u,v>.
    \end{align*}
    Notice how the recursive call in $P$ swaps $x$ and $y$.
    Let us see what happens if we unfold the recursion in $P$ and $Q$:
    \begin{align*}
        P &\equiv \pOut* x[a] \cdot \pIn x(b) ; \pOut* y[c] \cdot \pIn y(d) ; \pRec X(y,x) ; \pOut* y[a] \cdot \pIn y(b) ; \pOut* x[c] \cdot \pIn x(d) ; \pCall X<x,y>
        \\
        Q &\equiv \pIn u(a) ; \pOut* u[b] \cdot \pIn v(c) ; \pOut* v[d] \cdot \pRec X(u,v) ; \pIn u(a) ; \pOut* u[b] \cdot \pIn v(c) ; \pOut* v[d] \cdot \pCall X<u,v>
    \end{align*}
    If we connect these processes on the names $x$ and $u$ and on $y$ and $v$, we can see that the second recursive definition of this process contains a deadlock: the second receive on $y$ is blocking the second send on $x$, while the second receive on $u$ (waiting for the second send on $x$) is blocking the second send on $v$ (for which the second receive on $y$ is waiting).

    Yet, $P$ is typable using the naive typing rules described above:
    \[
        \begin{bussproof}
            \bussAx[\ruleLabel{typ-var-naive}]{
                X:2 \vdash \pCall X<y,x> :: x:X, y:X
            }
            \doubleLine
            \bussUn[$\cdots$]{
                X:2 \vdash \pOut* x[a] \cdot \pIn x(b) ; \pOut* y[c] \cdot \pIn y(d) ; \pCall X<y,x> :: x:\bullet \tensor^0 \bullet \parr^1 X, y:\bullet \tensor^2 \bullet \parr^3 X
            }
            \bussUn[\ruleLabel{typ-rec-naive}]{
                \emptyset \vdash \pRec X(x,y) ; \pOut* x[a] \cdot \pIn x(b) ; \pOut* y[c] \cdot \pIn y(d) ; \pCall X<y,x> :: \begin{array}[t]{@{}l@{}}
                    x:\mu X . \bullet \tensor^0 \bullet \parr^1 X, \\
                    y:\mu X . \bullet \tensor^2 \bullet \parr^3 X
                \end{array}
            }
        \end{bussproof}
    \]
    Thus, these naive rules prevent the type system of \APCP from guaranteeing deadlock freedom.
\end{exa}

The solution is for Rule~\ruleLabel{typ-rec} to unfold all types.
While unfolding, the priorities in these types are lifted by a common value, denoted $t$ in the rule, that must be greater than the top priority occurring in the original types (cf.\ \Cref{d:APCP:top}).
This makes sure that any priority requirements that come up in the typing of the recursive body of the process remain valid.
The recursive context is used to record the bodies of the original folded types.
Rule~\ruleLabel{typ-rec} then requires that the types of names are recursive on the recursion variable used for the call.
It checks that the bodies of the types concur with the types recorded in the recursive context, up to a lift by a common value $t$ (i.e., the lifter used in the application of Rule~\ruleLabel{typ-rec}).

\begin{exa}
    To see how the unfolding of types and the common lifter in Rule~\ruleLabel{typ-rec} prevents $P$ from \Cref{x:APCP:naiveRec} from being typable, let us attempt to find a typing derivation for $P$:
    \[
        \begin{bussproof}
            \bussAx[\ruleLabel{typ-var}]{
                X:(\bullet \tensor^0 \bullet \parr^1 X, \bullet \tensor^2 \bullet \parr^3 X) \vdash \pCall X<y,x> :: \begin{array}[t]{@{}l@{}}
                    x:\mu X . \lift{t}(\bullet \tensor^0 \bullet \parr^1 X), \\
                    y:\mu X . \lift{t}(\bullet \tensor^2 \bullet \parr^3 X)
                \end{array}
            }
            \doubleLine
            \bussUn[$\cdots$]{
                X:(\bullet \tensor^0 \bullet \parr^1 X, \bullet \tensor^2 \bullet \parr^3 X) \begin{array}[t]{@{}r@{}l@{}}
                    {} \vdash {}
                    & \pOut* x[a] \cdot \pIn x(b) ; \pOut* y[c] \cdot \pIn y(d) ; \pCall X<y,x>
                    \\
                    {} :: {}
                    & x:\bullet \tensor^0 \bullet \parr^1 (\mu X . \lift{t}( \bullet \tensor^0 \bullet \parr^1 X)),
                    \\
                    & y:\bullet \tensor^2 \bullet \parr^3 (\mu X . \lift{t}( \bullet \tensor^2 \bullet \parr^3 X))
                \end{array}
            }
            \bussUn[\ruleLabel{typ-rec}]{
                \emptyset \vdash \pRec X(x,y) ; \pOut* x[a] \cdot \pIn x(b) ; \pOut* y[c] \cdot \pIn y(d) ; \pCall X<y,x> :: \begin{array}[t]{@{}l@{}}
                    x:\mu X . \bullet \tensor^0 \bullet \parr^1 X, \\
                    y:\mu X . \bullet \tensor^2 \bullet \parr^3 X
                \end{array}
            }
        \end{bussproof}
    \]
    The application of Rule~\ruleLabel{typ-var} at the top here is invalid: it is impossible to find a lifter~$t$ that matches the priorities in the type of $x$ with those in the second type assigned to $X$, while simultaneously doing the same for the type of $y$ and the first type assigned to $X$.
    However, if the call were $\pCall X<x,y>$, the application of Rule~\ruleLabel{typ-var} would be valid.
\end{exa}

\begin{rem}[Comparison to \protect\PCP]
    \label{r:APCP:simplerPrios}
    Consider the typing rules for sending and selection in \PCP, in which both are blocking prefixes and do not involve continuation passing.
    Note that \PCP does not have recursion, and we thus omit the recursive context~$\Omega$.
    \begin{mathpar}
        \begin{bussproof}[typ-send-PCP]
            \bussAssume{
                \vdash P :: \Gamma, y:A, x:B
            }
            \bussAssume{
                \pri < \pr(\Gamma)
            }
            \bussBin{
                \vdash \pOut* x[y] ; P :: \Gamma, x:A \tensor^\pri B
            }
        \end{bussproof}
        \and
        \begin{bussproof}[typ-sel-PCP]
            \bussAssume{
                \vdash P :: \Gamma, x:A_j
            }
            \bussAssume{
                j \in I
            }
            \bussAssume{
                \pri < \pr(\Gamma)
            }
            \bussTern{
                \vdash \pSel* x < j ; P :: \Gamma, x:\oplus^\pri \{ i : A_i \}_{i \in I}
            }
        \end{bussproof}
    \end{mathpar}
    These rules are similar to the Rules~\ruleLabel{typ-send$\ast$} and~\ruleLabel{typ-sel$\ast$} in \Cref{f:APCP:typingRules} for the sugared syntax introduced in \Cref{n:APCP:sugar}; the differences are twofold.
    First, the semicolons `\,$;$\,' indicate that sends and selections are indeed blocking.
    Second, the rules compare the priority of the send/selection to the priorities in the context, whereas our rules compare this priority to the priorities of the continuation of the send/selection.
\end{rem}

As anticipated, the
binding of sends and selections to continuation processes (\Cref{n:APCP:sugar}) is consistent with typing in \APCP.
The corresponding typing rules in \Cref{f:APCP:typingRules} (bottom) are admissible using Rules~\ruleLabel{typ-par} and~\ruleLabel{typ-res}.
Note that rules for the sugared receive and branching in \Cref{n:APCP:sugar} are not necessary, because the sugared input constructs rely on $\alpha$ renaming only.
\Cref{f:APCP:typingRules}~(bottom) also includes an admissible Rule~\ruleLabel{typ-lift} that lifts a process' priorities.

\begin{thm}
    \label{t:APCP:derivable}
    Rules~\ruleLabel{typ-send$\ast$} and~\ruleLabel{typ-sel$\ast$} in \Cref{f:APCP:typingRules}~(bottom) are derivable, and Rule~\ruleLabel{typ-lift} in \Cref{f:APCP:typingRules}~(bottom) is admissible.
\end{thm}

\begin{figure}
    \begin{align*}
        & \begin{bussproof}[typ-send$\ast$]
            \bussAssume{
                \vdash P :: \Gamma, y:A, x:B
            }
            \bussAssume{
                \pri < \pr(A),\pr(B)
            }
            \bussBin{
                \vdash \pOut* x[y] \cdot P :: \Gamma, x:A \tensor^\pri B
            }
        \end{bussproof}
        \\
        & \Rightarrow
        \\
        & \begin{bussproof}
            \bussAssume{
                \pri < \pr(A),\pr(B)
            }
            \bussUn[\ruleLabel{typ-send}]{
                \vdash \pOut x[a, b] :: x:A \tensor^\pri B, a:\ol{A}, b:\ol{B}
            }
            \bussAssume{
                \vdash P \{ z/x \} :: \Gamma, y:A, z:B
            }
            \bussBin[\ruleLabel{typ-par}]{
                \vdash \pOut x[a, b] \| P \{ z/x \} :: \Gamma, x:A \tensor^\pri B, y:A, a:\ol{A}, z:B, b:\ol{B}
            }
            \doubleLine
            \bussUn[$2\times$\ruleLabel{typ-res}]{
                \vdash \underbrace{
                    \pRes{ya} \pRes{zb} ( \pOut x[a, b] \| P \{ z/x \} )
                }_{\pOut* x[y] \cdot P ~\text{(cf.\ \Cref{n:APCP:sugar})}} :: \Gamma, x:A \tensor^\pri B
            }
        \end{bussproof}
        \\[5pt]
        & \begin{bussproof}[typ-sel$\ast$]
            \bussAssume{
                \vdash P :: \Gamma, x:A_j
            }
            \bussAssume{
                j \in I
            }
            \bussAssume{
                \pri < \pr(A_j)
            }
            \bussTern{
                \vdash \pSel* x < j \cdot P :: \Gamma, x:\oplus^\pri \{ i : A_i \}_{i \in I}
            }
        \end{bussproof}
        \\
        & \Rightarrow
        \\
        & \begin{bussproof}
            \bussAssume{
                j \in I
            }
            \bussUn[\ruleLabel{typ-sel}]{
                \vdash \pSel x[b] < j :: x:\oplus^\pri \{ i : A_i \}_{i \in I}, b:\ol{A_j}
            }
            \bussAssume{
                \vdash P \{ z/x \} :: \Gamma, z:A_j
            }
            \bussBin[\ruleLabel{typ-par}]{
                \vdash \pSel x[b] < j \| P \{ z/x \} :: \Gamma, x:\oplus^\pri \{ i : A_i \}_{i \in I}, z:A_j, b:\ol{A_j}
            }
            \bussUn[\ruleLabel{typ-res}]{
                \vdash \underbrace{
                    \pRes{zb} ( \pSel x[b] < j \| P \{ z/x \} )
                }_{\pSel* x < j \cdot P ~\text{(cf.\ \Cref{n:APCP:sugar})}} :: \Gamma, x:\oplus^\pri \{ i : A_i \}_{i \in I}
            }
        \end{bussproof}
    \end{align*}

    \caption{Proof that Rules~\protect\ruleLabel{typ-send$\ast$} and~\protect\ruleLabel{typ-sel$\ast$} are derivable (cf.\ \Cref{t:APCP:derivable}).}
    \label{f:APCP:derivable}
\end{figure}

\begin{sketch}
    To show the derivability of Rules~\ruleLabel{typ-send$\ast$} and~\ruleLabel{typ-sel$\ast$}, we give their derivations in \Cref{f:APCP:derivable} (omitting the recursive context~$\Omega$).
    Rule~\ruleLabel{typ-lift} is admissible: $\Omega \vdash P :: \Gamma$ implies $\Omega \vdash P :: \lift{t} \Gamma$ (cf.\ Dardha and Gay~\cite{conf/fossacs/DardhaG18}),  simply by increasing all priorities in the derivation of $P$ by~$t$.
\end{sketch}

\noindent
\Cref{t:APCP:derivable} highlights how \APCP's asynchrony uncovers a more primitive, lower-level view of message passing.
In the next subsection we discuss deadlock freedom, which follows from a correspondence between reduction and the transformation of typing derivations to eliminate applications of Rule~\ruleLabel{typ-res}.
In the case of \APCP, this requires care: binding sends and selections to continuation processes leads to applications of Rule~\ruleLabel{typ-res} that do not immediately correspond to reductions.

\begin{rem}[Comparison To DeYoung \etal]
    Our rules for sending and selection are axiomatic, whereas DeYoung \etal's are in the form of Rules~\ruleLabel{typ-send$\ast$} and~\ruleLabel{typ-sel$\ast$}, even though their sends and selections are parallel atomic prefixes as well~\cite{conf/csl/DeYoungCPT12}.
    While our type system is based on classical linear logic (with single-sided judgments), their type system is based on \emph{intuitionistic} linear logic.
    As a result, their typing judgments are two sided, restriction involves a single name, and there are two rules (right and left) per connective.
    This way, for instance, their right rules for sending and selection are as follows (again omitting the recursive context $\Omega$ for a lack of recursion):
    \begin{mathpar}
        \begin{bussproof}[typ-send-R]
            \bussAssume{
                \Gamma \vdash P :: y:A
            }
            \bussAssume{
                \Delta \vdash Q :: x':B
            }
            \bussBin{
                \Gamma, \Delta \vdash \pRes{y} \pRes{x'} ( \pOut x[y,x'] \| P \| Q ) :: x:A \tensor B
            }
        \end{bussproof}
        \and
        \phantom{\qedsymbol}\hfill
        \begin{bussproof}[typ-sel-R]
            \bussAssume{
                \Gamma \vdash P :: x':A_j
            }
            \bussAssume{
                j \in I
            }
            \bussBin{
                \Gamma \vdash \pRes{x'} ( \pSel x[x'] < j \| P ) :: x:\oplus \{ i : A_i \}_{i \in I}
            }
            \hfill
        \end{bussproof}
        \hfill\qedsymbol
    \end{mathpar}
\end{rem}
\noindent 
The following result is important: it shows that the type system of \APCP is complete with respect to types, i.e., every syntactical type has a well-typed process.

\begin{prop}
    \label{p:APCP:charProc}
    Given a type $A$, there exists $\Omega \vdash P :: x:A$.
\end{prop}

\begin{sketch}
    By constructing $P$ from the structure of $A$.
    To this end, we define characteristc processes: a function $\chr^x(A)$ that constructs a process that performs the behavior described by $A$ on the name $x$.
    \begin{align*}
        \chr^x(A \tensor^\pri B)
        &\deq
        \pOut* x[y] \cdot ( \chr^y(A) \| \chr^x(B) )
        &
        \chr^x(\bullet)
        &\deq
        \0
        \\
        \chr^x(A \parr^\pri B)
        &\deq
        \pIn x(y) ; ( \chr^y(A) \| \chr^x(B) )
        &
        \chr^x(\mu X . A)
        &\deq
        \pRec X(x) ; \chr^x(A)
        \\
        \chr^x(\oplus^\pri \{ i : A_i \}_{i \in I})
        &\deq
        \pSel* x < j \cdot \chr^x(A_j)
        ~ \text{(any $j \in I$)}
        &
        \chr^x(X)
        &\deq
        \pCall X<x>
        \\
        \chr^x(\&^\pri \{ i : A_i \}_{i \in I})
        &\deq
        \pBra* x > {\{ i : \chr^x(A_i) \}_{i \in I}}
    \end{align*}
    For finite types, it is obvious that $\emptyset \vdash \chr^x(A) :: x:A$.
    For simplicity, we omit details about recursive types, which require unfolding.
    For closed, recursive types, the thesis is obvious as well: $\emptyset \vdash \chr^x(\mu X . A) :: x: \mu X . A$.
\end{sketch}

\subsection{Type Preservation and Deadlock Freedom}
\label{s:APCP:APCP:results}

Well-typed processes satisfy protocol fidelity, communication safety, and deadlock freedom.
The former two of these properties follow from \emph{type preservation}, which ensures that \APCP's semantics preserves typing.
In contrast to Caires and Pfenning~\cite{conf/concur/CairesP10} and Wadler~\cite{conf/icfp/Wadler12}, where type preservation corresponds to the elimination of (top-level) applications of Rule~\ruleLabel{typ-cut}, in \APCP it corresponds to the elimination of (top-level) applications of Rule~\ruleLabel{type-res}.

\APCP's semantics consists of reduction and structural congruence.
Since the former relies on the latter, we first need to show type preservation for structural congruence, i.e., \emph{subject congruence}.
The structural congruence rule that unfolds recursive definitions requires care, because the types of the unfolded process are also unfolded:

\begin{exa}
    \label{x:APCP:unfrel}
    Consider the following typed recursive definition:
    \[
        \emptyset \vdash \pRec X(x,y) ; \pIn x(a) ; \pIn y(b) ; \pCall X<x,y> :: x:\mu X . \bullet \parr^0 X, y:\mu X . \bullet \parr^1 X
    \]
    Let us derive the typing of the unfolding of this process:
    \[
        \begin{bussproof}
            \bussAx[\ruleLabel{typ-var}]{
                X:(\bullet \parr^2 X, \bullet \parr^3 X) \vdash \pCall X<x,y> :: x:\mu X . \underbrace{\bullet \parr^6 X}_{\lift{4} (\bullet \parr^2 X)}, y:\mu X . \underbrace{\bullet \parr^7 X}_{\lift{4} (\bullet \parr^3 X)}
            }
            \bussUn[\ruleLabel{typ-end}]{
                X:(\bullet \parr^2 X, \bullet \parr^3 X) \vdash \pCall X<x,y> :: x:\mu X . \bullet \parr^6 X, y:\mu X . \bullet \parr^7 X, b':\bullet
            }
            \bussUn[\ruleLabel{typ-recv}]{
                X:(\bullet \parr^2 X, \bullet \parr^3 X) \vdash \pIn y(b') ; \pCall X<x,y> :: x:\mu X . \bullet \parr^6 X, y:\bullet \parr^3 \mu X . \bullet \parr^7 X
            }
            \bussUn[\ruleLabel{typ-end}]{
                X:(\bullet \parr^2 X, \bullet \parr^3 X) \vdash \pIn y(b') ; \pCall X<x,y> :: x:\mu X . \bullet \parr^6 X, y:\bullet \parr^3 \mu X . \bullet \parr^7 X, a':\bullet
            }
            \bussUn[\ruleLabel{typ-recv}]{
                X:(\bullet \parr^2 X, \bullet \parr^3 X)
                \begin{array}[t]{@{}r@{}l@{}}
                    {} \vdash {}
                    & \pIn x(a') ; \pIn y(b') ; \pCall X<x,y>
                    \\
                    {} :: {}
                    & x:\underbrace{
                        \bullet \parr^2 \mu X . \bullet \parr^6 X
                    }_{\unfold^4(\mu X . \bullet \parr^2 X)}, y:\underbrace{
                        \bullet \parr^3 \mu X . \bullet \parr^7 X
                    }_{\unfold^4(\mu X . \bullet \parr^3 X)}
                \end{array}
            }
            \bussUn[\ruleLabel{typ-rec}]{
                \emptyset \vdash \pRec X(x,y) ; \pIn x(a') ; \pIn y(b') ; \pCall X<x,y> :: x:\mu X . \bullet \parr^2 X, y:\mu X . \bullet \parr^3 X
            }
            \bussUn[\ruleLabel{typ-end}]{
                \emptyset \vdash \pRec X(x,y) ; \pIn x(a') ; \pIn y(b') ; \pCall X<x,y> :: x:\mu X . \bullet \parr^2 X, y:\mu X . \bullet \parr^3 X, b:\bullet
            }
            \bussUn[\ruleLabel{typ-recv}]{
                \emptyset \vdash \pIn y(b) ; \pRec X(x,y) ; \pIn x(a') ; \pIn y(b') ; \pCall X<x,y> :: x:\mu X . \bullet \parr^2 X, y:\bullet \parr^1 \mu X . \bullet \parr^3 X
            }
            \bussUn[\ruleLabel{typ-end}]{
                \emptyset \vdash \pIn y(b) ; \pRec X(x,y) ; \pIn x(a') ; \pIn y(b') ; \pCall X<x,y> :: x:\mu X . \bullet \parr^2 X, y:\bullet \parr^1 \mu X . \bullet \parr^3 X, a:\bullet
            }
            \bussUn[\ruleLabel{typ-recv}]{
                \emptyset
                \begin{array}[t]{@{}r@{}l@{}}
                    {} \vdash {}
                    & \pIn x(a) ; \pIn y(b) ; \pRec X(x,y) ; \pIn x(a') ; \pIn y(b') ; \pCall X<x,y>
                    \\
                    {} :: {}
                    & x:\underbrace{
                        \bullet \parr^0 \mu X . \bullet \parr^2 X
                    }_{\unfold^2(\mu X . \bullet \parr^0 X)}, y:\underbrace{
                        \bullet \parr^1 \mu X . \bullet \parr^3 X
                    }_{\unfold^2(\mu X . \bullet \parr^1 X)}
                \end{array}
            }
        \end{bussproof}
    \]
    Clearly, the typing of the unfolded process is not the same as the initial type, but they are equal up to unfolding of types.
    Note that the application of Rule~\ruleLabel{typ-rec} unfolds the types with a common lifter of 4, because it needs to be larger than the top priority in the types before unfolding which is 3.
\end{exa}

Hence, type preservation holds \emph{up to unfolding}.
To formalize this, we define the relation $(\vdash P :: \Gamma) \unfrel \Gamma'$, which says that $\Gamma$ and $\Gamma'$ are equal up to (un)folding of recursive types consistent with the typing of $P$ under $\Gamma$:

\begin{defi}
    \label{d:APCP:unfrel}
    We define an asymmetric relation between a typed process $(\Omega \vdash P :: \Gamma)$ and a typing context $\Gamma'$, denoted $(\Omega \vdash P :: \Gamma) \unfrel \Gamma'$.
    The relation is defined by the inference rules in \Cref{f:APCP:unfrel}, where each rule implicitly requires that $\Omega \vdash P :: \Gamma$ is a valid typing derivation by the rules in \Cref{f:APCP:typingRules}.

    We write $(\Omega \vdash P :: \Gamma) \unfrel* (\Omega' \vdash Q :: \Gamma')$ if $(\Omega \vdash P :: \Gamma) \unfrel \Gamma'$ and $(\Omega' \vdash Q :: \Gamma') \unfrel \Gamma$.
\end{defi}

\begin{figure}[t!]
    \begin{mathpar}
        \begin{bussproof}[unf-fold]
            \bussAssume{
                (\Omega \vdash P \big\{ \big( \pRec X(\tilde{y}) ; P \{ \tilde{y}/\tilde{z} \} \big) / \pCall X<\tilde{y}> \big\} :: \tilde{z}:\tilde{U}) \unfrel \tilde{z}:\tilde{U'}
            }
            \bussAssume{
                \forall U'_i \in \tilde{U'};~ U'_i = \unfold^t(\mu X . A'_i)
            }
            \bussBin{
                (\Omega \vdash P \big\{ \big( \pRec X(\tilde{y}) ; P \{ \tilde{y}/\tilde{z} \} \big) / \pCall X<\tilde{y}> \big\} :: \tilde{z}:\tilde{U}) \unfrel \tilde{z}:\widetilde{\mu X . A'}
            }
        \end{bussproof}
        \and
        \begin{bussproof}[unf-unf]
            \bussAssume{
                (\Omega \vdash \pRec X(\tilde{z}) ; P :: \tilde{z}:\widetilde{\mu X . A}) \unfrel \tilde{z}:\widetilde{\mu X . A'}
            }
            \bussAssume{
                \forall U'_i \in \tilde{U'}.~ U'_i = \unfold^t(\mu X . A'_i)
            }
            \bussBin{
                (\Omega \vdash \pRec X(\tilde{z}) ; P :: \tilde{z}:\widetilde{\mu X . A}) \unfrel \tilde{z}:\tilde{U'}
            }
        \end{bussproof}
        \\ \phantom{.}\dashes \\ 
        \begin{bussproof}[unf-send]
            \bussAx{
                (\Omega \vdash \pOut x[y,z] :: x:A \tensor^\pri B, y:\ol{A}, z:\ol{B}) \unfrel x:A \tensor^\pri B, y:\ol{A}, z:\ol{B}
            }
        \end{bussproof}
        \and
        \begin{bussproof}[unf-recv]
            \bussAssume{
                (\Omega \vdash P :: \Gamma, y:A, z:B) \unfrel \Gamma', y:A', z:B'
            }
            \bussUn{
                (\Omega \vdash \pIn x(y, z) ; P :: \Gamma, x:A \parr^\pri B) \unfrel \Gamma', x:A' \parr^\pri B'
            }
        \end{bussproof}
        \and
        \begin{bussproof}[unf-sel]
            \bussAx{
                (\Omega \vdash \pSel x[z] < j :: x:\oplus^\pri \{ i : A_i \}_{i \in I}, z:\ol{A_j}) \unfrel x:\oplus^\pri \{ i : A_i \}_{i \in I}, z:\ol{A_j}
            }
        \end{bussproof}
        \and
        \begin{bussproof}[unf-bra]
            \bussAssume{
                \forall i \in I.~ (\Omega \vdash P_i :: \Gamma, z:A_i) \unfrel \Gamma', z:A'_i
            }
            \bussUn{
                (\Omega \vdash \pBra x(z) > {\{ i : P_i \}_{i \in I}} :: \Gamma, x:\&^\pri \{ i : A_i \}_{i \in I}) \unfrel \Gamma', x:\&^\pri \{ i : A'_i \}_{i \in I}
            }
        \end{bussproof}
        \and
        \begin{bussproof}[unf-end]
            \bussAssume{
                (\Omega \vdash P :: \Gamma) \unfrel \Gamma'
            }
            \bussUn{
                (\Omega \vdash P :: \Gamma, x:\bullet) \unfrel \Gamma', x:\bullet
            }
        \end{bussproof}
        \and
        \begin{bussproof}[unf-par]
            \bussAssume{
                (\Omega \vdash P :: \Gamma) \unfrel \Gamma'
            }
            \bussAssume{
                (\Omega \vdash Q :: \Delta) \unfrel \Delta'
            }
            \bussBin{
                (\Omega \vdash P \| Q :: \Gamma, \Delta) \unfrel \Gamma' , \Delta'
            }
        \end{bussproof}
        \and
        \begin{bussproof}[unf-res]
            \bussAssume{
                (\Omega \vdash P :: \Gamma, x:A, y:\ol{A}) \unfrel \Gamma', x:A', y:\ol{A'}
            }
            \bussUn{
                (\Omega \vdash \pRes{xy} P :: \Gamma) \unfrel \Gamma'
            }
        \end{bussproof}
        \and
        \begin{bussproof}[unf-inact]
            \bussAx{
                (\Omega \vdash \0 :: \emptyset) \unfrel \emptyset
            }
        \end{bussproof}
        \and
        \begin{bussproof}[unf-fwd]
            \bussAx{
                (\Omega \vdash \pFwd [x<>y] :: x:\ol{A}, y:A) \unfrel x:\ol{A}, y:A
            }
        \end{bussproof}
        \and
        \begin{bussproof}[unf-rec]
            \bussAssume{
                (\Omega, X:{\tilde{A}} \vdash P :: \tilde{z}:\tilde{U}) \unfrel \tilde{z}:\tilde{U'}
            }
            \bussUn{
                (\Omega \vdash \pRec X(\tilde{z}) ; P :: \tilde{z}:\widetilde{\mu X . A}) \unfrel \tilde{z}:\widetilde{\mu X . A'}
            }
        \end{bussproof}
        \and
        \begin{bussproof}[unf-var]
            \bussAx{
                (\Omega, X:\tilde{A} \vdash \pCall X<\tilde{z}> :: \tilde{z}:\tilde{U}) \unfrel \tilde{z}:\tilde{U}
            }
        \end{bussproof}
    \end{mathpar}

    \caption{Inference rules for \Cref{d:APCP:unfrel}.}
    \label{f:APCP:unfrel}
\end{figure}

The most important rules of \Cref{f:APCP:unfrel} are Rules~\ruleLabel{unf-unf} and~\ruleLabel{unf-fold} (above the line), as they relate unfolding and folding; the other rule (below the line) follow the typing rules in \Cref{f:APCP:typingRules}.
Note that the rules in \Cref{f:APCP:unfrel} require no priority requirements, as they are covered by the implicit validity of the derivation of $\Omega \vdash P :: \Gamma$.

\begin{prop}
    \label{p:APCP:unfrelTrans}
    If (i)~$(\Omega_1 \vdash P_1 :: \Gamma_1) \unfrel* (\Omega_2 \vdash P_2 :: \Gamma_2)$ and (ii)~$(\Omega_2 \vdash P_2 :: \Gamma_2) \unfrel* (\Omega_3 \vdash P_3 :: \Gamma_3)$, then $(\Omega_1 \vdash P_1 :: \Gamma_1) \unfrel* (\Omega_3 \vdash P_3 :: \Gamma_3)$.
\end{prop}

\begin{sketch}
    Since $\unfrel*$ is reflexive by definition, it suffices to show that (i)~$(\Omega_1 \vdash P_1 :: \Gamma_1) \unfrel \Gamma_2$ and (ii)~$(\Omega_2 \vdash P_2 :: \Gamma_2) \unfrel \Gamma_3$ imply $(\Omega_1 \vdash P_1 :: \Gamma_1) \unfrel \Gamma_3$.
    Assumptions~(i) and~(ii) relate $\Gamma_1,\Gamma_2,\Gamma_3$ by applications of Rules~\ruleLabel{unf-fold} and~\ruleLabel{unf-unf}.
    Clearly, the relation between $\Gamma_1$ and $\Gamma_3$ can then be derived directly by combining the applications of these rules.
\end{sketch}
\noindent 
Subject congruence additionally requires the following property of typing derivations involving recursion variables:

\begin{prop}
    \label{p:APCP:recPreserved}
    Suppose given a process $P$ and a derivation $\pi$ of $\Omega, X:{(A_l)}_{l \in L} \vdash P :: \Gamma$.
    In every step in $\pi$, the assignment $X:{(A_l)}_{l \in L}$ in the recursive context is not modified nor removed, and no other assignments on the recursion variable $X$ are added.
\end{prop}

\begin{sketch}
    By induction on the height of the derivation $\pi$: no typing rule eliminates assignments from the recursive context, and no typing rule changes the types assigned to a recursion variable in the recursive context.
    Moreover, in our type system, any rule adding an assignment to a context implicitly assumes that the newly introduced name or recursion variable is not in the context yet.
    Hence, an assignment on $X$ cannot be added by any rule.
\end{sketch}
\noindent 
Now we can state and prove subject congruence, which holds up to the relation in \Cref{d:APCP:unfrel}:

\begin{restatable}[Subject Congruence]{theorem}{tAPCPSC}
    \label{t:APCP:sc}
    If $\Omega \vdash P :: \Gamma$ and $P \equiv Q$, then there exists $\Gamma'$ such that $\Omega \vdash Q :: \Gamma'$ and $(\Omega \vdash P :: \Gamma) \unfrel* (\Omega \vdash Q :: \Gamma')$.
\end{restatable}

\begin{proof}
    By induction on the derivation of $P \equiv Q$.
    The cases for the structural rules follow from the IH directly.
    All base cases are straightforward, except the cases of unfolding and folding.
    We detail unfolding, where $P = \pRec X(\tilde{z}) ; P'$ and $Q = P' \big\{ \big( \pRec X(\tilde{y}) ; P' \{ \tilde{y}/\tilde{z} \} \big) / \pCall X<\tilde{y}> \big\}$.
    The other direction, folding, is similar.

    By type inversion of Rule~\ruleLabel{typ-rec},
    \begin{align*}
        \Omega &\vdash P :: \Omega; \tilde{z}:\widetilde{\mu X . A}
        \\
        \Omega, X:\tilde{A} &\vdash P' :: \tilde{z}:\tilde{U}
    \end{align*}
    where there exists $t \in \mbb{N} > \prtop(\widetilde{\mu X . A})$ such that each
    \[
        U_i = \unfold^t(\mu X . A_i) = A_i \{ \big( \mu X . \lift{t} A_i \big) / X \}.
    \]
    Let $\deriv$ denote the typing derivation of $P'$.

    $P'$ may contain $m \geq 0$ recursive calls $\pCall X<\tilde{y}>$.
    For each call, there is an application of Rule~\ruleLabel{typ-var} in $\deriv$.
    We uniquely identify each such application of Rule~\ruleLabel{typ-var} as $\deriv*_j$ for $0 < j \leq m$.
    Since \APCP only allows tail recursion, these multiple recursive calls can only occur inside branches on names not in $\tilde{z}$, so the common lifter of each $\deriv*_j$ must be~$t$.
    \begin{align*}
        \deriv*_j \left\{ \qquad
            \begin{bussproof}
                \bussAx[\ruleLabel{typ-var}]{
                    \Omega'_j, X:\tilde{A} \vdash \pCall X<\tilde{y}_j> :: \tilde{y}_j:\widetilde{\mu X . \lift{t} A}
                }
            \end{bussproof}
        \right.
    \end{align*}
    where $\Omega'_j \supseteq \Omega$, because rules applied in $\deriv$ can only add assignments to the recursive context.
    Recall that $\tilde{y}_j:\widetilde{\mu X . \lift{t} A}$ is notation for $y_{j_1}:\mu X . \lift{t} A_1 , \ldots , y_{j_k}:\mu X . \lift{t} A_k$, as introduced in \Cref{s:APCP:APCP:types}.

    The unfolding of a recursive process replaces each recursive call by a copy of the recursive definition.
    Hence, to find a typing derivation for $Q$ we proceed as follows, for each $0 < j \leq m$:
    \begin{enumerate}

        \item
            We obtain a derivation $\deriv'_j$ of $P' \{ \tilde{y}_j/\tilde{z} \}$ from $\deriv$ by substituting $\tilde{z}$ for $\tilde{y}_j$ (while avoiding capturing names) and by lifting all priorities by $t$ (including the priorities of $X$ in the recursive context of $\deriv*_j$).

        \item
            We apply Rule~\ruleLabel{typ-rec} on $X$ in the conclusion of $\deriv'_j$, resulting in a new typing derivation $\deriv*'_j$:
            \begin{align*}
                \deriv*'_j \left\{ \qquad
                    \begin{bussproof}(c)
                        \bussAssume{
                            \deriv'_j
                        }
                        \noLine
                        \bussUn{
                            \Omega'_j, X:\widetilde{\lift{t} A} \vdash P' \{ \tilde{y}_j/\tilde{z} \} :: \tilde{y}_j:\widetilde{\lift{t} U}
                        }
                        \bussUn[\ruleLabel{typ-rec}]{
                            \Omega'_j \vdash \pRec X(\tilde{y}_j) ; P' \{ \tilde{y}_j/\tilde{z} \} :: \tilde{y}_j:\widetilde{\mu X . \lift{t} A}
                        }
                    \end{bussproof}
                \right.
            \end{align*}

    \end{enumerate}

    For every $0 < j \leq m$, the context of the conclusion of $\deriv*'_j$ coincides with the context of the conclusion of $\deriv*_j$, up to the assignment $X:\widetilde{\lift{t} A}$ in the recursive context not present in~$\deriv*'_j$.
    We intend to obtain from $\deriv$ a new derivation $\deriv'$ by replacing each $\deriv*_j$ with $\deriv*'_j$.
    By \Cref{p:APCP:recPreserved}, the fact that each $\deriv*'_j$ is missing the assignment to $X$ in the recursive context does not influence the steps in $\deriv$.
    Hence, we can indeed obtain such a derivation~$\deriv'$:
    \[
        \begin{bussproof}
            \bussAssume{
                \deriv'
            }
            \noLine
            \bussUn{
                \Omega \vdash \underbrace{
                    P' \big\{ \big( \pRec X(\tilde{y}) ; P' \{ \tilde{y}/\tilde{z} \} \big) / \pCall X<\tilde{y}> \big\}
                }_{Q} :: \tilde{z}: \tilde{U}
            }
        \end{bussproof}
    \]
    where $\forall U_i \in \tilde{U}.~ U_i = \unfold^t(A_i)$.

    Let $\Gamma' = \tilde{z}:\tilde{U}$.
    We have $\Omega \vdash Q :: \Gamma'$.
    To prove the thesis, we have to show that $\Gamma$ and $\Gamma'$ are equal up to unfolding.
    Following the typing rules applied in the derivation of $\Omega \vdash Q :: \tilde{z}:\tilde{U}$,  we have $(\Omega \vdash Q :: \tilde{z}:\tilde{U}) \unfrel \tilde{z}:\tilde{U}$ (cf.\ \Cref{d:APCP:unfrel}).
    Then, by Rule~\ruleLabel{unf-fold}, $(\Omega \vdash Q :: \tilde{z}:\tilde{U}) \unfrel \tilde{z}:\widetilde{\mu X . A}$.
    Similarly, we have $(\Omega \vdash P :: \tilde{z}:\widetilde{\mu X . A}) \unfrel \tilde{z}:\widetilde{\mu X . A}$, so by Rule~\ruleLabel{unf-unf}, $(\Omega \vdash P :: \tilde{z}:\widetilde{\mu X . A}) \unfrel \tilde{z}:\tilde{U}$.
    Hence, $(\Omega \vdash P :: \Gamma) \unfrel* (\Omega \vdash Q :: \Gamma')$, proving the thesis.
\end{proof}

\begin{figure}[t]
    \begin{align*}
        & \begin{bussproof}
            \def\defaultHypSeparation{\hskip1ex}
            \bussAx[\ruleLabel{typ-send}]{
                \vdash \pOut x[a,b] :: x:A \tensor^\pri B, a:\ol{A}, b:\ol{B}
            }
            \bussAssume{
                \vdash P :: \Gamma, z:\ol{A}, y':\ol{B}
            }
            \bussUn[\ruleLabel{typ-recv}]{
                \vdash \pIn y(z,y') ; P :: \Gamma, y:\ol{A} \parr^\pri \ol{B}
            }
            \doubleLine
            \bussBin[\ruleLabel{typ-par}+\ruleLabel{typ-res}]{
                \vdash \pRes{xy} ( \pOut x[a,b] \| \pIn y(z,y') ; P ) :: \Gamma, a:\ol{A}, b:\ol{B}
            }
        \end{bussproof}
        \\
        & \redd
        \\
        & \vdash P \{ a/z,b/y' \} :: \Gamma, a:\ol{A}, b:\ol{B}
    \end{align*}

    \caption{
        Example of subject reduction (cf.\ \Cref{t:APCP:sr}) in Rule~\protect\ruleLabel{red-send-recv}.
        The well typedness of the process before reduction allows us to infer its typing derivation, also giving us the typing of $P$.
        Typing the process after reduction is then a matter of inductively substituting names in the typing derivation of $P$.
    }
    \label{f:APCP:tp}
\end{figure}
\noindent 
Having established subject congruence, we can prove that reduction preserves typing, i.e., \emph{subject reduction}:

\begin{thm}[Subject Reduction]
    \label{t:APCP:sr}
    If $\Omega \vdash P :: \Gamma$ and $P \redd Q$, then there exists $\Gamma'$ such that $\Omega \vdash Q :: \Gamma'$ and ${(\Omega \vdash P :: \Gamma) \unfrel* (\Omega \vdash Q :: \Gamma')}$.
\end{thm}

\begin{sketch}
    By induction on the derivation of $P \redd Q$ (\Cref{f:procdef} (bottom)), we find a $\Gamma'$ such that $\Omega \vdash Q :: \Gamma'$ and $(\Omega \vdash P :: \Gamma) \unfrel* (\Omega \vdash Q :: \Gamma')$.

    The cases of Rules~\ruleLabel{red-res} and~\ruleLabel{red-par} follow directly from the IH.
    Consider, e.g., $\pRes{xy} ( P \| R ) \redd \pRes{xy} ( Q \| R )$ derived from $P \redd Q$.
    By inversion of typing on the process before reduction, e.g., $\Omega \vdash P :: \Gamma , x:A$.
    Then, by the IH, $(\Omega \vdash P :: \Gamma, x:A) \unfrel* (\Omega \vdash Q :: \Gamma', x:A')$.
    The thesis then follows from an application of Rules~\ruleLabel{unf-par} and~\ruleLabel{unf-res}.

    The case of Rule~\ruleLabel{red-cong} follows from the IH and subject congruence (\Cref{t:APCP:sc}).
    To be precise, the rule says that (i)~$P \equiv P'$, (ii)~$P' \redd Q'$, and (iii)~$Q' \equiv Q$ imply $P \redd Q$.
    By \Cref{t:APCP:sc} (subject congruence) on assumptions~(i) and~(iii), we have $(\Omega \vdash P :: \Gamma) \unfrel* (\Omega \vdash P' :: \Gamma')$ and $(\Omega' \vdash Q' :: \Delta') \unfrel* (\Omega' \vdash Q :: \Delta)$.
    Also, by the IH on assumption~(ii), $(\Omega \vdash P' :: \Gamma') \unfrel* (\Omega' \vdash Q' :: \Delta')$ where $\Omega = \Omega'$.
    Then, by \Cref{p:APCP:unfrelTrans} (transitivity of $\unfrel*$), $(\Omega \vdash P :: \Gamma) \unfrel* (\Omega \vdash Q :: \Delta)$, proving the thesis.

    Key cases are Rules~\ruleLabel{red-send-recv}, \ruleLabel{red-sel-bra}, and~\ruleLabel{red-fwd}.
    \Cref{f:APCP:tp} shows the representative instance of Rule~\ruleLabel{red-send-recv}, and an example where Rules~\ruleLabel{red-res}, \ruleLabel{red-par}, and~\ruleLabel{red-cong} are used.
\end{sketch}
\noindent 
Protocol fidelity ensures that processes respect their intended (session) protocols.
Communication safety ensures the absence of communication errors and mismatches in processes.
Correct typability gives a static guarantee that a process conforms to its ascribed session protocols; type preservation gives a dynamic guarantee.
Because session types describe the intended protocols and error-free exchanges, type preservation entails both protocol fidelity and communication safety.
For a detailed account, we refer the curious reader to the early work by Honda \etal~\cite{conf/esop/HondaVK98}, which defines error processes
and shows by contradiction that well-typed processes do not reduce to an error.

In what follows, we consider a process to be deadlocked if it is not the inactive process and cannot reduce.
Our deadlock-freedom result for \APCP adapts that for \PCP~\cite{conf/fossacs/DardhaG18}.
The equivalent of our Rule~\ruleLabel{typ-res} in \PCP is Rule~\ruleLabel{cycle}.
Deadlock freedom for \PCP involves three steps to eliminate applications of Rule~\ruleLabel{cycle}:
\begin{enumerate}

    \item
        First, \ruleLabel{cycle} elimination states that we can remove all applications of \ruleLabel{cycle} in a typing derivation without affecting the derivation's assumptions and conclusion.

    \item
        Only the removal of \emph{top-level} applications of \ruleLabel{cycle} captures the intended process semantics; the removal of other applications of \ruleLabel{cycle} corresponds to reductions behind prefixes, which is not allowed~\cite{conf/icfp/Wadler12,conf/fossacs/DardhaG18}.
        Therefore, the second step is \emph{top-level deadlock freedom} (referred to here as \emph{progress}), which states that a process with a top-level application of \ruleLabel{cycle} reduces until there are no top-level applications of \ruleLabel{cycle} left.
        This step requires commuting conversions: a process with a top-level application of \ruleLabel{cycle} may not have reductions ready, so commuting conversions are used to remove top-level applications of \ruleLabel{cycle} by blocking them with prefixes.

    \item
        Third, deadlock freedom follows for processes typable under empty contexts.

\end{enumerate}
Here, we adapt and address \ruleLabel{typ-res} elimination and progress in one proof.

As mentioned before, binding \APCP's asynchronous sends and selections to continuations involves additional, low-level uses of \ruleLabel{typ-res}, which we cannot eliminate through process reduction.
Therefore, we establish progress for \emph{live processes} (\Cref{t:APCP:progress}).
A process is live if it is equivalent to a restriction on \emph{active names} used for unguarded prefixes.
This way, e.g., in $\pOut x[y,z]$ the name $x$ is active, but $y$ and $z$ are not.
We additionally need a notion of \emph{evaluation context}, under which reducible forwarders may occur.

\def\dAPCPan{
    \begin{defi}[Active Names]
        \label{d:APCP:an}
        The \emph{set of active names} of $P$, denoted $\an(P)$, contains the (free) names that are used for non-blocked prefixed:
        \begin{align*}
            \an(\pOut x[y,z])
            &\deq
            \{x\}
            &
            \an(\pIn x(y,z) ; P)
            &\deq
            \{x\}
            &
            \an(\0)
            &\deq
            \emptyset
            \\
            \an(\pSel x[z] < \ell)
            &\deq
            \{x\}
            &
            \an(\pBra x(z) > {\{i: P_i\}_{i \in I}})
            &\deq
            \{x\}
            &
            \an(\pFwd [x<>y])
            &\deq
            \emptyset
            \\
            \an(P \| Q)
            &\deq
            \an(P) \cup \an(Q)
            &
            \an(\pRec X(\tilde{x}) ; P)
            &\deq
            \an(P)
            \\
            \an(\pRes{xy} P)
            &\deq
            \an(P) \setminus \{x,y\}
            &
            \an(\pCall X<\tilde{x}>)
            &\deq
            \emptyset
        \end{align*}
    \end{defi}
}

\dAPCPan

\begin{defi}[Evaluation Context]
    \label{d:APCP:evalCtx}
    Evaluation contexts ($\evalCtx{E}$) are defined by the following grammar:
    \[
        \evalCtx{E} ::= \evalHole \sepr \evalCtx{E} \| P \sepr \pRes{xy} \evalCtx{E} \sepr \pRec X(\tilde{x}) ; \evalCtx{E}
    \]
    We write $\evalCtx{E}[P]$ to denote the process obtained by replacing the hole $\evalHole$ in $\evalCtx{E}$ by $P$.
\end{defi}

\def\dAPCPLive{
    \begin{defi}[Live Process]
        \label{d:APCP:live}
        A process $P$ is \emph{live}, denoted $\live(P)$, if
        \begin{enumerate}

            \item
                there are names $x,y$ and process $P'$ such that $P \equiv \pRes{xy} P'$ with $x,y \in \an(P')$, or

            \item
                there are names $x,y,z$ and process $P'$ such that $P \equiv \evalCtx[\big]{E}[\pRes{yz} ( \pFwd [x<>y] \| P' )]$ and $z \neq x$ (i.e., the forwarder is independent).

        \end{enumerate}
    \end{defi}
}

\dAPCPLive
\noindent 
We additionally need to account for recursion: as recursive definitions do not directly entail reductions, we must fully unfold them before eliminating applications of \ruleLabel{typ-res}:

\begin{restatable}[Unfolding]{lemma}{lAPCPUnfold}
    \label{l:APCP:unfold}
    If $\Omega \vdash P :: \Gamma$, then there is a process $P^\star$ such that $P^\star \equiv P$ and $P^\star$ is not of the form $\pRec X(\tilde{z}) ; Q$.
\end{restatable}

\begin{proof}
    By induction on the number $n$ of consecutive recursive definitions prefixing $P$, such that $P$ is of the form $\pRec X_1(\tilde{z}) ; \ldots ; \pRec X_n(\tilde{z}) ; Q$.
    If $n = 0$, the thesis follows immediately, as $P \equiv P$.
    Otherwise, $n \geq 1$.
    Then there are $X,Q$ such that $P = \pRec X(\tilde{z}) ; Q$, where $Q$ starts with $n-1$ consecutive recursive definitions.
    Let $R \deq Q \big\{ \big( \pRec X(\tilde{y}) ; Q \{ \tilde{y}/\tilde{z} \} \big) / \pCall X<\tilde{y}> \big\}$.
    Clearly, $R \equiv P$.
    Then, because $R$ starts with $n-1$ consecutive recursive definitions, the thesis follows by appealing to the IH.
\end{proof}
\noindent 
Dardha and Gay's progress result concerns a sequence of reduction steps that reaches a process that is not live anymore~\cite{conf/fossacs/DardhaG18}.
In our case, progress concerns a single reduction step only, because recursive processes might stay live across reductions forever.
Moreover, because of our definition of liveness, we do not need commuting conversions for this step.

\begin{restatable}[Progress]{theorem}{tAPCPProgress}
    \label{t:APCP:progress}
    If $\emptyset \vdash P :: \Gamma$ and $\live(P)$, then there is a process $Q$ such that $P \redd Q$.
\end{restatable}

\begin{proof}
    We distinguish the two cases of $\live(P)$: $P$ contains a restriction on a pair of active names, or $P$ contains a restriction on a forwarded name under a reduction context.
    In both cases, we first unfold any recursive definitions preceding the involved prefixes/forwarders, resulting in $P^\star \equiv P$.
    By subject congruence (\Cref{t:APCP:sc}), there exists $\Gamma'$ such that $\emptyset \vdash P^\star :: \Gamma'$ and $(\vdash P^\star :: \Gamma') \unfrel \Gamma$.
    By \Cref{d:APCP:unfrel}, the only difference between $\Gamma'$ and $\Gamma$ is the unfolding of recursive types.
    \begin{enumerate}

        \item
            $P^\star$ contains a restriction on a pair of active names.
            That is, $P^\star \equiv \pRes{xy} {P^\star}'$ and $x,y \in \an({P^\star}')$.
            The rest of the analysis depends on how $x$ and $y$ occur as active names in ${P^\star}'$.
            As a representative case, we consider that $x$ occurs as the subject of a send.
            By inversion of typing, then $y$ occurs as the subject of a receive.
            Hence, we have $P^\star \equiv \evalCtx[\big]{E}[\pRes{xy} ( \pOut x[a,b] \| \pIn y(v,z) ; {P^\star}'' )]$.
            Let $Q \deq \evalCtx{E}[{P^\star}'' \{ a/v,b/z \}]$.
            Then, by Rule~\ruleLabel{red-send-recv}, $P^\star \redd Q$.
            Hence, by Rule~\ruleLabel{red-cong}, $P \redd Q$, proving the thesis.

        \item
            $P^\star$ contains a restriction on a forwarded name under a reduction context.
            That is, $P^\star \equiv \evalCtx[\big]{E}[\pRes{yz} ( \pFwd [x<>y] \| {P^\star}' )]$ where $z \neq x$.
            Let $Q = \evalCtx{E}[{P^\star}' \{ x/z \}]$.
            Then, by Rule~\ruleLabel{red-fwd}, $P^\star \redd Q$.
            Hence, by Rule~\ruleLabel{red-cong}, $P \redd Q$, proving the thesis.
            \qedhere

    \end{enumerate}
\end{proof}
\noindent 
Our deadlock-freedom result concerns processes typable under empty contexts (as in, e.g., Caires and Pfenning~\cite{conf/concur/CairesP10} and Dardha and Gay~\cite{conf/fossacs/DardhaG18}).
We first need a lemma which ensures that non-live processes typable under empty contexts do not contain prefixes (sends/receives/selections/branches) or \emph{independent} forwarders (whose endpoints are not bound together using restriction).
This lemma is in fact the crux of our deadlock-freedom result, as it relies on the priority checks induced by our typing system:

\begin{restatable}{lemma}{lAPCPNotLiveNoAction}
    \label{l:APCP:notLiveNoAction}
    If $\emptyset \vdash P :: \emptyset$ and $P$ is not live, then $P$ contains no prefixes or independent forwarders.
\end{restatable}

\begin{proof}
    W.l.o.g., assume all recursion in $P$ is unfolded.
    Take $P^\star \equiv P$ such that $P^\star = \pRes{x_i x'_i}_{i \in I} \prod_{j \in J} P_j$ where, for every $j \in J$, $P_j$ is a prefix or forwarder.
    That is, every $P_j$ is a \emph{thread} which cannot be broken down further into parallel components or restrictions.

    By abuse of notation, we write $\pr(P_j)$ to denote the priority of the type of the subject of the prefix of $P_j$.
    Also, we write $\pr(x)$ to denote the priority of the type of $x$.

    Towards a contradiction, assume that there is at least one prefix or independent forwarder in $P$.
    We apply induction on the size of $J$.
    \begin{itemize}

        \item
            In the base case, $J = \emptyset$.
            Then there cannot be any prefixes or independent forwarders in $P$: a contradiction.

        \item
            In the inductive case, $J = J' \cup \{j\}$.
            W.l.o.g., assume that, for every $j' \in J'$, $\pr(P_j) \leq \pr(P_{j'})$ (i.e., pick $P_j$ as one of the prefixes/forwarders with the least priority of all threads).
            By assumption, $P_j$ denotes a prefix or forwarder on some endpoint $x_i$.
            By well typedness, $x_i$ is connected through restriction to $x'_i$.
            By Rule~\ruleLabel{typ-res}, $\pr(x_i) = \pr(x'_i)$.
            Further analysis depends on $P_j$: a receive (or, analogously, a branch), a send (or, analogously, a selection), or a forwarder.
            \begin{itemize}

                \item
                    Suppose $P_j = \pIn x_i(y,z) ; P'_j$.
                    By Rule~\ruleLabel{typ-recv}, $x'_i \notin \fn(P'_j)$, because the rule requires $\pr(x_i) < \pr(x'_i)$: otherwise, this would contradict the fact that $\pr(x_i) = \pr(x'_i)$.
                    Hence, there exists $j' \in J'$ such that $x'_i \in \fn(P_{j'})$.

                    If $J' = \emptyset$, the contradiction is immediate.
                    Otherwise, the analysis depends on the prefix of $P_{j'}$.
                    Since $P$ is not live, this cannot be a send with subject $x'_i$ or an independent forwarder on $x'_i$.
                    The prefix can also not be a dependent forwarder, for $x_i$ already appears in $P_j$.
                    This leaves us with two possibilities: a receive (or, analogously, a branch) on another endpoint, or a send (or, analogously, a selection) with object $x'_i$.
                    \begin{itemize}

                        \item
                            If $P_{j'} = \pIn x_k(v,w) ; P'_{j'}$ where $x'_i \in \fn(P'_{j'})$, then, by Rule~\ruleLabel{typ-recv}, $\pr(x_k) < \pr(x'_i)$.
                            Hence, $\pr(P_{j'}) = \pr(x_k) < \pr(x'_i) = \pr(x_i) = \pr(P_j)$: this contradicts the assumption that $\pr(P_j) \leq \pr(P_{j'})$.

                        \item
                            If $P_{j'} = \pOut x[a,b]$ where $x'_i \in \{a,b\}$, w.l.o.g., assume $x'_i = a$.
                            Then, by Rule~\ruleLabel{typ-send}, $\pr(x_k) < \pr(x'_i)$.
                            The contradiction follows as in the previous case.

                    \end{itemize}

                \item
                    Suppose $P_j = \pOut x_i[a,b]$.
                    By Rule~\ruleLabel{typ-send}, $x'_i \notin \{a,b\}$, because the rule requires $\pr(x_i) < \pr(x'_i)$: otherwise, this would contradict the fact that $\pr(x_i) = \pr(x'_i)$.
                    Hence, there exists $j' \in J'$ such that $x'_i \in \fn(P_{j'})$.
                    The contradiction follows as in the previous case.

                \item
                    Suppose $P_j = \pFwd [x_i<>x_k]$.
                    By well typedness, $x_i$ is connected through restriction to $x'_i$.
                    Since $P$ is not live, it must be that $x'_i = x_k$: $P_j$ is not an independent forwarder.
                    Then there must be a $j' \in J'$ where $P_{j'}$ fulfills the assumption that $P$ contains at least one prefix or independent forwarder.
                    The contradiction then follows from the IH.
                    \qedhere

            \end{itemize}

    \end{itemize}
\end{proof}

We now state our deadlock-freedom result:

\begin{restatable}[Deadlock Freedom]{theorem}{tAPCPDF}
    \label{t:APCP:df}
    If $\emptyset \vdash P :: \emptyset$, then either $P \equiv \0$ or $P \redd Q\mkern2mu$ for some $Q$.
\end{restatable}

\begin{proof}
    The analysis depends on whether $P$ is live or not.
    \begin{itemize}
        \item
            If $P$ is not live, then, by \Cref{l:APCP:notLiveNoAction}, it does not contain any prefixes or independent forwarders.
            Any recursive definitions in $P$ are thus of the form $\pRec X_1() ; \ldots ; \pRec X_n() ; \0$: contractiveness requires recursive calls to be prefixed by receives/branches or bound to parallel sends/selections/forwarders, of which there are none.
            Hence, we can use structural congruence to rewrite each recursive definition in $P$ to $\0$ by unfolding, yielding $P' \equiv P$.
            Any dependent forwarders in $P$ are of the form $\pRes{xy} \pFwd [x<>y]$, and can be rewritten to $\0$ using Rule~\ruleLabel{cong-res-fwd} (cf. \Cref{f:procdef}).
            The remaining derivation of $P'$ only contains applications of Rule~\ruleLabel{typ-inact}, \ruleLabel{typ-par}, \ruleLabel{typ-end}, or~\ruleLabel{typ-res} on closed names.
            It follows straightforwardly that $P \equiv P' \equiv \0$.

        \item
            If $P$ is live, by \Cref{t:APCP:progress} there is $Q$ s.t.\ $P \redd Q$.
            \qedhere
    \end{itemize}
\end{proof}

\subsection{Reactivity}
\label{s:APCP:APCP:react}

Progress (\Cref{t:APCP:progress}) is an important liveness property, as it defines precisely the conditions under which processes can reduce (namely, liveness in \Cref{d:APCP:live}).
As such, progress plays a key role in deadlock freedom (\Cref{t:APCP:df}).
However, in the presence of recursion, the strength of progress is limited: even if the majority of subprocesses is stuck, progress holds if at least one subprocess is live.
Nonetheless, \APCP's type system allows us to prove a stronger result: \emph{reactivity}, which essentially states that infinite recursion in \APCP cannot block sessions from progressing.

Processes typable under empty contexts are not only deadlock free, they are \emph{reactive}, in the following sense: for each name in the process, we can eventually observe a reduction involving that name.
To formalize this property, we define \emph{labeled reductions}, which expose details about synchronizations:

\begin{defi}[Labeled Reductions]
    \label{d:procLred}
    Consider the labels
    \begin{align*}
        \alpha ::= \pFwd [x<>y] \sepr \rLbl x>y:a \sepr \rLbl x>y:\ell
        \qquad\qquad
        \text{(forwarding, send/receive, selection/branching)}
    \end{align*}
    where each label has subjects $x$ and $y$.
    The \emph{labeled reduction} $P \lredd{\alpha} Q$ is defined by the following rules:
    \begin{mathpar}
        \begin{bussproof}[lred-send-recv]
            \bussAx{
                \pRes{xy} ( \pOut x[a,b] \| \pIn y(v,z) ; P )
                \lredd{\rLbl x>y:a}
                P \{ a/v,b/z \}
            }
        \end{bussproof}
        \and
        \begin{bussproof}[lred-sel-bra]
            \bussAssume{
                j \in I
            }
            \bussUn{
                \pRes{xy} ( \pSel x[b] < j \| \pBra y(z) > {\{i: P_i\}_{i \in I}} )
                \lredd{\rLbl x>y:j}
                P_j \{ b/z \}
            }
        \end{bussproof}
        \and
        \begin{bussproof}[lred-fwd]
            \bussAx{
                \pRes{yz} ( \pFwd [x<>y] \| P )
                \lredd{\pFwd [x<>y]}
                P \{ x/z \}
            }
        \end{bussproof}
        \and
        \begin{bussproof}[lred-cong]
            \bussAssume{
                P \equiv P'
            }
            \bussAssume{
                P' \lredd{\alpha} Q'
            }
            \bussAssume{
                Q' \equiv Q
            }
            \bussTern{
                P \lredd{\alpha} Q
            }
        \end{bussproof}
        \and
        \begin{bussproof}[lred-res]
            \bussAssume{
                P \lredd{\alpha} Q
            }
            \bussUn{
                \pRes{xy} P \lredd{\alpha} \pRes{xy} Q
            }
        \end{bussproof}
        \and
        \begin{bussproof}[lred-par]
            \bussAssume{
                P \lredd{\alpha} Q
            }
            \bussUn{
                P \| R \lredd{\alpha} Q \| R
            }
        \end{bussproof}
    \end{mathpar}
\end{defi}

\def\makePLredd{
    For any $P$ and $P'$, $P \redd P'$ if and only if there exists a label $\alpha$ such that~$P \lredd{\alpha} P'$.
}
\begin{restatable}{proposition}{pLredd}
    \makePLredd
\end{restatable}

\begin{proof}
    Immediate by definition, because each reduction in \Cref{f:procdef} (bottom) corresponds to a labeled reduction, and vice versa.
\end{proof}

Our reactivity result states that processes typable under empty contexts have at least one finite reduction sequence (denoted $\reddF$) that enables a labeled reduction involving a \emph{pending} name---a name that occurs as the subject of a prefix and is not bound by an input (see below).
Clearly, the typed process may have other reduction sequences, not necessarily finite.

\begin{defi}[Pending Names]
    \label{d:APCP:pn}
    Given a process $P$, we define the set of \emph{pending names} of $P$, denoted $\pn(P)$, as follows:
    \begin{align*}
        \pn(\pOut x[y,z])
        &\deq
        \{x\}
        &
        \pn(\pIn x(y,z) ; P)
        &\deq
        \{x\} \cup (\pn(P) \setminus \{y,z\})
        \span\span
        \\
        \pn(\pSel x[z] < \ell)
        &\deq
        \{x\}
        &
        \pn(\pBra x(z) > {\{i: P_i\}_{i \in I}})
        &\deq
        \{x\} \cup (\bigcup_{i \in I} \pn(P_i) \setminus \{z\})
        \span\span
        \\
        \pn(P \| Q)
        &\deq
        \pn(P) \cup \pn(Q)
        &
        \pn(\pRec X(\tilde{x}) ; P)
        &\deq
        \pn(P)
        &
        \pn(\0)
        &\deq
        \emptyset
        \\
        \pn(\pRes{xy} P)
        &\deq
        \pn(P)
        &
        \pn(\pCall X<\tilde{x}>)
        &\deq
        \emptyset
        &
        \pn(\pFwd [x<>y])
        &\deq
        \{x,y\}
    \end{align*}
\end{defi}
\noindent 
Note that the proof of reactivity below does not rely on deadlock freedom: suppose we are observing a blocked pending prefix in a process with a parallel recursive definition; deadlock freedom ensures a reduction from the recursive definition which would not unblock the pending prefix we are observing.
Instead, the proof relies on a priority analysis (similar to the one in the proof of \Cref{l:APCP:notLiveNoAction}) to unblock pending prefixes.

\def\makeTAPCPReact{
    Suppose given a process $\emptyset \vdash P :: \emptyset$.
    Then, for every $x \in \pn(P)$ there exists a process $P'$ such that $P \reddF P'$ and $P' \lredd{\alpha} Q$, for some process $Q$ and label $\alpha$ with subject $x$.
}
\begin{restatable}[Reactivity]{theorem}{tAPCPReact}
    \label{t:APCP:react}
    \makeTAPCPReact
\end{restatable}

\begin{proof}
    Take any $x \in \pn(P)$.
    Because $P$ is typable under empty contexts, $x$ is bound to some $y \in \pn(P)$ by restriction.
    By typing, in $P$ there is exactly one prefix on $x$ and one prefix on $y$ (they may also occur in forwarder processes).
    Following the restrictions on priorities in the typing of $x$ and $y$ in $P$, the prefixes on $x$ and $y$ cannot occur sequentially in $P$ (cf.\ the proof of \Cref{l:APCP:notLiveNoAction} for details on this reasoning).
    By typability, the prefix on $y$ is dual to the prefix on $x$.

    We apply induction on the number of receives, branches, and recursive definitions in $P$ blocking the prefixes on $x$ and $y$, denoted $n$ and $m$, respectively.
    Because $P$ is typable under empty contexts, the blocking receives and branches that are on names in $\pn(P)$ also have to be bound to pending names by restriction.
    The prefixes on these connected names may also be prefixed by receives, branches, and recursive definitions, so we may need to unblock those prefixes as well.
    Since there can only be a finite number of names in any given process, we also apply induction on the number of prefixes blocking these connected prefixes.
    \begin{itemize}

        \item
            If $n = 0$ and $m = 0$, then the prefixes on $x$ and $y$ occur at the top level; because they do not occur sequentially, the synchronization between $x$ and $y$ can take place immediately.
            Hence, $P \lredd{\alpha} Q$ where $x$ and $y$ are the subjects of $\alpha$.
            This proves the thesis, with $P' = P$.

        \item
            If $n > 0$ or $m > 0$, the analysis depends on the foremost prefixes blocking the prefixes on $x$ and $y$.

            If the either of these blocking prefixes is a recursive definition ($\pRec X(\tilde{y})$), we unfold the recursion.
            Because a corresponding recursive call ($\pCall X<\tilde{z}>$) cannot occur as a prefix, the effect of unfolding either (i)~triggers prefixes that occur in parallel to those on $x$ and $y$, or (ii)~the prefixes on $x$ or $y$ precede the punfolded recursive call.
            In either case, the number of prefixes decreases, and the thesis follows from the IH.

            Otherwise, if neither foremost prefix is a recursive definition, then the foremost prefixes must be on names in $\pn(P)$.
            Consider the prefix that is typable with the least priority.
            W.l.o.g.\ assume that this is the foremost prefix of $x$.
            Suppose this prefix is on some name $w$ connected to another name $z \in \pn(P)$ by restriction.
            By typability, the priority of $w$ is less than that of $x$ and all of the prefixes in between.
            This means that the number of prefixes blocking the prefix on $z$ strictly decreases.
            Hence, by the IH, $P \reddF P'' \lredd{\alpha'} Q'$ in a finite number of steps, where $w$ and $z$ are the subjects of $\alpha'$.
            The synchronization between $w$ and $z$ can be performed, and $n$ decreases.
            By type preservation (\Cref{t:APCP:sr}), $\emptyset \vdash Q' :: \emptyset$.
            The thesis then follows from the IH: $P \reddF P'' \lredd{\alpha'} Q' \reddF P' \lredd{\alpha} Q$ in finite steps, where $x$ and $y$ are the subjects of $\alpha$.
            \qedhere

    \end{itemize}
\end{proof}
\noindent 
Note that deadlock freedom (\Cref{t:APCP:df}) can be derived directly from reactivity: liveness implies the existence of pending names, so reduction is guaranteed by \Cref{t:APCP:react}.
In the following, deadlock freedom is strong enough, but in~\cite{journal/scico/vdHeuvelP22} where we apply \APCP for the analysis of distributed implementations of multiparty session types we do rely on the stronger reactivity.

\subsection{Typing Milner's  Cyclic Scheduler}
\label{s:APCP:APCP:milner}

\begin{figure}[t]
    \[
        \begin{bussproof}
            \bussAx[\ruleLabel{typ-var}]{
                X: (T_{\aChan{a_1}}, T_{\cChan{c_n}}, T_{\dChan{d_1}}) \vdash \begin{array}[t]{@{}l@{}}
                    \aChan{a_1}: \mu X . (\lift{t_1} T_{\aChan{a_1}}),
                    \cChan{c_n}: \mu X . (\lift{t_1} T_{\cChan{c_n}}),
                    \\
                    \dChan{d_1}: \mu X . (\lift{t_1} T_{\dChan{d_1}})
                \end{array}
                \quad
                \eqref{eq:APCP:pri:bra1}
            }
            \bussUn[\ruleLabel{typ-bra}]{
                X: T_X \vdash \begin{array}[t]{@{}l@{}}
                    \aChan{a_1}: \mu X . (\lift{t_1} T_{\aChan{a_1}}),
                    \cChan{c_n}: \&^{\rho_n} \{ \sff{next} : \mu X . (\lift{t_1} T_{\cChan{c_n}}) \},
                    \\
                    \dChan{d_1}: \mu X . (\lift{t_1} T_{\dChan{d_1}})
                \end{array}
                \quad
                \eqref{eq:APCP:pri:bra2}
            }
            \bussUn[\ruleLabel{typ-bra}]{
                X: T_X \vdash \begin{array}[t]{@{}l@{}}
                    \aChan{a_1}: \mu X . (\lift{t_1} T_{\aChan{a_1}}),
                    \cChan{c_n}: \&^{\pi_n} \{ \sff{start} : \&^{\rho_n} \{ \sff{next} : \mu X . (\lift{t_1} T_{\cChan{c_n}}) \} \},
                    \\
                    \dChan{d_1}: \mu X . (\lift{t_1} T_{\dChan{d_1}})
                \end{array}
                \quad
                \eqref{eq:APCP:pri:sel1}
            }
            \bussUn[\ruleLabel{typ-sel$\ast$}]{
                X: T_X \vdash
                \aChan{a_1}: \mu X . (\lift{t_1} T_{\aChan{a_1}}), \begin{array}[t]{@{}l@{}}
                    \cChan{c_n}: \&^{\pi_n} \{ \sff{start} : \&^{\rho_n} \{ \sff{next} : \mu X . (\lift{t_1} T_{\cChan{c_n}}) \} \},
                    \\
                    \dChan{d_1}: \oplus^{\rho_1} \{ \sff{next} : \mu X . (\lift{t_1} T_{\dChan{d_1}}) \}
                \end{array}
                \quad
                \eqref{eq:APCP:pri:bra3}
            }
            \bussUn[\ruleLabel{typ-bra}]{
                X: T_X \vdash \begin{array}[t]{@{}l@{}}
                    \aChan{a_1}: \&^{\kappa_1} \{ \sff{ack} : \mu X . (\lift{t_1} T_{\aChan{a_1}}) \},
                    \\
                    \cChan{c_n}: \&^{\pi_n} \{ \sff{start} : \&^{\rho_n} \{ \sff{next} : \mu X . (\lift{t_1} T_{\cChan{c_n}}) \} \},
                    \\
                    \dChan{d_1}: \oplus^{\rho_1} \{ \sff{next} : \mu X . (\lift{t_1} T_{\dChan{d_1}}) \}
                \end{array}
                \quad
                \eqref{eq:APCP:pri:sel2}
            }
            \bussUn[\ruleLabel{typ-sel$\ast$}]{
                X: T_X \vdash \begin{array}[t]{@{}l@{}}
                    \aChan{a_1}: \oplus^{\pri_1} \{ \sff{start} : \&^{\kappa_1} \{ \sff{ack} : \mu X . (\lift{t_1} T_{\aChan{a_1}}) \} \},
                    \\
                    \cChan{c_n}: \&^{\pi_n} \{ \sff{start} : \&^{\rho_n} \{ \sff{next} : \mu X . (\lift{t_1} T_{\cChan{c_n}}) \} \},
                    \\
                    \dChan{d_1}: \oplus^{\rho_1} \{ \sff{next} : \mu X . (\lift{t_1} T_{\dChan{d_1}}) \}
                \end{array}
                \quad
                \eqref{eq:APCP:pri:sel3}
            }
            \bussUn[\ruleLabel{typ-sel$\ast$}]{
                X: \underbrace{
                    (T_{\aChan{a_1}}, T_{\cChan{c_n}}, T_{\dChan{d_1}})
                }_{T_X} \vdash \begin{array}[t]{@{}l@{}}
                    \aChan{a_1}: \oplus^{\pri_1} \{ \sff{start} : \&^{\kappa_1} \{ \sff{ack} : \mu X . (\lift{t_1} T_{\aChan{a_1}}) \} \},
                    \\
                    \cChan{c_n}: \&^{\pi_n} \{ \sff{start} : \&^{\rho_n} \{ \sff{next} : \mu X . (\lift{t_1} T_{\cChan{c_n}}) \} \},
                    \\
                    \dChan{d_1}: \oplus^{\pi_1} \{ \sff{start} : \oplus^{\rho_1} \{ \sff{next} : \mu X . (\lift{t_1} T_{\dChan{d_1}}) \} \}
                \end{array}
                \quad
                \eqref{eq:APCP:pri:rec}
            }
            \bussUn[\ruleLabel{typ-rec}]{
                \emptyset \vdash \begin{array}[t]{@{}l@{}}
                    \aChan{a_1}: \mu X . \underbrace{
                        \oplus^{\pri_1} \{ \sff{start} : \&^{\kappa_1} \{ \sff{ack} : X \} \}
                    }_{T_{\aChan{a_1}}},
                    \cChan{c_n}: \mu X . \underbrace{
                        \&^{\pi_n} \{ \sff{start} : \&^{\rho_n} \{ \sff{next} : X \} \}
                    }_{T_{\cChan{c_n}}},
                    \\
                    \dChan{d_1}: \mu X . \underbrace{
                        \oplus^{\pi_1} \{ \sff{start} : \oplus^{\rho_1} \{ \sff{next} : X \} \}
                    }_{T_{\dChan{d_1}}}
                \end{array}
            }
        \end{bussproof}
    \]

    \caption{Typing derivation of the leader scheduler $A_1$ of Milner's cyclic scheduler (processes omitted).}
    \label{f:APCP:milnerLeader}
\end{figure}

Here we show that our specification of Milner's cyclic scheduler from \Cref{s:APCP:milner} is typable in \APCP, and thus deadlock free (cf.\ \Cref{t:APCP:df}).
Let us recall the process definitions of the leader and followers, omitting braces `$\{\ldots\}$' for branches with one option:
\cyclicAsyncMilnerScheds
\Cref{f:APCP:milnerLeader} gives the typing derivation of $A_1$, omitting processes from judgments, with the following priority requirements:
\begin{align}
    t_1 &> \max(\pri_1,\kappa_1,\pi_n,\rho_n,\pi_1,\rho_1) \label{eq:APCP:pri:rec}
    \\
    \pi_1 &< \rho_1 \label{eq:APCP:pri:sel3}
    \\
    \pri_1 &< \kappa_1 \label{eq:APCP:pri:sel2}
    \\
    \kappa_1 &< \pi_n,\rho_1 \label{eq:APCP:pri:bra3}
    \\
    \rho_1 &< \pi_1+t_1 \label{eq:APCP:pri:sel1}
    \\
    \pi_n &< \pri_1+t_1,\pi_1+t_1 \label{eq:APCP:pri:bra2}
    \\
    \rho_n &< \pri_1+t_1,\pi_1+t_1 \label{eq:APCP:pri:bra1}
\end{align}
Each process $A_{i+1}$ for $0 \leq i < n$---thus including the leader---is typable as follows, assuming \cChan{$c_i$} is \cChan{$c_n$} for $i=0$:
\begin{align*}
    \emptyset &\vdash
    A_{i+1}
    :: \aChan{a_{i+1}}: \mu X . \oplus^{\pri_{i+1}} \{ \sff{start} : \&^{\kappa_{i+1}} \{ \sff{ack} : X \} \},
    \begin{array}[t]{@{}l@{}}
        \cChan{c_i}: \mu X . \&^{\pi_i} \{ \sff{start} : \&^{\rho_i} \{ \sff{next} : X \} \},
        \\
        \dChan{d_{i+1}}: \mu X . \oplus^{\pi_{i+1}} \{ \sff{start} : \oplus^{\rho_{i+1}} \{ \sff{next} : \hspace{-2pt} X \} \} 
    \end{array}
\end{align*}
Note how, for each $1 \leq i \leq n$, the types for \cChan{$c_i$} and \dChan{$d_i$} are duals and are thus assigned equal priorities.

The priority requirements in the typing derivation of each $A_i$ are satisfiable.
The derivations of these processes have the following constraints:
\begin{itemize}

    \item
        For $A_1$ we require the inequalities listed above;

    \item
        For each $1 \leq i < n$, for $A_{i+1}$ we require $\pi_i < \pri_{i+1},\pi_{i+1}$, $\pri_{i+1} < \kappa_{i+1}$, $\pi_{i+1} < \rho_{i+1}$, $\kappa_{i+1} < \rho_i,\rho_{i+1}$, $\rho_i < \pri_{i+1}+t_{i+1},\rho_{i+1}$, and $\rho_{i+1} < \pi_{i+1}+t_{i+1}$.

\end{itemize}
We can satisfy these requirements by assigning $\pi_i \deq i$, $\pri_i \deq i+1$, $\kappa_i \deq i+2$, and $\rho_i \deq i+4$ for each $1 \leq i \leq n$, except with $\pi_n \deq n+3$ (to satisfy~\eqref{eq:APCP:pri:bra3} for $1 \leq n < 4$).
For the application of \ruleLabel{typ-rec}, each derivation also requires a common lifter $t_{i+1} > \max(\pri_{i+1}, \kappa_{i+1}, \pi_i, \rho_i, \pi_{i+1}, \rho_{i+1})$ for $0 \leq i < n$.
The priority requirements involving $t_{i+1}$ always require the priority lifted by $t_{i+1}$ to be higher than the priority not lifted by $t_{i+1}$, so the common lifter requirement easily satisfies these requirements.

Recall from \Cref{s:APCP:milner}:
\cyclicAsyncMilnerSched
Assuming given workers
\[
    \emptyset \vdash P_i :: \bChan{b_i}: \mu X . \&^{\pri_i} \{ \sff{start} : \oplus^{\kappa_i} \{ \sff{ack} : X \} \}
\]
for each $1 \leq i \leq n$, we have $\emptyset \vdash Sched_n :: \emptyset$.
Hence, it follows from \Cref{t:APCP:df} that $Sched_n$ is deadlock free for each $n \geq 1$.

\subsection{Extensions: Explicit Session Closing and Replicated Servers}
\label{s:APCP:APCP:exts}

\begin{figure}[t]
    \begin{mathpar}
        \begin{bussproof}[typ-close]
            \bussAx{
                \Omega \vdash \pClose x[] :: x: \1^\pri
            }
        \end{bussproof}
        \and
        \begin{bussproof}[typ-wait]
            \bussAssume{
                \Omega \vdash P :: \Gamma
            }
            \bussAssume{
                \pri < \pr(\Gamma)
            }
            \bussBin{
                \Omega \vdash \pWait x() ; P :: \Gamma, x: \bot^\pri
            }
        \end{bussproof}
        \and
        \begin{bussproof}[typ-cli]
            \bussAssume{
                \pri < \pr(A)
            }
            \bussUn{
                \Omega \vdash \pCli x[y] :: x: {?}^\pri A, y: \ol{A}
            }
        \end{bussproof}
        \and
        \begin{bussproof}[typ-srv]
            \bussAssume{
                \Omega \vdash P :: {?}\Gamma, y: A
            }
            \bussAssume{
                \pri < \pr({?}\Gamma)
            }
            \bussBin{
                \Omega \vdash \pSrv x(y) ; P :: {?}\Gamma, x: {!}^\pri A
            }
        \end{bussproof}
        \and
        \begin{bussproof}[typ-cli$\ast$]
            \bussAssume{
                \Omega \vdash P :: \Gamma, y: A
            }
            \bussAssume{
                \pri < \pr(A)
            }
            \bussBin{
                \Omega \vdash \pCli* x[y] \cdot P :: \Gamma, x: {?}^\pri A
            }
        \end{bussproof}
        \and
        \begin{bussproof}[typ-weaken]
            \bussAssume{
                \Omega \vdash P :: \Gamma
            }
            \bussUn{
                \Omega \vdash P :: \Gamma, x: {?}^\pri A
            }
        \end{bussproof}
        \and
        \begin{bussproof}[typ-contract]
            \bussAssume{
                \Omega \vdash P :: \Gamma, x: {?}^\pri A, x': {?}^\kappa A
            }
            \bussAssume{
                \pi = \min(\pri,\kappa)
            }
            \bussBin{
                \Omega \vdash P \{ x/x' \} :: \Gamma, x: {?}^\pi A
            }
        \end{bussproof}
        \\ \dashes \\
        \begin{bussproof}[red-close-wait]
            \bussAx{
                \pRes{xy} ( \pClose x[] \| \pWait y() ; P) \redd P
            }
        \end{bussproof}
        \and
        \begin{bussproof}[red-cli-srv]
            \bussAx{
                \pRes{xy} ( \pCli x[a] \| \pSrv y(v) ; P \| Q ) \redd P \{ a/v \} \| \pRes{xy} ( \pSrv y(v) ; P \| Q )
            }
        \end{bussproof}
    \end{mathpar}

    \caption{Typing rules (top) and reductions (bottom) for explicit closing and replicated servers.}
    \label{f:APCP:exrules}
\end{figure}

As already mentioned, our presentation of \APCP does not include explicit closing and replicated servers.
Here we briefly discuss what \APCP would look like if we were to include these constructs.

Explicit closing is useful in programming to be sure that all resources are cleaned up correctly.
There are several ways of integrating explicit closing in a calculus like \APCP.
Following, e.g., \cite{conf/concur/CairesP10,conf/icfp/Wadler12}, here we achieve explicit closing by adding closes (empty sends) $\pClose x[]$ and waits (empty receives) $\pWait x() ; P$ to the syntax in \Cref{f:procdef} (top).
We also add the Rule~\ruleLabel{red-close-wait} to \Cref{f:procdef} (bottom).
At the level of types, we replace the conflated type $\bullet$ with $\1^\pri$ and $\bot^\pri$, associated to closes and waits, respectively.
Note that we do need priority annotations on types for closed names now, because wait is blocking and thus requires priority checks.
In the type system in \Cref{f:APCP:typingRules} (top), we replace Rule~\ruleLabel{typ-end} with Rules~\ruleLabel{type-close} and~\ruleLabel{typ-wait} in \Cref{f:APCP:exrules} (top).

For replicated servers, we add (asynchronous) client requests $\pCli x[y]$ and servers $\pSrv x(y) ; P$, with types ${?}^\pri A$ and ${!}^\pri A$, respectively.
We include syntactic sugar for binding client requests to continuations as in \Cref{n:APCP:sugar}: $\pCli* x[y] \cdot P \deq \pRes{ya} ( \pCli x[a] \| P )$.
Rule~\ruleLabel{red-cli-srv} is in \Cref{f:APCP:exrules} (bottom): it connects a client and a server and forks a copy of the server.
Also, we add a structural congruence rule to clean up unused servers: $\pRes{xz} ( \pSrv x(y) ; P ) \equiv \0$.
In the type system, we add Rules~\ruleLabel{typ-cli}, \ruleLabel{typ-srv}, \ruleLabel{typ-weaken}, and~\ruleLabel{typ-contract} in \Cref{f:APCP:exrules} (top); the former two are for typing client requests and servers, respectively, and the latter two are for connecting to a server without requests and for multiple requests, respectively.
In Rule~\ruleLabel{typ-srv}, notation ${?}\Gamma$ means that every type in $\Gamma$ is of the form ${?}^\pri A$.
\Cref{f:APCP:exrules} (top) also includes a derivable Rule~\ruleLabel{typ-cli$\ast$} which types the syntactic sugar for bound client requests.

\section{An Intermezzo: From \APCP to \texorpdfstring{\protect\LAST}{LAST}}
\label{s:LAST}

As attested by its expressivity and
meta-theoretical results (\Cref{t:APCP:sr,t:APCP:df}), \APCP provides a convenient framework for analyzing asynchronous message passing between cyclically connected processes.
In particular, \APCP provides a firm basis for designing languages with session-typed concurrency, asynchronous communication, and cyclic structures.
Ideally, we would like to faithfully compile any such language into \APCP, in order to transfer its correctness guarantees.

As discussed in the Introduction, we look for answers in the realm of functional programming, in the form of variants of the \lamcalc with session-typed, message-passing concurrency.
In this context, Gay and Vasconcelos's \LAST~\cite{journal/jfp/GayV10} appears nicely positioned: \LAST is a call-by-value language in which programs consist of threads that are cyclically connected on channels that provide asynchronous message passing (through buffers).
The calculus \LAST has forked several variants that connect message-passing processes with message-passing functions, as we set out to do here.
Most notably, Walder~\cite{conf/icfp/Wadler12,journal/jfp/Wadler14} introduced \GV, a variant of \LAST with synchronous communication and non-cyclic thread connections, and gave a translation into his \CP (the synchronous ancestor of \APCP, without cyclic connections).
Subsequently, in the same spirit, Kokke and Dardha~\cite{conf/forte/KokkeD21} presented \PGV, a variant of \GV with cyclic  connections, and a translation from Dardha and Gay's \PCP (the synchronous ancestor of \APCP~\cite{conf/fossacs/DardhaG18}).

Hence, \LAST is a natural choice for a core programming language that can be studied via a translation into \APCP.
Notice that both \GV and \PGV enjoy deadlock freedom by typing, whereas \LAST does not.
This strengthens our motivation for designing a translation into \APCP, as transferring the deadlock-freedom property from \APCP to \LAST would address a significant gap.
To our knowledge, such a \emph{translational} approach to ensuring deadlock freedom for \LAST programs has not been achieved before.

In this section we gently recall \LAST and gradually introduce the key ingredients for its translation into \APCP.
For presentational purposes, we find it useful to present a variant of  \LAST that is more convenient towards a translation, denoted \LAST* (\Cref{s:LAST:syntaxSemantics}).
The type system for \LAST*, described in \Cref{s:LAST:typeSystem}, closely follows the one for \LAST in~\cite{journal/jfp/GayV10}.
Then, in \Cref{s:LAST:trans}, we discuss the potential design of a translation from \LAST* into \APCP.
We purposefully use the word ``potential'': we seek a translation that is \emph{faithful}, i.e., that preserves and reflects behaviors in a precise sense.
Given this focus, we shall argue that the call-by-value semantics of \LAST* is not well suited for inducing a faithful translation.
As such, this section serves as motivation for introducing \FIRST, a  variant of \LAST based on a call-by-name semantics with constructs for session closing and explicit substitutions, which enjoys a faithful translation into \APCP and admits the translational approach to the deadlock-freedom property (\Cref{s:FIRST}).

\subsection{The Syntax and Semantics of \texorpdfstring{\protect\LAST*}{LAST*}}
\label{s:LAST:syntaxSemantics}

In \LAST*, programs consist of two layers: while \emph{terms} ($\term{M,N,\ldots}$) define the behavior of threads, \emph{configurations} ($\term{C,D,\ldots}$) are obtained by composing threads in parallel, so as to enable their interaction by  exchanging messages on buffered channels.
Before detailing these syntactic elements (and their semantics), we point out the differences between \LAST and \LAST*.
While in \LAST endpoints are created by synchronization on shared names, \LAST* features a dedicated construct for endpoint creation (denoted~$\term{\tNew}$).
Also, thread creation in \LAST*  involves an explicit continuation, not present in \LAST.
Moreover,  in \LAST buffers run next to threads, whereas they are integrated in endpoint restrictions in \LAST*.
Finally, for simplicity, \LAST* accounts for linear sessions only.

\paragraph{Terms}
\begin{figure}[t!]
    Terms ($\term{M},\term{N},\ldots$), values ($\term{v}$), and reduction contexts ($\tCtx{R}$):
    \begin{align*}
        \term{M},\term{N}
        & \begin{array}[t]{@{}l@{\kern.5ex}lr@{\kern2ex}l@{\kern.5ex}lr@{}}
            {} ::= &
            \term{x} & \text{variable}
            & \sepr &
            \term{\tNew} & \text{create new channel}
            \\ \sepr* &
            \term{()} & \text{unit value}
            & \sepr &
            \term{\tFork M ; N} & \text{fork $\term{M}$ in parallel to $\term{N}$}
            \\ \sepr* &
            \term{\lam x . M} & \text{abstraction}
            & \sepr &
            \term{(M,N)} & \text{pair construction}
            \\ \sepr* &
            \term{M~N} & \text{application}
            & \sepr &
            \term{\tLet (x,y)=M \tIn N} & \text{pair deconstruction}
            \\ \sepr* &
            \term{\tSend M \, N} & \text{send $\term{M}$ along $\term{N}$}
            & \sepr &
            \term{\tSel \ell \, M} & \text{select label $\ell$ along $\term{M}$}
            \\ \sepr* &
            \term{\tRecv M} & \text{receive along $\term{M}$}
            & \sepr &
            \term{\tCase M \tOf \{ i : N \}_{i \in I}} & \text{offer labels in $I$ along $\term{M}$}
        \end{array}
        \\
        \term{v} &::=
        \term{x}
        \sepr
        \term{\lam x . M}
        \sepr
        \term{(v,v)}
        \sepr
        \term{()}
        \\
        \term{\tCtx{R}}
        & \begin{array}[t]{@{}l@{\kern.5ex}l@{}}
            {} ::= &
            \term{\tHole}
            \sepr
            \term{\tCtx{R}~M}
            \sepr
            \term{v~\tCtx{R}}
            \sepr
            \term{(\tCtx{R},M)}
            \sepr
            \term{(v,\tCtx{R})}
            \sepr
            \term{\tLet (x,y) = \tCtx{R} \tIn M}
            \\ \sepr* &
            \term{\tSend \tCtx{R} \, M}
            \sepr
            \term{\tSend v \, \tCtx{R}}
            \sepr
            \term{\tRecv \tCtx{R}}
            \sepr
            \term{\tSel \ell\, \tCtx{R}}
            \sepr
            \term{\tCase \tCtx{R} \tOf \{ i : M \}_{i \in I}}
        \end{array}
    \end{align*}

    \dashes

    Term reduction ($\reddM$):
    \begin{mathpar}
        \begin{bussproof}[red-lam]
            \bussAx{
                \term{(\lam x . M)~v}
                \reddM
                \term{M \{ v/x \}}
            }
        \end{bussproof}
        \and
        \begin{bussproof}[red-pair]
            \bussAx{
                \term{\tLet (x,y) = (v_1,v_2) \tIn M}
                \reddM
                \term{M \{ v_1/x,v_2/y \}}
            }
        \end{bussproof}
        \and
        \begin{bussproof}[red-lift]
            \bussAssume{
                \term{M}
                \reddM
                \term{N}
            }
            \bussUn{
                \tCtx{R}[M]
                \reddM
                \tCtx{R}[N]
            }
        \end{bussproof}
    \end{mathpar}

    \caption{The \protect\LAST* term language.}\label{f:LAST:terms}
\end{figure}

\Cref{f:LAST:terms} (top) gives the term syntax for \LAST*.
The functional behavior of terms is defined by standard \lamcalc constructs for variables $\term{x}$, the unit value $\term{()}$, abstraction $\term{\lam x . M}$ and application $\term{M\ N}$, and pair construction $\term{(M,N)}$ and deconstruction $\term{\tLet (x,y) = M \tIn N}$.
As usual, to improve readability, we often write $\term{\tLet x = M \tIn N}$ to denote $\term{(\lam x . N)~M}$.

The remaining constructs define the thread and message-passing behavior of terms; their exact semantics will be defined for configurations, so we briefly describe them here.
The construct $\term{\tNew}$ creates a new buffered channel with two endpoints (referred to with variables).
The construct $\term{\tFork M ; N}$ forks a new thread running term $\term{M}$ and continues as $\term{N}$.
By involving an explicit continuation $\term{N}$, this is slightly different than the corresponding construct in \LAST.
We can think of $\term{N}$ as the continuation to be run immediately after $\term{M}$ is spawned.
This is a mild generalization, as the construct in \LAST corresponds to the case in which $\term{N} = \term{()}$; it will be convenient for the correct translation into \APCP.

The constructs $\term{\tSend N M}$ and $\term{\tRecv M}$ denote sending and receiving messages along $\term{M}$ once it has reduced to a variable referring to a buffer endpoint, respectively; that is, sending entails placing the message $\term{N}$ at the end of the buffer, and receiving entails taking a message from the start of the buffer (if there).
The constructs $\term{\tSel \ell \, M}$ and $\term{\tCase M \tOf \{ i : N_i \}_{i \in I}}$ denote selecting and offering labels along $\term{M}$ once it has reduced to an endpoint variable, respectively; that is, selection entails placing the label $\term{\ell}$ at the end of the buffer, and offering entails taking a label $\term{j} \in I$ from the start of the buffer (if there) and continuing in the corresponding branch $\term{N_j}$.

Following~\cite{journal/jfp/GayV10}, the term reduction semantics of \LAST* employs a call-by-value (CbV) approach; \Cref{f:LAST:terms} (center) defines values $\term{v}$, i.e., terms that cannot further reduce on their own: variables, abstractions, pairs, and the unit value.
\Cref{f:LAST:terms} (center) also defines reduction contexts $\tCtx{R}$ that define under which positions subterms may reduce.
Finally, \Cref{f:LAST:terms} (bottom) defines term reduction ($\reddM$).
Rule~\ruleLabel{red-lam} reduces an abstraction applied to a value; the substitution of a value $\term{v}$ for a (free) variable $\term{x}$ is denoted $\term{\{ v/x \}}$, as usual.
Rule~\ruleLabel{red-pair} reduces the pair deconstruction of a pair of values to two substitutions.
Rule~\ruleLabel{red-lift} closes term reduction under reduction contexts.

\begin{exa}
    \label{x:CbV}
    To illustrate the CbV semantics of \LAST*, consider the following term and its behavior:
    \[
        \term{\big(\lam x . x~(\lam y . y)\big)~\big((\lam w . w)~(\lam z . z)\big)}
        \reddM
        \term{\big(\lam x . x~(\lam y . y)\big)~(\lam z . z)}
        \reddM
        \term{(\lam z . z)~(\lam y . y)}
        \reddM
        \term{\lam y . y}
        \tag*{\qedhere}
    \]
\end{exa}

\noindent
We will illustrate the message-passing behavior of \LAST* after introducing configurations below.

\paragraph{Configurations}

\begin{figure}[t!]
    \def\MathparLineskip{\lineskip=3pt}
    Markers ($\term{\phi}$), messages ($\term{m},\term{n}$), configurations ($\term{C},\term{D},\term{E}$), thread contexts ($\tCtx{F}$), configuration  contexts ($\tCtx{G}$):
    \begin{align*}
        \term{\phi} ::= {}
        & \term{\tMain} \sepr \term{\tChild}
        &
        \term{m},\term{n} ::= {}
        & \term{v} \sepr \term{\ell}
        \\
        \term{C},\term{D},\term{E} ::= {}
        & \term{\phi\, M}  \sepr \term{\pRes{x\tBfr{\vec{m}}y} C} \sepr \term{C \prl D}
        \\
        \tCtx{F} ::= {}
        & \term{\phi\, \tCtx{R}}
        &
        \term{\tCtx{G}} ::= {}
        & \term{\tHole} \sepr \term{\tCtx{G} \prl C} \sepr \term{\pRes{x\tBfr{\vec{m}}y} \tCtx{G}}
    \end{align*}

    \phantom{.}\dashes 

    Structural congruence for configurations ($\equivC$):\\

    \begin{mathpar}
        \begin{bussproof}[sc-res-swap]
            \bussAx{
                \term{\pRes{x\tBfr{\epsi}y} C}
                \equivC
                \term{\pRes{y\tBfr{\epsi}x} C}
            }
        \end{bussproof}
        \and
        \begin{bussproof}[sc-res-comm]
            \bussAx{
                \term{\pRes{x\tBfr{\vec{m}}y} \pRes{z\tBfr{\vec{n}}w} C}
                \equivC
                \term{\pRes{z\tBfr{\vec{n}}w} \pRes{x\tBfr{\vec{m}}y} C}
            }
        \end{bussproof}
        \and
        \begin{bussproof}[sc-res-ext]
            \bussAssume{
                \term{x},\term{y} \notin \fv(\term{C})
            }
            \bussUn{
                \term{\pRes{x\tBfr{\vec{m}}y} ( C \prl D )}
                \equivC
                \term{C \prl \pRes{x\tBfr{\vec{m}}y} D}
            }
        \end{bussproof}
        \and
        \begin{bussproof}[sc-par-comm]
            \bussAx{
                \term{C \prl D}
                \equivC
                \term{D \prl C}
            }
        \end{bussproof}
        \and
        \begin{bussproof}[sc-par-assoc]
            \bussAx{
                \term{C \prl (D \prl E)}
                \equivC
                \term{(C \prl D) \prl E}
            }
        \end{bussproof}
    \end{mathpar}

    \phantom{.}\dashes 

    Configuration reduction ($\reddC$):

    \begin{mathpar}
        \begin{bussproof}[red-new]
            \bussAx{
                \tCtx{F}[\tNew]
                \reddC
                \term{\pRes{x\tBfr{\epsi}y} ( \tCtx{F}[(x,y)] )}
            }
        \end{bussproof}
        \and
        \begin{bussproof}[red-fork]
            \bussAx{
                \tCtx{F}[\tFork M ; N]
                \reddC
                \term{\tCtx{F}[N] \prl \tChild\,M}
            }
        \end{bussproof}
        \and
        \begin{bussproof}[red-send]
            \bussAx{
                \term{\pRes{x\tBfr{\vec{m}}y} ( \tCtx{F}[\tSend v \, x] \prl C )}
                \reddC
                \term{\pRes{x\tBfr{v,\vec{m}}y} ( \tCtx{F}[x] \prl C )}
            }
        \end{bussproof}
        \and
        \begin{bussproof}[red-recv]
            \bussAx{
                \term{\pRes{x\tBfr{\vec{m},v}y} ( \tCtx{F}[\tRecv y] \prl C )}
                \reddC
                \term{\pRes{x\tBfr{\vec{m}}y} ( \tCtx{F}[(v,y)] \prl C )}
            }
        \end{bussproof}
        \and
        \begin{bussproof}[red-select]
            \bussAx{
                \term{\pRes{x\tBfr{\vec{m}}y} ( \tCtx{F}[\tSel \ell\, x] \prl C )}
                \reddC
                \term{\pRes{x\tBfr{\ell,\vec{m}}y} ( \tCtx{F}[x] \prl C )}
            }
        \end{bussproof}
        \and
        \begin{bussproof}[red-case]
            \bussAssume{
                j \in I
            }
            \bussUn{
                \term{\pRes{x\tBfr{\vec{m},j}y} ( \tCtx{F}[\tCase y \tOf \{ i : M_i \}_{i \in I}] \prl C )}
                \reddC
                \term{\pRes{x\tBfr{\vec{m}}y} ( \tCtx{F}[M_j~y] \prl C )}
            }
        \end{bussproof}
        \and
        \begin{bussproof}[red-res-nil]
            \bussAssume{
                \term{x},\term{y} \notin \fv(\term{C})
            }
            \bussUn{
                \term{\pRes{x\tBfr{\epsi}y} C} \reddC \term{C}
            }
        \end{bussproof}
        \and
        \begin{bussproof}[red-par-nil]
            \bussAx{
                \term{C \prl \tChild \, ()} \reddC \term{C}
            }
        \end{bussproof}
        \and
        \begin{bussproof}[red-lift-C]
            \bussAssume{
                \term{C}
                \reddC
                \term{C'}
            }
            \bussUn{
                \tCtx{G}[C]
                \reddC
                \tCtx{G}[C']
            }
        \end{bussproof}
        \and
        \begin{bussproof}[red-lift-M]
            \bussAssume{
                \term{M}
                \reddM
                \term{M'}
            }
            \bussUn{
                \tCtx{F}[M]
                \reddC
                \tCtx{F}[M']
            }
        \end{bussproof}
        \and
        \begin{bussproof}[red-conf-lift-sc]
            \bussAssume{
                \term{C}
                \equivC
                \term{C'}
            }
            \bussAssume{
                \term{C'}
                \reddC
                \term{D'}
            }
            \bussAssume{
                \term{D'}
                \equivC
                \term{D}
            }
            \bussTern{
                \term{C}
                \reddC
                \term{D}
            }
        \end{bussproof}
    \end{mathpar}
    \caption{The \protect\LAST* configuration language: syntax and semantics.}\label{f:LAST:confs}
\end{figure}

Functional calculi such as \LAST* feature a clear distinction between the static and dynamic parts of their languages.
That is, a \LAST* program starts as a closed functions (the static part) and evolves into several threads operating in parallel and communicating through message passing (the dynamic part).
We refer to the static and dynamic parts of \LAST* as terms and configurations, respectively.
In general, one writes a \LAST* program as a single main thread containing a term that forks and connects child threads.
To contrast, process calculi such as \APCP blur the lines between such static and dynamic parts, as \APCP ``programs'' are immediately written as configurations of parallel subprocesses connected on channels.

\Cref{f:LAST:confs} gives the configuration syntax for \LAST*.
Given a term $\term{M}$, the configuration $\term{\phi \, M}$ denotes a corresponding thread, where the marker $\term{\phi}$ is useful to distinguish the main thread $\term{\tMain}$ from child threads $\term{\tChild}$---this distinction will be useful for typing.
Buffered channels are denoted $\term{\pRes{x\tBfr{\vec{m}}y} C}$.
Here, $\term{C}$ has access to the endpoints $\term{x}$ and $\term{y}$.
The buffer itself $\term{x\tBfr{\vec{m}}y}$ is an ordered sequence of messages (values and labels, denoted $\term{\vec{m}}$) sent on $\term{x}$ and to be received on $\term{y}$.
This means that $\term{C}$ may send/select on $\term{x}$ and receive/offer on $\term{y}$.
Once the buffer is empty (i.e., $\term{\vec{m}} = \term{\epsi}$), $\term{x}$ and $\term{y}$ may switch roles.
Configuration $\term{C \prl D}$ denotes the parallel composition of $\term{C}$ and $\term{D}$.
\Cref{f:LAST:confs} (top) also defines thread contexts $\tCtx{F}$ as term reduction contexts inside threads, and configuration contexts $\tCtx{G}$.

The reduction semantics for configurations is defined on specific arrangements of buffers and threads.
To ensure such arrangements, we define a structural congruence for configurations ($\equivC$), the least congruence on configurations satisfying the rules in \Cref{f:LAST:confs} (center).
Rule~\ruleLabel{sc-res-swap} swaps the direction of buffered channels and thus the input/output roles of the channel's endpoints.
Rule~\ruleLabel{sc-res-comm} defines commutativity of buffered channels.
Rule~\ruleLabel{sc-res-ext} defines scope extrusion/inclusion for buffered channels.
Rules~\ruleLabel{sc-par-comm} and~\ruleLabel{sc-par-assoc} define commutativity and associativity for parallel composition.

\Cref{f:LAST:confs} (bottom) gives the reduction semantics for configurations ($\reddC$).
It defines how terms in threads interact with each other by exchanging messages through buffered channels.
Rule~\ruleLabel{red-new} reduces a $\term{\tNew}$ construct in a thread by creating a new buffered channel and returning the endpoints $\term{x}$ and $\term{y}$.
Rule~\ruleLabel{red-fork} reduces a $\term{\tFork}$ construct in a thread by creating a new child thread.
Rule~\ruleLabel{red-send} reduces a $\term{\tSend}$ by placing the value at the end of the enclosing buffer and returning the endpoint.
Dually, Rule~\ruleLabel{red-recv} reduces a $\term{\tRecv}$ by retrieving the value at the start of the enclosing buffer and returning it along with the endpoint.
Rule~\ruleLabel{red-select} reduces a $\term{\tSel}$ by placing the label at the end of the enclosing buffer and returning the endpoint.
Dually, Rule~\ruleLabel{red-case} reduces a $\term{\tCase}$ by retrieving the label at the start of the enclosing buffer and applying the label's corresponding continuation to the endpoint.
Rule~\ruleLabel{red-res-nil} garbage collects buffers that are no longer used, and Rule~\ruleLabel{red-par-nil} garbage collects child threads that have reduced to unit.
Rules~\ruleLabel{red-lift-C} and~\ruleLabel{red-lift-M} close configuration reduction under configuration contexts and enable terms in threads to reduce, respectively.
Rule~\ruleLabel{red-conf-lift-SC} closes configuration reduction under structural congruence.

\begin{exa}[The Bookshop Scenario, Revisited]
    \label{x:LAST:Shop}
    We illustrate the message-passing behavior of \LAST* by considering the bookshop example from \Cref{s:FIRST:example}.
    Note that \LAST* does not have $\term{\tClose}$ constructs: we will motivate them in \Cref{s:FIRST}; here, for simplicity, we consider the system in \Cref{s:FIRST:example} without these constructs.

    First, let us explore how $\term{\tMain \, \sff{Sys}}$ sets up some channels and threads:
    \begin{align*}
        &\hphantom{{}\reddC{}}
        \term{\tMain \, \tLet (s,s') = \tNew \tIn \tFork \sff{Shop}(s) ; \tLet (m,m') = \tNew \tIn \tFork \sff{Mother}(m) ; \sff{Son}(s',m')}
        \\
        &\reddC
        \term{\pRes{y\tBfr{\epsi}y'} \big( \tMain \, \begin{array}[t]{@{}l@{}}
            \tLet (s,s') = (y,y') \tIn \tFork \sff{Shop}(s) ; \\
            \tLet (m,m') = \tNew \tIn \tFork \sff{Mother}(m) ; \sff{Son}(s',m') \big)
        \end{array}}
        \\
        &\reddC
        \term{\pRes{y\tBfr{\epsi}y'} \big( \tMain \, \tFork \sff{Shop}(y) ; \tLet (m,m') = \tNew \tIn \tFork \sff{Mother}(m) ; \sff{Son}(y',m') \big)}
        \\
        &\reddC
        \term{\pRes{y\tBfr{\epsi}y'} \big( \tMain \, \tLet (m,m') = \tNew \tIn \tFork \sff{Mother}(m) ; \sff{Son}(y',m') \prl \tChild \, \sff{Shop}(y) \big)}
        \\
        &\reddC
        \term{\pRes{y\tBfr{\epsi}y'} \big( \pRes{z\tBfr{\epsi}z'} \big( \tMain \, \tLet (m,m') = (z,z') \tIn \tFork \sff{Mother}(m) ; \sff{Son}(y',m') \big) \prl \tChild \, \sff{Shop}(y) \big)}
        \\
        &\reddC
        \term{\pRes{y\tBfr{\epsi}y'} \big( \pRes{z\tBfr{\epsi}z'} \big( \tMain \, \tFork \sff{Mother}(z) ; \sff{Son}(y',z') \big) \prl \tChild \, \sff{Shop}(y) \big)}
        \\
        &\reddC
        \term{\pRes{y\tBfr{\epsi}y'} \big( \pRes{z\tBfr{\epsi}z'} \big( \tMain \, \sff{Son}(y',z') \prl \tChild \, \sff{Mother}(z) \big) \prl \tChild \, \sff{Shop}(y) \big)}
    \end{align*}

    \noindent
    Next, let us see how the son sends the book title and his choice to buy to the shop, and then his connection with the shop to his mother, without waiting for any of his messages to be received:
    \begin{align*}
        &\equivC
        \term{\pRes{y'\tBfr{\epsi}y} \big( \pRes{z\tBfr{\epsi}z'} \big( \tMain \, \tLet s'_1 = \tSend \text{``Dune''} \, y' \tIn \ldots \prl \tChild \, \sff{Mother}(z) \big) \prl \tChild \, \sff{Shop}(y) \big)}
        \\ \displaybreak[1]
        &\reddC
        \term{\pRes{y'\tBfr{\text{``Dune''}}y} \big( \pRes{z\tBfr{\epsi}z'} \big( \tMain \, \tLet s'_1 = y' \tIn \ldots \prl \tChild \, \sff{Mother}(z) \big) \prl \tChild \, \sff{Shop}(y) \big)}
        \\ \displaybreak[1]
        &\reddC
        \term{\pRes{y'\tBfr{\text{``Dune''}}y} \big( \pRes{z\tBfr{\epsi}z'} \big( \tMain \, \tLet s'_2 = \tSel \sff{buy} \, y' \tIn \ldots \prl \tChild \, \sff{Mother}(z) \big) \prl \tChild \, \sff{Shop}(y) \big)}
        \\ \displaybreak[1]
        &\reddC
        \term{\pRes{y'\tBfr{\sff{buy},\text{``Dune''}}y} \big( \pRes{z\tBfr{\epsi}z'} \big( \tMain \, \tLet s'_2 = y' \tIn \ldots \prl \tChild \, \sff{Mother}(z) \big) \prl \tChild \, \sff{Shop}(y) \big)}
        \\ \displaybreak[1]
        &\reddC
        \term{\pRes{y'\tBfr{\sff{buy},\text{``Dune''}}y} \big( \pRes{z\tBfr{\epsi}z'} \big( \tMain \, \tLet m'_1 = \tSend y' \, z' \tIn \ldots \prl \tChild \, \sff{Mother}(z) \big) \prl \tChild \, \sff{Shop}(y) \big)}
        \\ \displaybreak[1]
        &\reddC
        \term{\pRes{y'\tBfr{\sff{buy},\text{``Dune''}}y} \big( \pRes{z'\tBfr{y'}z} \big( \tMain \, \tLet m'_1 = z' \tIn \ldots \prl \tChild \, \sff{Mother}(z) \big) \prl \tChild \, \sff{Shop}(y) \big)}
        \\ \displaybreak[1]
        &\reddC
        \term{\pRes{y'\tBfr{\sff{buy},\text{``Dune''}}y} \big( \pRes{z'\tBfr{y'}z} \big( \tMain \, \tLet (\textit{book},m'_2) = \ldots \prl \tChild \, \sff{Mother}(z) \big) \prl \tChild \, \sff{Shop}(y) \big)}
    \end{align*}

    \noindent
    Now, we can see how the mother receives the shop's connection and sends her credit card information:
    \begin{align*}
        &=
        \term{
            \pRes{y'\tBfr{\sff{buy},\text{``Dune''}}y} \big(
                \pRes{z'\tBfr{y'}z} \big(
                    \begin{array}[t]{@{}l@{}}
                        \tMain \, \tLet (\textit{book},m'_2) = \tRecv z' \tIn \textit{book}
                        \\
                        \prl \tChild \, \tLet (x,m_1) = \tRecv z \tIn \ldots
                    \big)
                    \prl \tChild \, \sff{Shop}(y)
                \big)
            \end{array}
        }
        \\
        &\reddC
        \term{
            \pRes{y'\tBfr{\sff{buy},\text{``Dune''}}y} \big(
                \pRes{z\tBfr{\epsi}z'} \big(
                    \begin{array}[t]{@{}l@{}}
                        \tMain \, \tLet (\textit{book},m'_2) = \tRecv z' \tIn \textit{book}
                        \\
                        \prl \tChild \, \tLet (x,m_1) = (y',z) \tIn \ldots
                    \big)
                    \prl \tChild \, \sff{Shop}(y)
                \big)
            \end{array}
        }
        \\
        &\reddC
        \term{
            \pRes{y'\tBfr{\sff{buy},\text{``Dune''}}y} \big(
                \pRes{z\tBfr{\epsi}z'} \big(
                    \begin{array}[t]{@{}l@{}}
                        \tMain \, \tLet (\textit{book},m'_2) = \tRecv z' \tIn \textit{book}
                        \\
                        \prl \tChild \, \tLet x_1 = \tSend \text{visa} \, y' \tIn \ldots
                    \big)
                    \prl \tChild \, \sff{Shop}(y)
                \big)
            \end{array}
        }
        \\
        &\reddC
        \term{
            \pRes{y'\tBfr{\text{visa},\sff{buy},\text{``Dune''}}y} \big(
                \pRes{z\tBfr{\epsi}z'} \big(
                    \begin{array}[t]{@{}l@{}}
                        \tMain \, \tLet (\textit{book},m'_2) = \tRecv z' \tIn \textit{book}
                        \\
                        \prl \tChild \, \tLet x_1 = y' \tIn \ldots
                    \big)
                    \prl \tChild \, \sff{Shop}(y)
                \big)
            \end{array}
        }
        \\
        &\reddC
        \term{
            \pRes{y'\tBfr{\text{visa},\sff{buy},\text{``Dune''}}y} \big(
                \pRes{z\tBfr{\epsi}z'} \big(
                    \begin{array}[t]{@{}l@{}}
                        \tMain \, \tLet (\textit{book},m'_2) = \tRecv z' \tIn \textit{book}
                        \\
                        \prl \tChild \, \tLet (\textit{book},x_2) = \tRecv y' \tIn \ldots
                    \big)
                    \prl \tChild \, \sff{Shop}(y)
                \big)
            \end{array}
        }
    \end{align*}

    \noindent
    Finally, the sequence of reductions in \Cref{f:x:LAST:Shop} shows how the shop reads  messages and sends the book, and how the mother forwards it to her son.
    \end{exa}
    \begin{figure}[!t]
        \begin{align*}
        &\hphantom{{}={}}
        \term{
            \pRes{y'\tBfr{\text{visa},\sff{buy},\text{``Dune''}}y} \big(
                \begin{array}[t]{@{}l@{}}
                    \pRes{z\tBfr{\epsi}z'} \big(
                        \begin{array}[t]{@{}l@{}}
                            \tMain \, \tLet (\textit{book},m'_2) = \tRecv z' \tIn \textit{book}
                            \\
                            \prl \tChild \, \tLet (\textit{book},x_2) = \tRecv y' \tIn \ldots
                        \big)
                    \end{array}
                    \\
                    \prl \tChild \, \tLet (\textit{title},s_1) = \tRecv y \tIn \ldots
                \big)
            \end{array}
        }
        \\
        &\reddC
        \term{
            \pRes{y'\tBfr{\text{visa},\sff{buy}}y} \big(
                \pRes{z\tBfr{\epsi}z'} \big(
                    \tMain \, \ldots
                    \prl \tChild \, \ldots
                \big)
                \prl \tChild \, \tLet (\textit{title},s_1) = (\text{``Dune''},y) \tIn \ldots
            \big)
        }
        \\
        &\reddC
        \term{
            \pRes{y'\tBfr{\text{visa},\sff{buy}}y} \big(
                \pRes{z\tBfr{\epsi}z'} \big(
                    \tMain \, \ldots
                    \prl \tChild \, \ldots
                \big)
                \prl \tChild \, \tCase y \tOf \{\ldots\}
            \big)
        }
        \\
        &\reddC
        \term{
            \pRes{y'\tBfr{\text{visa}}y} \big(
                \pRes{z\tBfr{\epsi}z'} \big(
                    \tMain \, \ldots
                    \prl \tChild \, \ldots
                \big)
                \prl \tChild \, ( \lam s_2 \ldots )~y
            \big)
        }
        \\
        &\reddC
        \term{
            \pRes{y'\tBfr{\text{visa}}y} \big(
                \pRes{z\tBfr{\epsi}z'} \big(
                    \tMain \, \ldots
                    \prl \tChild \, \ldots
                \big)
                \prl \tChild \, \tLet (\textit{card},s_3) = \tRecv y \tIn \ldots
            \big)
        }
        \\
        &\reddC
        \term{
            \pRes{y\tBfr{\epsi}y'} \big(
                \pRes{z\tBfr{\epsi}z'} \big(
                    \tMain \, \ldots
                    \prl \tChild \, \ldots
                \big)
                \prl \tChild \, \tLet (\textit{card},s_3) = (\text{visa},y) \tIn \ldots
            \big)
        }
        \\
        &\reddC
        \term{
            \pRes{y\tBfr{\epsi}y'} \big(
                \pRes{z\tBfr{\epsi}z'} \big(
                    \tMain \, \ldots
                    \prl \tChild \, \ldots
                \big)
                \prl \tChild \, \tLet s_4 = \tSend \text{book}(\text{``Dune''}) \, y \tIn ()
            \big)
        }
        \\
        &\reddC
        \term{
            \pRes{y\tBfr{\text{book}(\text{``Dune''})}y'} \big(
                \pRes{z\tBfr{\epsi}z'} \big(
                    \tMain \, \ldots
                    \prl \tChild \, \ldots
                \big)
                \prl \tChild \, \tLet s_4 = y \tIn ()
            \big)
        }
        \\
        &\reddC
        \term{
            \pRes{y\tBfr{\text{book}(\text{``Dune''})}y'} \big(
                \pRes{z\tBfr{\epsi}z'} \big(
                    \tMain \, \ldots
                    \prl \tChild \, \ldots
                \big)
                \prl \tChild \, ()
            \big)
        }
        \\
        &\reddC
        \term{
            \pRes{y\tBfr{\text{book}(\text{``Dune''})}y'} \big(
                \pRes{z\tBfr{\epsi}z'} \big(
                    \tMain \, \ldots
                    \prl \tChild \, \tLet (\textit{book},x_2) = \tRecv y' \tIn \ldots
                \big)
            \big)
        }
        \\
        &\reddC
        \term{
            \pRes{y\tBfr{\epsi}y'} \big(
                \pRes{z\tBfr{\epsi}z'} \big(
                    \tMain \, \ldots
                    \prl \tChild \, \tLet (\textit{book},x_2) = (\text{book}(\text{``Dune''}),y') \tIn \ldots
                \big)
            \big)
        }
        \\ \displaybreak[1]
        &\reddC
        \term{
            \pRes{y\tBfr{\epsi}y'} \big(
                \pRes{z\tBfr{\epsi}z'} \big(
                    \tMain \, \ldots
                    \prl \tChild \, \tLet m_2 = \tSend \text{book}(\text{``Dune''}) \, z \tIn ()
                \big)
            \big)
        }
        \\ \displaybreak[1]
        &\reddC
        \term{
            \pRes{z\tBfr{\epsi}z'} \big(
                \tMain \, \ldots
                \prl \tChild \, \tLet m_2 = \tSend \text{book}(\text{``Dune''}) \, z \tIn ()
            \big)
        }
        \\ \displaybreak[1]
        &\reddC
        \term{
            \pRes{z\tBfr{\text{book}(\text{``Dune''})}z'} \big(
                \tMain \, \ldots
                \prl \tChild \, \tLet m_2 = z \tIn ()
            \big)
        }
        \\ \displaybreak[1]
        &\reddC
        \term{
            \pRes{z\tBfr{\text{book}(\text{``Dune''})}z'} \big(
                \tMain \, \ldots
                \prl \tChild \, ()
            \big)
        }
        \\ \displaybreak[1]
        &\reddC
        \term{
            \pRes{z\tBfr{\text{book}(\text{``Dune''})}z'} \big(
                \tMain \, \tLet (\textit{book},m'_2) = \tRecv z' \tIn \textit{book}
            \big)
        }
        \\ \displaybreak[1]
        &\reddC
        \term{
            \pRes{z\tBfr{\epsi}z'} \big(
                \tMain \, \tLet (\textit{book},m'_2) = (\text{book}(\text{``Dune''}),z') \tIn \textit{book}
            \big)
        }
        \\ \displaybreak[1]
        &\reddC
        \term{
            \pRes{z\tBfr{\epsi}z'} \big(
                \tMain \, \text{book}(\text{``Dune''})
            \big)
        }
        \reddC
        \term{
            \tMain \, \text{book}(\text{``Dune''})
        }
        \tag*{\qedhere}
    \end{align*}
    \caption{
        Reduction sequence from \Cref{x:LAST:Shop}.
        Recall that notation $\term{\text{book}(\textit{title})}$ is syntactic sugar for a lookup function.
        We abbreviate unchanged threads using ``$\term{\ldots}$''.
    }\label{f:x:LAST:Shop}
\end{figure}

\subsection{The Type System of \texorpdfstring{\protect\LAST*}{LAST*}}
\label{s:LAST:typeSystem}

The type system for \LAST* includes functional types for functions and pairs and session types for message passing.
The syntax and meaning of functional types ($\type{T},\type{U}$) and session types ($\type{S}$) are as follows:
\begin{align*}
    \begin{array}[t]{@{}r@{}lr@{\kern.5ex}l@{\kern2ex}lr@{\kern.5ex}l@{\kern2ex}lr@{\kern.5ex}l@{\kern2ex}lr@{}}
        \type{T},\type{U} ::= {} &
        \type{T \times U} & \text{pair}
        & \sepr &
        \type{T \lolli U} & \text{function}
        & \sepr &
        \type{\1} & \text{unit}
        & \sepr &
        \type{S} & \text{session}
        \\
        \type{S} ::= {} &
        \type{{!}T . S} & \text{send}
        & \sepr &
        \type{{?}T . S} & \text{receive}
        & \sepr &
        \type{\oplus \{ i : T \}_{i \in I}} & \text{select}
        & \sepr &
        \type{\& \{ i : T \}_{i \in I}} & \text{branch}
        \\ \sepr* &
        \type{\tEnd}
    \end{array}
\end{align*}

\noindent
Session type duality ($\type{\ol{S}}$) is defined as usual; note that only the continuations, and not the messages, of send and receive types are dualized.
\begin{align*}
    \type{\ol{{!}T . S}}
    &= \type{{?}T . \ol{S}}
    &
    \type{\ol{\oplus \{ i : S_i \}_{i \in I}}}
    &= \type{\& \{ i : \ol{S_i} \}_{i \in I}}
    &
    \type{\ol{\tEnd}}
    &= \type{\tEnd}
    \\
    \type{\ol{{?}T . S}}
    &= \type{{!}T . \ol{S}}
    &
    \type{\ol{\& \{ i : S_i \}_{i \in I}}}
    &= \type{\oplus \{ i : \ol{S_i} \}_{i \in I}}
\end{align*}

\begin{figure}[t!]
    \def\defaultHypSeparation{\hskip1ex}
    \begin{mathpar}
        \begin{bussproof}[typ-var]
            \bussAx{
                \term{x}:\type{T} \vdashM \term{x}:\type{T}
            }
        \end{bussproof}
        \and
        \begin{bussproof}[typ-abs]
            \bussAssume{
                \type{\Gamma}, \term{x}:\type{T} \vdashM \term{M} :\type{U}
            }
            \bussUn{
                \type{\Gamma} \vdashM \term{\lam x . M} : \type{T \lolli U}
            }
        \end{bussproof}
        \and
        \begin{bussproof}[typ-app]
            \bussAssume{
                \type{\Gamma} \vdashM \term{M}: \type{T \lolli U}
            }
            \bussAssume{
                \type{\Delta} \vdashM \term{N}: \type{T}
            }
            \bussBin{
                \type{\Gamma}, \type{\Delta} \vdashM \term{M~N}: \type{U}
            }
        \end{bussproof}
        \and
        \begin{bussproof}[typ-unit]
            \bussAx{
                \type{\emptyset} \vdashM \term{()}: \type{\1}
            }
        \end{bussproof}
        \and
        \begin{bussproof}[typ-pair]
            \bussAssume{
                \type{\Gamma} \vdashM \term{M}: \type{T}
            }
            \bussAssume{
                \type{\Delta} \vdashM \term{N}: \type{U}
            }
            \bussBin{
                \type{\Gamma}, \type{\Delta} \vdashM \term{(M,N)}: \type{T \times U}
            }
        \end{bussproof}
        \and
        \begin{bussproof}[typ-split]
            \bussAssume{
                \type{\Gamma} \vdashM \term{M}: \type{T \times T'}
            }
            \bussAssume{
                \type{\Delta}, \term{x}:\type{T}, \term{y}:\type{T'} \vdashM \term{N}: \type{U}
            }
            \bussBin{
                \type{\Gamma}, \type{\Delta} \vdashM \term{\tLet (x,y) = M \tIn N}: \type{U}
            }
        \end{bussproof}
        \and
        \begin{bussproof}[typ-new]
            \bussAx{
                \type{\emptyset} \vdashM \term{\tNew}: \type{S \times \ol{S}}
            }
        \end{bussproof}
        \and
        \begin{bussproof}[typ-fork]
            \bussAssume{
                \type{\Gamma} \vdashM \term{M}: \type{\1}
            }
            \bussAssume{
                \type{\Delta} \vdashM \term{N} : \type{T}
            }
            \bussBin{
                \type{\Gamma} , \type{\Delta} \vdashM \term{\tFork M ; N}: \type{T}
            }
        \end{bussproof}
        \and
        \begin{bussproof}[typ-end]
            \bussAssume{
                \type{\Gamma} \vdashM \term{M} : \type{T}
            }
            \bussUn{
                \type{\Gamma} , \term{x}:\type{\tEnd} \vdashM \term{M} : \type{T}
            }
        \end{bussproof}
        \and
        \begin{bussproof}[typ-send]
            \bussAssume{
                \type{\Gamma} \vdashM \term{M}: \type{T}
            }
            \bussAssume{
                \type{\Delta} \vdashM \term{N}: \type{{!}T . S}
            }
            \bussBin{
                \type{\Gamma} ,  \type{\Delta} \vdashM \term{\tSend M \, N}: \type{S}
            }
        \end{bussproof}
        \and
        \begin{bussproof}[typ-recv]
            \bussAssume{
                \type{\Gamma} \vdashM \term{M}: \type{{?}T . S}
            }
            \bussUn{
                \type{\Gamma} \vdashM \term{\tRecv M}: \type{T \times S}
            }
        \end{bussproof}
        \and
        \begin{bussproof}[typ-sel]
            \bussAssume{
                \type{\Gamma} \vdashM \term{M}: \type{\oplus \{ i : S_i \}_{i \in I}}
            }
            \bussAssume{
                j \in I
            }
            \bussBin{
                \type{\Gamma} \vdashM \term{\tSel j\, M}: \type{S_j}
            }
        \end{bussproof}
        \and
        \begin{bussproof}[typ-case]
            \bussAssume{
                \type{\Gamma} \vdashM \term{M}: \type{{\&}\{ i : S_i \}_{i \in I}}
            }
            \bussAssume{
                \forall i \in I.~ \type{\Delta} \vdashM \term{N_i}: \type{S_i \lolli U}
            }
            \bussBin{
                \type{\Gamma}, \type{\Delta} \vdashM \term{\tCase M \tOf \{ i : N_i \}_{i \in I}}: \type{U}
            }
        \end{bussproof}
        \\ \phantom{.}\dashes \\ 
        \begin{bussproof}[typ-buf]
            \bussAx{
                \type{\emptyset} \vdashB \term{\tBfr{\epsilon}}: \type{S'} \> \type{S'}
            }
        \end{bussproof}
        \and
        \begin{bussproof}[typ-buf-send]
            \bussAssume{
                \type{\Gamma} \vdashM \term{M}: \type{T}
            }
            \bussAssume{
                \type{\Delta} \vdashB \term{\tBfr{\vec{m}}}: \type{S'} \> \type{S}
            }
            \bussBin{
                \type{\Gamma}, \type{\Delta} \vdashB \term{\tBfr{\vec{m},M}}: \type{S'} \> \type{{!}T . S}
            }
        \end{bussproof}
        \and
        \begin{bussproof}[typ-buf-sel]
            \bussAssume{
                \type{\Gamma} \vdashB \term{\tBfr{\vec{m}}}: \type{S'} \> \type{S_j}
            }
            \bussAssume{
                j \in I
            }
            \bussBin{
                \type{\Gamma} \vdashB \term{\tBfr{\vec{m},j}}: \type{S'} \> \type{\oplus \{ i : S_i \}_{i \in I}}
            }
        \end{bussproof}
        \\ \phantom{.}\dashes \\ 
        \begin{bussproof}[typ-main]
            \bussAssume{
                \type{\Gamma} \vdashM \term{M}: \type{\hat{T}}
            }
            \bussUn{
                \type{\Gamma} \vdashC{\tMain} \term{\tMain\, M}: \type{\hat{T}}
            }
        \end{bussproof}
        \and
        \begin{bussproof}[typ-child]
            \bussAssume{
                \type{\Gamma} \vdashM \term{M}: \type{\1}
            }
            \bussUn{
                \type{\Gamma} \vdashC{\tChild} \term{\tChild\, M}: \type{\1}
            }
        \end{bussproof}
        \and
        \begin{bussproof}[typ-par]
            \bussAssume{
                \type{\Gamma} \vdashC{\phi_1} \term{C}: \type{T_1}
            }
            \bussAssume{
                \type{\Delta} \vdashC{\phi_2} \term{D}: \type{T_2}
            }
            \bussBin{
                \type{\Gamma}, \type{\Delta} \vdashC{\phi_1 + \phi_2} \term{C \prl D}: \type{T_1 + T_2}
            }
        \end{bussproof}
        \and
        \begin{bussproof}[typ-res]
            \bussAssume{
                \type{\Gamma} \vdashB \term{\tBfr{\vec{m}}}: \type{S'} \> \type{S}
            }
            \bussAssume{
                \type{\Delta}, \term{x}: \type{S'} \vdashC{\phi} \term{C}: \type{T}
            }
            \bussAssume{
                \type{\Gamma'} , \term{y}:\type{\ol{S}} = \type{\Gamma} , \type{\Delta}
            }
            \bussTern{
                \type{\Gamma'} \vdashC{\phi} \term{\pRes{x\tBfr{\vec{m}}y} C}: \type{T}
            }
        \end{bussproof}
    \end{mathpar}
    \caption{\protect\LAST* typing rules for terms (top), buffers (center), and configurations (bottom).}\label{f:LAST:type}
\end{figure}

The type system for \LAST* has three layers: typing for terms, for buffers, and for configurations.
Typing for terms uses judgments of the form $$\type{\Gamma} \vdashM \term{M} : \type{T}$$ which decrees that $\term{M}$ has a behavior described by $\type{T}$ using the typing context $\type{\Gamma}$, which is defined as a list of variable-type assignments $\term{x}:\type{T}$.

\Cref{f:LAST:type} (top) gives the typing rules for terms.
Rules often combine typing contexts $\type{\Gamma}$ and $\type{\Delta}$ to form $\type{\Gamma} , \type{\Delta}$; this implicitly assumes that the domains of $\type{\Gamma}$ and $\type{\Delta}$ are disjoint. We briefly comment on them:
\begin{itemize}
	\item
Rules~\ruleLabel{typ-var}, \ruleLabel{typ-abs}, \ruleLabel{typ-app}, \ruleLabel{typ-unit}, \ruleLabel{typ-pair}, and~\ruleLabel{typ-split} are standard typing rules for functional terms.
\item Rule~\ruleLabel{typ-new} types the $\term{\tNew}$ construct as a pair of dual session types.
\item Rule~\ruleLabel{typ-fork} takes a term $\term{M}$ of type $\type{\1}$ and a term $\term{N}$ of type $\type{T}$ to type the $\term{\tFork}$ construct as $\type{T}$; as we will see in the typing of configurations, child threads must be typed $\type{\1}$, which explains the type of $\term{M}$.
\item Rule~\ruleLabel{typ-end} allows weakening typing contexts with $\type{\tEnd}$-typed variables, as closed sessions are not used.
\item Rule~\ruleLabel{typ-send} takes a term $\term{M}$ of type $\type{T}$ and a term $\term{N}$ of type $\type{{!}T.S}$ to type a $\term{\tSend}$ of $\term{M}$ along $\term{N}$ as the continuation type $\type{S}$.
Dually, Rule~\ruleLabel{typ-recv} takes a term $\term{M}$ of type $\type{{?}T.S}$ to type a $\term{\tRecv}$ along $\term{M}$ as a pair of the message's payload type and continuation $\type{T \times S}$.
\item Rule~\ruleLabel{typ-sel} takes a term $\term{M}$ of type $\type{\oplus \{ i : S_i \}_{i \in I}}$ to type the selection of some $\term{j} \in I$ along $\term{M}$ as the continuation $\type{S_j}$.
Dually, Rule~\ruleLabel{typ-case} takes a term $\term{M}$ of type $\type{\& \{ i : S_i \}_{i \in I}}$ and for every $\term{i} \in I$ a term $\term{N_i}$ typed $\type{S_i \lolli U}$ (i.e., a function from the label $\term{i}$'s continuation type $\type{S_i}$ to some common but arbitrary type $\type{U}$) to type a $\term{\tCase}$ along $\term{M}$ as $\type{U}$.
\end{itemize}

\begin{exa}
    \label{x:LAST:typTerm}
    To illustrate the typing rules, let us derive the typing of term $\term{\sff{Son}(s',m')}$ from \Cref{s:FIRST:example}.
    As in \Cref{x:LAST:Shop}, we omit the $\term{\tClose}$ construct.
    To type the sugared terms $\term{\tLet x = M \tIn N}$ we use a sugared Rule~\ruleLabel{typ-let} derivable from Rules~\ruleLabel{typ-abs} and~\ruleLabel{typ-app}.
    We consider $\type{\sff{Str}}$ (string), $\type{\sff{B}}$ (book), and $\type{\sff{P}}$ (payment) to be primitive non-linear types that can be weakened/contracted at will and are self dual.
    Below, $\type{S} = \type{{!}\sff{P}.{?}\sff{B}.\tEnd}$ and $\type{S'} = \type{{?}\sff{B}.\tEnd}$.
    We omit ``\textsc{typ-}'' from rule labels.
    \begin{align*}
        &
        \pi \deq
        \begin{bussproof}
            \bussAx[\ruleLabel{var}]{
                \term{m'_1}:\type{{?}\sff{B}.\tEnd} \vdashM \term{m'_1}:\type{{?}\sff{B}.\tEnd}
            }
            \bussUn[\ruleLabel{recv}]{
                \term{m'_1}:\type{{?}\sff{B}.\tEnd} \vdashM \term{\tRecv m'_1}:\type{\sff{B} \times \tEnd}
            }
            \bussAx[\ruleLabel{var}]{
                \term{\textit{book}}:\type{\sff{B}} \vdashM \term{\textit{book}} : \type{\sff{B}}
            }
            \bussUn[\ruleLabel{end}]{
                \term{\textit{book}}:\type{\sff{B}} , \term{m'_2}:\type{\tEnd} \vdashM \term{\textit{book}} : \type{\sff{B}}
            }
            \bussBin[\ruleLabel{split}]{
                \term{m'_1}:\type{{?}\sff{B}.\tEnd} \vdashM \term{\tLet (\textit{book},m'_2) = \tRecv m'_1 \tIn \textit{book}} : \type{\sff{B}}
            }
        \end{bussproof}
        \\[5pt]
        &
        \begin{bussproof}
            \def\defaultHypSeparation{\hskip1ex}
            \bussAx[\ruleLabel{var}]{
                \term{s'_2}:\type{S} \vdashM \term{s'_2} : \type{S}
            }
            \bussAx[\ruleLabel{var}]{
                \term{m'}:\type{{!}S.{?}\sff{B}.\tEnd} \vdashM \term{m'} : \type{{!}S.{?}\sff{B}.\tEnd}
            }
            \bussBin[\ruleLabel{send}]{
                \term{s'_2}:\type{S} , \term{m'}:\type{{!}S.{?}\sff{B}.\tEnd} \vdashM \term{\tSend s'_2 \, m'} : \type{{?}\sff{B}.\tEnd}
            }
            \bussAssume{
                \pi
            }
            \noLine
            \bussUn{
                \term{m'_1}:\type{{?}\sff{B}.\tEnd} \vdashM \term{\ldots} : \type{\sff{B}}
            }
            \bussBin[\ruleLabel{let}]{
                \term{s'_2}:\type{S} , \term{m'}:\type{{!}S.{?}\sff{B}.\tEnd} \vdashM \term{\tLet m'_1 = \tSend s'_2 \, m' \tIn \ldots} : \type{\sff{B}}
            }
            \doubleLine
            \bussUn[\dots]{
                \term{s'}:\type{{!}\sff{Str}.\oplus\{\sff{buy}:S,\sff{blurb}:S'\}} , \term{m'}:\type{{!}S.{?}\sff{B}.\tEnd} \vdashM \term{\sff{Son}(s',m')} : \type{\sff{B}}
            }
        \end{bussproof}
    \end{align*}

    Similarly, the typings of the shop and the mother are as follows:
    \begin{align*}
        \term{s}:\type{{?}\sff{Str}.\&\{\sff{buy}:\ol{S},\sff{blurb}:\ol{S'}\}} &\vdashM \term{\sff{Shop}(s)} : \type{\1}
        \\
        \term{m}:\type{{?}S.{!}\sff{B}.\tEnd} &\vdashM \term{\sff{Mother}(m)} : \type{\1}
    \end{align*}
    As such, the types of $\term{s},\term{s'}$ and $\term{m},\term{m'}$ are pairwise dual.
    Hence, the typing of the entire system is simply
    \[
        \type{\emptyset} \vdashM \term{\sff{Sys}} : \type{\sff{B}}.
        \tag*{\qedhere}
    \]
\end{exa}
\noindent 
The typing for  buffered channels  uses judgments of the form $$\type{\Gamma} \vdashB \term{\tBfr{\vec{m}}} : \type{S'}\>\type{S}$$
The difference between $\type{S'}$ and $\type{S}$ is determined by the messages in the buffer $\term{\vec{m}}$: $\type{S}$ denotes a sequence of sends and selections corresponding to the values and labels in $\term{\vec{m}}$, after which the type continues as $\type{S'}$.
Thus, $\type{S}$ denotes the type of a buffered channel's output endpoint before it sent the messages in $\term{\vec{m}}$, and $\type{S'}$ denotes this endpoint's current type.
This way, $\type{\ol{S}}$ signifies the type of the buffered channel's input endpoint, corresponding to a sequence of receives and offers corresponding to the values and labels in $\term{\vec{m}}$.

\Cref{f:LAST:type} (center) gives the three typing rules for buffers.
The typing context $\type{\Gamma}$ is used in the typing of the values in the buffer.
Rule~\ruleLabel{typ-buf} types an empty buffer; as such, $\type{S'} = \type{S}$.
Rule~\ruleLabel{typ-buf-send} takes a value $\term{v}$ of type $\type{T}$ and a buffer typed $\type{S'}\>\type{S}$ to insert $\term{v}$ at the start of the buffer now typed $\type{S'}\>\type{{!}T.S}$.
Rule~\ruleLabel{typ-buf-sel} takes a buffer typed $\type{S'}\>\type{S_j}$ (for some $\term{j} \in I$) to insert $\term{j}$ at the start of the buffer, now typed as $\type{S'}\>\type{\oplus \{ i : S_i \}_{i \in I}}$.

Typing for configurations uses judgments of the form $$\type{\Gamma} \vdashC{\phi} \term{C} : \type{T}$$ where the  marker $\term{\phi}$  indicates whether $\term{C}$ contains the main thread ($\term{\phi} = \term{\tMain}$) or only child threads ($\term{\phi} = \term{\tChild}$).
When composing  configurations marked $\term{\phi_1}$ and $\term{\phi_2}$, we  compute a new marker $\term{\phi_1 + \phi_2}$, as follows:
\begin{mathpar}
    \term{\tMain + \tChild} \deq \term{\tChild + \tMain} \deq \term{\tMain}
    \and
    \term{\tChild + \tChild} \deq \term{\tChild}
    \and
    \text{($\term{\tMain + \tMain}$ is undefined)}
\end{mathpar}
\Cref{f:LAST:type} (bottom) gives the typing rules for configurations.
Rule~\ruleLabel{typ-main} takes a term of non-session type (denoted $\type{\hat{T}}$) and turns it into a main thread.
Rule~\ruleLabel{typ-child} takes a term of type $\type{\1}$ and turns it into a child thread.
Rule~\ruleLabel{typ-par} composes two configurations typed $\type{T_1}$ and $\type{T_2}$ in parallel; the rule requires one of the configurations to be typed $\type{\1}$ and uses the other configuration's type to type the composition $\type{T_1 + T_2}$:
\[
    \type{T_1 + T_2} \deq \begin{cases}
        \type{T_1} & \text{if $\type{T_2} = \type{\1}$} \\
        \type{T_2} & \text{if $\type{T_1} = \type{\1}$} \\
        \text{undefined} & \text{otherwise}
    \end{cases}
\]
As such, Rule~\ruleLabel{typ-par} is not defined for configurations both not typed $\type{\1}$; moreover, if both configurations contain a main thread,  the sum of their markers is undefined, and the rule cannot be applied.
Rule~\ruleLabel{typ-res} types a configuration under a buffered channel with output endpoint $\term{x}$ and input endpoint $\term{y}$.
The rule uses typing for buffers to type the buffer $\type{S'}\>\type{S}$.
As such, the configuration should use $\term{x}$ of type $\type{S'}$.
Since $\type{S}$ is the type of $\term{x}$ before it sent the messages already in the buffer, $\term{y}$ should be of type $\type{\ol{S}}$.
Note that $\term{y}$ may be used in the configuration, but may also appear in a message in the buffer.

\subsection{Towards a Faithful Translation of \texorpdfstring{\protect\LAST*}{LAST*} into \APCP}
\label{s:LAST:trans}

As already discussed, we are interested in faithfully translating \LAST* into \APCP.
We first discuss what we mean precisely by `faithful'.
Then, we consider some existing translations of (variants of) the $\lambda$-calculus with call-by-value semantics into (variants of) the \picalc; we briefly discuss their status with respect to our definition of faithfulness.

\paragraph{Faithfulness}
We shall aim for correct translations in the sense of Gorla~\cite{journal/ic/Gorla10}, a well-established set of criteria whereby faithful translations should satisfy \emph{operational correspondence}, a criterion that is divided into
\emph{completeness} and \emph{soundness} properties.
Intuitively, completeness says that the target language    \emph{does enough} to represent the behavior of the source language, whereas soundness says that the target language \emph{does not do too much}.
Alternatively, one may see completeness and soundness as properties that ensure that reduction steps are \emph{preserved} and \emph{reflected} by the translation, respectively.

We formulate these requirements in the specific setting of \LAST* and \APCP, by considering a translation of configurations into processes, denoted `$\encc{z}{\cdot}$', where, as usual, $z$ denotes a name on which the resulting process exhibits the behavior of the source term:
\begin{description}

    \item[Completeness]
        If $\term{C} \reddC \term{D}$, then $\encc{z}{C} \redd* \encc{z}{D}$.

    \item[Soundness]
        If $\encc{z}{C} \redd* Q$, then there exists $\term{D}$ such that $\term{C} \reddC^\ast \term{D}$ and $Q \redd* \encc{z}{D}$.
\end{description}

Soundness and completeness come in different flavors of strength; see, e.g.,~\cite{DBLP:journals/corr/abs-1908-08633}.
Our definitions can be seen as being strictly stronger than Gorla's, in that we do not consider a behavioral relation on target processes (typically used to abstract away from `junk processes' in comparisons with $\encc{z}{D}$).

\paragraph{Existing Translations are not Sound}
We are not aware of translations of  $\lambda$-calculi with call-by-value semantics into the \picalc that satisfy soundness as stated above.
To illustrate the problem, let us consider the well-known translation by Milner~\cite{conf/icalp/Milner90,journal/mscs/Milner92}.
This is a translation into a \picalc with synchronous communication, while we seek a translation into \APCP, which is asynchronous. This discrepancy is not crucial: the unsound reductions we will show next are not enabled by the asynchrony of \APCP (i.e., they are also enabled in synchronous processes).

It is actually sufficient to consider Milner's approach to translating variables and applications.
By adapting this approach to \APCP, we obtain the translations given next.
Below, we  write `$\_$' to denote a fresh name of type $\bullet$\,; when sending names denoted `$\_$', we omit binders `$\pRes{\_\_}$'.
\begin{align*}
    \encc{z}{x} &\deq \pRes{ab} ( \pOut z[\_,a] \| \pIn b(\_,c) ; \pOut x[\_,c] )
    \\
    \encc{z}{M~N} &\deq \pRes{ab} \pRes{cd} \big( \pIn a(\_,e) ; \pRes{fg} ( \pOut e[\_,f] \| \pIn c(\_,h) ; \pOut g[h,z] ) \| \encc{b}{M} \| \encc{d}{N} \big)
\end{align*}
Consider the very simple term $\term{x~y}$, which cannot reduce.
Yet, its translation $\encc{z}{x~y}$ does reduce, as shown next  (each step underlines the send and receive that synchronize):
\begin{align*}
    \encc{z}{x~y} &= \pRes{a_1b_1} \pRes{c_1d_1} \big(
        \begin{array}[t]{@{}l@{}}
            \underline{\pIn a_1(\_,e_1)} ; \pRes{f_1g_1} ( \pOut e_1[\_,f_1] \| \pIn c_1(\_,h_1) ; \pOut g_1[h_1,z] )
            \\
            {} \| \pRes{a_2b_2} ( \underline{\pOut b_1[\_,a_2]} \| \pIn b_2(\_,c_2) ; \pOut x[\_,c_2] )
            \\
            {} \| \pRes{a_3b_3} ( \pOut d_1[\_,a_3] \| \pIn b_3(\_,c_3) ; \pOut y[\_,c_3] )
        \big)
    \end{array}
    \\
    &\redd \pRes{c_1d_1} \pRes{a_2b_2} \big(
        \begin{array}[t]{@{}l@{}}
            \pRes{f_1g_1} ( \underline{\pOut a_2[\_,f_1]} \| \pIn c_1(\_,h_1) ; \pOut g_1[h_1,z] )
            \\
            {} \| \underline{\pIn b_2(\_,c_2)} ; \pOut x[\_,c_2]
            \\
            {} \| \pRes{a_3b_3} ( \pOut d_1[\_,a_3] \| \pIn b_3(\_,c_3) ; \pOut y[\_,c_3] )
        \big)
    \end{array}
    \\
    &\redd \pRes{c_1d_1} \pRes{f_1g_1} \big(
        \begin{array}[t]{@{}l@{}}
            \underline{\pIn c_1(\_,h_1)} ; \pOut g_1[h_1,z]
            \\
            {} \| \pOut x[\_,f_1]
            \\
            {} \| \pRes{a_3b_3} ( \underline{\pOut d_1[\_,a_3]} \| \pIn b_3(\_,c_3) ; \pOut y[\_,c_3] )
        \big)
    \end{array}
    \\
    &\redd \pRes{f_1g_1} \pRes{a_3b_3} ( \pOut g_1[a_3,z] \| \pOut x[\_,f_1] \| \pIn b_3(\_,c_3) ; \pOut y[\_,c_3] ) \nredd
\end{align*}
It is clear that this final term, which cannot reduce any further, cannot be reconciled with any source \LAST* term.
Hence, the translation is unsound.

Besides Milner's, other translations of  $\lambda$-calculi with call-by-value semantics into (variants of) the \picalc have been proposed by Lindley and Morris~\cite{conf/esop/LindleyM15} (a precise formalization of Walder's~\cite{conf/icfp/Wadler12,journal/jfp/Wadler14}), by Vasconcelos~\cite{report/Vasconcelos00}, and by Fowler \etal~\cite{DBLP:journals/lmcs/FowlerKDLM23}.
One of the most significant differences between them is how they translate variables: Milner uses sends; Lindley and Morris and Fowler \etal use forwarders; Vasconcelos uses substitutions.
None of them satisfy soundness as defined above, although Vasconcelos' and Fowler \etal's enjoy a weaker form of soundness that holds up to an appropriate behavioral equivalence.

By examining these three translations, we observe that the call-by-value semantics is overly contextual, in the sense that determining whether a subterm may reduce depends on the context.
This way, e.g., Rule~\ruleLabel{red-lam} (\Cref{f:LAST:terms}) only applies to abstractions applied to values.
The semantic rules for the \picalc are much less contextual,
so translations require additional machinery to prevent unwanted reductions.
We are not aware of  translations of call-by-value $\lambda$-calculi into $\pi$ that are sound, which in our view suggests that such a sound translation may not exist.

Based on this discussion, we conclude that the call-by-value semantics of \LAST* does not lend itself for a faithful representation in \APCP.
To address this issue, in the next section we will propose a variant of \LAST with a call-by-name semantics.
This variant, denoted \FIRST, will admit a faithful (i.e., sound and complete)   translation into \APCP.

\section{\FIRST and a Faithful Translation into \APCP}
\label{s:FIRST}

Here we present \FIRST, a variant of \LAST with call-by-name semantics.
Our presentation proceeds gradually, based on the presentation we gave for  \LAST* (\Cref{s:LAST}).
We then give a typed translation of \FIRST into \APCP, and prove that it is faithful as discussed in \Cref{s:LAST:trans}.
Finally, we show how the translation can help us identify a fragment of deadlock-free \FIRST programs.

\subsection{The Language of \FIRST}
\label{s:FIRST:language}

\begin{figure}[t]
    Terms ($\term{M},\term{N},\ldots$) and reduction contexts ($\tCtx{R}$):
    \begin{align*}
        \term{M},\term{N}
        & \begin{array}[t]{@{}l@{\kern.5ex}lr@{\kern1.2ex}l@{\kern.5ex}lr@{}}
            {} ::= &
            \term{x} & \text{variable}
            & \sepr &
            \term{\tNew} & \text{create new channel}
            \\ \sepr* &
            \term{()} & \text{unit value}
            & \sepr &
            \term{\tFork M ; N} & \text{fork $\term{M}$ in parallel to $\term{N}$}
            \\ \sepr* &
            \term{\lam x . M} & \text{abstraction}
            & \sepr &
            \term{(M,N)} & \text{pair construction}
            \\ \sepr* &
            \term{M~N} & \text{application}
            & \sepr &
            \term{\tLet (x,y)=M \tIn N} & \text{pair deconstruction}
            \\ \sepr* &
            \term{\tSend M \, N} & \text{send $\term{M}$ along $\term{N}$}
            & \sepr &
            \term{\tSel \ell \, M} & \text{select label $\ell$ along $\term{M}$}
            \\ \sepr* &
            \term{\tRecv M} & \text{receive along $\term{M}$}
            & \sepr &
            \term{\tCase M \tOf \{ i : M \}_{i \in I}} & \text{offer labels in $I$ along $\term{M}$}
            \\ \sepr* &
            \term{\tClose M ; N} & \text{close $\term{M}$}
            & \sepr &
            \term{M \tSub{ N/x }} & \text{explicit substitution}
        \end{array}
        \\
        \term{\tCtx{R}}
        & \begin{array}[t]{@{}l@{\kern.5ex}l@{}}
            {} ::= &
            \term{\tHole}
            \sepr
            \term{\tCtx{R}~M}
            \sepr
            \term{\tSend M \, \tCtx{R}}
            \sepr
            \term{\tRecv \tCtx{R}}
            \sepr
            \term{\tLet (x,y) = \tCtx{R} \tIn M}
            \\ \sepr* &
            \term{\tSel \ell\, \tCtx{R}}
            \sepr
            \term{\tCase \tCtx{R} \tOf \{ i : M \}_{i \in I}}
            \sepr
            \term{\tClose \tCtx{R} ; M}
            \sepr
            \term{\tCtx{R} \tSub{ M/x }}
        \end{array}
    \end{align*}

    \caption{Term syntax and reduction contexts for \FIRST.}\label{f:LASTn:termCtx}
\end{figure}

To define \FIRST, we start from \LAST* and modify its semantics.
We only need to change the definitions of reduction contexts and reduction in \Cref{f:LAST:terms,f:LAST:confs}.
For the reader's convenience, \Cref{f:LASTn:termCtx} gives the syntax of \FIRST terms and reduction contexts; \Cref{as:FIRST:lang} contains a full, self-contained definition of the language and type system of \FIRST.

The crux of the required changes is that call-by-name semantics applies abstractions more eagerly, and not only when arguments are values.
To be precise, term reduction changes Rules~\ruleLabel{red-lam} and~\ruleLabel{red-pair} as follows:
\begin{mathpar}
    \begin{bussproof}[red-lam]
        \bussAx{
            \term{(\lam x . M)~\underline{N}} \reddM \term{M \{ \underline{N}/x \}}
        }
    \end{bussproof}
    \and
    \begin{bussproof}[red-pair]
        \bussAx{
            \term{\tLet (x,y) = (\underline{N_1},\underline{N_2}) \tIn M} \reddM \term{M \{ \underline{N_1}/x,\underline{N_2}/y \}}
        }
    \end{bussproof}
\end{mathpar}
Above, we underline modified parts: instead of requiring values, the rules allow arbitrary terms.
Moreover, to enforce the eager application of these rules, we remove from the definition of reduction contexts the clauses $\term{v~\tCtx{R}}$, $\term{(\tCtx{R},M)}$, and $\term{(v,\tCtx{R})}$.
That is, \FIRST disallows terms in parameter positions to reduce.

\begin{exa}
    \label{x:CbN}
    We consider the term from \Cref{x:CbV}, now using the call-by-name semantics of \FIRST:
    \[
        \term{\big(\lam x . x~(\lam y . y)\big)~\big((\lam w . w)~(\lam z . z)\big)}
        \reddM
        \term{\big((\lam w . w)~(\lam z . z)\big)~(\lam y . y)}
        \reddM
        \term{(\lam z . z)~(\lam y . y)}
        \reddM
        \term{\lam y . y}
    \]
    Notice how here the function on $\term{x}$ is applied \emph{before} its parameter is evaluated as in \Cref{x:CbV}.
\end{exa}

Following the same call-by-name spirit, message passing in \FIRST transmits unevaluated terms, instead of only values as in \LAST*.
To accommodate this, we replace the two reduction-context clauses for $\term{\tSend}$ with the clause $\term{\tSend M \, \tCtx{R}}$, and modify Rule~\ruleLabel{red-send} accordingly:
\[
    \begin{bussproof}[red-send]
        \bussAx{
            \term{\nu{x\tBfr{\vec m}y} ( \tCtx{F}[\tSend M \, x] \prl C )} \reddC \term{\nu{x\tBfr{\vec m , M}y} ( \tCtx{F}[x] \prl C )}
        }
    \end{bussproof}
\]

\smallskip

As is to be expected, there are subtle differences between the semantics of \FIRST and \APCP.
In order to deal with these discrepancies, while enabling the desired completeness and soundness results, we enrich \FIRST with \emph{explicit substitutions} and \emph{closed sessions}.

\paragraph{Explicit Substitutions}

Term reductions~\ruleLabel{red-lam} and~\ruleLabel{red-pair} substitute variables for terms, regardless of the position of these variables.
To mimic this in \APCP, we would have to be able to substitute the translations of variables for translations of terms.
Although this is possible when those variables occur in reduction contexts, there is no mechanism for representing this in \APCP when they occur in different contexts (where their translations are blocked by prefixes).

One way of dealing with this issue is to equate translations up to substitutions, using the so-called  \emph{substitution lifting}~\cite{conf/fossacs/ToninhoCP12}.
Here, we opt for an alternative, more direct treatment based on \emph{explicit substitutions} (see, e.g., \cite{conf/fsttcs/LevyM99}).
Intuitively, explicit substitutions delay variable substitution until those variables occur in reduction contexts.
To incorporate explicit substitutions, we proceed as follows:
\begin{itemize}

    \item
        We add explicit substitutions to the syntax of terms $\term{M \tSub{ N/x }}$ and configurations $\term{C \tSub{ N/x }}$.
        These are not meant to be used when writing programs, instead appearing and disappearing as programs reduce ({runtime} syntax).
        That is, they appear when abstractions are applied and pairs deconstructed:
        \begin{mathpar}
            \begin{bussproof}[red-lam]
                \bussAx{
                    \term{(\lam x . M)~N} \reddM \term{M \tSub{ N/x }}
                }
            \end{bussproof}
            \and
            \begin{bussproof}[red-pair]
                \bussAx{
                    \term{\tLet (x,y) = (N_1,N_2) \tIn M} \reddM \term{M \tSub{ N_1/x,N_2/y }}
                }
            \end{bussproof}
        \end{mathpar}
        Explicit substitutions disappear when they meet the substituted variable, as per the following  rule:
        \[
            \begin{bussproof}[red-subst]
                \bussAx{
                    \term{x \tSub{ N/x }} \reddM \term{N}
                }
            \end{bussproof}
        \]

    \item
        We add reduction contexts $\term{\tCtx{R} \tSub{ N/x }}$ and configuration contexts $\term{\tCtx{G} \tSub{ N/x }}$.
        We then define a structural congruence for terms, denoted $\equivM$, with a single rule that enables extruding the scope of explicit substitutions across reduction contexts:
        \[
            \begin{bussproof}[sc-sub-ext]
                \bussAssume{
                    \term{x} \notin \fv(\tCtx{R})
                }
                \bussUn{
                    \term{(\tCtx{R}[M]) \tSub{ N/x }} \equivM \tCtx{R}[M \tSub{ N/x }]
                }
            \end{bussproof}
        \]
        We close term reduction under this structural congruence:
        \[
            \begin{bussproof}[red-lift-sc]
                \bussAssume{
                    \term{M} \equivM \term{M'}
                }
                \bussAssume{
                    \term{M'} \reddM \term{N'}
                }
                \bussAssume{
                    \term{N'} \equivM \term{N}
                }
                \bussTern{
                    \term{M} \reddM \term{N}
                }
            \end{bussproof}
        \]
        We also add rules to the structural congruence for configurations that lift explicit substitutions in terms and enable scope extrusion on the level of configurations, respectively:
        \begin{mathpar}
            \begin{bussproof}[sc-conf-sub]
                \bussAx{
                    \term{\phi \, ( M \tSub{ N/x } )} \equivC \term{(\phi \, M) \tSub{ N/x }}
                }
            \end{bussproof}
            \and
            \begin{bussproof}[sc-conf-sub-ext]
                \bussAssume{
                    \term{x} \notin \fv(\tCtx{G})
                }
                \bussUn{
                    \term{(\tCtx{G}[C]) \tSub{ N/x }} \equivC \tCtx{G}[C \tSub{ N/x }]
                }
            \end{bussproof}
        \end{mathpar}

    \item
        When moving terms between threads, we need to make sure that we do not affect variables that are bound by explicit substitutions.
        As such we define a specific form of thread context, denoted~$\tCtx{\hat{F}}$, that disallows the hole to appear under explicit substitutions.
        Configuration reductions~\ruleLabel{red-fork}, \ruleLabel{red-send}, and~\ruleLabel{red-recv} (cf.\ \Cref{f:LAST:confs}) then use these specific contexts.

    \item
        To type explicit substitutions, we add the following typing rules for terms and configurations:
        \begin{mathpar}
            \begin{bussproof}[typ-sub]
                \bussAssume{
                    \type{\Gamma} , \term{x}:\type{T} \vdashM \term{M} : \type{U}
                }
                \bussAssume{
                    \type{\Delta} \vdashM \term{N} : \type{T}
                }
                \bussBin{
                    \type{\Gamma} , \type{\Delta} \vdashM \term{M \tSub{ N/x }} : \type{U}
                }
            \end{bussproof}
            \and
            \begin{bussproof}[typ-conf-sub]
                \bussAssume{
                    \type{\Gamma} , \term{x}:\type{T} \vdashC{\phi} \term{C} : \type{U}
                }
                \bussAssume{
                    \type{\Delta} \vdashM \term{N} : \type{T}
                }
                \bussBin{
                    \type{\Gamma} , \type{\Delta} \vdashC{\phi} \term{C \tSub{ N/x }} : \type{U}
                }
            \end{bussproof}
        \end{mathpar}

\end{itemize}
\begin{exa}
    \label{x:explSub}
    We revisit \Cref{x:CbN}, this time using explicit substitutions:
    \begin{align*}
        \term{\big(\lam x . x~(\lam y . y)\big)~\big((\lam w . w)~(\lam z . z)\big)}
        &\reddM
        \term{\big(x~(\lam y . y)\big) \tSub{ \big((\lam w . w)~(\lam z . z)\big)/x }}
        \\
        &\equivM
        \term{(x \tSub{ \big((\lam w . w)~(\lam z . z)\big)/x })~(\lam y . y)}
        \\
        &\reddM
        \term{\big((\lam w . w)~(\lam z . z)\big)~(\lam y . y)}
        \\
        &\reddM
        \term{(w \tSub{ (\lam z . z)/w })~(\lam y . y)}
        \reddM
        \term{(\lam z . z)~(\lam y . y)}
        \\
        &\reddM
        \term{z \tSub{(\lam y . y)/z}}
        \reddM
        \term{\lam y . y}
        \tag*{\qedhere}
    \end{align*}
\end{exa}

\begin{exa}[The Bookshop Scenario in \FIRST]
\label{x:FIRST:Shop}
    Recall the bookshop scenario introduced in \Cref{s:FIRST:example}, already illustrated by \Cref{x:LAST:Shop} for \LAST*.
    Although the end result is the same, the CbN semantics and explicit substitutions of \FIRST do change the behavior of the system ($\term{\tMain \, \sff{Sys}}$) compared to its behavior under \LAST*.

    We start to illustrate this behavior by reconsidering how the system starts by setting up channels; as we will see, not all explicit substitutions can be immediately resolved (we sometimes abbreviate unchanged threads using ``$\term{\ldots}$''):
    \begin{align*}
        \term{\tMain \, \sff{Sys}}
        &= \term{\tMain \, \tLet (s,s') = \tNew \tIn \ldots}
        \\
        &\reddC \term{\pRes{y\tBfr{\epsi}y'} \tMain \, \tLet (s,s') = (y,y') \tIn \ldots}
        \\
        &\reddC \term{\pRes{y\tBfr{\epsi}y'} \tMain \, \tFork \sff{Shop}(s) ; \ldots \tSub{y/s,y'/s'}}
        \\
        &\equivC \term{\pRes{y\tBfr{\epsi}y'} \big( ( \tMain \, \tFork \sff{Shop}(s) ; \ldots ) \tSub{y/s,y'/s'} \big)}
        \\
        &\reddC \term{\pRes{y\tBfr{\epsi}y'} \big( ( \tMain \, \tLet (m,m') = \tNew \tIn \ldots \prl \tChild \, \sff{Shop}(s) ) \tSub{y/s,y'/s'} \big)}
        \\
        &\equivC \term{\pRes{y\tBfr{\epsi}y'} \big( ( \tMain \, \ldots \prl \tChild \, \tLet (\textit{title},s_1) = \tRecv (s \tSub{y/s}) \tIn \ldots ) \tSub{y'/s'} \big)}
        \\
        &\reddC \term{\pRes{y\tBfr{\epsi}y'} \big( ( \tMain \, \ldots \prl \tChild \, \tLet (\textit{title},s_1) = \tRecv y \tIn \ldots ) \tSub{y'/s'} \big)}
        \\
        &\reddC^2 \term{\pRes{y\tBfr{\epsi}y'} \big( \pRes{z\tBfr{\epsi}z'} ( \tMain \, \tFork \sff{Mother}(m) ; \sff{Son}(s',m') ) \tSub{z/m,z'/m',y'/s'} \prl \tChild \, \sff{Shop}(y) \big)}
        \\
        &\reddC \term{\pRes{y\tBfr{\epsi}y'} \big( \pRes{z\tBfr{\epsi}z'} \big( \tMain \, \sff{Son}(s',m') \tSub{z'/m',y'/s'} \prl \tChild \, \sff{Mother}(m) \tSub{z/m} \big) \prl \tChild \, \sff{Shop}(y) \big)}
        \\
        &\equivC \term{\pRes{y\tBfr{\epsi}y'} \big(
            \pRes{z\tBfr{\epsi}z'} \big( \tMain \, \ldots
            \prl \tChild \, \tLet (x,m_1) = \tRecv (m \tSub{z/m}) \tIn \ldots \big) \prl \tChild \, \sff{Shop}(y)
        \big) }
        \\
        &\reddC \term{\pRes{y\tBfr{\epsi}y'} \big( \pRes{z\tBfr{\epsi}z'} \big( \tMain \, \ldots \prl \tChild \, \tLet (x,m_1) = \tRecv z \tIn \ldots \big) \prl \tChild \, \sff{Shop}(y) \big)}
        \\
        &= \term{\pRes{y\tBfr{\epsi}y'} \big( \pRes{z\tBfr{\epsi}z'} \big( \tMain \, \sff{Son}(s',m') \tSub{z'/m',y'/s'} \prl \tChild \, \sff{Mother}(z) \big) \prl \tChild \, \sff{Shop}(y) \big)} \deq \term{\sff{Sys}^1}
    \end{align*}
    At this point, the explicit substitutions on the main thread cannot be resolved, as neither of $\term{m'},\term{s'}$ appears under a reduction context in $\term{\sff{Son}_{s',m'}}$.
    We continue the example by examining the behavior of the son at this point (abbreviating some substitutions using ``$\term{\ldots}$''):
    \begin{align*}
        & \term{\sff{Son}(s',m') \tSub{z'/m',y'/s'}}
        \\
        &= \term{\tLet s'_1 = \tSend \text{``Dune''} \, s' \tIn \ldots \tSub{z'/m',y'/s'}}
        \\
        &\reddM \term{\tLet s'_2 = \tSel \sff{buy} \, s'_1 \tIn \ldots \tSub{\tSend \text{``Dune''} \, s'/s'_1,z'/m',y'/s'}}
        \\
        &\reddM \term{\tLet m'_1 = \tSend s'_2 \, (m' \tSub{z'/m'}) \tIn \ldots \tSub{\tSel \sff{buy} \, s'_1/s'_2,\tSend \text{``Dune''} \, s'/s'_1,y'/s'}}
        \\
        &\reddM \term{\tLet m'_1 = \tSend s'_2 \, z' \tIn \ldots \tSub{\ldots}}
        \\
        &\reddM \term{\tLet (\textit{book},m'_2) = \tRecv (m'_1 \tSub{\tSend s'_2 \, z'/m'_1}) \tIn \ldots \tSub{\ldots}}
        \\
        &\reddM \term{\tLet (\textit{book},m'_2) = \tRecv (\tSend s'_2 \, z') \tIn \ldots \tSub{\ldots}}
        \deq \term{\sff{Son}^1(s',z')}
    \end{align*}
    This chain of explicit substitutions cannot be resolved further: $\term{s'_2}$ does not appear under a reduction context.
    In fact, they will not be resolved until the son has delegated his (pending) access to $\term{z'}$ to his mother, along with his pending messages in the form of explicit substitutions, as shown by the sequence leading to process $\sff{Sys}^2$ in \Cref{f:FIRST:Shop}.

    \begin{figure}[!t]
        \begin{align*}
            &\hphantom{{}\reddC{}} \term{\sff{Sys}^1}
            \\
            &\reddC^5 \term{\pRes{y\tBfr{\epsi}y'} \big( \pRes{z\tBfr{\epsi}z'} \big( \tMain \, \sff{Son}^1(s',z') \prl \tChild \, \sff{Mother}(z) \big) \prl \tChild \, \sff{Shop}(y) \big)}
            \\
            &\equivC \term{
                \pRes{y\tBfr{\epsi}y'} \big(
                    \pRes{z'\tBfr{\epsi}z} \big( \tMain \, \tLet (\textit{book},m'_2) = \tRecv (\tSend s'_2 \, z') \tIn \ldots \prl \tChild \, \ldots \big) \tSub{\ldots}
                    \prl \tChild \, \ldots
                \big)
            }
            \\
            &\reddC \term{
                \pRes{y\tBfr{\epsi}y'} \big(
                    \begin{array}[t]{@{}l@{}}
                        \pRes{z'\tBfr{s'_2}z} \big( \tMain \, \tLet (\textit{book},m'_2) = \tRecv z' \tIn \ldots \prl \tChild \, \tLet (x,m_1) = \tRecv z \tIn \ldots \big)
                        \\
                        \tSub{\ldots} \prl \tChild \, \ldots
                    \big)
                \end{array}
            }
            \\
            &\reddC \term{
                \pRes{y\tBfr{\epsi}y'} \big(
                    \pRes{z'\tBfr{\epsi}z} \big(
                        \tMain \, \ldots
                        \prl \tChild \, \tLet (x,m_1) = (s'_2,z) \tIn \ldots \tSub{\ldots}
                    \big)
                    \prl \tChild \, \ldots
                \big)
            }
            \\
            &\reddC \term{
                \pRes{y\tBfr{\epsi}y'} \big(
                    \pRes{z'\tBfr{\epsi}z} \big(
                        \begin{array}[t]{@{}l@{}}
                            \tMain \, \ldots
                            \prl \tChild \, \begin{array}[t]{@{}l@{}}
                                \tLet x_1 = \tSend \text{visa} \, (x \tSub{s'_2/x,\ldots})
                                \tIn \ldots \tSub{z/m_1}
                            \big)
                            \prl \tChild \, \ldots
                        \big)
                    \end{array}
                \end{array}
            }
            \\
            &\reddC \term{
                \pRes{y\tBfr{\epsi}y'} \big(
                    \pRes{z'\tBfr{\epsi}z} \big(
                        \tMain \, \ldots
                        \prl \tChild \, \tLet x_1 = \ldots (s'_2 \tSub{\tSel \sff{buy} \, s'_1/s'_2,\ldots})
                        \tIn \ldots
                    \big)
                    \prl \tChild \, \ldots
                \big)
            }
            \\
            &\reddC \term{
                \pRes{y\tBfr{\epsi}y'} \big(
                    \pRes{z'\tBfr{\epsi}z} \big(
                        \tMain \, \ldots
                        \prl \tChild \,
                        \begin{array}[t]{@{}l@{}}
                            \tLet x_1 = \ldots \, (\tSel \sff{buy} \, (s'_1 \tSub{\tSend \text{``Dune''} \, s'/s'_1,y'/s'}))
                            \\
                            \tIn \ldots
                        \big)
                        \prl \tChild \, \ldots
                    \big)
                \end{array}
            }
            \\
            &\reddC \term{
                \pRes{y\tBfr{\epsi}y'} \big(
                    \pRes{z'\tBfr{\epsi}z} \big(
                        \tMain \, \ldots
                        \prl \tChild \,
                        \tLet x_1 = \ldots (\ldots (\tSend \text{``Dune''} \, (s' \tSub{y'/s'})))
                        \tIn \ldots
                    \big)
                    \prl \tChild \, \ldots
                \big)
            }
            \\
            &\reddC \term{
                \pRes{y\tBfr{\epsi}y'} \big(
                    \pRes{z'\tBfr{\epsi}z} \big(
                        \tMain \, \ldots
                        \prl \tChild \,
                        \tLet x_1 = \ldots (\ldots (\ldots y'))
                        \tIn \ldots
                    \big)
                    \prl \tChild \, \ldots
                \big) \mathrel{\normalcolor\deq} \sff{Sys}^2
            }
        \end{align*}
        \caption{
            The Bookshop Scenario in \FIRST (cf.\ \Cref{x:FIRST:Shop}).
            We abbreviate unchanged terms and substitutions using ``$\term{\ldots}$''.
        }\label{f:FIRST:Shop}
    \end{figure}

    At this point, the system behaves roughly as in \Cref{x:LAST:Shop}: there is no further session delegation (as we just witnessed between the son and the mother), so upcoming explicit substitutions can be resolved straightforwardly.
    \begin{align*}
        \term{\sff{Sys}^2}
        &\reddC^3 \term{
            \pRes{y'\tBfr{\text{visa},\sff{buy},\text{``Dune''}}y} \big(
                \pRes{z'\tBfr{\epsi}z} \big(
                    \begin{array}[t]{@{}l@{}}
                        \tMain \, \tLet (\textit{book},m'_2) = \tRecv z' \tIn \ldots
                        \\
                        \prl \tChild \,
                        \tLet x_1 = y' \tIn \ldots \tSub{z/m_1}
                    \big)
                    \prl \tChild \, \sff{Shop}(y)
                \big)
            \end{array}
        }
    \end{align*}
    From here, the thread representing the mother can finally perform her son's two original outputs, followed by her own (omitted for brevity).
\end{exa}

\paragraph{Closed Sessions}

Both in \FIRST and \APCP, closed sessions have no associated behavior.
This entails some management on both sides to garbage collect unused variables with type $\type{\tEnd}$.
As a result, translating terms with variables requires two forms: one for regular variables, and another for variables typed~$\type{\tEnd}$.
Because such a translation can get burdensome, we opt to add a form of explicit closing to \FIRST.
Once both endpoints of a channel have been closed, the channel and its buffer can be garbage collected.
This treatment relieves us from translating unused variables, in such a way that the translation of terms with variables comes in one intuitive form only.
To formalize this idea, we proceed as follows:
\begin{itemize}

    \item
        We add terms for endpoint closing $\term{\tClose M ; N}$ and a reduction context $\term{\tClose \tCtx{R} ; N}$.
        Similar to $\term{\tFork}$, the term $\term{N}$ is an explicit continuation for when the endpoint is closed.
        We define behavior on the level of configurations by adding a new reduction rule:
        \[
            \begin{bussproof}[red-close]
                \bussAx{
                    \term{\pRes{x\tBfr{\vec{m}}y} ( \tCtx{F}[\tClose x ; M] \prl C )} \reddC \term{\pRes{\tNil\tBfr{\vec{m}}y} ( \tCtx{F}[M] \prl C )}
                }
            \end{bussproof}
        \]
        Here, `$\term{\tNil}$' is a special runtime endpoint variable: it indicates that the endpoint  $x$  has been closed and so the variable is not used anywhere inside the enclosed configuration.
        Then, buffered channels can be garbage collected when both endpoints have been closed, as enabled by the following reduction:
        \[
            \begin{bussproof}[red-res-nil]
                \bussAx{
                    \term{\pRes{\tNil\tBfr{\epsi}\tNil} C} \reddC \term{C}
                }
            \end{bussproof}
        \]

    \item
        We type the session closing construct by replacing Rule~\ruleLabel{typ-end} with the following:
        \[
            \begin{bussproof}[typ-close]
                \bussAssume{
                    \type{\Gamma} \vdashM \term{M} : \type{\tEnd}
                }
                \bussAssume{
                    \type{\Delta} \vdashM \term{N} : \type{T}
                }
                \bussBin{
                    \type{\Gamma} , \type{\Delta} \vdashM \term{\tClose M ; N} : \type{T}
                }
            \end{bussproof}
        \]

    \item
        We add a new session type for endpoints that have been  already closed.
        With a slight abuse of notation, we denote it as $\type{\tNil}$.
        We disallow using this type in the typing of terms, leaving it only to appear in the typing of buffered channels.
        To ensure that closed endpoint variables are not used in configurations, we add side conditions to Rule~\ruleLabel{typ-res}: $\term{x} = \term{\tNil}$ if and only if $\type{S'} = \type{\tNil}$ and similarly for $\term{y}$.
        The following two new rules enable typing partially closed buffered channels:
        \begin{mathpar}
            \begin{bussproof}[typ-buf-end-L]
                \bussAx{
                    \type{\emptyset} \vdashB \term{\tBfr{\epsi}} : \type{\tEnd}\>\type{\tNil}
                }
            \end{bussproof}
            \and
            \begin{bussproof}[typ-buf-end-R]
                \bussAx{
                    \type{\emptyset} \vdashB \term{\tBfr{\epsi}} : \type{\tNil}\>\type{\tEnd}
                }
            \end{bussproof}
        \end{mathpar}

\end{itemize}

\begin{exa}
    We illustrate the explicit closing of sessions in \FIRST in the context of the bookstore scenario introduced in \Cref{s:FIRST:example}.
    Consider the system where all session interactions have taken place, and all three threads are ready to close their sessions:
    \begin{align*}
        \term{\sff{Sys}}
        &\reddC^\ast \term{\pRes{y\tBfr{\epsi}y'} \big( \pRes{z\tBfr{\epsi}z'} \big( \tMain \, \tClose z' ; \text{book}(\text{``Dune''}) \prl \tChild \, \tClose z ; \tClose y' \big) \prl \tChild \tClose y \big)}
        \\
        &\deq \term{\sff{Sys}^c}
    \end{align*}
    Once again, notation $\term{\text{book}(\textit{title})}$ is syntactic sugar for a lookup function.
    The order in which endpoints are closed is unimportant; here, we execute the $\term{\tClose}$ primitives from left to right.
    \begin{align*}
        \term{\sff{Sys}^c}
        &\equivC \term{\pRes{y\tBfr{\epsi}y'} \big( \pRes{z'\tBfr{\epsi}z} \big( \tMain \, \tClose z' ; \text{book}(\text{``Dune''}) \prl \tChild \, \tClose z ; \tClose y' \big) \prl \tChild \tClose y \big)}
        \\
        &\reddC \term{\tMain \, \text{book}(\text{``Dune''}) \prl \pRes{y\tBfr{\epsi}y'} \big( \pRes{\tNil\tBfr{\epsi}z} \tChild \, \tClose z ; \tClose y' \prl \tChild \tClose y \big)}
        \\
        &\equivC \term{\tMain \, \text{book}(\text{``Dune''}) \prl \pRes{y\tBfr{\epsi}y'} \big( \pRes{z\tBfr{\epsi}\tNil} \tChild \, \tClose z ; \tClose y' \prl \tChild \tClose y \big)}
        \\
        &\reddC \term{\tMain \, \text{book}(\text{``Dune''}) \prl \pRes{y\tBfr{\epsi}y'} \big( \pRes{\tNil\tBfr{\epsi}\tNil} \tChild \, \tClose y' \prl \tChild \tClose y \big)}
        \\
        &\reddC \term{\tMain \, \text{book}(\text{``Dune''}) \prl \pRes{y\tBfr{\epsi}y'} \big( \tChild \, \tClose y' \prl \tChild \tClose y \big)}
        \\
        &\equivC \term{\tMain \, \text{book}(\text{``Dune''}) \prl \pRes{y'\tBfr{\epsi}y} \big( \tChild \, \tClose y' \prl \tChild \tClose y \big)}
        \\
        &\reddC \term{\tMain \, \text{book}(\text{``Dune''}) \prl \tChild \, () \prl \pRes{\tNil\tBfr{\epsi}y} \tChild \tClose y}
        \\
        &\reddC \term{\tMain \, \text{book}(\text{``Dune''}) \prl \pRes{\tNil\tBfr{\epsi}y} \tChild \tClose y}
        \\
        &\equivC \term{\tMain \, \text{book}(\text{``Dune''}) \prl \pRes{y\tBfr{\epsi}\tNil} \tChild \tClose y}
        \tag{$\ast$}
        \\
        &\reddC \term{\tMain \, \text{book}(\text{``Dune''}) \prl \pRes{\tNil\tBfr{\epsi}\tNil} \tChild ()}
        \\
        &\reddC \term{\tMain \, \text{book}(\text{``Dune''}) \prl \tChild ()}
        \\
        &\reddC \term{\tMain \, \text{book}(\text{``Dune''})}
    \end{align*}
\noindent 
    Additionally, we illustrate the typing of half-closed sessions on the configuration marked~($\ast$).
    Recall from \Cref{x:LAST:typTerm} that we consider $\type{\sff{B}}$ (book) a primitive non-linear type that can be weakened/contracted at will and is self dual.
    We have (omitting ``\textsc{typ-}'' from rule labels):
    \[
        \begin{bussproof}
            \def\defaultHypSeparation{\hskip1ex}
            \def\ScoreOverhang{1pt}
            \bussAx{
                \type{\emptyset} \vdashM \term{\text{book}(\text{``Dune''})} : \type{\sff{B}}
            }
            \bussUn[\ruleLabel{main}\hspace{-2em}]{ 
                \type{\emptyset} \vdashC{\tMain} \term{\tMain \, \text{book}(\text{``Dune''})} : \type{\sff{B}}
            }
            \bussAx[\ruleLabel{var}]{
                \term{y}:\type{\tEnd} \vdashM \term{y} : \type{\tEnd}
            }
            \bussAx[\ruleLabel{unit}\hspace{-4em}]{
                \type{\emptyset} \vdashM \term{()} : \type{\1}
            }
            \bussBin[\ruleLabel{close}\hspace{-6em}]{
                \term{y}:\type{\tEnd} \vdashM \term{\tClose y ; ()} : \type{\1}
            }
            \bussUn[\ruleLabel{child}]{
                \term{y}:\type{\tEnd} \vdashC{\tChild} \term{\tChild \, \tClose y ; ()} : \type{\1}
            }
            \bussAx[\ruleLabel{buf-end-L}]{
                \type{\emptyset} \vdashB \term{\tBfr{\epsi}} : \type{\tEnd} \> \type{\tNil}
            }
            \bussBin[\ruleLabel{res}]{
                \type{\emptyset} \vdashC{\tChild} \term{\pRes{y\tBfr{\epsi}\tNil} \tChild \, \tClose y ; ()} : \type{\1}
            }
            \bussBin[\ruleLabel{par}]{
                \type{\emptyset} \vdashC{\tMain} \term{\tMain \, \text{book}(\text{``Dune''}) \prl \pRes{y\tBfr{\epsi}\tNil} \tChild \, \tClose y ; ()} : \type{\sff{B}}
            }
        \end{bussproof}
        \tag*{\qedhere}
    \]
\end{exa}

\paragraph{Type Preservation}

\FIRST satisfies the expected correctness properties for session-typed languages: protocol fidelity and communication safety.
Both properties follow directly from \emph{type preservation}.

\begin{restatable}[Type Preservation for \protect\FIRST]{theorem}{tFIRSTTp}
\label{t:FIRST:tp}
    Given $\type{\Gamma} \vdashC{\phi} \term{C} : \type{T}$, if $\term{C} \equivC \term{D}$ or $\term{C} \reddC \term{D}$, then $\type{\Gamma} \vdashC{\phi} \term{D} : \type{T}$.
\end{restatable}

\begin{proof}
    By proving subject congruence ($\equivC$ preserves typing) and subject reduction ($\reddC$ preserves typing) separately.
    In both cases, we first prove separately subject congruence and subject reduction on the term level ($\equivM$ and $\reddM$, respectively).
    These results follow by induction on the derivation of the structural congruence or reduction.
    Below we give a few interesting cases; \Cref{as:FIRST:tp} contains detailed proofs.
    \begin{itemize}

        \item
            For $\equivM$, the only interesting case is the only base case, i.e.,%
            \global\def\proofScTermsSubExt{%
                Rule~\ruleLabel{sc-sub-ext}:
                \[
                    \term{x} \notin \fv(\tCtx{R}) \implies \term{(\tCtx{R}[M]) \tSub{ N/x }} \equivM \tCtx{R}[M \tSub{ N/x }]
                \]
                \noindent 
                We apply induction on the structure of the reduction context $\tCtx{R}$.
                As an interesting, representative case, consider $\tCtx{R} = \term{L \tSub{ \tCtx{R'}/y }}$.
                Assuming $\term{x} \notin \fv(\tCtx{R})$, we have $\term{x} \notin \fv(\term{L}) \cup \fv(\tCtx{R'})$.
                We apply inversion of typing:
                \[
                    \begin{bussproof}
                        \bussAssume{
                            \type{\Gamma}, \term{y}:\type{U} \vdashM \term{L}: \type{T}
                        }
                        \bussAssume{
                            \type{\Delta}, \term{x}:\type{U'} \vdashM \tCtx{R'}[M]: \type{U}
                        }
                        \bussBin[\ruleLabel{typ-sub}]{
                            \type{\Gamma}, \type{\Delta}, \term{x}:\type{U'} \vdashM \term{L \tSub{ (\tCtx{R'}[M])/y }}: \type{T}
                        }
                        \bussAssume{
                            \type{\Delta'} \vdashM \term{N}: \type{U'}
                        }
                        \bussBin[\ruleLabel{typ-sub}]{
                            \type{\Gamma}, \type{\Delta}, \type{\Delta'} \vdashM \term{(L \tSub{ (\tCtx{R'}[M])/y }) \tSub{ N/x }}: \type{T}
                        }
                    \end{bussproof}
                \]
                We can derive:
                \[
                    \begin{bussproof}
                        \bussAssume{
                            \type{\Delta}, \term{x}:\type{U'} \vdashM \tCtx{R'}[M]: \type{U}
                        }
                        \bussAssume{
                            \type{\Delta'} \vdashM \term{N}: \type{U'}
                        }
                        \bussBin[\ruleLabel{typ-sub}]{
                            \type{\Delta}, \type{\Delta'} \vdashM \term{(\tCtx{R'}[M]) \tSub{ N/x }}: \type{U}
                        }
                    \end{bussproof}
                \]
                Since $\term{x} \notin \fv(\tCtx{R'})$, by Rule~\ruleLabel{sc-sub-ext}, $\term{(\tCtx{R'}[M]) \tSub{ N/x }} \equivM \tCtx{R'}[M \tSub{ N/x }]$.
                Then, by the IH, $\type{\Delta}, \type{\Delta'} \vdashM \tCtx{R'}[M \tSub{ N/x }]: \type{U}$.
                Hence, we can conclude the following:
                \[
                    \begin{bussproof}
                        \bussAssume{
                            \type{\Gamma}, \term{y}:\type{U} \vdashM \term{L}: \type{T}
                        }
                        \bussAssume{
                            \type{\Delta}, \type{\Delta'} \vdashM \tCtx{R'}[M \tSub{ N/x }]: \type{U}
                        }
                        \bussBin[\ruleLabel{typ-sub}]{
                            \type{\Gamma}, \type{\Delta}, \type{\Delta'} \vdashM \term{L \tSub{ (\tCtx{R'}[M \tSub{ N/x }])/y }}: \type{T}
                        }
                    \end{bussproof}
                \]
                It is straightforward to see that this reasoning works in opposite direction as well.%
            }
            \proofScTermsSubExt

        \item
            For $\reddM$, all cases are straightforward.

        \item
            For $\equivC$, we detail the base case%
            \global\def\proofScResSwap{%
                Rule~\ruleLabel{sc-res-swap}: $\term{\pRes{x\tBfr{\epsi}y} C} \equivC \term{\pRes{y\tBfr{\epsi}x} C}$.

                The analysis depends on whether exactly one of $\term{x},\term{y}$ is $\term{\tNil}$ or not.
                We discuss both cases.
                \begin{itemize}

                    \item
                        Exactly one of $\term{x},\term{y}$ is $\term{\tNil}$; w.l.o.g., assume $\term{x} = \term{\tNil}$.
                        We have the following:
                        \begin{align*}
                            & \begin{bussproof}
                                \bussAx[\ruleLabel{typ-buf-end-R}]{
                                    \type{\emptyset} \vdashB \term{\tBfr{\epsi}} : \type{\tNil}\>\type{\tEnd}
                                }
                                \bussAssume{
                                    \type{\Gamma} , \term{x} : \type{\tEnd} \vdashC{\phi} \term{C} : \type{T}
                                }
                                \bussBin[\ruleLabel{typ-res}]{
                                    \type{\Gamma} \vdashC{\phi} \term{\pRes{\tNil\tBfr{\epsi}y} C} : \type{T}
                                }
                            \end{bussproof}
                            \\
                            & \equivC
                            \\
                            & \begin{bussproof}
                                \bussAx[\ruleLabel{typ-buf-end-L}]{
                                    \type{\emptyset} \vdashB \term{\tBfr{\epsi}} : \type{\tEnd}\>\type{\tNil}
                                }
                                \bussAssume{
                                    \type{\Gamma} , \term{x} : \type{\tEnd} \vdashC{\phi} \term{C} : \type{T}
                                }
                                \bussBin[\ruleLabel{typ-res}]{
                                    \type{\Gamma} \vdashC{\phi} \term{\pRes{y\tBfr{\epsi}\tNil} C} : \type{T}
                                }
                            \end{bussproof}
                        \end{align*}

                    \item
                        Neither or both of $\term{x},\term{y}$ are $\term{\tNil}$.
                        We have the following:
                        \begin{align*}
                            & \begin{bussproof}
                                \bussAx[\ruleLabel{typ-buf}]{
                                    \type{\emptyset} \vdashB \term{\tBfr{\epsi}} : \type{S'}\>\type{S'}
                                }
                                \bussAssume{
                                    \type{\Gamma} , \term{x}:\type{S'} , \term{y}:\type{\ol{S'}} \vdashC{\phi} \term{C} : \type{T}
                                }
                                \bussBin[\ruleLabel{typ-res}]{
                                    \type{\Gamma} \vdashC{\phi} \term{\pRes{x\tBfr{\epsi}y} C} : \type{T}
                                }
                            \end{bussproof}
                            \\
                            & \equivC
                            \\
                            & \begin{bussproof}
                                \bussAx[\ruleLabel{typ-buf}]{
                                    \type{\emptyset} \vdashB \term{\tBfr{\epsi}} : \type{\ol{S'}}\>\type{\ol{S'}}
                                }
                                \bussAssume{
                                    \type{\Gamma} , \term{x}:\type{S'} , \term{y}:\type{\ol{S'}} \vdashC{\phi} \term{C} : \type{T}
                                }
                                \bussBin[\ruleLabel{typ-res}]{
                                    \type{\Gamma} \vdashC{\phi} \term{\pRes{y\tBfr{\epsi}x} C} : \type{T}
                                }
                            \end{bussproof}
                        \end{align*}

                \end{itemize}%
            }
            \proofScResSwap

        \item
            For $\reddC$, we detail the base case
            \global\def\proofSrSend{%
                Rule~\ruleLabel{red-send}: $\term{\pRes{x\tBfr{\vec{m}}y}(\tCtx{\hat{F}}[\tSend M \, x] \prl C)} \reddC \term{\pRes{x\tBfr{M,\vec{m}}y}(\tCtx{\hat{F}}[x] \prl C)}$.

                This case follows by induction on the structure of $\tCtx{\hat{F}}$.
                The inductive cases follow from the IH straightforwardly.
                The fact that the hole in $\tCtx{\hat{F}}$ does not occur under an explicit substitution guarantees that we can move $\term{M}$ out of the context of $\tCtx{\hat{F}}$ and into the buffer.
                We consider the base case ($\tCtx{\hat{F}} = \term{\phi\,\tCtx{R}}$).
                By well typedness, it must be that $\tCtx{R} = \tCtx{R_1}[\tClose \tCtx{R_2} ; M]$.
                We apply induction on the structures of $\tCtx{R_1},\tCtx{R_2}$ and consider the base cases: $\tCtx{R_1} = \tCtx{R_2} = \tHole$.
                We apply inversion of typing, w.l.o.g.\ assuming that $\term{\phi} = \term{\tMain}$ and $\term{y} \in \fv(\term{C})$ (omitting ``\textsc{typ-}'' from rule labels):
                \[
                    \def\defaultHypSeparation{\hskip1ex}
                    \def\ScoreOverhang{.9pt}
                    \begin{bussproof}
                        \bussAssume{
                            \type{\Gamma} \vdashB \term{\tBfr{\vec{m}}}: \mathrlap{\type{{!}T . \tEnd} \> \type{S}}
                        }
                        \bussAssume{
                            \type{\Delta_1} \vdashM \term{M}: \type{T}
                        }
                        \bussAx[\ruleLabel{var}]{
                            \term{x}:\type{{!}T . \tEnd} \vdashM \term{x}: \type{{!}T . \tEnd}
                        }
                        \bussBin[\ruleLabel{send}]{
                            \type{\Delta_1}, \term{x}:\type{{!}T . \tEnd} \vdashM \term{\tSend M \, x}: \type{\tEnd}
                        }
                        \bussAssume{
                            \type{\Delta_2} \vdashM \term{N} : \type{U}
                        }
                        \bussBin[\ruleLabel{close}]{
                            \type{\Delta_1} , \type{\Delta_2} , \term{x}:\type{{!}T . \tEnd} \vdashM \term{\tClose (\tSend M \, x) ; N} : \type{U}
                        }
                        \bussUn[\ruleLabel{main}]{
                            \type{\Delta_1}, \type{\Delta_2} , \term{x}:\type{{!}T . \tEnd} \vdashC{\tMain} \term{\tMain\, (\tClose (\tSend M \, x) ; N)}: \type{U}
                        }
                        \bussAssume{
                            \hspace{-4.5ex} 
                            \type{\Lambda}, \term{y}:\type{\ol{S}} \vdashC{\tChild} \term{C}: \type{\1}
                        }
                        \bussBin[\hspace{-4.2pt}\ruleLabel{par}]{ 
                            \type{\Delta_1} , \type{\Delta_2} , \type{\Lambda}, \term{x}:\type{{!}T . \tEnd}, \term{y}:\type{\ol{S}} \vdashC{\tMain} \term{\tMain\, (\tClose (\tSend M \, x) ; N) \prl C}: \type{U}
                        }
                        \bussBin[\ruleLabel{res}]{
                            \type{\Gamma}, \type{\Delta_1} , \type{\Delta_2}, \type{\Lambda} \vdashC{\tMain} \term{\pRes{x\tBfr{\vec{m}}y}(\tMain\, (\tClose (\tSend M \, x) ; N) \prl C)}: \type{U}
                        }
                    \end{bussproof}
                \]

                Note that the derivation of $\type{\Gamma} \vdashB \term{\tBfr{\vec{m}}}: \type{{!}T . \tEnd} \> \type{S}$ depends on the size of $\term{\vec{m}}$.
                By induction on the size of $\term{\vec{m}}$ (\ih{2}), we derive $\type{\Gamma}, \type{\Delta_1} \vdashB \term{\tBfr{M,\vec{m}}}: \type{\tEnd} \> \type{S}$:
                \begin{itemize}

                    \item
                        If $\term{\vec{m}}$ is empty, it follows by inversion of typing that $\type{\Gamma} = \type{\emptyset}$ and $\type{S} = \type{{!}T . \tEnd}$:
                        \[
                            \begin{bussproof}
                                \bussAx[\ruleLabel{typ-buf}]{
                                    \type{\emptyset} \vdashB \term{\tBfr{\epsi}}: \type{{!}T . \tEnd} \> \type{{!}T . \tEnd}
                                }
                            \end{bussproof}
                        \]
                        Then, we derive the following:
                        \[
                            \begin{bussproof}
                                \bussAssume{
                                    \type{\Delta_1} \vdashM \term{M}: \type{T}
                                }
                                \bussAx[\ruleLabel{typ-buf}]{
                                    \type{\emptyset} \vdashB \term{\tBfr{\epsi}}: \type{\tEnd} \> \type{\tEnd}
                                }
                                \bussBin[\ruleLabel{typ-buf-send}]{
                                    \type{\Delta_1} \vdashB \term{\tBfr{M}}: \type{\tEnd} \> \type{{!}T . \tEnd}
                                }
                            \end{bussproof}
                        \]

                    \item
                        If $\term{\vec{m}} = \term{\vec{m}',L}$, it follows by inversion of typing that $\type{\Gamma} = \type{\Gamma'},\type{\Gamma''}$ and $\type{S} = \type{{!}T' . \tEnd'}$:
                        \[
                            \begin{bussproof}
                                \bussAssume{
                                    \type{\Gamma'} \vdashM \term{L}: \type{T'}
                                }
                                \bussAssume{
                                    \type{\Gamma''} \vdashB \term{\tBfr{\vec{m}'}}: \type{{!}T . \tEnd} \> \type{\tEnd'}
                                }
                                \bussBin[\ruleLabel{typ-buf-send}]{
                                    \type{\Gamma'}, \type{\Gamma''} \vdashB \term{\tBfr{\vec{m}',L}}: \type{{!}T . \tEnd} \> \type{{!}T' . \tEnd'}
                                }
                            \end{bussproof}
                        \]
                        By \ih{2}, $\type{\Gamma''}, \type{\Delta_1} \vdashB \term{\tBfr{M,\vec{m}'}}: \type{\tEnd} \> \type{\tEnd'}$, allowing us to derive the following:
                        \[
                            \begin{bussproof}
                                \bussAssume{
                                    \type{\Gamma'} \vdashM \term{L}: \type{T'}
                                }
                                \bussAssume{
                                    \type{\Gamma''}, \type{\Delta_1} \vdashB \term{\tBfr{M,\vec{m}'}}: \type{\tEnd} \> \type{\tEnd'}
                                }
                                \bussBin[\ruleLabel{typ-buf-send}]{
                                    \type{\Gamma'}, \type{\Gamma''}, \type{\Delta_1} \vdashB \term{\tBfr{M,\vec{m}',L}}: \type{\tEnd} \> \type{{!}T' . \tEnd'}
                                }
                            \end{bussproof}
                        \]

                    \item
                        If $\term{\vec{m}} = \term{\vec{m}', j}$, it follows by inversion of typing that there exist types $\type{S_i}$ for each $i$ in a set of labels $I$, where $j \in I$, such that $\type{S} = \type{\oplus \{ i : S_i \}_{i \in I}}$:
                        \[
                            \begin{bussproof}
                                \bussAssume{
                                    \type{\Gamma} \vdashB \term{\tBfr{\vec{m}'}}: \type{{!}T . \tEnd} \> \type{S_j}
                                }
                                \bussUn[\ruleLabel{typ-buf-sel}]{
                                    \type{\Gamma} \vdashB \term{\tBfr{\vec{m},j}}: \type{{!}T . \tEnd} \> \type{\oplus \{ i : S_i \}_{i \in I}}
                                }
                            \end{bussproof}
                        \]
                        By \ih{2}, $\type{\Gamma}, \type{\Delta_1} \vdashB \term{\tBfr{M,\vec{m}'}}: \type{\tEnd} \> \type{S_j}$, so we derive the following:
                        \[
                            \begin{bussproof}
                                \bussAssume{
                                    \type{\Gamma}, \type{\Delta_1} \vdashB \term{\tBfr{M,\vec{m}'}}: \type{\tEnd} \> \type{S_j}
                                }
                                \bussUn[\ruleLabel{typ-buf-sel}]{
                                    \type{\Gamma}, \type{\Delta_1} \vdashB \term{\tBfr{M,\vec{m}',j}}: \type{\tEnd} \> \type{\oplus \{ i : S_i \}_{i \in I}}
                                }
                            \end{bussproof}
                        \]

                \end{itemize}
                \noindent 
                Now, we can derive the typing of the structurally congruent configuration (omitting ``\textsc{typ-}'' from rule labels):
                \[
                    \def\defaultHypSeparation{\hskip1ex}
                    \def\ScoreOverhang{1pt}
                    \begin{bussproof}
                        \bussAssume{
                            \type{\Gamma}, \type{\Delta_1} \vdashB \term{\tBfr{M,\vec{m}}}: \type{\tEnd} \> \type{S}
                        }
                        \bussAx[\ruleLabel{var}]{
                            \term{x}:\type{\tEnd} \vdashM \term{x}: \type{\tEnd}
                        }
                        \bussAssume{
                            \type{\Delta_2} \vdashM \term{N} : \type{U}
                        }
                        \bussBin[\ruleLabel{close}]{
                            \type{\Delta_2} , \term{x}:\type{\tEnd} \vdashM \term{\tClose x ; N} : \type{U}
                        }
                        \bussUn[\ruleLabel{main}]{
                            \type{\Delta_2} , \term{x}:\type{\tEnd} \vdashC{\tMain} \term{\tMain\, (\tClose x ; N)}: \type{U}
                        }
                        \bussAssume{
                            \type{\Lambda}, \term{y}:\type{\ol{S}} \vdashC{\tChild} \term{C}: \type{\1}
                        }
                        \bussBin[\ruleLabel{par-R}]{
                            \type{\Delta_2} , \type{\Lambda}, \term{x}:\type{\tEnd}, \term{y}:\type{\ol{S}} \vdashC{\tMain} \term{\tMain\, (\tClose x ; N) \prl C}: \type{U}
                        }
                        \bussBin[\ruleLabel{res}]{
                            \type{\Gamma}, \type{\Delta_1} , \type{\Delta_2} , \type{\Lambda} \vdashC{\tMain} \term{\pRes{x\tBfr{M,\vec{m}}y} ( \tMain\, (\tClose x ; N) \prl C )}: \type{U}
                        }
                    \end{bussproof}
                \]
            }
            \proofSrSend
            \vspace{-2.7\baselineskip} 

    \end{itemize}

\end{proof}

\subsection{Faithfully Translating \FIRST into \APCP}
\label{s:FIRST:trans}

We now define a typed translation from \FIRST into \APCP.
Then, we show that this translation is operationally complete and sound, in the sense of \Cref{s:LAST:trans}.

\paragraph{The Translation: Key Ideas}

Before we define the translation, let us discuss its key idea, inspired by Milner's translation of the \emph{lazy} \lamcalc~\cite{journal/mscs/Milner92}.
The most important design decision is how to translate variables.
Variables serve two purposes: (i)~as placeholders for future substitutions and (ii)~as an access point to buffered channels.
Accordingly, our translation uses variables as sends that enable (i)~an explicit substitution or (ii)~interactions with a buffer.
Another important design decision is the translation of buffers: the buffer's types guide the translation to form a sequence of inputs/outputs that are explicitly forwarded (i.e., not using the forwarder process) between the buffer endpoints.

\paragraph{The Translation: Types}

Equipped with these ideas about translating variables and buffers, let us give a deeper insight into our translation by taking a look at how it turns functional and session types in \FIRST into session types in \APCP.

Before moving on, note that \FIRST does not guarantee deadlock freedom by typing.
As such, there is no point in annotating types with priorities, as the priority requirements induced by typing translated processes may not always be satisfiable.
We therefore use a special \APCP typing judgment
$$~ \vdash* P ::\Gamma~$$
which indicates well typedness of $P$ modulo priority annotations and requirements in $\Gamma$ (we omit $\Omega$, as \FIRST is a finite language).
We will recover priorities and deadlock freedom in \Cref{s:FIRST:df}.

\Cref{d:FIRST:transTypes} below introduces two forms of type translation: while we use $\enct{-}$ to translate the types of terms/configurations, we use $\enct*{-}$ to translate the types of variables.
We write $\enct*{\Gamma}$ to denote a component-wise translation, where each assignment $\term{x}:\type{T}$ translates to an assignment $x:\enct*{T}$.
As such, our \emph{typed} translation takes, e.g., a typed term $\type{\Gamma} \vdashM \term{M} : \type{T}$ and returns a typed process
\[
    \vdash* \encc{z}{M} :: \enct*{\Gamma} , z:\enct{T}.
\]
Above, $\encc{z}{M}$ is the process that models the behavior of $\term{M}$ on a fresh name $z$ (defined hereafter).

Let us now define and explain the two forms of translation of types (that are mutually recursive):
\begin{defi}[Translation of Types]
    \label{d:FIRST:transTypes}
    \begin{align*}
        \enct*{T} &\deq \bullet \tensor \ol{\enct{T}} \quad \text{(if $\type{T} \neq \type{\tNil}$)}
        \span\span
        \\
        \enct{T \times U} &\deq \ol{\enct*{T}} \tensor \ol{\enct*{U}}
        &
        \enct{T \lolli U} &\deq \enct*{T} \parr \enct{U}
        &
        \enct{\1} &\deq \bullet
        \\
        \enct{{!}T.S} &\deq \bullet \tensor \enct*{T} \parr \ol{\enct*{S}}
        &
        \enct{\oplus \{ i : S_i \}_{i \in I}} &\deq \bullet \tensor \& \{ i : \ol{\enct*{S_i}} \}_{i \in I}
        &
        \enct{\tEnd} &\deq \bullet \tensor \bullet
        \\
        \enct{{?}T.S} &\deq \ol{\enct*{T}} \tensor \ol{\enct*{S}}
        &
        \enct{\& \{ i : S_i \}_{i \in I}} &\deq \oplus \{ i : \ol{\enct*{S_i}} \}_{i \in I}
        &
        \enct{\tNil} &\deq \enct*{\tNil} \deq \bullet
    \end{align*}
\end{defi}
Some intuitions follow:
\begin{itemize}

    \item
        Following the intuitions above, the translation $\enct*{-}$ (row~1) codifies a variable as a send action that enables further behaviors.

    \item
        The translation $\enct{-}$ translates functional types straightforwardly (row~2): pairs send access to their components as variables, abstractions receive their parameter as a variable and then provide the return type, and units have no behavior.

    \item
        The translation of session types (rows~3 and~4) is more interesting: at a first glance they seem to be translated dually.
        This is because these are types that belong to buffered channel endpoints.
        As such, when a variable is typed $\type{{!}T.S}$ and we send a term on this variable, the translation sends something along the variable and thus the variable receives.
        Additionally, the translation of sends and selections have an extra send: this signifies a handshake between the variable and the buffer, indicating that both are ready to perform the actual send/selection.

\end{itemize}

\begin{exa}
    \label{x:FIRST:transTypes}
    We illustrate the translation of types by means of an example.
    After giving the translation of the type $\type{(T \times S) \lolli {!}T.S}$, we break the resulting \APCP type down and explain it in terms of the associated behavior of a process translated from a term implementing the type.
    \begin{align*}
        \enct{(T \times S) \lolli {!}T.S}
        &= \enct*{T \times S} \parr \enct{{!}T.S}
        \\
        &= ( \bullet \tensor \ol{\enct{T \times S}} ) \parr \bullet \tensor \enct*{T} \parr \ol{\enct*{S}}
        \\
        &= ( \bullet \tensor \ol{\ol{\enct*{T}} \tensor \ol{\enct*{S}}} ) \parr \bullet \tensor ( \bullet \tensor \ol{\enct{T}} ) \parr \ol{\bullet \tensor \ol{\enct{S}}}
        \\
        &= ( \bullet \tensor \enct*{T} \parr \enct*{S} ) \parr \bullet \tensor ( \bullet \tensor \ol{\enct{T}} ) \parr \bullet \parr \enct{S}
        \\
        &= ( \bullet \tensor ( \bullet \tensor \ol{\enct{T}} ) \parr \bullet \tensor \ol{\enct{S}} ) \parr \bullet \tensor ( \bullet \tensor \ol{\enct{T}} ) \parr \bullet \parr \enct{S}
    \end{align*}
         \Cref{f:FIRST:transTypes} gives a detailed explanation for this translated type.
     \end{exa}

\begin{figure}[!t]
    \begin{align*}
        & ( \bullet \tensor {}
        &&
        \tikzmark{x:FIRST:transTypes:1s}
        \tikzmark{x:FIRST:transTypes:11s}
        \tikzmark{x:FIRST:transTypes:11e}
        \tag*{announce pair ready}
        \\
        & \q[1] ( \bullet \tensor {}
        &&
        \tikzmark{x:FIRST:transTypes:12s}
        \tikzmark{x:FIRST:transTypes:121s}
        \tikzmark{x:FIRST:transTypes:121e}
        \tag*{announce substitution ready}
        \\
        & \q[2] \ol{\enct{T}}
        &&
        \tikzmark{x:FIRST:transTypes:122s}
        \tikzmark{x:FIRST:transTypes:122e}
        \tag*{actual first component}
        \\
        & \q[1] ) \parr {}
        &&
        \tikzmark{x:FIRST:transTypes:12e}
        \tag*{receive first component}
        \\
        & \q[1] \bullet \tensor {}
        &&
        \tikzmark{x:FIRST:transTypes:13s}
        \tikzmark{x:FIRST:transTypes:13e}
        \tag*{announce substitution ready}
        \\
        & \q[1] \ol{\enct{S}}
        &&
        \tikzmark{x:FIRST:transTypes:14s}
        \tikzmark{x:FIRST:transTypes:14e}
        \tag*{actual second component}
        \\
        & ) \parr {}
        &&
        \tikzmark{x:FIRST:transTypes:1e}
        \tag*{receive parameter}
        \\
        & \bullet \tensor {}
        &&
        \tikzmark{x:FIRST:transTypes:2s}
        \tikzmark{x:FIRST:transTypes:2e}
        \tag*{trigger buffer}
        \\
        & ( \bullet \tensor {}
        &&
        \tikzmark{x:FIRST:transTypes:3s}
        \tikzmark{x:FIRST:transTypes:31s}
        \tikzmark{x:FIRST:transTypes:31e}
        \tag*{announce substitution ready}
        \\
        & \q[1] \ol{\enct{T}}
        &&
        \tikzmark{x:FIRST:transTypes:32s}
        \tikzmark{x:FIRST:transTypes:32e}
        \tag*{actual payload}
        \\
        & ) \parr {}
        &&
        \tikzmark{x:FIRST:transTypes:3e}
        \tag*{receive payload provider}
        \\
        & \bullet \parr {}
        &&
        \tikzmark{x:FIRST:transTypes:4s}
        \tikzmark{x:FIRST:transTypes:4e}
        \tag*{await substitution ready}
        \\
        & \enct{S}
        &&
        \tikzmark{x:FIRST:transTypes:5s}
        \tikzmark{x:FIRST:transTypes:5e}
        \tag*{actual continuation}
    \end{align*}
    \begin{tikzpicture}[overlay,remember picture]
        \draw ([yshift=2ex]pic cs:x:FIRST:transTypes:1s) to [square left brace] (pic cs:x:FIRST:transTypes:1e);
        \draw ([yshift=1ex,xshift=.1cm]pic cs:x:FIRST:transTypes:1e) to ([yshift=1ex,xshift=.4cm]pic cs:x:FIRST:transTypes:1e);
        \draw ([yshift=2ex]pic cs:x:FIRST:transTypes:2s) to [square left brace] (pic cs:x:FIRST:transTypes:2e);
        \draw ([yshift=1ex,xshift=.1cm]pic cs:x:FIRST:transTypes:2e) to ([yshift=1ex,xshift=.4cm]pic cs:x:FIRST:transTypes:2e);
        \draw ([yshift=2ex]pic cs:x:FIRST:transTypes:3s) to [square left brace] (pic cs:x:FIRST:transTypes:3e);
        \draw ([yshift=1ex,xshift=.1cm]pic cs:x:FIRST:transTypes:3e) to ([yshift=1ex,xshift=.4cm]pic cs:x:FIRST:transTypes:3e);
        \draw ([yshift=2ex]pic cs:x:FIRST:transTypes:4s) to [square left brace] (pic cs:x:FIRST:transTypes:4e);
        \draw ([yshift=1ex,xshift=.1cm]pic cs:x:FIRST:transTypes:4e) to ([yshift=1ex,xshift=.4cm]pic cs:x:FIRST:transTypes:4e);
        \draw ([yshift=2ex]pic cs:x:FIRST:transTypes:5s) to [square left brace] (pic cs:x:FIRST:transTypes:5e);
        \draw ([yshift=1ex,xshift=.1cm]pic cs:x:FIRST:transTypes:5e) to ([yshift=1ex,xshift=.4cm]pic cs:x:FIRST:transTypes:5e);
        \draw ([yshift=2ex,xshift=.2cm]pic cs:x:FIRST:transTypes:11s) to [square left brace] ([xshift=.2cm]pic cs:x:FIRST:transTypes:11e);
        \draw ([yshift=1ex,xshift=.3cm]pic cs:x:FIRST:transTypes:11e) to ([yshift=1ex,xshift=.6cm]pic cs:x:FIRST:transTypes:11e);
        \draw ([yshift=2ex,xshift=.2cm]pic cs:x:FIRST:transTypes:12s) to [square left brace] ([xshift=.2cm]pic cs:x:FIRST:transTypes:12e);
        \draw ([yshift=1ex,xshift=.3cm]pic cs:x:FIRST:transTypes:12e) to ([yshift=1ex,xshift=.6cm]pic cs:x:FIRST:transTypes:12e);
        \draw ([yshift=2ex,xshift=.2cm]pic cs:x:FIRST:transTypes:13s) to [square left brace] ([xshift=.2cm]pic cs:x:FIRST:transTypes:13e);
        \draw ([yshift=1ex,xshift=.3cm]pic cs:x:FIRST:transTypes:13e) to ([yshift=1ex,xshift=.6cm]pic cs:x:FIRST:transTypes:13e);
        \draw ([yshift=2ex,xshift=.2cm]pic cs:x:FIRST:transTypes:14s) to [square left brace] ([xshift=.2cm]pic cs:x:FIRST:transTypes:14e);
        \draw ([yshift=1ex,xshift=.3cm]pic cs:x:FIRST:transTypes:14e) to ([yshift=1ex,xshift=.6cm]pic cs:x:FIRST:transTypes:14e);
        \draw ([yshift=2ex,xshift=.2cm]pic cs:x:FIRST:transTypes:31s) to [square left brace] ([xshift=.2cm]pic cs:x:FIRST:transTypes:31e);
        \draw ([yshift=1ex,xshift=.3cm]pic cs:x:FIRST:transTypes:31e) to ([yshift=1ex,xshift=.6cm]pic cs:x:FIRST:transTypes:31e);
        \draw ([yshift=2ex,xshift=.2cm]pic cs:x:FIRST:transTypes:32s) to [square left brace] ([xshift=.2cm]pic cs:x:FIRST:transTypes:32e);
        \draw ([yshift=1ex,xshift=.3cm]pic cs:x:FIRST:transTypes:32e) to ([yshift=1ex,xshift=.6cm]pic cs:x:FIRST:transTypes:32e);
        \draw ([yshift=2ex,xshift=.4cm]pic cs:x:FIRST:transTypes:121s) to [square left brace] ([xshift=.4cm]pic cs:x:FIRST:transTypes:121e);
        \draw ([yshift=1ex,xshift=.5cm]pic cs:x:FIRST:transTypes:121e) to ([yshift=1ex,xshift=.8cm]pic cs:x:FIRST:transTypes:121e);
        \draw ([yshift=2ex,xshift=.4cm]pic cs:x:FIRST:transTypes:122s) to [square left brace] ([xshift=.4cm]pic cs:x:FIRST:transTypes:122e);
        \draw ([yshift=1ex,xshift=.5cm]pic cs:x:FIRST:transTypes:122e) to ([yshift=1ex,xshift=.8cm]pic cs:x:FIRST:transTypes:122e);
    \end{tikzpicture}
\caption{\Cref{x:FIRST:transTypes}: the translation for $\type{(T \times S) \lolli {!}T.S}$, in detail.\label{f:FIRST:transTypes}}
\end{figure}

\paragraph{The Translation: Terms, Configurations, and Buffers}

Now we complete the definition of our translation.
We proceed inductively on the typing derivations of terms, configurations, and buffers, obtaining typing derivations of \APCP processes.
Hence, the definition considers the typing rules in \Cref{f:LAST:type}, including the changes discussed in \Cref{s:FIRST:language}.
\begin{itemize}
	\item
The translation of a well-typed term, denoted $\encc{z}{\normalcolor \type{\Gamma} \vdashM \term{M} : \type{T}}$, corresponds to a judgment ${\vdash* P :: \Delta , z:A}$ in \APCP, where $z$ is a fresh name that executes the behavior of $\term{M}$, as usual, and the shape of the process $P$, context $\Delta$, and session type $A$ will become precise shortly.
  Similarly, the translation of a well-typed configuration, denoted $\encc{z}{\normalcolor \type{\Gamma} \vdashC{\phi} \term{C} : \type{T}}$, corresponds to the judgment ${\vdash* P :: \Delta , z:A}$, for some $P$, $\Delta$, and $A$.
\item Also, $\encc{a\>b}{\normalcolor \type{\Gamma} \vdashB \term{\tBfr{\vec{m}}} : \type{S'}\>\type{S}}$, the translation of a well-typed buffer holding messages $\term{\vec{m}}$, corresponds to the judgment $ {\vdash* P :: \Delta , a:A , b:B}$, for some $P$, $\Delta$, $A$, and $B$.
The translated buffer stores the translations of the messages in $\term{\vec{m}}$ and forwards them on the fresh name $b$; the  buffer receives new messages on the fresh name $a$.
Note that, if the buffer is empty, the roles of $a$ and $b$ may be reversed (depending on $\type{S}$), and the buffer may receive new messages on $b$ instead.
\end{itemize}

\begin{figure}[p]
    \begin{align*}
        \arrayrulecolor{gray!50}
        \setlength{\arrayrulewidth}{.1pt}
        & \tinyRL{typ-var}
        &
        \encc{z}{x}
        &\deq
        \pOut x[\_,z]
        \\[1pt] \hline \\[-14pt]
        & \tinyRL{typ-abs}
        &
        \encc{z}{\lam x . M}
        &\deq
        \pIn z(x,a) ; \encc{a}{M}
        \tag*{receive $x$, then run body}
        \\[1pt] \hline \\[-14pt]
        & \tinyRL{typ-app}
        &
        \encc{z}{M~N}
        &\deq
        \pRes{ab}\pRes{cd} (
        \encc{a}{M}
        \tag*{run abstraction}
        \\ &&&\hphantom{{}\deq{}}
        ~~ \| \pOut b[c,z]
        \tag*{trigger function body}
        \\ &&&\hphantom{{}\deq{}}
        ~~ \| \pIn d(\_,e) ; \encc{e}{N} )
        \tag*{parameter as future substitution}
        \\[1pt] \hline \\[-14pt]
        & \tinyRL{typ-unit}
        &
        \encc{z}{()}
        &\deq \0
        \\[1pt] \hline \\[-14pt]
        & \tinyRL{typ-pair}
        &
        \encc{z}{(M,N)}
        &\deq
        \pRes{ab} \pRes{cd} (
        \pOut z[a,c]
        \tag*{announce pair is ready}
        \\ &&&\hphantom{{}\deq{}}
        ~~ \| \pIn b(\_,e) ; \encc{e}{M} \| \pIn d(\_,f) ; \encc{f}{N} )
        \tag*{components as future substitutions}
        \\[1pt] \hline \\[-14pt]
        & \tinyRL{typ-split}
        &
        \encc{z}{\tLet (x,y) = M \tIn N}
        &\deq
        \pRes{ab} (
        \pIn a(x,y) ; \encc{z}{N}
        \tag*{block body until pair ready}
        \\ &&&\hphantom{{}\deq{}}
        ~~ \| \encc{b}{M} )
        \tag*{run pair}
        \\[1pt] \hline \\[-14pt]
        & \tinyRL{typ-new}
        &
        \encc{z}{\tNew}
        &\deq
        \pRes{ab} (
        \pOut a[\_,z]
        \tag*{activate buffer}
        \\ &&&\hphantom{{}\deq{}}
        ~~ \| \pIn b(\_,c) ; \pRes{dx} \pRes{ey} (
        \tag*{block until activated}
        \\ &&&\hphantom{{}\deq{}}
        ~~~~ \encc{d\>e}{\tBfr{\epsi}}
        \tag*{prepare buffer}
        \\ &&&\hphantom{{}\deq{}}
        ~~~~ \| \encc{c}{(x,y)} ) )
        \tag*{return pair of endpoints}
        \\[1pt] \hline \\[-14pt]
        & \tinyRL{typ-fork}
        &
        \encc{z}{\tFork M ; N}
        &\deq
        \pRes{ab} (
        \pOut a[\_,z]
        \tag*{activate bodies}
        \\ &&&\hphantom{{}\deq{}}
        ~~ \| \pIn b(\_,c) ; (
        \tag*{block until activated}
        \\ &&&\hphantom{{}\deq{}}
        ~~~~ \pRes{\_\_} \encc{\_}{M} \| \encc{c}{N} ) )
        \tag*{run bodies}
        \\[1pt] \hline \\[-14pt]
        & \tinyRL{typ-close}
        &
        \encc{z}{\tClose M ; N}
        &\deq
        \pRes{ab} (
        \encc{a}{M}
        \tag*{run argument to activate buffer}
        \\ &&&\hphantom{{}\deq{}}
        ~~ \| \pIn b(\_,\_) ; \encc{z}{N} )
        \tag*{wait for buffer to close}
        \\[1pt] \hline \\[-14pt]
        & \tinyRL{typ-send}
        &
        \encc{z}{\tSend M \, N}
        &\deq
        \pRes{ab} \pRes{cd} (
        \pIn a(\_,e) ; \encc{e}{M}
        \tag*{block payload until received}
        \\ &&&\hphantom{{}\deq{}}
        ~~ \| \encc{c}{N}
        \tag*{run channel term to activate buffer}
        \\ &&&\hphantom{{}\deq{}}
        ~~ \| \pIn d(\_,f) ; \pRes{gh} (
        \tag*{wait for buffer to activate}
        \\ &&&\hphantom{{}\deq{}}
        ~~~~ \pOut f[b,g]
        \tag*{send to buffer}
        \\ &&&\hphantom{{}\deq{}}
        ~~~~ \| \pOut h[\_,z] ) )
        \tag*{prepare returned endpoint variable}
        \\[1pt] \hline \\[-14pt]
        & \tinyRL{typ-recv}
        &
        \encc{z}{\tRecv M}
        &\deq
        \pRes{ab} (
        \encc{a}{M}
        \tag*{run channel term to activate buffer}
        \\ &&&\hphantom{{}\deq{}}
        ~~ \| \pIn b(c,d) ;
        \tag*{receive from buffer}
        \\ &&&\hphantom{{}\deq{}}
        ~~~~ \pRes{ef} ( \pOut z[c,e] \| \pIn f(\_,g) ; d[\_,g] ) )
        \tag*{returned pair}
        \\[1pt] \hline \\[-14pt]
        & \tinyRL{typ-sel}
        &
        \encc{z}{\tSel j \, M}
        &\deq
        \pRes{ab} (
        \encc{a}{M}
        \tag*{run channel term to activate buffer}
        \\ &&&\hphantom{{}\deq{}}
        ~~ \| \pIn b(\_,c) ; \pRes{de} (
        \tag*{wait for buffer to activate}
        \\ &&&\hphantom{{}\deq{}}
        ~~~~ \pSel c[d] < j
        \tag*{select with buffer}
        \\ &&&\hphantom{{}\deq{}}
        ~~~~ \| e[\_,z] ) )
        \tag*{prepare returned endpoint variable}
    \end{align*}

    \caption{Translating \protect\FIRST into \APCP, Part 1/3.}
    \label{f:FIRST:trans1}
\end{figure}

\begin{figure}[p]
    \begin{align*}
        \arrayrulecolor{gray!50}
        \setlength{\arrayrulewidth}{.1pt}
        & \tinyRL{typ-case}
        &
        \encc{z}{\tCase M \tOf \{ i : N_i \}_{i \in I}}
        &\deq
        \pRes{ab} (
        \encc{a}{M}
        \tag*{run channel term to activate buffer}
        \\ &&&\hphantom{{}\deq{}}
        ~~ \| \pBra b(c) > \{ i :
        \tag*{branch on buffer}
        \\ &&&\hphantom{{}\deq{}}
        ~~~~ \encc{z}{N_i\ c} \}_{i \in I} )
        \tag*{apply continuation to endpoint}
        \\[1pt] \hline \\[-14pt]
        & \tinyRL{typ-sub}
        &
        \encc{z}{M \tSub{ N/x }}
        &\deq
        \pRes{xa} (
        \encc{z}{M}
        \tag*{run body}
        \\ &&&\hphantom{{}\deq{}}
        ~~ \pIn a(\_,b) ; \encc{b}{N} )
        \tag*{block until body is variable}
        \\[1pt] \hline \\[-14pt]
        & \tinyRL{typ-main}
        &
        \encc{z}{\tMain \, M}
        &\deq
        \encc{z}{M}
        \\[1pt] \hline \\[-14pt]
        & \tinyRL{typ-child}
        &
        \encc{z}{\tChild \, M}
        &\deq
        \encc{z}{M}
        \\[1pt] \hline \\[-14pt]
        & \tinyRL{typ-par}~(\type{T_1} = \type{\1})
        &
        \encc{z}{C \prl D}
        &\deq
        \pRes{\_\_} \encc{\_}{C} \| \encc{z}{D}
        \\[1pt] \hline \\[-14pt]
        & \tinyRL{typ-par}~(\type{T_2} = \type{\1})
        &
        \encc{z}{C \prl D}
        &\deq
        \encc{z}{C} \| \pRes{\_\_} \encc{\_}{D}
        \\[1pt] \hline \\[-14pt]
        & \tinyRL{typ-res}
        &
        \encc{z}{\pRes{x\tBfr{\vec{m}}y} C}
        &\deq
        \pRes{ax} \pRes{by} (
        \encc{a\>b}{\tBfr{\vec m}}
        \tag*{run buffer}
        \\ &&&\hphantom{{}\deq{}}
        ~~ \| \encc{z}{C} )
        \tag*{run configuration}
        \\[1pt] \hline \\[-14pt]
        & \tinyRL{typ-conf-sub}
        &
        \encc{z}{C \tSub{ N/x }}
        &\deq
        \pRes{xa} (
        \encc{z}{C}
        \tag*{run configuration}
        \\ &&&\hphantom{{}\deq{}}
        ~~ \| \pIn a(\_,b) ; \encc{b}{N}
        \tag*{block until configuration is variable}
        \\[1pt] \hline \\[-14pt]
        & \tinyRL{typ-buf}~(\type{S'} = \type{{!}T.S})
        &
        \encc{a\>b}{\tBfr{\epsi}}
        &\deq
        \pIn a(\_,c) ; \pRes{de} (
        \tag*{wait for activation}
        \\ &&&\hphantom{{}\deq{}}
        ~~ \pOut c[\_,d]
        \tag*{activate send}
        \\ &&&\hphantom{{}\deq{}}
        ~~ \| \pIn e(f,g) ; \pRes{hk} (
        \tag*{receive from send}
        \\ &&&\hphantom{{}\deq{}}
        ~~~~ \pIn b(\_,l) ;
        \tag*{wait for activation}
        \\ &&&\hphantom{{}\deq{}}
        ~~~~~~ \pOut l[f,h]
        \tag*{send to receive}
        \\ &&&\hphantom{{}\deq{}}
        ~~~~ \| \enc{g\>k}{\term{\tBfr{\epsi}} : \type{S}\>\type{S}} ) )
        \\[1pt] \hline \\[-14pt]
        & \tinyRL{typ-buf}~(\type{S'} = \type{\oplus \{ i : S_i \}_{i \in I}})
        &
        \encc{a\>b}{\tBfr{\epsi}}
        &\deq
        \pIn a(\_,c) ; \pRes{de} (
        \tag*{wait for activation}
        \\ &&&\hphantom{{}\deq{}}
        ~~ \pOut c[\_,d]
        \tag*{activate select}
        \\ &&&\hphantom{{}\deq{}}
        ~~ \| \pBra e(f) > \{ i : \pRes{gh} (
        \tag*{receive selection}
        \\ &&&\hphantom{{}\deq{}}
        ~~~~ \pIn b(\_,k) ;
        \tag*{wait for activation}
        \\ &&&\hphantom{{}\deq{}}
        ~~~~~~ \pSel k[g] < i
        \tag*{make selection}
        \\ &&&\hphantom{{}\deq{}}
        ~~~~ \| \enc{f\>h}{\term{\tBfr{\epsi}} : \type{S_i}\>\type{S_i}} ) \} )
    \end{align*}

    \caption{Translating \protect\FIRST into \APCP, Part 2/3.}
    \label{f:FIRST:trans2}
\end{figure}

\begin{figure}[t]
    \begin{align*}
        \arrayrulecolor{gray!50}
        \setlength{\arrayrulewidth}{.1pt}
        & \tinyRL{typ-buf}~(\type{S'} \in \{\type{{?}T.S},\type{\& \{ i : S_i \}_{i \in I}}\})
        &
        \encc{a\>b}{\tBfr{\epsi}}
        &\deq
        \enc{b\>a}{\term{\tBfr{\epsi}} : \type{\ol{S'}}\>\type{\ol{S'}}}
        \\[1pt] \hline \\[-14pt]
        & \tinyRL{typ-buf}~(\type{S'} = \type{\tEnd})
        &
        \encc{a\>b}{\tBfr{\epsi}}
        &\deq
        \pIn a(\_,c) ; \pOut c[\_,\_] \| \pIn b(\_,d) ; \pOut d[\_,\_]
        \tag*{close concurrently}
        \\[1pt] \hline \\[-14pt]
        & \tinyRL{typ-buf}~(\type{S'} = \type{\tNil})
        &
        \encc{a\>b}{\tBfr{\epsi}}
        &\deq
        \0
        \\[1pt] \hline \\[-14pt]
        & \tinyRL{typ-buf-send}
        &
        \encc{a\>b}{\tBfr{\vec{m},M}}
        &\deq
        \pRes{cd} \pRes{ef} (
        \pIn c(\_,g) ; \encc{g}{M}
        \tag*{block payload until received}
        \\ &&&\hphantom{{}\deq{}}
        ~~ \| \pIn b(\_,h) ;
        \tag*{wait for activation}
        \\ &&&\hphantom{{}\deq{}}
        ~~~~ \pOut h[d,e]
        \tag*{send to receive}
        \\ &&&\hphantom{{}\deq{}}
        ~~ \| \encc{a\>f}{\tBfr{\vec{m}}} )
        \tag*{run buffer}
        \\[1pt] \hline \\[-14pt]
        & \tinyRL{typ-buf-sel}
        &
        \encc{a\>b}{\tBfr{\vec{m},j}}
        &\deq
        \pRes{cd} (
        \pIn b(\_,e) ;
        \tag*{wait for activation}
        \\ &&&\hphantom{{}\deq{}}
        ~~~~ \pSel e[c] < j
        \tag*{make selection}
        \\ &&&\hphantom{{}\deq{}}
        ~~ \| \encc{a\>d}{\tBfr{\vec{m}}} )
        \tag*{run buffer}
        \\[1pt] \hline \\[-14pt]
        & \tinyRL{typ-buf-end-L}
        &
        \encc{a\>b}{\tBfr{\epsi}}
        &\deq
        \pIn a(\_,c) ; \pOut c[\_,\_]
        \\[1pt] \hline \\[-14pt]
        & \tinyRL{typ-buf-end-R}
        &
        \encc{a\>b}{\tBfr{\epsi}}
        &\deq
        \pIn b(\_,c) ; \pOut c[\_,\_]
    \end{align*}

    \caption{Translating \protect\FIRST into \APCP, Part 3/3.}
    \label{f:FIRST:trans3}
\end{figure}
\noindent 
\Cref{f:FIRST:trans1,f:FIRST:trans2,f:FIRST:trans3} give the translation of term, configuration and buffer typing rules.
In the figures (and in the rest of the paper), for the sake of readability, we often omit typing information: we write  $\encc{z}{\term{M}}$ to refer to the typed process associated to $\encc{z}{\type{\Gamma} \vdashM \term{M} : \type{T}}$, and similarly for the translations of typed configurations and buffers. The omitted typing information/derivations is made precise by \Cref{t:FIRST:transTp}, given below.

Instead of explaining the translation separately, the figures include line-by-line explanations.
Recall that we often write `$\_$' to denote a fresh name of type $\bullet$; when sending names denoted `$\_$', we omit binders `$\pRes{\_\_}$'.

Importantly, the translation is \emph{type preserving} using the translations of types introduced in \Cref{d:FIRST:transTypes}.
\Cref{as:FIRST:transTp} contains a detailed proof.
\newline\newline 
\vspace{-5\baselineskip}
\begin{restatable}[Type Preservation for the Translation]{theorem}{tFIRSTTransTp}
\label{t:FIRST:transTp}
    \leavevmode
    \begin{itemize}
        \item
            If $\type{\Gamma} \vdashM \term{M} : \type{T}$, then $\vdash* \encc{z}{M} :: \enct*{\Gamma} , z:\enct{T}$;
        \item
            If $\type{\Gamma} \vdashC{\phi} \term{C} : \type{T}$, then $\vdash* \encc{z}{C} :: \enct*{\Gamma} , z:\enct{T}$;
        \item
            If $\type{\Gamma} \vdashB \term{\tBfr{\vec{m}}} : \type{S'}\>\type{S}$, then $\vdash* \encc{a\>b}{\tBfr{\vec m}} :: \enct*{\Gamma} , a:\ol{\enct*{S'}} , b:\ol{\enct*{\ol{S}}}$.

    \end{itemize}
\end{restatable}

\begin{exa}
    \label{x:FIRST:trans}
    We illustrate our translation by translating a subterm of the shop defined in \Cref{s:FIRST:example} (leaving the translation of the syntactic sugar $\term{\text{book}(\textit{title})}$ unspecified):
    \begin{align*}
        & \encc{z}{
            \lam s_2 . \tLet (\textit{card},s_3) = \tRecv s_2 \tIn
            \tLet s_4 = \tSend \text{book}(\textit{title}) \, s_3 \tIn
            \tClose s_4 ; ()
        }
        \span\span
        \\
        \mathllap{{}={}}& \pIn z(s_2,a_0) ;
        &&
        \tikzmark{x:FIRST:trans:1s}
        \tag*{$\encc{z}{\lam s_2 \ldots}$}
        \\
        & \q[1] \pRes{a_1b_1} ( \pIn a_1(\textit{card},s_3) ;
        &&
        \tikzmark{x:FIRST:trans:11s}
        \tag*{$\encc{a_0}{\tLet (\textit{card},s_3) = \ldots}$}
        \\
        & \q[2] \pRes{a_2b_2} \pRes{c_2d_2} (
        &&
        \tikzmark{x:FIRST:trans:111s}
        \tag*{$\encc{a_0}{\tLet s_4 = \ldots}$}
        \\
        & \q[3] \pIn a_2(s_4,a_3) ;
        &&
        \tikzmark{x:FIRST:trans:1111s}
        \tag*{$\encc{a_2}{\lam s_4 . \ldots}$}
        \\
        & \q[4] \pRes{a_4b_4} (
        &&
        \tikzmark{x:FIRST:trans:11111s}
        \tag*{$\encc{a_3}{\tClose \ldots}$}
        \\
        & \q[5] \pOut s_4[\_,a_4]
        &&
        \tikzmark{x:FIRST:trans:111111s}
        \tikzmark{x:FIRST:trans:111111e}
        \tag*{$\encc{a_4}{s_4}$}
        \\
        & \q[5] \| \pIn b_4(\_,\_) ;
        \\
        & \q[6] \0 )
        &&
        \tikzmark{x:FIRST:trans:111112s}
        \tikzmark{x:FIRST:trans:111112e}
        \tikzmark{x:FIRST:trans:11111e}
        \tikzmark{x:FIRST:trans:1111e}
        \tag*{$\encc{a_3}{()}$}
        \\
        & \q[3] \pOut b_2[c_2,a_0] \| \pIn d_2(\_,e_2) ;
        \\
        & \q[4] \pRes{a_5b_5} \pRes{c_5d_5} ( \pIn a_5(\_,e_5) ;
        &&
        \tikzmark{x:FIRST:trans:1112s}
        \tag*{$\encc{e_2}{\tSend \ldots}$}
        \\
        & \q[5] \encc{e_5}{\text{book}(\textit{title})}
        \\
        & \q[5] \| \pOut s_3[\_,c_5]
        &&
        \tikzmark{x:FIRST:trans:11121s}
        \tikzmark{x:FIRST:trans:11121e}
        \tag*{$\encc{c_5}{s_3}$}
        \\
        & \q[5] \| \pIn d_5(\_,f_5) ; \pRes{g_5h_5} ( \pOut f_5[b_5,g_5] \| \pOut h_5[\_,e_2] ) ) )
        &&
        \tikzmark{x:FIRST:trans:1112e}
        \tikzmark{x:FIRST:trans:111e}
        \\
        & \q[2] \| \pRes{a_6b_6} (
        &&
        \tikzmark{x:FIRST:trans:112s}
        \tag*{$\encc{b_1}{\tRecv \ldots}$}
        \\
        & \q[3] \pOut s_2[\_,a_6]
        &&
        \tikzmark{x:FIRST:trans:1121s}
        \tikzmark{x:FIRST:trans:1121e}
        \tag*{$\encc{a_6}{s_2}$}
        \\
        & \q[3] \| \pIn b_6(c_6,d_6) ; \pRes{e_6f_6} ( \pOut b_1[c_6,e_6] \| \pIn f_6(\_,g_6) ; \pOut d_6[\_,g_6] ) ) )
        \hspace*{-8mm}
        &&
        \tikzmark{x:FIRST:trans:1e}
        \tikzmark{x:FIRST:trans:112e}
        \tikzmark{x:FIRST:trans:11e}
        \q[10]
    \end{align*}

    \begin{tikzpicture}[overlay,remember picture]
        \draw ([yshift=2ex]pic cs:x:FIRST:trans:1s) to [square left brace] (pic cs:x:FIRST:trans:1e);
        \draw ([yshift=1ex,xshift=.1cm]pic cs:x:FIRST:trans:1s) to ([yshift=1ex,xshift=.4cm]pic cs:x:FIRST:trans:1s);
        \draw ([yshift=2ex,xshift=.2cm]pic cs:x:FIRST:trans:11s) to [square left brace] ([xshift=.2cm]pic cs:x:FIRST:trans:11e);
        \draw ([yshift=1ex,xshift=.3cm]pic cs:x:FIRST:trans:11s) to ([yshift=1ex,xshift=.6cm]pic cs:x:FIRST:trans:11s);
        \draw ([yshift=2ex,xshift=.4cm]pic cs:x:FIRST:trans:111s) to [square left brace] ([xshift=.4cm]pic cs:x:FIRST:trans:111e);
        \draw ([yshift=1ex,xshift=.5cm]pic cs:x:FIRST:trans:111s) to ([yshift=1ex,xshift=.8cm]pic cs:x:FIRST:trans:111s);
        \draw ([yshift=2ex,xshift=.4cm]pic cs:x:FIRST:trans:112s) to [square left brace] ([xshift=.4cm]pic cs:x:FIRST:trans:112e);
        \draw ([yshift=1ex,xshift=.5cm]pic cs:x:FIRST:trans:112s) to ([yshift=1ex,xshift=.8cm]pic cs:x:FIRST:trans:112s);
        \draw ([yshift=2ex,xshift=.6cm]pic cs:x:FIRST:trans:1111s) to [square left brace] ([xshift=.6cm]pic cs:x:FIRST:trans:1111e);
        \draw ([yshift=1ex,xshift=.7cm]pic cs:x:FIRST:trans:1111s) to ([yshift=1ex,xshift=1.0cm]pic cs:x:FIRST:trans:1111s);
        \draw ([yshift=2ex,xshift=.6cm]pic cs:x:FIRST:trans:1112s) to [square left brace] ([xshift=.6cm]pic cs:x:FIRST:trans:1112e);
        \draw ([yshift=1ex,xshift=.7cm]pic cs:x:FIRST:trans:1112s) to ([yshift=1ex,xshift=1.0cm]pic cs:x:FIRST:trans:1112s);
        \draw ([yshift=2ex,xshift=.6cm]pic cs:x:FIRST:trans:1121s) to [square left brace] ([xshift=.6cm]pic cs:x:FIRST:trans:1121e);
        \draw ([yshift=1ex,xshift=.7cm]pic cs:x:FIRST:trans:1121s) to ([yshift=1ex,xshift=1.0cm]pic cs:x:FIRST:trans:1121s);
        \draw ([yshift=2ex,xshift=.8cm]pic cs:x:FIRST:trans:11111s) to [square left brace] ([xshift=.8cm]pic cs:x:FIRST:trans:11111e);
        \draw ([yshift=1ex,xshift=.9cm]pic cs:x:FIRST:trans:11111s) to ([yshift=1ex,xshift=1.2cm]pic cs:x:FIRST:trans:11111s);
        \draw ([yshift=2ex,xshift=.8cm]pic cs:x:FIRST:trans:11121s) to [square left brace] ([xshift=.8cm]pic cs:x:FIRST:trans:11121e);
        \draw ([yshift=1ex,xshift=.9cm]pic cs:x:FIRST:trans:11121s) to ([yshift=1ex,xshift=1.2cm]pic cs:x:FIRST:trans:11121s);
        \draw ([yshift=2ex,xshift=1.0cm]pic cs:x:FIRST:trans:111111s) to [square left brace] ([xshift=1.0cm]pic cs:x:FIRST:trans:111111e);
        \draw ([yshift=1ex,xshift=1.1cm]pic cs:x:FIRST:trans:111111s) to ([yshift=1ex,xshift=1.4cm]pic cs:x:FIRST:trans:111111s);
        \draw ([yshift=2ex,xshift=1.0cm]pic cs:x:FIRST:trans:111112s) to [square left brace] ([xshift=1.0cm]pic cs:x:FIRST:trans:111112e);
        \draw ([yshift=1ex,xshift=1.1cm]pic cs:x:FIRST:trans:111112s) to ([yshift=1ex,xshift=1.4cm]pic cs:x:FIRST:trans:111112s);
    \end{tikzpicture}

    We also give the translation of the outer buffer in the final state of the system in \Cref{x:FIRST:Shop}:

    \begin{align*}
        & \encc{z}{\pRes{y'\tBfr{\text{visa},\sff{buy},\text{``Dune''}}y} \big( \pRes{z'\tBfr{\epsi}z} \ldots \prl \tChild \, \sff{Shop}(y) \big)}
        \span\span
        \\
        \mathllap{{}={}}
        & \pRes{a_1y'} \pRes{b_1y} (
        &&
        \tikzmark{x:FIRST:trans2:1s}
        \tag*{$\encc{z}{\pRes{y'\tBfr{\ldots}y} \ldots}$}
        \\
        & \q[1] \pRes{c_1d_1} \pRes{e_1f_1} ( \pIn c_1(\_,g_1) ;
        &&
        \tikzmark{x:FIRST:trans2:11s}
        \tag*{$\encc{a_1\>b_1}{\tBfr{\ldots,\text{``Dune''}}}$}
        \\
        & \q[2] \encc{g_1}{\text{``Dune''}}
        \\
        & \q[2] {} \| \pIn b_1(\_,h_1) ; \pOut h_1[d_1,e_1]
        \\
        & \q[2] {} \| \pRes{c_2d_2} ( \pIn f_1(\_,e_2) ; \pSel e_2[c_2] < \sff{buy}
        &&
        \tikzmark{x:FIRST:trans2:111s}
        \tag*{$\encc{a_1\>f_1}{\tBfr{\ldots,\sff{buy}}}$}
        \\
        & \q[3] {} \| \pRes{c_3d_3} \pRes{e_3f_3} ( \pIn c_3(\_,g_3) ;
        &&
        \tikzmark{x:FIRST:trans2:1111s}
        \tag*{$\encc{a_1\>d_2}{\tBfr{\text{visa}}}$}
        \\
        & \q[4] \encc{g_3}{\text{visa}}
        \\
        & \q[4] {} \| \pIn d_2(\_,h_3) ; \pOut h_3[d_3,e_3]
        \\
        & \q[4] {} \| \pIn f_3(\_,c_4) ; \pRes{d_4e_4} ( \pOut c_4[\_,d_4] \| \pIn e_4(f_4,g_4) ;
        &&
        \tikzmark{x:FIRST:trans2:11111s}
        \tag*{$\enc{f_3\>a_1}{\term{\tBfr{\epsi}} : \type{{!}\sff{B}.\tEnd} \> \type{{!}\sff{B}.\tEnd}}$}
        \\
        & \q[5] \pRes{h_4k_4} ( \pIn a_1(\_,l_4) ; \pOut l_4[f_4,h_4]
        \\
        & \q[6] {} \| \pIn g_4(\_,c_5) ; \pOut c_5[\_,\_] \| \pIn k_4(\_,d_5) ; \pOut d_5[\_,\_] )\!)\!)\!)\!)
        \hspace{-2em}
        &&
        \tikzmark{x:FIRST:trans2:111111s}
        \tikzmark{x:FIRST:trans2:111111e}
        \tikzmark{x:FIRST:trans2:11111e}
        \tikzmark{x:FIRST:trans2:1111e}
        \tikzmark{x:FIRST:trans2:111e}
        \tikzmark{x:FIRST:trans2:11e}
        \hspace{4em}
        \tag*{$\enc{g_4\>k_4}{\term{\tBfr{\epsi}} : \type{\tEnd} \> \type{\tEnd}}$}
        \\
        & \q[1] {} \| \encc{z}{\pRes{z'\tBfr{\epsi}z} \ldots \prl \tChild \, \sff{Shop}(y)}
        &&
        \tikzmark{x:FIRST:trans2:1e}
        \tag*{\qedhere}
    \end{align*}

    \begin{tikzpicture}[overlay,remember picture]
        \draw ([yshift=2ex]pic cs:x:FIRST:trans2:1s) to [square left brace] (pic cs:x:FIRST:trans2:1e);
        \draw ([yshift=1ex,xshift=.1cm]pic cs:x:FIRST:trans2:1s) to ([yshift=1ex,xshift=.4cm]pic cs:x:FIRST:trans2:1s);
        \draw ([yshift=2ex,xshift=.2cm]pic cs:x:FIRST:trans2:11s) to [square left brace] ([xshift=.2cm]pic cs:x:FIRST:trans2:11e);
        \draw ([yshift=1ex,xshift=.3cm]pic cs:x:FIRST:trans2:11s) to ([yshift=1ex,xshift=.6cm]pic cs:x:FIRST:trans2:11s);
        \draw ([yshift=2ex,xshift=.4cm]pic cs:x:FIRST:trans2:111s) to [square left brace] ([xshift=.4cm]pic cs:x:FIRST:trans2:111e);
        \draw ([yshift=1ex,xshift=.5cm]pic cs:x:FIRST:trans2:111s) to ([yshift=1ex,xshift=.8cm]pic cs:x:FIRST:trans2:111s);
        \draw ([yshift=2ex,xshift=.6cm]pic cs:x:FIRST:trans2:1111s) to [square left brace] ([xshift=.6cm]pic cs:x:FIRST:trans2:1111e);
        \draw ([yshift=1ex,xshift=.7cm]pic cs:x:FIRST:trans2:1111s) to ([yshift=1ex,xshift=1.0cm]pic cs:x:FIRST:trans2:1111s);
        \draw ([yshift=2ex,xshift=.8cm]pic cs:x:FIRST:trans2:11111s) to [square left brace] ([xshift=.8cm]pic cs:x:FIRST:trans2:11111e);
        \draw ([yshift=1ex,xshift=.9cm]pic cs:x:FIRST:trans2:11111s) to ([yshift=1ex,xshift=1.2cm]pic cs:x:FIRST:trans2:11111s);
        \draw ([yshift=2ex,xshift=1.0cm]pic cs:x:FIRST:trans2:111111s) to [square left brace] ([xshift=1.0cm]pic cs:x:FIRST:trans2:111111e);
        \draw ([yshift=1ex,xshift=1.1cm]pic cs:x:FIRST:trans2:111111s) to ([yshift=1ex,xshift=1.4cm]pic cs:x:FIRST:trans2:111111s);
    \end{tikzpicture}
\end{exa}

\paragraph{Design Decisions: Explicit Substitutions and Closing}
Having presented our typed translation, we reflect on our design decision to enrich \FIRST with explicit substitutions and closing.

As we will see next, adopting   explicit   substitutions leads to a direct operational correctness result.
To see why,
suppose we were to apply substitutions immediately.
As an example, consider function application (\Cref{f:FIRST:trans1}), which entails substituting a variable in the body of the function.
The translation would need to encode the substitution of the translation of this variable.
However, if the variable does not occur under an evaluation context, there is no way in \APCP to perform such a substitution immediately.
Hence, the translation would still need to encode such \emph{implicit} substitutions in \FIRST \emph{explicitly} in \APCP.
This discrepancy would then have to be handled by means of the aforementioned substitution lifting in our operational correspondence results.
Although this is a perfectly viable approach, we prefer to have a more direct operational correspondence, which entails more direct proofs that are not affected by an asymmetric treatment of substitutions.

Our choice for explicit closing of sessions is more pragmatic: it leads to a more compact translation.
Suppose we were to treat closed sessions by silently weakening them.
Consider, e.g., Rule~\ruleLabel{red-send} (\Cref{f:LAST:confs}): the $\term{\tSend}$ primitive is replaced by a variable pointing to the buffered channel, even if the session ends after the send.
Hence, the translation would need a separate case for translating variables for closed sessions.
Consequently, the translation would need similar such separate cases anywhere variables/closed sessions may occur.
This would lead to a translation where the key ideas are unnecessarily obfuscated.
We find it then preferable to treat closed sessions explicitly in favor of a more streamlined translation.

\paragraph{Faithfulness: Operational Correspondence}
Following the discussion in \Cref{s:LAST:trans}, here we finally show that the translation presented above satisfies the operational correctness criterion (completeness and soundness) as proposed by Gorla~\cite{journal/ic/Gorla10}.
We first state both results, and then their proofs.

\begin{restatable}[Completeness]{theorem}{tFIRSTTransCompl}\label{t:FIRST:transCompl}
    Given $\type{\Gamma} \vdashC{\phi} \term{C}: \type{T}$, if $\term{C} \reddC \term{D}$, then $\encc{z}{C} \redd^\ast \encc{z}{D}$.
\end{restatable}

\begin{restatable}[Soundness]{theorem}{tFIRSTTransSound}
\label{t:FIRST:transSound}
    Given $\type{\Gamma} \vdashC{\phi} \term{C} : \type{T}$, if $\encc{z}{C} \redd* Q$, then there exists $\term{D}$ such that $\term{C} \reddC^\ast \term{D}$ and $Q \redd* \encc{z}{D}$.
\end{restatable}

\noindent 
Both results rely on the following lemma, which decomposes translations of terms/con\-figurations under contexts into some evaluation context containing the translation of the respective term/configuration, and transfers free variables/names and substitutions in and out of the translation (cf.\ \Cref{as:FIRST:oc}):

\begin{restatable}{lemma}{lFIRSTTransCtxs}
\label{l:FIRST:transCtxs}
    \leavevmode
    \begin{itemize}

        \item
            $\encc{z}{\tCtx{R}[M]} = \evalCtx{E}[\encc{z'}{M}]$ for some $\evalCtx{E},z'$;

        \item
            $\encc{z}{\tCtx{F}[M]} = \evalCtx{E}[\encc{z'}{M}]$ for some $\evalCtx{E},z'$;

        \item
            $\encc{z}{\tCtx{G}[C]} = \evalCtx{E}[\encc{z'}{C}]$ for some $\evalCtx{E},z'$;

        \item
            $\term{x} \notin \fv(\term{C})$ implies $x \notin \fn(\encc{z}{C})$;

        \item
            $\term{x} \in \fv(\term{C})$ implies $\encc{z}{C \{ y/x \}} = \encc{z}{C} \{ y/x \}$.

    \end{itemize}
\end{restatable}

\begin{sketch}
    The first three items follow by induction on the structure of the contexts.
    In the base case, the context is a hole and the thesis follows immediately.
    The inductive cases follow by construction of the translation and IH.
    The last two items follow by induction on the structure of the configuration and construction of the translation.
\end{sketch}
\noindent 
Completeness additionally relies on the following lemma, that decomposes the translation of buffers of several shapes into an evaluation context containing the translation of a continuation of the respective buffer:

\begin{restatable}{lemma}{lFIRSTTransBuf}
    \label{l:FIRST:transBuf}
    \leavevmode
    \begin{itemize}

        \item
            $\type{S'} \neq \type{\tNil}$ implies $\encc{a\>b}{\type{\Gamma} \vdashB \term{\tBfr{\vec{m}}} : \type{S'}\>\type{S}} = \evalCtx[\big]{E}[\encc{a\>c}{\type{\emptyset} \vdashB \term{\tBfr{\epsi}} : \type{S'}\>\type{S'}}]$
            for some $\evalCtx{E},c$;

        \item
            $\type{S'} \neq \type{\tNil}$ implies $\encc{a\>b}{\type{\Gamma},\type{\Delta} \vdashB \term{\tBfr{\mbb{M},\vec{m}}} : \type{S'}\>\type{S}} = \evalCtx[\big]{E}[\encc{a\>c}{\type{\Delta} \vdashB \term{\tBfr{\mbb{M}}} : \type{S'}\>\type{{!}T.S'}}]$
            for some $\evalCtx{E},c$;

        \item
            $\type{S'} \neq \type{\tNil}$ implies $\encc{a\>b}{\type{\Gamma} \vdashB \term{\tBfr{j,\vec{m}}} : \type{S'}\>\type{S}} = \evalCtx[\big]{E}[\encc{a\>c}{\type{\emptyset} \vdashB \term{\tBfr{j}} : \type{S'}\>\type{\oplus \{ i : S_i \}_{i \in I} \cup \{ j : S' \}}}]$
            for some $\evalCtx{E},c$;

        \item
            $\type{S} \neq \type{\tNil}$ implies $\encc{a\>b}{\type{\Gamma} \vdashB \term{\tBfr{\vec{m}}} : \type{\tEnd} \> \type{S}} = \evalCtx[\big]{E}[\encc{a\>c}{\type{\emptyset} \vdashB \term{\tBfr{\epsi}} : \type{\tEnd}\>\type{\tEnd}}]$
            for some $\evalCtx{E},c$;

        \item
            $\type{S} \neq \type{\tNil}$ implies $\encc{a\>b}{\type{\Gamma} \vdashB \term{\tBfr{\vec{m}}} : \type{\tNil} \> \type{S}} = \evalCtx[\big]{E}[\encc{a\>c}{\type{\emptyset} \vdashB \term{\tBfr{\epsi}} : \type{\tNil}\>\type{\tEnd}}]$
            for some $\evalCtx{E},c$.

    \end{itemize}
\end{restatable}

\begin{sketch}
    Each item follows by induction on the size of $\term{\vec m}$, and a case analysis on the shape of $\type{S}$.
    In the base case, $\term{\vec m} = \term{\epsi}$ and the thesis follows immediately.
    In the inductive cases, the thesis follows by IH and the construction of the translation.
\end{sketch}

With these prerequisites in place, we move on to discuss the proofs of \Cref{t:FIRST:transCompl,t:FIRST:transSound}.

\begin{proof}[Proof of completeness (\Cref{t:FIRST:transCompl})]
    By induction on the derivation of $\term{C} \reddC \term{D}$.
    We discuss one representative rule for message passing, as well as the structural rules; other rules are detailed in \Cref{as:FIRST:oc:compl}.
    \begin{itemize}

        \item
            \global\def\proofTransComplSend{%
                Rule~\ruleLabel{red-send}:
                $\term{\pRes{x\tBfr{\vec{m}}y} ( \tCtx{\hat{F}}[\tSend M \, x] \prl C )} \reddC \term{\pRes{x\tBfr{M,\vec{m}}y} ( \tCtx{\hat{F}}[x] \prl C )}$.
                W.l.o.g., assume $\term{C}$ is a child thread.
                By \Cref{l:FIRST:transCtxs}, for any $\term{L}$, $\encc{z}{\tCtx{F}[L]} = \evalCtx[\big]{E_1}[\encc{z'}{L}]$ for some $\evalCtx{E_1},z'$ ($\ast_1$).
                Moreover, since $\tCtx{\hat{F}}$ does not have its hole under an explicit substitution, it does not capture any free variables of $M$; hence, by \Cref{l:FIRST:transCtxs}, $\evalCtx{E_1}$ does not capture any free names of $\encc{u}{M}$ for any $u$ ($\ast_2$).
                By inversion of typing, $\type{\Gamma} \vdashB \term{\tBfr{\vec{m}}} : \type{S_1}\>\type{S}$ where $\type{S_1} = \type{{!}T.S_2}$ ($\ast_3$).
                By \Cref{l:FIRST:transBuf}, $\encc{a\>b}{\type{\Gamma} \vdashB \term{\tBfr{\vec{m}}} : \type{S_1}\>\type{S}} = \evalCtx[\big]{E_2}[\encc{a\>c}{\type{\emptyset} \vdashB \term{\tBfr{\epsi}} : \type{S_1}\>\type{S_1}}]$ ($\ast_4$) and $\encc{a\>b}{\type{\Gamma},\type{\Delta} \vdashB \term{\tBfr{M,\vec{m}}} : \type{S_1}\>\type{S}} = \evalCtx[\big]{E_2}[\encc{a\>c}{\type{\Delta} \vdashB \term{\tBfr{M}} : \type{S_1}\>\type{S_1}}]$ ($\ast_5$) for some $\evalCtx{E_2},c$.
                Below, we omit types from translations of buffers.
                The thesis holds as follows:
                \begin{align*}
                    & \encc{z}{\pRes{x\tBfr{\vec{m}}y} ( \tCtx{\hat{F}}[\tSend M \, x] \prl C )}
                    \\
                    &= \pRes{ax} \pRes{by} ( \encc{a\>b}{\tBfr{\vec{m}}} \| \encc{z}{\tCtx{\hat{F}}[\tSend M \, x]} \| \pRes{\_\_} \encc{\_}{C} )
                    \\
                    &= \pRes{ax} \pRes{by} ( \evalCtx[\big]{E_2}[\encc{a\>c}{\tBfr{\epsi}}] \| \evalCtx[\big]{E_1}[\encc{z'}{\tSend M \, x}] \| \pRes{\_\_} \encc{\_}{C} )
                    \tag{$\ast_1$,$\ast_4$}
                    \\
                    &= \begin{array}[t]{@{}l@{}}
                        \pRes{ax} \pRes{by} (
                        \\ ~
                        \evalCtx[\big]{E_2}[\pIn a(\_,c_1) ; \pRes{d_1e_1} \big( \pOut c_1[\_,d_1] \| \pIn e_1(f_1,g_1) ; \pRes{h_1k_1} ( \pIn c(\_,l_1) ; \pOut l_1[f_1,h_1] \| \encc{g_1\>k_1}{\tBfr{\epsi}} ) \big)]
                        \\ ~
                        {} \| \evalCtx[\big]{E_1}[\pRes{a_2b_2} \pRes{c_2d_2} \big( \pIn a_2(\_,e_2) ; \encc{e_2}{M} \| \pOut x[\_,c_2] \| \pIn d_2(\_,f_2) ; \pRes{g_2h_2} ( \pOut f_2[b_2,g_2] \| \pOut h_2[\_,z'] ) \big)]
                        \\ ~
                        {} \| \pRes{\_\_} \encc{\_}{C} )
                    \end{array}
                    \tag{$\ast_3$}
                    \\
                    &\redd \begin{array}[t]{@{}l@{}}
                        \pRes{c_2d_2} \pRes{by} (
                        \\ ~
                        \evalCtx[\big]{E_2}[\pRes{d_1e_1} \big( \pOut c_2[\_,d_1] \| \pIn e_1(f_1,g_1) ; \pRes{h_1k_1} ( \pIn c(\_,l_1) ; \pOut l_1[f_1,h_1] \| \encc{g_1\>k_1}{\tBfr{\epsi}} ) \big)]
                        \\ ~
                        {} \| \evalCtx[\big]{E_1}[\pRes{a_2b_2} \big( \pIn a_2(\_,e_2) ; \encc{e_2}{M} \| \pIn d_2(\_,f_2) ; \pRes{g_2h_2} ( \pOut f_2[b_2,g_2] \| \pOut h_2[\_,z'] ) \big)]
                        \\ ~
                        {} \| \pRes{\_\_} \encc{\_}{C} )
                    \end{array}
                    \\
                    &\redd \begin{array}[t]{@{}l@{}}
                        \pRes{e_1d_1} \pRes{by} (
                        \\ ~
                        \evalCtx[\big]{E_2}[\pIn e_1(f_1,g_1) ; \pRes{h_1k_1} ( \pIn c(\_,l_1) ; \pOut l_1[f_1,h_1] \| \encc{g_1\>k_1}{\tBfr{\epsi}} )]
                        \\ ~
                        {} \| \evalCtx[\big]{E_1}[\pRes{a_2b_2} \big( \pIn a_2(\_,e_2) ; \encc{e_2}{M} \| \pRes{g_2h_2} ( \pOut d_1[b_2,g_2] \| \pOut h_2[\_,z'] ) \big)]
                        \\ ~
                        {} \| \pRes{\_\_} \encc{\_}{C} )
                    \end{array}
                    \\
                    &\redd \begin{array}[t]{@{}l@{}}
                        \pRes{b_2a_2} \pRes{g_2h_2} \pRes{by} (
                        \\ ~
                        \evalCtx[\big]{E_2}[\pRes{h_1k_1} ( \pIn c(\_,l_1) ; \pOut l_1[b_2,h_1] \| \encc{g_2\>k_1}{\tBfr{\epsi}} )]
                        \\ ~
                        {} \| \evalCtx[\big]{E_1}[\pIn a_2(\_,e_2) ; \encc{e_2}{M} \| \pOut h_2[\_,z']]
                        \\ ~
                        {} \| \pRes{\_\_} \encc{\_}{C} )
                    \end{array}
                    \\
                    &\equiv \begin{array}[t]{@{}l@{}}
                        \pRes{ax} \pRes{by} (
                        \\ ~
                        \evalCtx[\big]{E_2}[\pRes{a_2b_2} \pRes{h_1k_1} ( \pIn a_2(\_,e_2) ; \encc{e_2}{M} \| \pIn c(\_,l_1) ; \pOut l_1[b_2,h_1] \| \encc{a\>k_1}{\tBfr{\epsi}} )]
                        \\ ~
                        {} \| \evalCtx[\big]{E_1}[\pOut x[\_,z']]
                        \\ ~
                        {} \| \pRes{\_\_} \encc{\_}{C} )
                    \end{array}
                    \tag{$\ast_2$}
                    \\
                    &= \pRes{ax} \pRes{by} ( \evalCtx[\big]{E_2}[\encc{a\>c}{\tBfr{M}}] \| \evalCtx[\big]{E_1}[\encc{z'}{x}] \| \pRes{\_\_} \encc{\_}{C} )
                    \\
                    &= \pRes{ax} \pRes{by} ( \encc{a\>b}{\tBfr{M,\vec{m}}} \| \encc{z}{\tCtx{\hat{F}}[x]} \| \pRes{\_\_} \encc{\_}{C} )
                    \tag{$\ast_1$,$\ast_5$}
                    \\
                    &= \encc{z}{\pRes{x\tBfr{M,\vec{m}}y} ( \tCtx{\hat{F}}[x] \prl C )}
                \end{align*}
            }
            \proofTransComplSend

        \item
            Rule~\ruleLabel{red-lift-C}: $\term{C} \reddC \term{C'}$ implies $\tCtx{G}[C] \reddC \tCtx{G}[C']$.
            This case follows from \Cref{l:FIRST:transCtxs} and the IH.

        \item
            Rule~\ruleLabel{red-lift-M}: $\term{M} \reddM \term{M'}$ implies $\tCtx{F}[M] \reddM \tCtx{F}[M']$.
            By \Cref{l:FIRST:transCtxs}, it suffices to show completeness on the level of terms.
            Hence, we apply induction on the derivation of $\term{M} \reddM \term{M'}$ (\ih2).
            In \Cref{as:FIRST:oc:compl}, this property is proven separately as \Cref{t:FIRST:transTermRed}.
            Here, we discuss one representative reduction for computation, as well as the structural rules:
            \begin{itemize}

                \item
                    \global\def\proofTransTermRedLam{%
                        Rule~\ruleLabel{red-lam}:
                        $\term{(\lam x . M)\ N} \reddM \term{M \tSub{ N/x }}$.
                        The thesis holds as follows:
                        \begin{align*}
                            \encc{z}{(\lam x . M)\ N} &= \pRes{a_1b_1} \pRes{c_1d_1} ( \pIn a_1(x,a_2) ; \encc{a_2}{M} \| \pOut b_1[c_1,z] \| \pIn d_1(\_,e_1) ; \encc{e_1}{N} )
                            \\
                            &\redd \pRes{c_1d_1} ( \encc{z}{M} \{ c_1/x \} \| \pIn d_1(\_,e_1) ; \encc{e_1}{N} )
                            \\
                            &\equiv \pRes{xd_1} ( \encc{z}{M} \| \pIn d_1(\_,e_1) ; \encc{e_1}{N} )
                            \\
                            &= \encc{z}{M \tSub{ N/x }}
                        \end{align*}
                    }
                    \proofTransTermRedLam

                \item
                    Rule~\ruleLabel{red-lift}: $\term{M} \reddM \term{N}$ implies $\tCtx{R}[M] \reddM \tCtx{R}[N]$.
                    This case follows from \Cref{l:FIRST:transCtxs} and \ih2.

                \item
                    Rule~\ruleLabel{red-lift-sc}: $\term{M} \equivM \term{M'}$, $\term{M'} \reddM \term{N'}$, and $\term{N'} \equivM \term{N}$ imply $\term{M} \reddM \term{N}$.
                    By \ih2, it suffices to show that the translation preserves structural congruence of terms as structural congruence of processes.
                    Hence, we apply induction on the derivation of $\term{M} \equivM \term{M'}$ (and similarly for $\term{N'} \equivM \term{N}$; \ih3).
                    In \Cref{as:FIRST:oc:compl}, this property is proven separately as \Cref{t:FIRST:transTermSc}.
                    \global\def\proofFIRSTTransTermSc#1#2{%
                        The inductive cases follow from \ih#1 and \Cref{l:FIRST:transCtxs} directly.
                        We detail the (only) base case of Rule~\ruleLabel{sc-sub-ext}:
                        $\term{x} \notin \fn(\tCtx{R})$ implies $\term{(\tCtx{R}[M]) \tSub{ N/x }} \equivM \tCtx{R}[M \tSub{ N/x }]$.

                        The analysis is by induction on the structure of $\tCtx{R}$ (\ih#2), assuming $\term{x} \notin \fn(\tCtx{R})$.
                        The base case where $\tCtx{R} = \tHole$ is immediate.
                        We detail one representative inductive case: $\tCtx{R} = \term{\tCtx{R'}\ M'}$.
                        The thesis holds as follows:
                        \begin{align*}
                            &\hphantom{{}={}}
                            \encc{z}{(\tCtx{R'}[M]\ M') \tSub{ N/x }}
                            \\
                            &= \pRes{xa_1} \big( \pRes{a_2b_2} \pRes{c_2d_2} ( \encc{a_2}{\tCtx{R'}[M]} \| \pOut b_2[c_2,z] \| \pIn d_2(\_,e_2) ; \encc{e_2}{M'} ) \| \pIn a_1(\_,b_1) ; \encc{b_1}{N} \big)
                            \\
                            &\equiv \pRes{a_2b_2} \pRes{c_2d_2} \big( \pRes{xa_1} ( \encc{a_2}{\tCtx{R'}[M]} \| \pIn a_1(\_,b_1) ; \encc{b_1}{N} ) \| \pOut b_2[c_2,z] \| \pIn d_2(\_,e_2) ; \encc{e_2}{M'} \big)
                            \\
                            &= \pRes{a_2b_2} \pRes{c_2d_2} \big( \encc{a_2}{(\tCtx{R'}[M]) \tSub{ N/x }} \| \pOut b_2[c_2,z] \| \pIn d_2(\_,e_2) ; \encc{e_2}{M'} \big)
                            \\
                            &\equiv \pRes{a_2b_2} \pRes{c_2d_2} \big( \encc{a_2}{\tCtx{R'}[M \tSub{ N/x }]} \| \pOut b_2[c_2,z] \| \pIn d_2(\_,e_2) ; \encc{e_2}{M'} \big)
                            \tag{\ih#2}
                            \\
                            &= \encc{z}{(\tCtx{R'}[M \tSub{ N/x }])\ M'} = \encc{z}{\tCtx{R}[M \tSub{ N/x }]}
                            \tag*{#1}
                        \end{align*}
                    }
                    \proofFIRSTTransTermSc34{}

            \end{itemize}

        \item
            Rule~\ruleLabel{red-conf-lift-sc}:
            $\term{C} \equivC \term{C'}$, $\term{C'} \reddC \term{D'}$, and $\term{D'} \equivC \term{D}$ imply $\term{C} \reddC \term{D}$.
            By the IH, it suffices to show that the translation preserves structural congruence of configurations as structural congruence of processes.
            Hence, we apply induction on the derivation of $\term{C} \equivC \term{C'}$ (and similarly for $\term{D'} \equivC \term{D}$; \ih2).
            In \Cref{as:FIRST:oc:compl}, this property is proven separately as \Cref{t:FIRST:transConfSc}.
            The inductive cases follow from \ih2 and \Cref{l:FIRST:transCtxs} straightforwardly.
            We detail the interesting base case of
            \global\def\proofFIRSTTransConfScResSwap#1#2{%
                Rule~\ruleLabel{sc-res-swap}:
                $\term{\pRes{x\tBfr{\epsi}y} C} \equivC \term{\pRes{y\tBfr{\epsi}x} C}$.
                Both directions are analogous; we detail the left to right direction.
                We first infer the typing of the left configuration:
                \[
                    \begin{bussproof}
                        \bussAssume{
                            \type{\emptyset} \vdashB \term{\tBfr{\epsi}} : \type{S'}\>\type{S}
                        }
                        \bussAssume{
                            \type{\Gamma} \vdashC{\phi} \term{C} : \type{T}
                        }
                        \bussBin[\ruleLabel{typ-res/-buf}]{
                            \type{\Gamma'} \vdashC{\phi} \term{\pRes{x\tBfr{\epsi}y} C} : \type{T}
                        }
                    \end{bussproof}
                \]
                Here, $\type{\Gamma'} = \type{\Gamma} \setminus \term{x}:\type{S'} , \term{y}:\type{\ol{S}}$.
                The analysis depends on whether $\term{x} = \term{\tNil}$ and/or $\term{y} = \term{\tNil}$.
                In each case, we show that
                \begin{align}
                    \encc{a\>b}{\type{\emptyset} \vdashB \term{\tBfr{\epsi}} : \type{S'}\>\type{S}} = \encc{b\>a}{\type{\emptyset} \vdashB \term{\tBfr{\epsi}} : \type{\ol{S}}\>\type{\ol{S'}}}:
                    \label{eq:CGV2:trans:confSC:resSwap#2}
                \end{align}
                \begin{itemize}

                    \item
                        Case $\term{x} = \term{y} = \term{\tNil}$, or $\term{x} \neq \term{\tNil}$ and $\term{x} \neq \term{\tNil}$.
                        Either way, $\type{S'} = \type{S}$.
                        If $\type{S'} = \type{\tNil}$, both translations are $\0$, from which the thesis follows immediately.
                        Otherwise, the thesis follows by induction on the structure of $\type{S'}$; clearly, the resulting translations are exactly the same.

                    \item
                        Case $\term{x} = \term{\tNil}$ and $\term{y} \neq \term{\tNil}$, or $\term{x} \neq \term{\tNil}$ and $\term{y} = \term{\tNil}$.
                        W.l.o.g., assume the former.
                        Then $\type{S'} = \type{\tNil}$ and $\type{S} = \type{\tEnd}$.
                        The thesis then holds as follows:
                        \begin{align*}
                            \encc{a\>b}{\type{\emptyset} \vdashB \term{\tBfr{\epsi}} : \type{\tNil}\>\type{\tEnd}} &= \pIn b(\_,c) ; \pOut c[\_,\_]
                            \\
                            &= \encc{b\>c}{\type{\emptyset} \vdashB \term{\tBfr{\epsi}} : \type{\tEnd}\>\type{\tNil}}
                        \end{align*}

                \end{itemize}
                The thesis then holds as follows:
                \begin{align*}
                    \encc{z}{\pRes{x\tBfr{\epsi}y} C} &= \pRes{ax} \pRes{by} ( \encc{a\>b}{\type{\emptyset} \vdashB \term{\tBfr{\epsi}} : \type{S'}\>\type{S}} \| \encc{z}{C} )
                    \\
                    &= \pRes{ax} \pRes{by} ( \encc{b\>a}{\type{\emptyset} \vdashB \term{\tBfr{\epsi}} : \type{\ol{S}}\>\type{\ol{S'}}} \| \encc{z}{C} )
                    \tag*{\eqref{eq:CGV2:trans:confSC:resSwap#2}}
                    \\
                    &\equiv \pRes{by} \pRes{ax} ( \encc{b\>a}{\type{\emptyset} \vdashB \term{\tBfr{\epsi}} : \type{\ol{S}}\>\type{\ol{S'}}} \| \encc{z}{C} )
                    \\
                    &= \encc{z}{\pRes{y\tBfr{\epsi}x} C}
                    \tag*{#1}
                \end{align*}
            }
            \proofFIRSTTransConfScResSwap{\qedhere}{}

    \end{itemize}
\end{proof}

\begin{proof}[Proof of soundness (\Cref{t:FIRST:transSound})]
    \global\def\proofFIRSTTransSoundIntro{%
        By induction on the number $k$ of steps $\encc{z}{C} \redd^k Q$ (\ih1).
        We distinguish cases on all possible initial reductions $\encc{z}{C} \redd Q_0$ and discuss all possible following reductions.
        Here, we rely on \APCP's confluence of independent reductions, allowing us to focus on a specific sequence of reductions, postponing other possibilities that eventually lead to the same result.

        We then use induction on the structure of $\term{C}$ (\ih2).
        The goal is to identify some $\term{D_0}$ such that we can isolate $k_0 \geq 0$ reductions such that $\term{C} \reddC \term{D_0}$ and $\encc{z}{C} \redd Q_0 \redd^{k_0} \encc{z}{D_0}$ (where $k_0$ may be different in each case).
        We then have $\encc{z}{D_0} \redd^{k-k_0} Q$, so it follows from \ih1 that there exists $\term{D}$ such that $\term{D_0} \reddC^\ast \term{D}$ and $\encc{z}{D_0} \redd* \encc{z}{D}$.
    }
    \proofFIRSTTransSoundIntro
    Here, we detail only the interesting case of an interaction between the translations of a buffer and a contained configuration; \Cref{as:FIRST:oc:sound} details the whole proof.%
    \global\def\proofFIRSTTransSoundBuf#1{%
        $\term{C} = \term{\pRes{x\tBfr{\vec{m}}y} C_1}$.
        We have
        \[
            \encc{z}{\pRes{x\tBfr{\vec{m}}y} C_1} = \pRes{a_1x} \pRes{b_1y} ( \encc{a_1\>b_1}{\tBfr{\vec{m}}} \| \encc{z}{C_1} ).
        \]
        The reduction may originate from (i)~$\encc{a_1\>b_1}{\tBfr{\vec{m}}}$, (ii)~$\encc{z}{C_1}$, (iii)~a synchronization between $a_1$ in $\encc{a_1\>b_1}{\tBfr{\vec{m}}}$ and $x$ in $\encc{z}{C_1}$, or (iv)~a synchronization between $b_1$ in $\encc{a_1\>b_1}{\tBfr{\vec{m}}}$ and $y$ in $\encc{z}{C_1}$.
        We detail each case:
\begin{enumerate}[label=(\roman*)] 

            \item
                The reduction originates from $\encc{a_1\>b_1}{\tBfr{\vec{m}}}$.
                No matter what $\term{\tBfr{\vec{m}}}$ is, no reduction is possible.

            \item
                The reduction originates from $\encc{z}{C_1}$.
                We thus have $\encc{z}{C_1} \redd Q_1$.
                By \ih2, there are $\term{D_1},k_1 \geq 0$ such that $\term{C_1} \reddC \term{D_1}$ and $\encc{z}{C_1} \redd Q_1 \redd^{k_1} \encc{z}{D_1}$.
                Let $\term{D_0} \deq \term{\pRes{x\tBfr{\vec{m}}y} D_1}$.
                We have $\term{C} \reddC \term{D_0}$.
                Moreover:
                \begin{align*}
                    \encc{z}{C}
                    &= \pRes{a_1x} \pRes{b_1y} ( \encc{a_1\>b_1}{\tBfr{\vec{m}}} \| \encc{z}{C_1} )
                    \\
                    &\redd^{k_1+1} \pRes{a_1x} \pRes{b_1y} ( \encc{a_1\>b_1}{\tBfr{\vec{m}}} \| \encc{z}{D_1} )
                    \\
                    &= \encc{z}{D_0}
                \end{align*}

            \item
                The reduction originates from a synchronization between $a_1$ in $\encc{a_1\>b_1}{\tBfr{\vec{m}}}$ and $x$ in $\encc{z}{C_1}$.
                By well typedness, $\type{\Delta} \vdashB \term{\tBfr{\vec{m}}} : \type{S'}\>\type{S}$.
                Note first that, by \Cref{l:FIRST:transBuf}, $\type{S'} \neq \type{\tNil}$ implies that there are $\evalCtx{E_2},c_1$ such that $\enc{a_1\>b_1}{\term{\tBfr{\vec{m}}} : \type{S'}\>\type{S}} = \evalCtx[\big]{E_2}[\enc{a_1\>c_1}{\term{\tBfr{\epsi}} : \type{S'}\>\type{S'}}]$.
                The analysis depends on $\type{S'}$, so we consider all possibilities.
                In each case, if the reduction is indeed possible, we show that the reduction is the first step in the execution of some rule such that $\term{C} \reddC \term{D_0}$.
                The corresponding reduction $\encc{z}{C} \redd Q_0 \redd^{k_0} \encc{z}{D_0}$ follows the corresponding case in the proof of \Cref{t:FIRST:transCompl} (\nameref{t:FIRST:transCompl}).
                \begin{itemize}

                    \item
                        Case $\type{S'} = \type{\tNil}$.
                        Then $\term{x} = \term{\tNil}$ is not free in $\term{C_1}$, and thus $x$ is not free in $\encc{z}{C_1}$: the reduction is not possible.

                    \item
                        Case $\type{S'} = \type{\tEnd}$.
                        The analysis depends on whether $\type{S} = \type{\tNil}$ or not; w.l.o.g., assume not.
                        We have
                        \[
                            \enc{a_1\>c_1}{\term{\tBfr{\epsi}} : \type{\tEnd}\>\type{\tEnd}} = \pIn a_1(\_,c_2) ; \ldots \| \ldots
                        \]
                        Thus, the synchronization is between the receive on $a_1$ and a send on $x$ in $\encc{z}{C_1}$.

                        A send on a variable $\term{x}$ can only occur in the translation of that variable directly, under some reduction context.
                        Since $\term{x}$ is of type $\type{\tEnd}$ and its translation appears under a reduction context, the only well-typed way for $\term{x}$ to appear in $\term{C_1}$ is if $\term{C_1} = \tCtx[\big]{G}[\tCtx{F}[\tClose x ; M_1]]$.
                        We then have
                        \[
                            \term{C} \equivC \tCtx[\big]{G'}[\pRes{x\tBfr{\vec{m}}y} ( \tCtx{F}[\tClose x ; M_1] \| \term{C_2} )].
                        \]
                        Hence, the observed reduction is the first step of executing Rule~\ruleLabel{red-close}.

                    \item
                        Case $\type{S'} = \type{{!}T_2.S'_2}$.
                        We have
                        \[
                            \enc{a_1\>c_1}{\term{\tBfr{\epsi}} : \type{{!}T_2.S'_2}\>\type{{!}T_2.S'_2}} = \pIn a_1(\_,c_2) ; \ldots
                        \]
                        Thus, the synchronization is between the receive on $a_1$ and a send on $x$ in $\encc{z}{C_1}$.

                        A send on a variable $\term{x}$ can only occur in the translation of that variable directly, under some reduction context.
                        Since $\term{x}$ is of type $\type{{!}T_2.S'_2}$ and its translation appears under a reduction context, the only well-typed way for $\term{x}$ to appear in $\term{C_1}$ is if $\term{C_1} = \tCtx[\big]{G}[\tCtx{F}[\tSend M_1 \, x]]$.
                        We then have
                        \[
                            \term{C} \equivC \tCtx[\big]{G'}[\pRes{x\tBfr{\vec{m}}y} ( \tCtx{\hat{F'}}[\tSend M_1 \, x] \| \term{C_2} )].
                        \]
                        Hence, the observed reduction is the first step of executing Rule~\ruleLabel{red-send}.

                    \item
                        Case $\type{S'} = \type{{?}T_2.S'_2}$.
                        We have
                        \[
                            \enc{a_1\>c_1}{\term{\tBfr{\epsi}} : \type{{?}T_2.S'_2}\>\type{{?}T_2.S'_2}} = \enc{c_1\>a_1}{\term{\tBfr{\epsi}} : \type{{!}T_2.\ol{S'_2}}\>\type{{!}T_2.\ol{S'_2}}} = \pIn c_1(\_,c_2) ; \ldots
                        \]
                        Thus, the reduction is not possible.

                    \item
                        Case $\type{S'} = \type{\oplus \{ i : S^i_2 \}_{i \in I}}$.
                        We have
                        \[
                            \enc{a_1\>c_1}{\term{\tBfr{\epsi}} : \type{\oplus \{ i : S^i_2 \}_{i \in I}}\>\type{\oplus \{ i : S^i_2 \}_{i \in I}}} = \pIn a_1(\_,c_2) ; \ldots
                        \]
                        Thus, the synchronization is between the receive on $a_1$ and a send on $x$ in $\encc{z}{C_1}$.

                        A send on a variable $\term{x}$ can only occur in the translation of that variable directly, under some reduction context.
                        Since $\term{x}$ is of type $\type{\oplus \{ i : S^i_2 \}_{i \in I}}$ and its translation appears under a reduction context, the only well-typed way for $\term{x}$ to appear in $\term{C_1}$ is if $\term{C_1} = \tCtx[\big]{G}[\tCtx{F}[\tSel j \, x]]$ where $j \in I$.
                        We then have
                        \[
                            \term{C} \equivC \tCtx[\big]{G'}[\pRes{x\tBfr{\vec{m}}y} ( \tCtx{F}[\tSel j \, x] \| \term{C_2} )].
                        \]
                        Hence, the observed reduction is the first step of executing Rule~\ruleLabel{red-select}.

                    \item
                        Case $\type{S'} = \type{\& \{ i : S^i_2 \}_{i \in I}}$.
                        We have
                        \begin{align*}
                            \enc{a_1\>c_1}{\term{\tBfr{\epsi}} : \type{\& \{ i : S^i_2 \}_{i \in I}}\>\type{\& \{ i : S^i_2 \}_{i \in I}}}
                            &= \enc{c_1\>a_1}{\term{\tBfr{\epsi}} : \type{\oplus \{ i : \ol{S^i_2} \}_{i \in I}}\>\type{\oplus \{ i : \ol{S^i_2} \}_{i \in I}}}
                            \\
                            &= \pIn c_1(\_,c_2) ; \ldots
                        \end{align*}
                        Thus, the reduction is not possible.

                \end{itemize}

            \item
                The reduction originates from a synchronization between $b_1$ in $\encc{a_1\>b_1}{\tBfr{\vec{m}}}$ and $y$ in $\encc{z}{C_1}$.
                By well typedness, $\type{\Delta} \vdashB \term{\tBfr{\vec{m}}} : \type{S'}\>\type{S}$.
                The analysis depends on $\type{S}$, so we consider all possibilities.
                In each case, if the reduction is indeed possible, we show that the reduction is the first step in the execution of some rule such that $\term{C} \reddC \term{D_0}$.
                The corresponding reduction $\encc{z}{C} \redd Q_0 \redd^{k_0} \encc{z}{D_0}$ follows the corresponding case in the proof of \Cref{t:FIRST:transCompl} (\nameref{t:FIRST:transCompl}).
                \begin{itemize}

                    \item
                        Case $\type{S} = \type{\tNil}$.
                        Then $\term{y} = \term{\tNil}$ is not free in $\term{C_1}$, and thus $y$ is not free in $\encc{z}{C_1}$: the reduction is not possible.

                    \item
                        Case $\type{S} = \type{\tEnd}$.
                        By well typedness, then $\term{\vec{m}} = \term{\epsi}$.
                        Let $\term{C'} \deq \term{\pRes{y\tBfr{\epsi}x} C_1}$; we have $\term{C} \equivC \term{C'}$ and $\encc{z}{C} \equiv \encc{z}{C'}$ (by \Cref{t:FIRST:transConfSc}).
                        The thesis then follows as in the analogous case under Subcase~(iii) above.

                    \item
                        Case $\type{S} = \type{{!}T_2.S_2}$.
                        By well typedness, then $\term{\vec{m}} = \term{\vec{m}',M_1}$.
                        We have
                        \[
                            \enc{a_1\>b_1}{\term{\tBfr{\vec{m}',M_1}} : \type{S'}\>\type{{!}T_2.S_2}} = \pRes{\ldots} \pRes{\ldots} ( \ldots \| \pIn b_1(\_,h_2) ; \ldots \| \ldots ).
                        \]
                        Thus, the synchronization is between the receive on $b_1$ and a send on $y$ in $\encc{z}{C_1}$.

                        A send on a variable $\term{y}$ can only occur in the translation of that variable directly, under some reduction context.
                        Since $\term{y}$ is of type $\type{\ol{S}} = \type{{?}T_2.\ol{S_2}}$ and its translation appears under a reduction context, the only well-typed way for $\term{y}$ to appear in $\term{C_1}$ is if $\term{C_1} = \tCtx[\big]{G}[\tCtx{F}[\tRecv y]]$.
                        We then have
                        \[
                            \term{C} \equivC \tCtx[\big]{G'}[\pRes{x\tBfr{\vec{m}',M_1}y} ( \tCtx{\hat{F'}}[\tRecv y] \| \term{C_2} )].
                        \]
                        Hence, the observed reduction is the first step of executing Rule~\ruleLabel{red-recv}.

                    \item
                        Case $\type{S} = \type{{?}T_2.S_2}$.
                        By well typedness, then $\term{\vec{m}} = \term{\epsi}$; this case is analogous to Case~$\type{S} = \type{\tEnd}$ above.

                    \item
                        Case $\type{S} = \type{\oplus \{ i : S_2^i \}_{i \in I}}$.
                        By well typedness, then $\term{\vec{m}} = \term{\vec{m}',j}$ where $j \in I$.
                        We have
                        \[
                            \enc{a_1\>b_1}{\term{\tBfr{\vec{m}',j}} : \type{S'}\>\type{\oplus \{ i : S_2^i \}_{i \in I}}} = \pRes{\ldots} ( \pIn b_1(\_,e_2) ; \ldots \| \ldots ).
                        \]
                        Thus, the synchronization is between the receive on $b_1$ and a send on $y$ in $\encc{z}{C_1}$.

                        A send on a variable $\term{y}$ can only occur in the translation of that variable directly, under some reduction context.
                        Since $\term{y}$ is of type $\type{\ol{S}} = \type{\& \{ i : \ol{S_2^i} \}_{i \in I}}$ and its translation appears under a reduction context, the only well-typed way for $\term{y}$ to appear in $\term{C_1}$ is if $\term{C_1} = \tCtx[\big]{G}[\tCtx{F}[\tCase y \tOf \{ i : M_{1.i} \}_{i \in I}]]$.
                        We then have
                        \[
                            \term{C} \equivC \tCtx[\big]{G'}[\pRes{x\tBfr{\vec{m}',j}y} ( \tCtx{F}[\tCase y \tOf \{ i : M_{1.i} \}_{i \in I}] \| \term{C_2} )].
                        \]
                        Hence, the observed reduction is the first step of executing Rule~\ruleLabel{red-case}.

                    \item
                        Case $\type{S} = \type{\& \{ i : S_2^i \}_{i \in I}}$.
                        By well typedness, then $\term{\vec{m}} = \term{\epsi}$; this case is analogous to Case~$\type{S} = \type{\tEnd}$ above.
                        #1

                \end{itemize}

        \end{enumerate}
    }
    That is, \proofFIRSTTransSoundBuf{\qedhere}
\end{proof}

\subsection{Deadlock Free \FIRST}
\label{s:FIRST:df}

By virtue of \Cref{t:APCP:df},
well-typed \APCP processes that are typable under empty contexts ($\vdash P :: \emptyset$) are deadlock free.
We may transfer this result to \FIRST configurations by appealing to the operational correctness of our translation (\Cref{t:FIRST:transCompl,t:FIRST:transSound} above).
Each deadlock-free configuration in \FIRST obtained via this transference satisfies two requirements:
\begin{itemize}

    \item
        The configuration is typable $\type{\emptyset} \vdashC{\tMain} \term{C} : \type{\1}$, i.e.,  it needs no external resources and has no external behavior.

    \item
        The typed translation of the configuration satisfies  priority requirements in \APCP: it is well typed under `$\vdash$', not only under `$\vdash*$' (cf.\ \Cref{s:FIRST:trans}).

\end{itemize}
\noindent 
We rely on soundness (\Cref{t:FIRST:transSound}) to transfer deadlock freedom to configurations:

\begin{thm}[Deadlock Freedom for \protect\FIRST]
\label{t:FIRST:df}
    Given $\type{\emptyset} \vdashC{\tMain} \term{C}: \type{\1}$, if $\vdash \encc{z}{C} :: \Gamma$ for some $\Gamma$, then $\term{C} \equiv \term{\tMain\,()}$ or $\term{C} \reddC \term{D}$ for some~$\term{D}$.
\end{thm}

\begin{proof}
    By \Cref{t:FIRST:transTp} (type preservation for the translation), $\Gamma = z:\enct{\1} = z:\bullet$.
    Then $\vdash \pRes{z\_} \encc{z}{C} :: \emptyset$.
    Hence, by \Cref{t:APCP:df} (deadlock freedom), either $\pRes{z\_} \encc{z}{C} \equiv \0$ or $\pRes{z\_} \encc{z}{C} \redd Q$ for some $Q$.
    The rest of the analysis depends on which possibility holds:
    \begin{itemize}

        \item
            We have $\pRes{z\_} \encc{z}{C} \equiv \0$.
            We straightforwardly deduce from the well typedness and translation of $\term{C}$ that $\term{C} \equivC \term{\tMain\,()}$, proving the thesis.

        \item
            We have $\pRes{z\_} \encc{z}{C} \redd Q$ for some $Q$.
            We argue that this implies that $\encc{z}{C} \redd Q_0$, for some $Q_0$.

            \begin{itemize}

                \item
                    The reduction involves $z$.
                    Since $z$ is of type $\bullet$ in $\encc{z}{C}$, then $z$ must occur in a forwarder, and the reduction involves a forwarder $\pFwd [z<>x]$ for some $x$.
                    Since $x$ is not free in $\encc{z}{C}$, there must be a restriction on $x$.
                    Hence, the forwarder $\pFwd [z<>x]$ can also reduce with this restriction, instead of with the restriction $\pRes{z\_}$.
                    This means that $\encc{z}{C} \redd Q_0$ for some $Q_0$.

                \item
                    The reduction does not involve  $z$, in which case the reduction must be internal to $\encc{z}{C}$.
                    That is, $\encc{z}{C} \redd Q_0$ for some $Q_0$ and $Q \equiv \pRes{z\_}Q_0$.
            \end{itemize}
\noindent 
            First, by contradiction, we show that $z$ is not involved in this reduction:
            Since $z$ is of type $\bullet$ in $\encc{z}{C}$, then $z$ must occur in a forwarder, and the reduction involves a forwarder $\pFwd [z<>x]$ for some $x$.
            However, our translation (\Cref{f:FIRST:trans1,f:FIRST:trans2,f:FIRST:trans3}) does not generate forwarders, so this is clearly a contradiction.
            Hence, the reduction does not involve $z$.

            Thus, the reduction must be internal to $\encc{z}{C}$.
            That is, $\encc{z}{C} \redd Q_0$ for some $Q_0$ and $Q \equiv \pRes{z\_}Q_0$.
            Then, by \Cref{t:FIRST:transSound} (soundness), there exists $\term{D}$ such that $\term{C} \reddC^\ast \term{D}$.
            Looking at the proof of \Cref{t:FIRST:transSound} (in \Cref{as:FIRST:oc:sound}), it is easy to see that in fact $\term{C} \reddC^+ \term{D}$.
            That is, there exists $\term{D_0}$ such that $\term{C} \reddC \term{D_0}$, proving the thesis.
            \qedhere
    \end{itemize}

\end{proof}

\begin{exa}
    \label{x:FIRST:deadlocked}
    To illustrate the insights gained from deadlock analysis in \APCP through translation from \FIRST, we consider a seemingly deadlock-free configuration that is not.
    To this end, we first discuss how we can assign priorities to the translated types of the configuration, and extract from the translation requirements on those priorities.
    Next, we show that our example leads to unsatisfiable priority requirements, indicating the elusive deadlock in the original configuration.
    We close the example by considering an alternative configuration that is indeed deadlock free.

    Our example is a classic showcase of deadlock due to cyclic connections under synchronous communication.
    Though one would expect this example to be deadlock free under asynchronous communication, it is not: there is a deadlock induced by the call-by-name semantics of \FIRST.
    Let
    \begin{align*}
        \term{M(x,y)} &\deq \term{
            \begin{array}[t]{@{}l@{}}
                \tLet x_1 = \tSend () \, x \tIn
                \\
                \tLet (v,y_1) = \tRecv y \tIn
                \\
                \tClose x_1 ; \tClose y_1 ; v \normalcolor,
            \end{array}
        }
        &
        \term{C} &\deq \term{
            \tMain \, \begin{array}[t]{@{}l@{}}
                \tLet (x,x') = \tNew \tIn
                \\
                \tLet (y,y') = \tNew \tIn
                \\
                \tFork M(x,y) ; M(y',x') \normalcolor.
            \end{array}
        }
    \end{align*}
    Before showing the deadlock and its source, we first analyze the deadlock through translation into \APCP.

    By \Cref{t:FIRST:transTp} and assigning priority variables to each connective, we have
    \begin{align*}
        &\vdash* \encc{z_1}{M(x,y)} :: \begin{array}[t]{@{}l@{}}
            y : \bullet \tensor^{\pri_1} (\bullet \tensor^{\pri_2} \bullet) \parr^{\pri_3} \bullet \tensor^{\pri_4} \bullet \parr^{\pri_5} \bullet , \\
            x : \bullet \tensor^{\pri_6} \bullet \parr^{\pri_7} (\bullet \parr^{\pri_8} \bullet) \tensor^{\pri_9} \bullet \tensor^{\pri_{10}} \bullet \parr^{\pri_{11}} \bullet , z_1 : \bullet
        \end{array}
        \\
        &\vdash* \encc{z_2}{M(y',x')} :: \begin{array}[t]{@{}l@{}}
            x' : \bullet \tensor^{\pi_1} (\bullet \tensor^{\pi_2} \bullet) \parr^{\pi_3} \bullet \tensor^{\pi_4} \bullet \parr^{\pi_5} \bullet , \\
            y' : \bullet \tensor^{\pi_6} \bullet \parr^{\pi_7} (\bullet \parr^{\pi_8} \bullet) \tensor^{\pi_9} \bullet \tensor^{\pi_{10}} \bullet \parr^{\pi_{11}} \bullet , z_2 : \bullet
        \end{array}
        \\
        &\vdash* \encc{z}{C} :: z : \bullet
    \end{align*}
    To determine requirements on these priorities, we analyze the translations of typing rules into \APCP derivations in \Cref{f:FIRST:trans1,f:FIRST:trans2,f:FIRST:trans3} and the priority requirements induced by them.
    Indeed, a deadlock is detected, because \APCP requires $\pri_6 < \pi_1 < \pi_6 < \pri_1 < \pri_6$, which is unsatisfiable.

    Before detailing the  origin of these requirements, we  first give an intuition behind this unsatisfiable chain of priority requirements, and how they reflect the deadlock that is indeed present in $\term{C}$.
    From left to right, the requirements denote that (1)~the send on $x$ occurs before the receive on $x'$, (2)~the receive on $x'$ occurs before the send on $y'$, (3)~the send on $y'$ occurs before the receive on $y$, (4)~the receive on $y$ occurs before the send on $x$.
    Steps~(2) and~(3) stand out: $\term{M(x,y)}$ and $\term{M(y',x')}$ seem to define an opposite order.
    This is because in, e.g., $\term{M(x,y)}$, the call-by-name semantics of \FIRST transforms the $\term{\tLet x_1}$ into an explicit substitution that can only be resolved---thus enabling the $\term{\tSend () \, x}$---once the $\term{\tClose x_1}$ is unblocked by the $\term{\tLet (v,y_1)}$, which in turn is waiting for the $\term{\tRecv y}$ to be resolved.
    To be precise:
    \begin{align*}
        \term{M(x,y)} &\reddM \term{\big( \tLet (v,y_1) = \tRecv y \tIn \ldots \big) \tSub{\tSend () \, x / x_1}} \deq \term{M'(x,y)} \not\mkern-6mu\reddM
        \\
        \term{M(y',x')} &\reddM \term{\big( \tLet (v',x'_1) = \tRecv x' \tIn \ldots \big) \tSub{\tSend () \, y' / y'_1}} \deq \term{M'(y',x')} \not\mkern-6mu\reddM
        \\
        \term{C} &\reddC^9 \term{\pRes{s\tBfr{\epsi}s'} \pRes{t\tBfr{\epsi}t'} \big( \tChild \, M'(x,t) \tSub{s/x} \prl \tMain \, M'(y',s') \tSub{t'/y'} \big) } \not\mkern-6mu\reddC
    \end{align*}

    We now detail the origin of the conflicting priority requirements; that is, we spell out the rules of the translation and the induced requirements.
    Note that many of these rules include names and priority variables that are not visible outside the derivation, being created from within certain parts of the translation.
    \begin{enumerate}

        \item
            Requirement $\pri_6 < \pi_1$ originates from the translation of $\term{\tNew}$ in the $\term{\tLet (x,x')}$.
            More precisely, the translation of $\term{\tNew}$ prepares the translation of the upcoming buffer, which follows the involved types.
            The requirement is induced by a step in the translation of this buffer (Rule~\ruleLabel{typ-buf} in \Cref{f:FIRST:trans2}):
            \[
                \begin{bussproof}
                    \def\defaultHypSeparation{\hskip1.5ex}
                    \bussAssume{
                        \vdash* \enc{g_3\>k_3}{\term{\tBfr{\epsi}} : \type{\tEnd}\>\type{\tEnd}} :: g_3 : \bullet \parr^{\rho_5} \bullet \tensor^{\pri_{11}} \bullet , k_3 : \bullet \parr^{\rho_6} \bullet \tensor^{\pi_5} \bullet
                    }
                    \bussAssume{
                        \begin{array}[b]{@{}l@{}}
                            \pri_8=\pi_2 , \underline{\pri_6<\pi_1} , \pri_9<\pi_1 ,
                            \\
                            \pi_1 < \pri_8 , \pi_1 < \pi_5 , \pri_7 < \pri_9 ,
                            \\
                            \pi_3 < \pi_2 , \pi_3 < \pi_4
                        \end{array}
                    }
                    \bussBin{
                        \vdash* \enc{d_2\>e_2}{\term{\tBfr{\epsi}} : \type{{!}\1.\tEnd}\>\type{{!}\1.\tEnd}} :: \begin{array}[t]{@{}l@{}}
                            d_2 : \bullet \parr^{\pri_6} \bullet \tensor^{\pri_7} (\bullet \tensor^{\pri_8} \bullet) \parr^{\pri_9} \bullet \parr^{\pri_{10}} \bullet \tensor^{\pri_{11}} \bullet ,
                            \\
                            e_2 : \bullet \parr^{\pi_1} (\bullet \parr^{\pi_2} \bullet) \tensor^{\pi_3} \bullet \parr^{\pi_4} \bullet \tensor^{\pi_5} \bullet
                        \end{array}
                    }
                \end{bussproof}
            \]

        \item
            Requirement $\pi_1 < \pi_6$ originates from four substeps, including intermediate priorities: $\pi_1 < \pi_3 < \pi_{14} < \pi_{12} < \pi_6$.
            \begin{enumerate}

                \item
                    $\pi_1 < \pi_3$ originates from the translation of the variable $\term{x'}$ (Rule~\ruleLabel{typ-var} in \Cref{f:FIRST:trans1}):
                    \[
                        \begin{bussproof}
                            \bussAssume{
                                \underline{\pi_1<\pi_3}
                            }
                            \bussUn{
                                \vdash* \encc{a'_4}{x'} :: x' : \bullet \tensor^{\pi_1} (\bullet \tensor^{\pi_2} \bullet) \parr^{\pi_3} \bullet \tensor^{\pi_4} \bullet \parr^{\pi_5} \bullet , a'_4 : (\bullet \parr^{\pi_2} \bullet) \tensor^{\pi_3} \bullet \parr^{\pi_4} \bullet \tensor^{\pi_5} \bullet
                            }
                        \end{bussproof}
                    \]

                \item
                   $\pi_3 < \pi_{14}$ originates from the translation of the $\term{\tRecv x'}$ (Rule~\ruleLabel{typ-recv} in \Cref{f:FIRST:trans1}):
                    \[
                        \begin{bussproof}
                            \def\defaultHypSeparation{\hskip1.4ex}
                            \bussAssume{
                                \vdash* \encc{a'_4}{x'} :: \begin{array}[b]{@{}l@{}}
                                    x' : \bullet \tensor^{\pi_1} (\bullet \tensor^{\pi_2} \bullet) \parr^{\pi_3} \bullet \tensor^{\pi_4} \bullet \parr^{\pi_5} \bullet ,
                                    \\
                                    a'_4 : (\bullet \parr^{\pi_2} \bullet) \tensor^{\pi_3} \bullet \parr^{\pi_4} \bullet \tensor^{\pi_5} \bullet
                                \end{array}
                            }
                            \bussAssume{
                                \pi_{14}<\pi_2 , \pi_{14}<\pi_{15} , \pi_4<\pi_5 ,  \underline{\pi_3<\pi_{14}}
                            }
                            \bussBin{
                                \vdash* \encc{b'_3}{\tRecv x'} :: x' : \bullet \tensor^{\pi_1} (\bullet \tensor^{\pi_2} \bullet) \parr^{\pi_3} \bullet \tensor^{\pi_4} \bullet \parr^{\pi_5} \bullet , b'_3 : (\bullet \parr^{\pi_2} \bullet) \tensor^{\pi_{14}} \bullet \parr^{\pi_{15}} \bullet \tensor^{\pi_5} \bullet
                            }
                        \end{bussproof}
                    \]

                \item
                     $\pi_{14} < \pi_{12}$ originates from the translation of the $\term{\tLet (v',x'_1)}$ (Rule~\ruleLabel{typ-split} in \Cref{f:FIRST:trans1}):
                    \[
                        \begin{bussproof}
                            \bussAssume{
                                \begin{array}[b]{@{}l@{}}
                                    \vdash* \encc{b'_3}{\tRecv x'}
                                    \\
                                    :: \begin{array}[t]{@{}l@{}}
                                        x' : \bullet \tensor^{\pi_1} (\bullet \tensor^{\pi_2} \bullet) \parr^{\pi_3} \bullet \tensor^{\pi_4} \bullet \parr^{\pi_5} \bullet ,
                                        \\
                                        b'_3 : (\bullet \parr^{\pi_2} \bullet) \tensor^{\pi_{14}} \bullet \parr^{\pi_{15}} \bullet \tensor^{\pi_5} \bullet
                                    \end{array}
                                \end{array}
                            }
                            \bussAssume{
                                \begin{array}[b]{@{}l@{}}
                                    \vdash* \encc{a'_2}{\ldots}
                                    \\
                                    :: \begin{array}[t]{@{}l@{}}
                                        y'_1 : \bullet \tensor^{\pi_{12}} \bullet \parr^{\pi_{11}} \bullet ,
                                        \\
                                        v' : \bullet \tensor^{\pi_2} \bullet ,
                                        \\
                                        x'_1 : \bullet \tensor^{\pi_{15}} \bullet \parr^{\pi_5} \bullet
                                    \end{array}
                                \end{array}
                            }
                            \bussAssume{
                                 \underline{\pi_{14}<\pi_{12}}
                            }
                            \bussTern{
                                \vdash* \encc{a'_2}{\tLet (v',x'_1) = \tRecv x' \tIn \ldots} :: \begin{array}[t]{@{}l@{}}
                                    x' : \bullet \tensor^{\pi_1} (\bullet \tensor^{\pi_2} \bullet) \parr^{\pi_3} \bullet \tensor^{\pi_4} \bullet \parr^{\pi_5} \bullet ,
                                    \\
                                    y'_1 : \bullet \tensor^{\pi_{12}} \bullet \parr^{\pi_{11}} \bullet , a'_2 : \bullet
                                \end{array}
                            }
                        \end{bussproof}
                    \]

                \item
                 Finally, $\pi_{12} < \pi_6$ originates from the translation of the $\term{\tLet y'_1}$.
                    To be precise, this syntactic sugar breaks down into an abstraction and an application; the requirement originates from the application (Rule~\ruleLabel{typ-app} in \Cref{f:FIRST:trans1}):
                    \[
                        \begin{bussproof}
                            \bussAssume{
                                \begin{array}[b]{@{}l@{}}
                                    \vdash* \encc{a'_1}{\lam y'_1 \ldots}
                                    \\
                                    :: \begin{array}[t]{@{}l@{}}
                                        x' : \bullet \tensor^{\pi_1} (\bullet \tensor^{\pi_2} \bullet) \parr^{\pi_3} \bullet \tensor^{\pi_4} \bullet \parr^{\pi_5} \bullet ,
                                        \\
                                        a'_1 : (\bullet \tensor^{\pi_{12}} \bullet \parr^{\pi_{11}} \bullet) \parr^{\pi_{13}} \bullet
                                    \end{array}
                                \end{array}
                            }
                            \bussAssume{
                                \begin{array}[b]{@{}l@{}}
                                    \vdash* \encc{e'_1}{\tSend () \, y'}
                                    \\
                                    :: \begin{array}[t]{@{}l@{}}
                                        y' : \begin{array}[t]{@{}l@{}}
                                            \bullet \tensor^{\pi_6} \bullet \parr^{\pi_7} (\bullet \parr^{\pi_8} \bullet)
                                            \\
                                            {} \tensor^{\pi_9} \bullet \tensor^{\pi_{10}} \bullet \parr^{\pi_{11}} \bullet ,
                                        \end{array}
                                        \\
                                        e'_1 : \bullet \tensor^{\pi_{11}} \bullet
                                    \end{array}
                                \end{array}
                            }
                            \bussAssume{
                                \begin{array}[b]{@{}l@{}}
                                    \underline{\pi_{12}<\pi_6} ,
                                    \\
                                    \pi_{13}<\pi_{12}
                                \end{array}
                            }
                            \bussTern{
                                \vdash* \encc{z_2}{\tLet y'_1 = \tSend y' \tIn \ldots} :: \begin{array}[t]{@{}l@{}}
                                    x' : \bullet \tensor^{\pi_1} (\bullet \tensor^{\pi_2} \bullet) \parr^{\pi_3} \bullet \tensor^{\pi_4} \bullet \parr^{\pi_5} \bullet ,
                                    \\
                                    y' : \bullet \tensor^{\pi_6} \bullet \parr^{\pi_7} (\bullet \parr^{\pi_8} \bullet) \tensor^{\pi_9} \bullet \tensor^{\pi_{10}} \bullet \parr^{\pi_{11}} \bullet ,
                                    \\
                                    z_2 : \bullet
                                \end{array}
                            }
                        \end{bussproof}
                    \]

            \end{enumerate}

        \item
            Requirement $\pi_6 < \pri_1$ originates from the buffer prepared by the translation of $\term{\tNew}$ in the $\term{\tLet (y,y')}$.
            It is derived similar to Requirement~(1).

        \item
            Requirement $\pri_1 < \pri_6$ originates from four similar substeps as in Requirement~(2).

    \end{enumerate}
    We conclude that the translation of $\term{C}$ cannot be typed under $\vdash$, because the priority requirements cannot be satisfied.
    Hence, \Cref{t:FIRST:df} does not apply, and so deadlock freedom cannot be guaranteed for $\term{C}$.

    \smallskip

    Note that it is straightforward to define variants of $\term{C}$ whose translations are well typed under $\vdash$, thus guaranteeing deadlock freedom through \Cref{t:FIRST:df}.
    An example is the variant of $ \term{M(x,y)}$ above in which the $\term{\tClose x_1}$ occurs before the $\term{\tLet (v,y_1)}$.
\end{exa}

\begin{rem}
    \label{r:FIRST:transDfComplex}
    Determining the deadlock freedom of well-typed \FIRST programs by means of translation into \APCP is of similar complexity as a direct, priority-based approach.
    It is easy to see that the translation is $\mcl{O}(n)$ in the size of the \FIRST typing derivation.
    We could add priorities to the types of \FIRST and derive priority requirements from the translation; this would boil down to roughly the same amount of checking.
    Hence, the indirect approach via \APCP is similarly complex.
    This justifies keeping the type system of \FIRST simple, compared to type systems that use priorities for deadlock-free functional programs such as those by, e.g., Padovani and Novara~\cite{conf/forte/PadovaniN15} and Kokke and Dardha~\cite{conf/forte/KokkeD21}.
\end{rem}

\section{Related Work}
\label{s:APCP:rw}

Closely related work on the two themes of the paper (deadlock freedom by typing and functional calculi with concurrency) has been already discussed throughout the paper. Here we comment on other related literature.

\paragraph{Asynchronous Communication}
Asynchronous communication has a longstanding history in process algebras (see, e.g., \cite{conf/concurrency/BergstraKT85,conf/ifip/HeJH90,conf/lics/deBoerKP92}).
The first accounts of asynchronous communication for the $\pi$-calculus were developed independently by Honda and Tokoro~\cite{conf/ecoop/HondaT91,conf/occ/HondaT91} and by Boudol~\cite{report/Boudol92}.
Palamidessi~\cite{journal/mscs/Palamidessi03} shows that the synchronous $\pi$-calculus is strictly more expressive than its asynchronous variant, due to \emph{mixed choices}: non-deterministic choices involving both inputs and outputs.
Beauxis \etal~\cite{chapter/BeauxisPV08} study the exact form of asynchronous communication modeled by the asynchronous $\pi$-calculus; they examine communication mediated by different mechanisms (bags, stacks, queues) in the synchronous $\pi$-calculus, and prove that bags lead to the strongest operational correspondence with the asynchronous $\pi$-calculus.

As discussed already, asynchrony is a relevant phenomenon in session $\pi$-calculi; the communication structures delineated by sessions strongly influence the expected ordering of messages.
The expressiveness gap between asynchronous and synchronous communication shown by Palamidessi in the untyped setting does not hold in this context, since session-typed $\pi$-calculi consider only deterministic choices and do not account for mixed choices.
A notable exception is the work on \emph{mixed sessions} by Casal \etal~\cite{conf/esop/VasconcelosCAM20,journal/tcs/CasalMV22}, which does not address deadlock freedom.
Nevertheless, fundamental differences between mixed choices in untyped and session-typed settings remain, as established by Peters and Yoshida~\cite{conf/express/PetersY22}.

In the context of session types, the first formal theory of asynchronous communication for the $\pi$-calculus is by Kouzapas \etal~\cite{conf/forte/KouzapasYH11}, using a buffered semantics.
Their focus is on the behavioral theory induced by asynchrony and the program transformations it enables.
Follow-up work~\cite{journal/mscs/KouzapasYHH16} goes beyond to consider asynchronous communication in combination with constructs for event-driven programming, and develops a corresponding type system and behavioral theory.
Unlike our work, these two works do not consider deadlock freedom for asynchronous session processes.

DeYoung \etal~\cite{conf/csl/DeYoungCPT12} give the first connection between linear logic and a session-typed $\pi$-calculus with asynchronous communication using a continuation-passing semantics, and show that this semantics is equivalent to a buffered semantics.
As mentioned before, \APCP's semantics is based on this work, except that our typing rules for sending and selection are axiomatic and that we consider recursion.
On a similar line, Jia \etal~\cite{conf/popl/JiaGP16} consider a session-typed language with asynchronous communication; their focus is on the dynamic monitoring of session behaviors, not on deadlock freedom.
Both these works are based on the correspondence with intuitionistic linear logic, which restricts the kind of process networks allowed by typing.
Pruiksma and Pfenning~\cite{conf/places/PruiksmaP19,journal/jlamp/PruiksmaP20} derive a ``propositions-as-sessions'' type system from adjoint logic, which combines several logics with different structural rules through modalities~\cite{report/Reed09,report/PruiksmaCPR18}.
Their process language features asynchronous communication with continuation passing and their type system treats asynchronous, non-blocking outputs via axiomatic typing rules, similar to Rules~\ruleLabel{typ-send} and~\ruleLabel{typ-sel} in \Cref{f:APCP:typingRules}.

Padovani's linear type system for the asynchronous $\pi$-calculus~\cite{conf/lics/Padovani14} has been already mentioned in the context of deadlock freedom for cyclic process networks.
His language is different from \APCP, as it lacks session constructs and does not have continuation passing baked into the type system.
While it should be possible to encode sessions in his typed framework (using communication of pairs to model continuation passing), it seems unclear how to transfer the analysis of deadlock freedom from this setting to a language such as \APCP via such a translation.

\paragraph{Deadlock Freedom for Cyclic Process Networks}

We have already discussed the related works by Kobayashi~\cite{conf/concur/Kobayashi06}, Padovani~\cite{conf/lics/Padovani14}, and Dardha and Gay~\cite{conf/fossacs/DardhaG18}.
The work of Kobayashi and Laneve~\cite{journal/ic/KobayashiL17} is related to \APCP in that it addresses deadlock freedom for \emph{unbounded} process networks.
 Toninho and Yoshida's work~\cite{journal/toplas/ToninhoY18} addresses deadlock freedom for cyclic process networks by generating global types from binary types.
The work by Balzer, Toninho and Pfenning~\cite{conf/icfp/BalzerP17,conf/esop/BalzerTP19} is also worth mentioning: it guarantees deadlock freedom for processes with shared, mutable resources by means of manifest sharing, i.e., explicitly acquiring and releasing access to resources.

\paragraph{Functional Languages with Sessions}
We have already discussed the related works by Gay and Vasconcelos~\cite{journal/jfp/GayV10}, Wadler~\cite{conf/icfp/Wadler12}, Kokke and Dardha~\cite{conf/forte/KokkeD21,report/KokkeD21}, and Padovani and Novara~\cite{conf/forte/PadovaniN15}.
Lindley and Morris~\cite{conf/esop/LindleyM15} formally define a semantics for Wadler's \GV, prove deadlock freedom and operational correspondence of the translation into Wadler's \CP, and give a translation from \CP into \GV.
Other related work in this line is Fowler \etal's \emph{Asynchronous} \GV with buffered asynchronous communication and a call-by-value semantics, and \emph{Exceptional} \GV (\EGV) which extends the former with exception handling~\cite{thesis/Fowler19,conf/popl/FowlerLMD19}.

\Cref{tbl:CGV:compare} summarizes the comparison of \FIRST to some of these related works.
Note that none of the mentioned works supports sending anything but values, unlike \FIRST.
Our work ensures deadlock freedom for configurations with cyclic topologies by means of a translation into a system with an established deadlock-freedom result (cf.\ \Cref{r:FIRST:transDfComplex}); the alternative approach is to enhance \GV's type system with priorities, as done by Padovani and Novara~\cite{conf/forte/PadovaniN15} and Kokke and Dardha~\cite{conf/forte/KokkeD21,report/KokkeD21}.

\begin{table}[!t]
    \begin{center}
        \begin{tabular}{cccccc}
            \toprule
            & \LAST
            & \GV
            & \EGV
            & \PGV
            & \textbf{\FIRST}
            \\
            & \cite{journal/jfp/GayV10}
            & \cite{conf/icfp/Wadler12}
            & \cite{conf/popl/FowlerLMD19}
            & \cite{conf/forte/KokkeD21,report/KokkeD21}
            & \textbf{(this paper)}
            \\ \midrule
            Communication
            & Async.
            & Sync.
            & Async.
            & Sync.
            & Async.
            \\ \midrule
            \makecell{Cyclic Topologies}
            & $\checkmark$
            & $\times$
            & $\times$
            & $\checkmark$
            & $\checkmark$
            \\ \midrule
            \makecell{Deadlock \\ Freedom}
            & $\times$
            & \makecell{$\checkmark$ \\ (by typing)}
            & \makecell{$\checkmark$ \\ (by typing)}
            & \makecell{$\checkmark$ \\ (by typing)}
            & \makecell{$\checkmark$ \\ (via \APCP)}
            \\ \bottomrule
        \end{tabular}
    \end{center}
    \caption{The features of \CGV compared to related works.}
    \label{tbl:CGV:compare}
\end{table}

\paragraph{Typed Encodings between (Concurrent) $\lambda$-calculi and $\pi$-calculi}
Several prior works develop typed encodings between $\lambda$- and $\pi$-calculi; to our knowledge, all of them consider call-by-value semantics, most do not consider $\lambda$ with message passing, and none translate variables as sends, as we do.
Some of these works have been already discussed  in \Cref{s:LAST:trans}, where we justified the notions of completeness and soundness relevant for our translation of \FIRST into \APCP.

Vasconcelos~\cite{report/Vasconcelos00} translates an untyped $\lambda$-calculus into an input/output-typed $\pi$-calculus; the translation is sound, up to barbed congruence.
The aforementioned work by Wadler~\cite{conf/icfp/Wadler12,journal/jfp/Wadler14} translates the session-typed \GV with synchronous message passing into the session-typed \CP, without giving a semantics for \GV nor associated completeness/soundness results.
Lindley and Morris~\cite{conf/esop/LindleyM15} formalize a semantics for \GV and its translation into~\CP, and prove completeness.
Similarly to our approach, they use closures/explicit substitutions in  \GV to obtain a direct, more controlled correspondence with cut reduction in linear logic; however, no soundness result is given.
Kokke and Dardha~\cite{conf/forte/KokkeD21} define \PGV, an extension of \GV with priority annotations to support deadlock-free cyclic thread configurations; they give a translation from \PCP into \PGV that is both sound and complete, but no translation in the other direction is studied.
Toninho \etal~\cite{conf/fossacs/ToninhoCP12} translate a linearly typed $\lambda$-calculus into a session-typed $\pi$-calculus derived from intuitionistic linear logic; they prove completeness, but for soundness they extend reduction in $\lambda$ to call by name.
Toninho and Yoshida~\cite{journal/toplas/ToninhoY21} translate a linearly typed $\lambda$-calculus without message passing into a session-typed polymorphic $\pi$-calculus. Their translation is fully abstract: it is complete and sound, with the latter result requiring an extension of reduction in $\pi$ with commuting conversions.
Fowler \etal~\cite{DBLP:journals/lmcs/FowlerKDLM23}
introduce a variant of (call-by-value) \GV based on \emph{hypersequents}, and
consider its relationship with hypersequent \CP: they exhibit translations that preserve and reflect reduction, up to weak bisimilarity.

\paragraph{Other Applications of \APCP}
As mentioned in the introduction, two salient aspects of Curry-Howard correspondences for session types, namely analysis of deadlock freedom and connections with functional calculi, define the main themes of this paper.
Yet another highlight of the logical correspondences is their suitability for the analysis of \emph{multiparty protocols}.
In separate work~\cite{journal/scico/vdHeuvelP22}, we have devised a methodology for the analysis of multiparty session types (MPST) based on \APCP~\cite{journal/acm/HondaYC16}.
In the multiparty context, two or more participants interact following a common protocol.
\APCP is well suited for the analysis of process implementations of MPST, which rely on asynchrony and recursion.
In our analysis, multiple separate processes implement the roles of one or more participants of a multiparty protocol.
The support for cyclic process networks in \APCP allows us to connect these implementations with each other directly, without requiring an additional process (as used in similar prior works~\cite{conf/forte/CairesP16,conf/concur/CarboneMSY15,conf/concur/CarboneLMSW16}).
The asynchrony in \APCP captures the asynchronous nature of MPST, and \APCP's recursion enables an expressive class of supported protocols.
This way, our methodology unlocks the transfer of (static) analysis of deadlock freedom and protocol conformance from \APCP to distributed implementations of MPST.
The key ideas underlying the methodology in~\cite{journal/scico/vdHeuvelP22} can be applied to the \emph{runtime} verification of such distributed implementations, as recently shown in~\cite{conf/rv/vdHeuvelPD23}.

\section{Closing Remarks}
\label{s:APCP:concl}

In this paper, we have presented two contributions to the challenging issue of ensuring deadlock-free, message-passing interactions in a session-typed setting.
Our \emph{first contribution} is \APCP: a typed framework for deadlock-free cyclic process networks with asynchronous communication and recursion.
The design of \APCP and its type system has a solid basis, with tight ties to logic (``propositions as sessions''): our syntax, semantics and type system harmoniously integrate insights separately presented in prior works for different typed variants of the $\pi$-calculus.

The step from synchronous communication (as in \PCP~\cite{conf/fossacs/DardhaG18}) to asynchronous communication (in \APCP) is significant.
In combination with cyclic process networks, asynchronous communication enables checking a larger class of deadlock-free processes: processes that under synchronous communication would be deadlocked due to blocking outputs may be deadlock free in \APCP (see, e.g., \Cref{x:APCP:redd}).
Perhaps more significantly, asynchronous communication simplifies the priority management involved in the detection of cyclic dependencies (cf.\ \Cref{r:APCP:simplerPrios}).

In our \emph{second contribution}, we move to consider cyclic process networks with asynchronous communication in the setting of a prototypical functional language.
We introduced \FIRST, a new functional language with asynchronous, session-typed communication.
Inspired by Gay and Vasconcelos's \LAST, the design of \FIRST combines buffered communication with a call-by-name reduction strategy, explicit substitutions, and explicit session closing.
In our opinion, this design makes \FIRST an interesting object of study on its own.
In the spirit of Gay and Vasconcelos's work, we defined a deliberately simple type system for \FIRST, whose type preservation property ensures protocol fidelity and communication safety, but not deadlock freedom.
This means that well-typed threads in \FIRST can perform session protocols that may lead to stuck terms/configurations.
To connect our contributions, and to transfer the deadlock-freedom guarantee to the functional realm, we presented a translation from \FIRST to \APCP that is \emph{operationally correct}.
In particular, operational correctness includes the \emph{soundness} property, which ensures that the translation reflects reduction steps and is critical to the transfer of deadlock freedom formalized by \Cref{t:FIRST:df}.
This way, we were able to analyze deadlock freedom in \FIRST by leaning on the established results for \APCP (\Cref{t:APCP:df}) without sacrificing complexity (cf.\ \Cref{r:FIRST:transDfComplex}).

In future work, it would be interesting to study the extension of \FIRST with recursion/recursive types, and to extend our translation into \APCP  (and its operational correspondence results) accordingly. We do not expect technical difficulties, in particular if \FIRST is enhanced with the kind of tail-recursive behavior present in \APCP. In the same spirit, it would be insightful to explore how to accommodate \APCP's support for priorities into a process language with inductive and coinductive types (least and greatest fixed points, respectively), such as the one studied by Rocha and Caires~\cite{DBLP:conf/esop/RochaC23}.
Finally, following~\cite{DBLP:journals/acta/Kobayashi05}, it would be interesting to study type inference algorithms for \APCP which could (automatically) determine priorities for typed processes.

\paragraph{Acknowledgements}
We are grateful to the anonymous reviewers for their careful reading of our paper and their useful feedback.
We also thank Ornela Dardha for clarifying the typing rules of \PCP to us, and Simon Fowler and Simon Gay for their helpful feedback.
We gratefully acknowledge the support of the Dutch Research Council (NWO) under project No.\,016.Vidi.189.046 (Unifying Correctness for Communicating Software).

\bibliographystyle{alphaurl}
\bibliography{refs}

\appendix
\newpage

\tableofcontents

\newpage
\section{\protect\FIRST: Detailed Definitions and Proofs}
\label{as:FIRST}

\subsection{Self-contained Definition of \protect\FIRST and its Type System}
\label{as:FIRST:lang}

\begin{figure}[t!]
    Terms ($\term{M},\term{N},\ldots$) and reduction contexts ($\tCtx{R}$):
    \begin{align*}
        \term{M},\term{N}
        & \begin{array}[t]{@{}l@{\kern.5ex}lr@{\kern1.2ex}l@{\kern.5ex}lr@{}}
            {} ::= &
            \term{x} & \text{variable}
            & \sepr &
            \term{\tNew} & \text{create new channel}
            \\ \sepr* &
            \term{()} & \text{unit value}
            & \sepr &
            \term{\tFork M ; N} & \text{fork $\term{M}$ in parallel to $\term{N}$}
            \\ \sepr* &
            \term{\lam x . M} & \text{abstraction}
            & \sepr &
            \term{(M,N)} & \text{pair construction}
            \\ \sepr* &
            \term{M~N} & \text{application}
            & \sepr &
            \term{\tLet (x,y)=M \tIn N} & \text{pair deconstruction}
            \\ \sepr* &
            \term{\tSend M \, N} & \text{send $\term{M}$ along $\term{N}$}
            & \sepr &
            \term{\tSel \ell \, M} & \text{select label $\ell$ along $\term{M}$}
            \\ \sepr* &
            \term{\tRecv M} & \text{receive along $\term{M}$}
            & \sepr &
            \term{\tCase M \tOf \{ i : M \}_{i \in I}} & \text{offer labels in $I$ along $\term{M}$}
            \\ \sepr* &
            \term{\tClose M ; N} & \text{close $\term{M}$}
            & \sepr &
            \term{M \tSub{ N/x }} & \text{explicit substitution}
        \end{array}
        \\
        \term{\tCtx{R}}
        & \begin{array}[t]{@{}l@{\kern.5ex}l@{}}
            {} ::= &
            \term{\tHole}
            \sepr
            \term{\tCtx{R}~M}
            \sepr
            \term{\tSend M \, \tCtx{R}}
            \sepr
            \term{\tRecv \tCtx{R}}
            \sepr
            \term{\tLet (x,y) = \tCtx{R} \tIn M}
            \\ \sepr* &
            \term{\tSel \ell\, \tCtx{R}}
            \sepr
            \term{\tCase \tCtx{R} \tOf \{ i : M \}_{i \in I}}
            \sepr
            \term{\tClose \tCtx{R} ; M}
            \sepr
            \term{\tCtx{R} \tSub{ M/x }}
        \end{array}
    \end{align*}

    \dashes

    Structural congruence for terms ($\equivM$) and term reduction ($\reddM$):
    \begin{mathpar}
        \begin{bussproof}[sc-sub-ext]
            \bussAssume{
                \term{x} \notin \fv(\tCtx{R})
            }
            \bussUn{
                \term{(\tCtx{R}[M]) \tSub{ N/x }}
                \equivM
                \term{\tCtx*{R}[M \tSub{ N/x }]}
            }
        \end{bussproof}
        \and
        \begin{bussproof}[red-lam]
            \bussAx{
                \term{(\lam x . M)~N}
                \reddM
                \term{M \tSub{ N/x }}
            }
        \end{bussproof}
        \and
        \begin{bussproof}[red-pair]
            \bussAx{
                \term{\tLet (x,y) = (M_1,M_2) \tIn N}
                \reddM
                \term{N \tSub{ M_1/x,M_2/y }}
            }
        \end{bussproof}
        \and
        \begin{bussproof}[red-name-sub]
            \bussAx{
                \term{x \tSub{ M/x }}
                \reddM
                \term{M}
            }
        \end{bussproof}
        \and
        \begin{bussproof}[red-lift]
            \bussAssume{
                \term{M}
                \reddM
                \term{N}
            }
            \bussUn{
                \tCtx{R}[M]
                \reddM
                \tCtx{R}[N]
            }
        \end{bussproof}
        \and
        \begin{bussproof}[red-lift-sc]
            \bussAssume{
                \term{M}
                \equivM
                \term{M'}
            }
            \bussAssume{
                \term{M'}
                \reddM
                \term{N'}
            }
            \bussAssume{
                \term{N'}
                \equivM
                \term{N}
            }
            \bussTern{
                \term{M}
                \reddM
                \term{N}
            }
        \end{bussproof}
    \end{mathpar}

    \caption{The \protect\FIRST term language.}\label{f:FIRST:terms}
\end{figure}

\begin{figure}[p]
    \def\MathparLineskip{\lineskip=3pt}
    Markers ($\term{\phi}$), messages ($\term{m},\term{n}$), configurations ($\term{C},\term{D},\term{E}$), thread contexts ($\tCtx{F}$) and configuration contexts ($\tCtx{G}$):
    \begin{align*}
        \term{\phi} ::= {}
        & \term{\tMain} \sepr \term{\tChild}
        &
        \term{m},\term{n} ::= {}
        & \term{M} \sepr \term{\ell}
        \\
        \term{C},\term{D},\term{E} ::= {}
        & \term{\phi\, M} \sepr \term{C \prl D} \sepr \term{\pRes{x\tBfr{\vec{m}}y} C} \sepr \term{C \tSub{ M/x }}
        &
        \tCtx{F} ::= {}
        & \term{\phi\, \tCtx{R}}
        \\
        \term{\tCtx{G}} ::= {}
        & \term{\tHole} \sepr \term{\tCtx{G} \prl C} \sepr \term{\pRes{x\tBfr{\vec{m}}y} \tCtx{G}} \sepr \term{\tCtx{G} \tSub{ M/x }}
    \end{align*}

    \phantom{.}\dashes

    Structural congruence for configurations ($\equivC$):

    \vspace{-2.7ex} 
    \begin{mathpar}
        \begin{bussproof}[sc-term-sc]
            \bussAssume{
                \term{M}
                \equivM
                \term{M'}
            }
            \bussUn{
                \term{\phi\, M}
                \equivC
                \term{\phi\, M'}
            }
        \end{bussproof}
        \and
        \begin{bussproof}[sc-res-swap]
            \bussAx{
                \term{\pRes{x\tBfr{\epsi}y} C}
                \equivC
                \term{\pRes{y\tBfr{\epsi}x} C}
            }
        \end{bussproof}
        \and
        \begin{bussproof}[sc-res-comm]
            \bussAx{
                \term{\pRes{x\tBfr{\vec{m}}y} \pRes{z\tBfr{\vec{n}}w} C}
                \equivC
                \term{\pRes{z\tBfr{\vec{n}}w} \pRes{x\tBfr{\vec{m}}y} C}
            }
        \end{bussproof}
        \and
        \begin{bussproof}[sc-res-ext]
            \bussAssume{
                \term{x},\term{y} \notin \fv(\term{C})
            }
            \bussUn{
                \term{\pRes{x\tBfr{\vec{m}}y} ( C \prl D )}
                \equivC
                \term{C \prl \pRes{x\tBfr{\vec{m}}y} D}
            }
        \end{bussproof}
        \and
        \begin{bussproof}[sc-par-comm]
            \bussAx{
                \term{C \prl D}
                \equivC
                \term{D \prl C}
            }
        \end{bussproof}
        \and
        \begin{bussproof}[sc-par-assoc]
            \bussAx{
                \term{C \prl (D \prl E)}
                \equivC
                \term{(C \prl D) \prl E}
            }
        \end{bussproof}
        \and
        \begin{bussproof}[sc-conf-sub]
            \bussAx{
                \term{\phi\,(M \tSub{ N/x })}
                \equivC
                \term{(\phi\,M) \tSub{ N/x }}
            }
        \end{bussproof}
        \and
        \begin{bussproof}[sc-conf-sub-ext]
            \bussAssume{
                \term{x} \notin \fv(\tCtx{G})
            }
            \bussUn{
                \term{(\tCtx{G}[C]) \tSub{ M/x }}
                \equivC
                \term{\tCtx{G}[C \tSub{ M/x }]}
            }
        \end{bussproof}
    \end{mathpar}

    \phantom{.}\dashes 

    Configuration reduction ($\reddC$):

    \vspace{-2.7ex} 
    \begin{mathpar}
        \begin{bussproof}[red-new]
            \bussAx{
                \tCtx{F}[\tNew]
                \reddC
                \term{\pRes{x\tBfr{\epsi}y} ( \tCtx{F}[(x,y)] )}
            }
        \end{bussproof}
        \and
        \begin{bussproof}[red-fork]
            \bussAx{
                \tCtx{\hat{F}}[\tFork M ; N]
                \reddC
                \term{\tCtx{\hat{F}}[N] \prl \tChild\,M}
            }
        \end{bussproof}
        \and
        \begin{bussproof}[red-send]
            \bussAx{
                \term{\pRes{x\tBfr{\vec{m}}y} ( \tCtx{\hat{F}}[\tSend M \, x] \prl C )}
                \reddC
                \term{\pRes{x\tBfr{M,\vec{m}}y} ( \tCtx{\hat{F}}[x] \prl C )}
            }
        \end{bussproof}
        \and
        \begin{bussproof}[red-recv]
            \bussAx{
                \term{\pRes{x\tBfr{\vec{m},M}y} ( \tCtx{\hat{F}}[\tRecv y] \prl C )}
                \reddC
                \term{\pRes{x\tBfr{\vec{m}}y} ( \tCtx{\hat{F}}[(M,y)] \prl C )}
            }
        \end{bussproof}
        \and
        \begin{bussproof}[red-select]
            \bussAx{
                \term{\pRes{x\tBfr{\vec{m}}y} ( \tCtx{F}[\tSel \ell\, x] \prl C )}
                \reddC
                \term{\pRes{x\tBfr{\ell,\vec{m}}y} ( \tCtx{F}[x] \prl C )}
            }
        \end{bussproof}
        \and
        \begin{bussproof}[red-case]
            \bussAssume{
                j \in I
            }
            \bussUn{
                \term{\pRes{x\tBfr{\vec{m},j}y} ( \tCtx{F}[\tCase y \tOf \{ i : M_i \}_{i \in I}] \prl C )}
                \reddC
                \term{\pRes{x\tBfr{\vec{m}}y} ( \tCtx{F}[M_j~y] \prl C )}
            }
        \end{bussproof}
        \and
        \begin{bussproof}[red-close]
            \bussAx{
                \term{\pRes{x\tBfr{\vec{m}}y} ( \tCtx{F}[\tClose x ; M] \prl C )}
                \reddC
                \term{\pRes{\tNil\tBfr{\vec{m}}y} ( \tCtx{F}[M] \prl C )}
            }
        \end{bussproof}
        \and
        \begin{bussproof}[red-res-nil]
            \bussAx{
                \term{\pRes{\tNil\tBfr{\epsilon}\tNil} C}
                \reddC
                \term{C}
            }
        \end{bussproof}
        \and
        \begin{bussproof}[red-par-nil]
            \bussAx{
                \term{C \prl \tChild\, ()}
                \reddC
                \term{C}
            }
        \end{bussproof}
        \and
        \begin{bussproof}[red-lift-C]
            \bussAssume{
                \term{C}
                \reddC
                \term{C'}
            }
            \bussUn{
                \tCtx{G}[C]
                \reddC
                \tCtx{G}[C']
            }
        \end{bussproof}
        \and
        \begin{bussproof}[red-lift-M]
            \bussAssume{
                \term{M}
                \reddM
                \term{M'}
            }
            \bussUn{
                \tCtx{F}[M]
                \reddC
                \tCtx{F}[M']
            }
        \end{bussproof}
        \and
        \begin{bussproof}[red-conf-lift-sc]
            \bussAssume{
                \term{C}
                \equivC
                \term{C'}
            }
            \bussAssume{
                \term{C'}
                \reddC
                \term{D'}
            }
            \bussAssume{
                \term{D'}
                \equivC
                \term{D}
            }
            \bussTern{
                \term{C}
                \reddC
                \term{D}
            }
        \end{bussproof}
    \end{mathpar}
    \caption{The \protect\FIRST configuration language.}\label{f:FIRST:confs}
\end{figure}

\begin{figure}[p]
    \def\defaultHypSeparation{\hskip1ex}
    \def\MathparLineskip{\lineskip=7pt}
    \begin{mathpar}
        \begin{bussproof}[typ-var]
            \bussAx{
                \term{x}:\type{T} \vdashM \term{x}:\type{T}
            }
        \end{bussproof}
        \and
        \begin{bussproof}[typ-abs]
            \bussAssume{
                \type{\Gamma}, \term{x}:\type{T} \vdashM \term{M} :\type{U}
            }
            \bussUn{
                \type{\Gamma} \vdashM \term{\lam x . M} : \type{T \lolli U}
            }
        \end{bussproof}
        \and
        \begin{bussproof}[typ-app]
            \bussAssume{
                \type{\Gamma} \vdashM \term{M}: \type{T \lolli U}
            }
            \bussAssume{
                \type{\Delta} \vdashM \term{N}: \type{T}
            }
            \bussBin{
                \type{\Gamma}, \type{\Delta} \vdashM \term{M~N}: \type{U}
            }
        \end{bussproof}
        \and
        \begin{bussproof}[typ-unit]
            \bussAx{
                \type{\emptyset} \vdashM \term{()}: \type{\1}
            }
        \end{bussproof}
        \and
        \begin{bussproof}[typ-pair]
            \bussAssume{
                \type{\Gamma} \vdashM \term{M}: \type{T}
            }
            \bussAssume{
                \type{\Delta} \vdashM \term{N}: \type{U}
            }
            \bussBin{
                \type{\Gamma}, \type{\Delta} \vdashM \term{(M,N)}: \type{T \times U}
            }
        \end{bussproof}
        \and
        \begin{bussproof}[typ-split]
            \bussAssume{
                \type{\Gamma} \vdashM \term{M}: \type{T \times T'}
            }
            \bussAssume{
                \type{\Delta}, \term{x}:\type{T}, \term{y}:\type{T'} \vdashM \term{N}: \type{U}
            }
            \bussBin{
                \type{\Gamma}, \type{\Delta} \vdashM \term{\tLet (x,y) = M \tIn N}: \type{U}
            }
        \end{bussproof}
        \and
        \begin{bussproof}[typ-new]
            \bussAx{
                \type{\emptyset} \vdashM \term{\tNew}: \type{S \times \ol{S}}
            }
        \end{bussproof}
        \and
        \begin{bussproof}[typ-fork]
            \bussAssume{
                \type{\Gamma} \vdashM \term{M}: \type{\1}
            }
            \bussAssume{
                \type{\Delta} \vdashM \term{N} : \type{T}
            }
            \bussBin{
                \type{\Gamma} , \type{\Delta} \vdashM \term{\tFork M ; N}: \type{T}
            }
        \end{bussproof}
        \and
        \begin{bussproof}[typ-close]
            \bussAssume{
                \type{\Gamma} \vdashM \term{M} : \type{\tEnd}
            }
            \bussAssume{
                \type{\Delta} \vdashM \term{N} : \type{T}
            }
            \bussBin{
                \type{\Gamma} , \type{\Delta} \vdashM \term{\tClose M ; N} : \type{T}
            }
        \end{bussproof}
        \and
        \begin{bussproof}[typ-send]
            \bussAssume{
                \type{\Gamma} \vdashM \term{M}: \type{T}
            }
            \bussAssume{
                \type{\Delta} \vdashM \term{N}: \type{{!}T . S}
            }
            \bussBin{
                \type{\Gamma} , \type{\Delta} \vdashM \term{\tSend M \, N}: \type{S}
            }
        \end{bussproof}
        \and
        \begin{bussproof}[typ-recv]
            \bussAssume{
                \type{\Gamma} \vdashM \term{M}: \type{{?}T . S}
            }
            \bussUn{
                \type{\Gamma} \vdashM \term{\tRecv M}: \type{T \times S}
            }
        \end{bussproof}
        \and
        \begin{bussproof}[typ-sel]
            \bussAssume{
                \type{\Gamma} \vdashM \term{M}: \type{\oplus \{ i : S_i \}_{i \in I}}
            }
            \bussAssume{
                j \in I
            }
            \bussBin{
                \type{\Gamma} \vdashM \term{\tSel j\, M}: \type{S_j}
            }
        \end{bussproof}
        \and
        \begin{bussproof}[typ-case]
            \bussAssume{
                \type{\Gamma} \vdashM \term{M}: \type{{\&}\{ i : S_i \}_{i \in I}}
            }
            \bussAssume{
                \forall i \in I.~ \type{\Delta} \vdashM \term{N_i}: \type{S_i \lolli U}
            }
            \bussBin{
                \type{\Gamma}, \type{\Delta} \vdashM \term{\tCase M \tOf \{ i : N_i \}_{i \in I}}: \type{U}
            }
        \end{bussproof}
        \and
        \begin{bussproof}[typ-sub]
            \bussAssume{
                \type{\Gamma}, \term{x}:\type{T} \vdashM \term{M}: \type{U}
            }
            \bussAssume{
                \type{\Delta} \vdashM \term{N}: \type{T}
            }
            \bussBin{
                \type{\Gamma}, \type{\Delta} \vdashM \term{M \tSub{ N/x }}: \type{U}
            }
        \end{bussproof}
        \\ \phantom{.}\dashes \\ 
        \begin{bussproof}[typ-buf]
            \bussAx{
                \type{\emptyset} \vdashB \term{\tBfr{\epsilon}}: \type{S'} \> \type{S'}
            }
        \end{bussproof}
        \and
        \begin{bussproof}[typ-buf-send]
            \bussAssume{
                \type{\Gamma} \vdashM \term{M}: \type{T}
            }
            \bussAssume{
                \type{\Delta} \vdashB \term{\tBfr{\vec{m}}}: \type{S'} \> \type{S}
            }
            \bussBin{
                \type{\Gamma}, \type{\Delta} \vdashB \term{\tBfr{\vec{m},M}}: \type{S'} \> \type{{!}T . S}
            }
        \end{bussproof}
        \and
        \begin{bussproof}[typ-buf-sel]
            \bussAssume{
                \type{\Gamma} \vdashB \term{\tBfr{\vec{m}}}: \type{S'} \> \type{S_j}
            }
            \bussAssume{
                j \in I
            }
            \bussBin{
                \type{\Gamma} \vdashB \term{\tBfr{\vec{m},j}}: \type{S'} \> \type{\oplus \{ i : S_i \}_{i \in I}}
            }
        \end{bussproof}
        \and
        \begin{bussproof}[typ-buf-end-L]
            \bussAx{
                \type{\emptyset} \vdashB \term{\tBfr{\epsi}} : \type{\tEnd}\>\type{\tNil}
            }
        \end{bussproof}
        \and
        \begin{bussproof}[typ-buf-end-R]
            \bussAx{
                \type{\emptyset} \vdashB \term{\tBfr{\epsi}} : \type{\tNil}\>\type{\tEnd}
            }
        \end{bussproof}
        \\ \phantom{.}\dashes \\ 
        \begin{bussproof}[typ-main]
            \bussAssume{
                \type{\Gamma} \vdashM \term{M}: \type{\hat{T}}
            }
            \bussUn{
                \type{\Gamma} \vdashC{\tMain} \term{\tMain\, M}: \type{\hat{T}}
            }
        \end{bussproof}
        \and
        \begin{bussproof}[typ-child]
            \bussAssume{
                \type{\Gamma} \vdashM \term{M}: \type{\1}
            }
            \bussUn{
                \type{\Gamma} \vdashC{\tChild} \term{\tChild\, M}: \type{\1}
            }
        \end{bussproof}
        \and
        \begin{bussproof}[typ-par]
            \bussAssume{
                \type{\Gamma} \vdashC{\phi_1} \term{C}: \type{T_1}
            }
            \bussAssume{
                \type{\Delta} \vdashC{\phi_2} \term{D}: \type{T_2}
            }
            \bussBin{
                \type{\Gamma}, \type{\Delta} \vdashC{\phi_1 + \phi_2} \term{C \prl D}: \type{T_1 + T_2}
            }
        \end{bussproof}
        \and
        \begin{bussproof}[typ-res]
            \bussAssume{
                \type{\Gamma} \vdashB \term{\tBfr{\vec{m}}}: \type{S'} \> \type{S}
            }
            \bussAssume{
                \type{\Delta}, \term{x}: \type{S'} \vdashC{\phi} \term{C}: \type{T}
            }
            \bussAssume{
                \type{\Gamma'} , \term{y}:\type{\ol{S}} = \type{\Gamma} , \type{\Delta}
            }
            \bussTern{
                \type{\Gamma'} \vdashC{\phi} \term{\pRes{x\tBfr{\vec{m}}y} C}: \type{T}
            }
        \end{bussproof}
        \and
        \begin{bussproof}[typ-conf-sub]
            \bussAssume{
                \type{\Gamma}, \term{x}:\type{T} \vdashC{\phi} \term{C}: \type{U}
            }
            \bussAssume{
                \type{\Delta} \vdashM \term{M}: \type{T}
            }
            \bussBin{
                \type{\Gamma}, \type{\Delta} \vdashC{\phi} \term{C \tSub{ M/x }}: \type{U}
            }
        \end{bussproof}
    \end{mathpar}
    \caption{\protect\FIRST typing rules for terms (top), buffers (center), and configurations (bottom).}
    \label{f:FIRST:type}
\end{figure}

\Cref{f:FIRST:terms} gives the term language of \FIRST, \Cref{f:FIRST:confs} the configuration language, and \Cref{f:FIRST:type} the type system.

\newpage
\subsection{Type Preservation}
\label{as:FIRST:tp}

Here, we prove type preservation for \FIRST:

\tFIRSTTp*
\noindent 
The proof is split into two parts: subject congruence and subject reduction.
These and intermediate results are organized as follows:
\begin{itemize}

    \item
        \Cref{t:FIRST:scTerms,t:FIRST:srTerms} prove subject congruence and subject reduction for terms, respectively.

    \item
        \Cref{t:FIRST:sc,t:FIRST:sr} then prove subject congruence and subject reduction for configurations, respectively, from which \Cref{t:FIRST:tp} follows.

\end{itemize}

\begin{thm}[Subject Congruence for Terms]
\label{t:FIRST:scTerms}
    If $\type{\Gamma} \vdashM \term{M} : \type{T}$ and $\term{M} \equivM \term{N}$, then $\type{\Gamma} \vdashM \term{N} : \type{T}$.
\end{thm}

\begin{proof}
    By induction on the derivation of $\term{M} \equivM \term{N}$.
    The inductive cases follow from the IH directly.
    We consider the only \proofScTermsSubExt
\end{proof}

\begin{thm}[Subject Reduction for Terms]
\label{t:FIRST:srTerms}
    If $\type{\Gamma} \vdashM \term{M} : \type{T}$ and $\term{M} \reddM \term{N}$, then $\type{\Gamma} \vdashM \term{N} : \type{T}$.
\end{thm}

\begin{proof}
    By induction on the derivation of $\term{M} \reddM \term{N}$ (\ih{1}).
    The case of Rule~\ruleLabel{red-lift} follows by induction on the structure of the reduction context $\tCtx{R}$, where the base case ($\tCtx{R} = \tHole$) follows from \ih{1}.
    The case of Rule~\ruleLabel{red-lift-sc} follows from \ih{1} and \Cref{t:FIRST:scTerms} (subject congruence for terms).
    We consider the other cases, applying inversion of typing and deriving the typing of the term after reduction:
    \begin{itemize}

        \item
            Rule~\ruleLabel{red-lam}: $\term{(\lam x . M)~N} \reddM \term{M \tSub{ N/x }}$.
            \begin{align*}
                & \begin{bussproof}
                    \bussAssume{
                        \type{\Gamma}, \term{x}:\type{T} \vdashM \term{M}: \type{U}
                    }
                    \bussUn[\ruleLabel{typ-abs}]{
                        \type{\Gamma} \vdashM \term{\lam x . M}: \type{T \lolli U}
                    }
                    \bussAssume{
                        \type{\Delta} \vdashM \term{N}: \type{T}
                    }
                    \bussBin[\ruleLabel{typ-app}]{
                        \type{\Gamma}, \type{\Delta} \vdashM \term{(\lam x . M)~N}: \type{U}
                    }
                \end{bussproof}
                \\
                & \reddM
                \\
                & \begin{bussproof}
                    \bussAssume{
                        \type{\Gamma}, \term{x}:\type{T} \vdashM \term{M}: \type{U}
                    }
                    \bussAssume{
                        \type{\Delta} \vdashM \term{N}: \type{T}
                    }
                    \bussBin[\ruleLabel{typ-sub}]{
                        \type{\Gamma}, \type{\Delta} \vdashM \term{M \tSub{ N/x }}: \type{U}
                    }
                \end{bussproof}
            \end{align*}

        \item
            Rule~\ruleLabel{red-pair}: $\term{\tLet (x,y) = (M_1,M_2) \tIn N} \reddM \term{N \tSub{ M_1/x,M_2/y }}$.
            \begin{align*}
                & \begin{bussproof}
                    \bussAssume{
                        \type{\Gamma} \vdashM \term{M_1}: \type{T}
                    }
                    \bussAssume{
                        \type{\Gamma'} \vdashM \term{M_2}: \type{T'}
                    }
                    \bussBin[\ruleLabel{typ-pair}]{
                        \type{\Gamma}, \type{\Gamma'} \vdashM \term{(M_1,M_2)}: \type{T \times T'}
                    }
                    \bussAssume{
                        \type{\Delta}, \term{x}:\type{T}, \term{y}:\type{T'} \vdashM \term{N}: \type{U}
                    }
                    \bussBin[\ruleLabel{typ-split}]{
                        \type{\Gamma}, \type{\Gamma'}, \type{\Delta} \vdashM \term{\tLet (x,y) = (M_1,M_2) \tIn N}: \type{U}
                    }
                \end{bussproof}
                \\
                & \reddM
                \\
                & \begin{bussproof}
                    \bussAssume{
                        \type{\Delta}, \term{x}:\type{T}, \term{y}:\type{T'} \vdashM \term{N}: \type{U}
                    }
                    \bussAssume{
                        \type{\Gamma} \vdashM \term{M_1}: \type{T}
                    }
                    \bussBin[\ruleLabel{typ-sub}]{
                        \type{\Gamma}, \type{\Delta}, \term{y}:\type{T'} \vdashM \term{N \tSub{ M_1/x }}: \type{U}
                    }
                    \bussAssume{
                        \type{\Gamma'} \vdashM \term{M_2}: \type{T'}
                    }
                    \bussBin[\ruleLabel{typ-sub}]{
                        \type{\Gamma}, \type{\Gamma'}, \type{\Delta} \vdashM \term{N \tSub{ M_1/x,M_2/y }}: \type{U}
                    }
                \end{bussproof}
            \end{align*}

        \item
            Rule~\ruleLabel{red-name-sub}: $\term{x \tSub{ M/x }} \reddM \term{M}$.
            \begin{align*}
                & \begin{bussproof}
                    \bussAx[\ruleLabel{typ-var}]{
                        \term{x}:\type{U} \vdashM \term{x}: \type{U}
                    }
                    \bussAssume{
                        \type{\Gamma} \vdashM \term{M}: \type{U}
                    }
                    \bussBin[\ruleLabel{typ-sub}]{
                        \type{\Gamma} \vdashM \term{x \tSub{ M/x }}: \type{U}
                    }
                \end{bussproof}
                \\
                & \reddM
                \\
                & \type{\Gamma} \vdashM \term{M}: \type{U}
                \tag*{\qedhere}
            \end{align*}
    \end{itemize}
\end{proof}

\begin{thm}[Subject Congruence for Configurations]
    \label{t:FIRST:sc}
    If $\type{\Gamma} \vdashC{\phi} \term{C}: \type{T}$ and $\term{C} \equivC \term{D}$, then $\type{\Gamma} \vdashC{\phi} \term{D} : \type{T}$.
\end{thm}

\begin{proof}
    By induction on the derivation of $\term{C} \equivC \term{D}$.
    The inductive cases follow from the IH directly.
    The case for Rule~\ruleLabel{sc-term-sc} follows from \Cref{t:FIRST:scTerms} (subject congruence for terms).
    The cases for Rules~\ruleLabel{sc-res-comm}, \ruleLabel{sc-par-nil}, \ruleLabel{sc-par-comm}, and~\ruleLabel{par-assoc} are straightforward.
    We consider the other cases:
    \begin{itemize}
        \item
            \proofScResSwap

        \item
            Rule~\ruleLabel{sc-res-ext}: $\term{x},\term{y} \notin \fv(\term{C}) \implies \term{\pRes{x\tBfr{\vec{m}}y}(C \prl D)} \equivC \term{C \prl \pRes{x\tBfr{\vec{m}}y}D}$.

            The analysis depends on whether $\term{C}$ or $\term{D}$ are child threads.
            W.l.o.g., we assume $\term{C}$ is a child thread.
            Assuming $\term{x},\term{y} \notin \fn(\term{C})$, we apply inversion of typing:
            \[
                \begin{bussproof}
                    \bussAssume{
                        \type{\Gamma} \vdashB \term{\tBfr{\vec{m}}}: \type{S'} \> \type{S}
                    }
                    \bussAssume{
                        \type{\Delta} \vdashC{\tChild} \term{C}: \type{\1}
                    }
                    \bussAssume{
                        \type{\Lambda}, \term{x}:\type{S'}, \term{y}:\type{\ol{S}} \vdashC{\phi} \term{D}: \type{T}
                    }
                    \bussBin[\ruleLabel{typ-par}]{
                        \type{\Delta}, \type{\Lambda}, \term{x}:\type{S'}, \term{y}:\type{\ol{S}} \vdashC{\tChild+\phi} \term{C \prl D}: \type{T}
                    }
                    \bussBin[\ruleLabel{typ-res}]{
                        \type{\Gamma}, \type{\Delta}, \type{\Lambda} \vdashC{\tChild+\phi} \term{\pRes{x\tBfr{\vec{m}}y}(C \prl D)}: \type{T}
                    }
                \end{bussproof}
            \]
            Then, we derive the typing of the structurally congruent configuration:
            \[
                \begin{bussproof}
                    \bussAssume{
                        \type{\Delta} \vdashC{\tChild} \term{C}: \type{\1}
                    }
                    \bussAssume{
                        \type{\Gamma} \vdashB \term{\tBfr{\vec{m}}}: \type{S'} \> \type{S}
                    }
                    \bussAssume{
                        \type{\Lambda}, \term{x}:\type{S'}, \term{y}:\type{\ol{S}} \vdashC{\phi} \term{D}: \type{T}
                    }
                    \bussBin[\ruleLabel{typ-res}]{
                        \type{\Gamma}, \type{\Lambda} \vdashC{\phi} \term{\pRes{x\tBfr{\vec{m}}y}D}: \type{T}
                    }
                    \bussBin[\ruleLabel{typ-par}]{
                        \type{\Gamma}, \type{\Delta}, \type{\Lambda} \vdashC{\tChild+\phi} \term{C \prl \pRes{x\tBfr{\vec{m}}y}D}: \type{T}
                    }
                \end{bussproof}
            \]
            The other direction is analogous.

        \item
            Rule~\ruleLabel{sc-conf-sub}: $\term{\phi\,(M \tSub{ N/x })} \equivC \term{(\phi\,M) \tSub{ N/x }}$.

            This case follows by a straightforward inversion of typing on both terms:
            \begin{align*}
                & \begin{bussproof}
                    \bussAssume{
                        \type{\Gamma}, \term{x}:\type{T} \vdashM \term{M}: \type{U}
                    }
                    \bussAssume{
                        \type{\Delta} \vdashM \term{N}: \type{T}
                    }
                    \bussBin[\ruleLabel{typ-sub}]{
                        \type{\Gamma}, \type{\Delta} \vdashM \term{M \tSub{ N/x }}: \type{U}
                    }
                    \bussUn[\ruleLabel{typ-main/child}]{
                        \type{\Gamma}, \type{\Delta} \vdashC{\phi} \term{\phi\,(M \tSub{ N/x })}: \type{U}
                    }
                \end{bussproof}
                \\
                & \equivC
                \\
                & \begin{bussproof}
                    \bussAssume{
                        \type{\Gamma}, \term{x}:\type{T} \vdashM \term{M}: \type{U}
                    }
                    \bussUn[\ruleLabel{typ-main/child}]{
                        \type{\Gamma}, \term{x}:\type{T} \vdashC{\phi} \term{\phi\,M}: \type{U}
                    }
                    \bussAssume{
                        \type{\Delta} \vdashM \term{N}: \type{T}
                    }
                    \bussBin[\ruleLabel{typ-conf-sub}]{
                        \type{\Gamma}, \type{\Delta} \vdashC{\phi} \term{(\phi\,M) \tSub{ N/x }}: \type{U}
                    }
                \end{bussproof}
            \end{align*}

        \item
            Rule~\ruleLabel{sc-conf-sub-ext}: $\term{x} \notin \fv(\tCtx{G}) \implies \term{(\tCtx{G}[C]) \tSub{ M/x }} \equivC \term{\tCtx{G}[C \tSub{ M/x }]}$.

            This case follows by induction on the structure of $\tCtx{G}$.
            The inductive cases follow from the IH straightforwardly.
            For the base case ($\tCtx{G} = \tHole$), the structural congruence is simply an equality.
            \qedhere

    \end{itemize}
\end{proof}

\begin{thm}[Subject Reduction for Configurations]
\label{t:FIRST:sr}
    If $\type{\Gamma} \vdashC{\phi} \term{C}: \type{T}$ and $\term{C} \reddC \term{D}$, then $\type{\Gamma} \vdashC{\phi} \term{D}: \type{T}$.
\end{thm}

\begin{proof}
    By induction on the derivation of $\term{C} \reddC \term{D}$ (\ih{1}).
    The case of Rule~\ruleLabel{red-lift-C} ($\term{C} \reddC \term{C'} \implies \tCtx{G}[C] \reddC \tCtx{G}[C']$) follows by induction on the structure of $\tCtx{G}$, directly from \ih{1}.
    The case of Rule~\ruleLabel{red-lift-M} ($\term{M} \reddM \term{M'} \implies \tCtx{F}[M] \reddC \tCtx{F}[M']$) follows by induction on the structure of $\tCtx{F}$, where the base case ($\tCtx{F} = \term{\phi\,\tHole}$) follows from \Cref{t:FIRST:srTerms} (subject reduction for terms).
    The case for Rule~\ruleLabel{red-conf-lift-sc} ($\term{C} \equivC \term{C'} \wedge \term{C'} \reddC \term{D'} \wedge \term{D'} \equivC \term{D} \implies \term{C} \reddC \term{D}$) follows from \ih{1} and \Cref{t:FIRST:sc} (subject congruence for configurations).
    We consider the other cases:
    \begin{itemize}

        \item
            Rule~\ruleLabel{red-new}: $\tCtx{F}[\tNew] \reddC \term{\pRes{x\tBfr{\epsilon}y} ( \tCtx{F}[(x,y)] )}$.

            This case follows by induction on the structure of $\tCtx{F}$.
            The inductive cases follow from the IH directly.
            For the base case ($\tCtx{F} = \term{\phi\,\tHole}$), we apply inversion of typing and derive the typing of the reduced configuration:
            \begin{align*}
                & \begin{bussproof}
                    \bussAx[\ruleLabel{typ-new}]{
                        \type{\emptyset} \vdashM \term{\tNew}: \type{S \times \ol{S}}
                    }
                    \bussUn[\ruleLabel{typ-main/child}]{
                        \type{\emptyset} \vdashC{\phi} \term{\phi\,\tNew}: \type{S \times \ol{S}}
                    }
                \end{bussproof}
                \\
                & \reddC
                \\
                & \begin{bussproof}
                    \bussAx[\ruleLabel{typ-buf}]{
                        \type{\emptyset} \vdashB \term{\tBfr{\epsi}}: \type{S} \> \type{S}
                    }
                    \bussAx[\ruleLabel{typ-var}]{
                        \term{x}:\type{S} \vdashM \term{x}: \type{S}
                    }
                    \bussAx[\ruleLabel{typ-var}]{
                        \term{y}:\type{\ol{S}} \vdashM \term{y}: \type{\ol{S}}
                    }
                    \bussBin[\ruleLabel{typ-pair}]{
                        \term{x}:\type{S}, \term{y}:\type{\ol{S}} \vdashM \term{(x,y)}: \type{S \times \ol{S}}
                    }
                    \bussUn[\ruleLabel{typ-main/child}]{
                        \term{x}:\type{S}, \term{y}:\type{\ol{S}} \vdashC{\phi} \term{\phi\,(x,y)}: \type{S \times \ol{S}}
                    }
                    \bussBin[\ruleLabel{typ-res}]{
                        \type{\emptyset} \vdashC{\phi} \term{\pRes{x\tBfr{\epsi}y}(\phi\,(x,y))}: \type{S \times \ol{S}}
                    }
                \end{bussproof}
            \end{align*}

        \item
            Rule~\ruleLabel{red-fork}:
            $\tCtx{\hat{F}}[\tFork M ; N] \reddC \term{\tCtx{\hat{F}}[N] \prl \tChild M}$.
            By induction on the structure of $\tCtx{\hat{F}}$, which excludes holes under explicit substitution and so no names of $\term{M}$ are captured by $\tCtx{\hat{F}}$.
            The inductive cases follows from the IH directly; we detail the base case ($\tCtx{\hat{F}} = \term{\phi \, \tHole}$):
            \begin{align*}
                & \begin{bussproof}
                    \bussAssume{
                        \type{\Gamma} \vdashM \term{M} : \type{\1}
                    }
                    \bussAssume{
                        \type{\Delta} \vdashM \term{N} : \type{T}
                    }
                    \bussBin[\ruleLabel{typ-fork}]{
                        \type{\Gamma} , \type{\Delta} \vdashM \term{\tFork M ; N} : \type{T}
                    }
                    \bussUn[\ruleLabel{typ-main/-child}]{
                        \type{\Gamma} , \type{\Delta} \vdashC{\phi} \term{\phi \, \tFork M ; N} : \type{T}
                    }
                \end{bussproof}
                \\
                & \reddC
                \\
                & \begin{bussproof}
                    \bussAssume{
                        \type{\Delta} \vdashM \term{N} : \type{T}
                    }
                    \bussUn[\ruleLabel{typ-main/-child}]{
                        \type{\Delta} \vdashC{\phi} \term{\phi \, N} : \type{T}
                    }
                    \bussAssume{
                        \type{\Gamma} \vdashM \term{M} : \type{\1}
                    }
                    \bussUn[\ruleLabel{typ-child}]{
                        \type{\Gamma} \vdashC{\tChild} \term{\tChild \, M} : \type{\1}
                    }
                    \bussBin[\ruleLabel{typ-par}]{
                        \type{\Gamma} , \type{\Delta} \vdashC{\phi+\tChild} \term{\phi \, N \prl \tChild \, M} : \type{T}
                    }
                \end{bussproof}
            \end{align*}
            Clearly, $\term{\phi} + \term{\tChild} = \term{\phi}$, proving the thesis.

        \item
            \proofSrSend

        \item
            Rule~\ruleLabel{red-recv}: $\term{\pRes{x\tBfr{\vec{m},M}y} ( \tCtx{\hat{F}}[\tRecv y] \prl C )} \reddC \term{\pRes{x\tBfr{\vec{m}}y} ( \tCtx{\hat{F}}[(M,y)] \prl C )}$.

            For this case, we apply induction on the structure of $\tCtx{\hat{F}}$.
            The inductive cases follow from the IH directly.
            We consider the base case ($\tCtx{\hat{F}} = \term{\phi\,\tHole}$).
            We apply inversion of typing, w.l.o.g.\ assuming that $\term{\phi} = \term{\tMain}$, and then derive the typing of the reduced configuration:
            \begin{align*}
                & \pi \deq \begin{bussproof}
                    \bussAx[\ruleLabel{typ-var}]{
                        \term{y}:\type{{?}T . \ol{S}} \vdashM \term{y}: \type{{?}T . \ol{S}}
                    }
                    \bussUn[\ruleLabel{typ-recv}]{
                        \term{y}:\type{{?}T . \ol{S}} \vdashM \term{\tRecv y}: \type{T \times \ol{S}}
                    }
                    \bussUn[\ruleLabel{typ-main}]{
                        \term{y}:\type{{?}T . \ol{S}} \vdashC{\tMain} \term{\tMain\,(\tRecv y)}: \type{T \times \ol{S}}
                    }
                    \bussAssume{
                        \type{\Lambda}, \term{x}:\type{S'} \vdashC{\tChild} \term{C}: \type{\1}
                    }
                    \bussBin[\ruleLabel{typ-par}]{
                        \type{\Lambda}, \term{x}:\type{S'}, \term{y}:\type{{?}T . \ol{S}} \vdashC{\tMain} \term{\tMain\,(\tRecv y) \prl C}: \type{T \times \ol{S}}
                    }
                \end{bussproof}
                \\
                & \begin{bussproof}
                    \bussAssume{
                        \type{\Gamma} \vdashM \term{M}: \type{T}
                    }
                    \bussAssume{
                        \type{\Delta} \vdashB \term{\tBfr{\vec{m}}}: \type{S'} \> \type{S}
                    }
                    \bussBin[\ruleLabel{typ-buf-send}]{
                        \type{\Gamma}, \type{\Delta} \vdashB \term{\tBfr{\vec{m},M}}: \type{S'} \> \type{{!}T . S}
                    }
                    \bussAssume{
                        \pi
                    }
                    \bussBin[\ruleLabel{typ-res}]{
                        \type{\Gamma}, \type{\Delta}, \type{\Lambda} \vdashC{\tMain} \term{\pRes{x\tBfr{\vec{m},M}y} ( \tMain\,(\tRecv y) \prl C )}: \type{T \times \ol{S}}
                    }
                \end{bussproof}
                \\
                & \reddC
                \\
                & \begin{bussproof}
                    \def\defaultHypSeparation{\hskip1ex}
                    \def\ScoreOverhang{1pt}
                    \bussAssume{
                        \type{\Delta} \vdashB \term{\tBfr{\vec{m}}}: \type{S'} \> \type{S}
                    }
                    \bussAssume{
                        \type{\Gamma} \vdashM \term{M}: \type{T}
                    }
                    \bussAx[\ruleLabel{typ-var}]{
                        \term{y}:\type{\ol{S}} \vdashM \term{y}: \type{\ol{S}}
                    }
                    \bussBin[\ruleLabel{typ-pair}]{
                        \type{\Gamma}, \term{y}:\type{\ol{S}} \vdashM \term{(M,y)}: \type{T \times \ol{S}}
                    }
                    \bussUn[\ruleLabel{typ-main}]{
                        \type{\Gamma}, \term{y}:\type{\ol{S}} \vdashC{\tMain} \term{\tMain\,(M,y)}: \type{T \times \ol{S}}
                    }
                    \bussAssume{
                        \type{\Lambda}, \term{x}:\type{S'} \vdashC{\tChild} \term{C}: \type{\1}
                    }
                    \bussBin[\ruleLabel{typ-par}]{
                        \type{\Gamma}, \type{\Lambda}, \term{x}:\type{S'}, \term{y}:\type{\ol{S}} \vdashC{\tMain} \term{\tMain\,(M,y) \prl C}: \type{T \times \ol{S}}
                    }
                    \bussBin[\ruleLabel{typ-res}]{
                        \type{\Gamma}, \type{\Delta}, \type{\Lambda} \vdashC{\tMain} \term{\pRes{x\tBfr{\vec{m}}y} ( \tMain\,(M,y) \prl C )}: \type{T \times \ol{S}}
                    }
                \end{bussproof}
            \end{align*}

        \item
            Rule~\ruleLabel{red-select} is similar to the case of Rule~\ruleLabel{red-send}:
            \[
                \term{\pRes{x\tBfr{\vec{m}}y} ( \tCtx{F}[\tSel \ell \, x] \prl C )} \reddC \term{\pRes{x\tBfr{\ell,\vec{m}}y} ( \tCtx{F}[x] \prl C )}.
            \]

        \item
            Rule~\ruleLabel{red-case} is similar to the case of Rule~\ruleLabel{red-recv}:
            \[
                j \in I \implies \term{\pRes{x\tBfr{\vec{m},j}y} ( \tCtx{F}[\tCase y \tOf \{ i : M_i \}_{i \in I}] \prl C )} \reddC \term{\pRes{x\tBfr{\vec{m}}y} ( \tCtx{F}[M_j~y] \prl C )}
            \]

        \item
            Rule~\ruleLabel{red-close}:
            $\term{\pRes{x\tBfr{\vec{m}}y} ( \tCtx{F}[\tClose x ; M] \prl C )} \reddC \term{\pRes{\tNil\tBfr{\vec{m}}y} ( \tCtx{F}[M] \prl C )}$.
            By induction on the structure of $\tCtx{F}$.
            The inductive cases follows from the IH straightforwardly; we detail the base case ($\tCtx{F} = \term{\phi \, \tHole}$).
            Assume, w.l.o.g., that $\term{\phi} = \term{\tMain}$.
            We first derive the typing of the configuration before reduction:
            \[
                \begin{bussproof}
                    \bussAssume{
                        \type{\Gamma} \vdashB \term{\tBfr{\vec{m}}} : \type{\tEnd}\>\type{S}
                    }
                    \bussAssume{
                        \term{x}:\type{\tEnd} \vdashM \term{x} : \type{\tEnd}
                    }
                    \bussAssume{
                        \type{\Delta} \vdashM \term{M} : \type{T}
                    }
                    \bussBin[\ruleLabel{typ-close}]{
                        \type{\Delta} , \term{x}:\type{\tEnd} \vdashM \term{\tClose x ; M} : \type{T}
                    }
                    \bussUn[\ruleLabel{typ-main}]{
                        \type{\Delta} , \term{x}:\type{\tEnd} \vdashC{\tMain} \term{\tMain \, \tClose x ; M} : \type{T}
                    }
                    \bussAssume{
                        \type{\Theta} \vdashC{\tChild} \term{C} : \type{\1}
                    }
                    \bussBin[\ruleLabel{typ-par}]{
                        \type{\Delta} , \type{\Theta} , \term{x}:\type{\tEnd} \vdashC{\tMain} \term{\tMain \, \tClose x ; M \prl C} : \type{T}
                    }
                    \bussBin[\ruleLabel{typ-res/-buf}]{
                        \type{\Gamma'} \vdashC{\tMain} \term{\pRes{x\tBfr{\vec{m}}y} ( \tMain \, \tClose x ; M \prl C )} : \type{T}
                    }
                \end{bussproof}
            \]
            We prove that $\type{\Gamma} \vdashB \term{\tBfr{\vec{m}}} : \type{\tNil}\>\type{S}$.
            The analysis depends on whether $\term{y} = \term{\tNil}$.
            \begin{itemize}

                \item
                    If $\term{y} = \term{\tNil}$, then $\term{\vec{m}} = \term{\epsi}$, $\type{S} = \type{\tNil}$, $\type{\Gamma'} = \type{\Gamma} , \type{\Delta} , \type{\Theta}$, and $\type{\Gamma} = \type{\emptyset}$, because $\type{\Gamma} \vdashB \term{\tBfr{\vec{m}}} : \type{\tEnd}\>\type{S}$ must be derived as follows:
                    \[
                        \begin{bussproof}
                            \bussAx[\ruleLabel{typ-buf-end-L}]{
                                \type{\emptyset} \vdashB \term{\tBfr{\epsi}} : \type{\tEnd}\>\type{\tNil}
                            }
                        \end{bussproof}
                    \]
                    The thesis holds as follows:
                    \[
                        \begin{bussproof}
                            \bussAx[\ruleLabel{typ-buf}]{
                                \type{\emptyset} \vdashB \term{\tBfr{\epsi}} : \type{\tNil}\>\type{\tNil}
                            }
                        \end{bussproof}
                    \]

                \item
                    If $\term{y} \neq \term{\tNil}$, then $\type{\Gamma'} , \term{y}:\type{\ol{S}} = \type{\Gamma} , \type{\Delta} , \type{\Theta}$.
                    We apply induction on the size of $\term{\vec{m}}$ (\ih2):
                    \begin{itemize}

                        \item
                            If $\term{\vec{m}} = \term{\epsi}$, then $\type{\Gamma} = \type{\emptyset}$, $\type{S} = \type{\tEnd}$, because $\type{\Gamma} \vdashB \term{\tBfr{\epsi}} : \type{\tEnd}\>\type{S}$ must be derived as follows:
                            \[
                                \begin{bussproof}
                                    \bussAx[\ruleLabel{typ-buf}]{
                                        \type{\emptyset} \vdashB \term{\tBfr{\epsi}} : \type{\tEnd}\>\type{\tEnd}
                                    }
                                \end{bussproof}
                            \]
                            The thesis then holds as follows:
                            \[
                                \begin{bussproof}
                                    \bussAx[\ruleLabel{typ-buf-end-R}]{
                                        \type{\emptyset} \vdashB \term{\tBfr{\epsi}} : \type{\tNil}\>\type{\tEnd}
                                    }
                                \end{bussproof}
                            \]

                        \item
                            If $\term{\vec{m}} = \term{\vec{m}',N}$, then $\type{\Gamma} = \type{\Gamma_1} , \type{\Gamma_2}$, and $\type{S} = \type{{!}U.S'}$, because $\type{\Gamma} \vdashB \term{\tBfr{\vec{m}',N}} : \type{\tEnd}\>\type{S}$ must be derived as follows:
                            \[
                                \begin{bussproof}
                                    \bussAssume{
                                        \type{\Gamma_1} \vdashM \term{N} : \type{U}
                                    }
                                    \bussAssume{
                                        \type{\Gamma_2} \vdashB \term{\tBfr{\vec{m}'}} : \type{\tEnd}\>\type{S'}
                                    }
                                    \bussBin[\ruleLabel{typ-buf-send}]{
                                        \type{\Gamma_1} , \type{\Gamma_2} \vdashB \term{\tBfr{\vec{m}',N}} : \type{\tEnd}\>\type{{!}U.S'}
                                    }
                                \end{bussproof}
                            \]
                            By \ih2, $\type{\Gamma_2} \vdashB \term{\tBfr{\vec{m}'}} : \type{\tNil}\>\type{S'}$.
                            The thesis then holds as follows:
                            \[
                                \begin{bussproof}
                                    \bussAssume{
                                        \type{\Gamma_1} \vdashM \term{N} : \type{U}
                                    }
                                    \bussAssume{
                                        \type{\Gamma_2} \vdashB \term{\tBfr{\vec{m}'}} : \type{\tNil}\>\type{S'}
                                    }
                                    \bussBin[\ruleLabel{typ-buf-send}]{
                                        \type{\Gamma_1} , \type{\Gamma_2} \vdashB \term{\tBfr{\vec{m}',N}} : \type{\tNil}\>\type{{!}U.S'}
                                    }
                                \end{bussproof}
                            \]
                    \end{itemize}

            \end{itemize}
            Now, we derive the typing of the reduced configuration:
            \[
                \begin{bussproof}
                    \bussAssume{
                        \type{\Gamma} \vdashB \term{\tBfr{\vec{m}}} : \type{\tNil}\>\type{S}
                    }
                    \bussAssume{
                        \type{\Delta} \vdashM \term{M} : \type{T}
                    }
                    \bussUn[\ruleLabel{typ-main}]{
                        \type{\Delta} \vdashC{\tMain} \term{\tMain \, M} : \type{T}
                    }
                    \bussAssume{
                        \type{\Theta} \vdashC{\tChild} \term{C} : \type{\1}
                    }
                    \bussBin[\ruleLabel{typ-par}]{
                        \type{\Delta} , \type{\Theta} \vdashC{\tMain} \term{\tMain \, M \prl C} : \type{T}
                    }
                    \bussBin[\ruleLabel{typ-res/-buf}]{
                        \type{\Gamma'} \vdashC{\tMain} \term{\pRes{\tNil\tBfr{\vec{m}}y} ( \tMain \, M \prl C )} : \type{T}
                    }
                \end{bussproof}
            \]

        \item
            Rule~\ruleLabel{red-res-nil}:
            $\term{\pRes{\tNil\tBfr{\epsi}\tNil} C} \reddC \term{C}$.
            We have
            \begin{align*}
                & \begin{bussproof}
                    \bussAx[\ruleLabel{typ-buf}]{
                        \type{\emptyset} \vdashB \term{\tBfr{\epsi}} : \type{\tNil}\>\type{\tNil}
                    }
                    \bussAssume{
                        \type{\Gamma} \vdashC{\phi} \term{C} : \type{T}
                    }
                    \bussBin[\ruleLabel{typ-res/-res-buf}]{
                        \type{\Gamma} \vdashC{\phi} \term{\pRes{\tNil\tBfr{\epsi}\tNil} C} : \type{T}
                    }
                \end{bussproof}
                \\
                & \reddC
                \\
                & \type{\Gamma} \vdashC{\phi} \term{C} : \type{T}
            \end{align*}

        \item
            Rule~\ruleLabel{red-par-nil}:
            $\term{C \prl \tChild \, ()} \reddC \term{C}$.
            We have the following:
            \begin{align*}
                & \begin{bussproof}
                    \bussAssume{
                        \type{\Gamma} \vdashC{\phi} \term{C} : \type{T}
                    }
                    \bussAx[\ruleLabel{typ-unit}]{
                        \type{\emptyset} \vdashM \term{()} : \type{\1}
                    }
                    \bussUn[\ruleLabel{typ-child}]{
                        \type{\emptyset} \vdashC{\tChild} \term{\tChild \, ()} : \type{\1}
                    }
                    \bussBin[\ruleLabel{typ-par}]{
                        \type{\Gamma} \vdashC{\phi} \term{C \prl \tChild \, ()} : \type{T}
                    }
                \end{bussproof}
                \\
                & \reddC
                \\
                & \type{\Gamma} \vdashC{\phi} \term{C} : \type{T}
                \tag*{\qedhere}
            \end{align*}

    \end{itemize}
\end{proof}

\subsection{Translation: Type Preservation}
\label{as:FIRST:transTp}

\tFIRSTTransTp*

\begin{proof}
    By induction on the \FIRST typing derivation.
    It is sufficient to give the typing of the translations of typing rules in \Cref{f:FIRST:trans1,f:FIRST:trans2,f:FIRST:trans3} as follows; checking the derivations is straightforward.
    \begin{align*}
        & \tinyRL{typ-var}
        && \vdash* \encc{z}{\normalcolor \term{x}:\type{T} \vdashM \term{x} : \type{T}}
        && :: x:\enct*{T} , z:\enct{T}
        \\ \displaybreak[1]
        & \tinyRL{typ-abs}
        && \vdash* \encc{z}{\normalcolor \type{\Gamma} \vdashM \term{\lam x . M} : \type{T \lolli U}}
        && :: \enct*{\Gamma} , z:\enct*{T} \parr \enct{U} = \enct{T \lolli U}
        \\ \displaybreak[1]
        & \tinyRL{typ-app}
        && \vdash* \encc{z}{\normalcolor \type{\Gamma} , \type{\Delta} \vdashM \term{M~N} : \type{U}}
        && :: \enct*{\Gamma} , \enct*{\Delta} , z:\enct{U}
        \\ \displaybreak[1]
        & \tinyRL{typ-unit}
        && \vdash* \encc{z}{\normalcolor \type{\emptyset} \vdashM \term{()} : \type{\1}}
        && :: z:\bullet = \enct{\1}
        \\ \displaybreak[1]
        & \tinyRL{typ-pair}
        && \vdash* \encc{z}{\normalcolor \type{\Gamma} , \type{\Delta} \vdashM \term{(M,N)} : \type{T \times U}}
        && :: \begin{array}[t]{@{}l@{}}
            \enct*{\Gamma} , \enct*{\Delta} ,
            \\
            z:\begin{array}[t]{@{}l@{}}
                (\bullet \parr \enct{T}) \tensor (\bullet \parr \enct{U})
                \\
                = \ol{\enct*{T}} \tensor \ol{\enct*{U}} = \enct{T \times U}
            \end{array}
        \end{array}
        \\ \displaybreak[1]
        & \tinyRL{typ-split}
        && \vdash* \encc{z}{\normalcolor \type{\Gamma} , \type{\Delta} \vdashM \term{\tLet (x,y) = M \tIn N} : \type{U}}
        && :: \enct*{\Gamma} , \enct*{\Delta} , z:\enct{U}
        \\ \displaybreak[1]
        & \tinyRL{typ-new}
        && \vdash* \encc{z}{\normalcolor \type{\emptyset} \vdashM \term{\tNew} : \type{S \times \ol{S}}}
        && :: z:\ol{\enct*{S}} \tensor \ol{\enct*{\ol{S}}} = \enct{S \times \ol{S}}
        \\ \displaybreak[1]
        & \tinyRL{typ-fork}
        && \vdash* \encc{z}{\normalcolor \type{\Gamma} , \type{\Delta} \vdashM \term{\tFork M ; N} : \type{T}}
        && :: \enct*{\Gamma} , \enct*{\Delta} , z:\enct{T}
        \\ \displaybreak[1]
        & \tinyRL{typ-close}
        && \vdash* \encc{z}{\normalcolor \type{\Gamma} , \type{\Delta} \vdashM \term{\tClose M ; N} : \type{T}}
        && :: \enct*{\Gamma} , \enct*{\Delta} , z:\enct{T}
        \\ \displaybreak[1]
        & \tinyRL{typ-send}
        && \vdash* \encc{z}{\normalcolor \type{\Gamma} \vdashM \term{\tSend M \, N} : \type{S}}
        && :: \enct*{\Gamma} , \enct*{\Delta} , z:\enct{S}
        \\ \displaybreak[1]
        & \tinyRL{typ-recv}
        && \vdash* \encc{z}{\normalcolor \type{\Gamma} \vdashM \term{\tRecv M} : \type{T \times S}}
        && :: \enct*{\Gamma} , z:\begin{array}[t]{@{}l@{}}
            (\bullet \parr \enct{T}) \tensor (\bullet \parr \enct{S})
            \\
            = \ol{\enct*{T}} \tensor \ol{\enct*{S}} = \enct{T \times S}
        \end{array}
        \\ \displaybreak[1]
        & \tinyRL{typ-sel}
        && \vdash* \encc{z}{\normalcolor \type{\Gamma} \vdashM \term{\tSel j \, M} : \type{S_j}}
        && :: \enct*{\Gamma} , z:\enct{S_j}
        \\ \displaybreak[1]
        & \tinyRL{typ-case}
        && \vdash* \encc{z}{\normalcolor \type{\Gamma} \vdashM \term{\tCase M \tOf \{ i : N_i \}_{i \in I}} : \type{U}}
        && :: \enct*{\Gamma} , \enct*{\Delta} , z:\enct{U}
        \\ \displaybreak[1]
        & \tinyRL{typ-sub}
        && \vdash* \encc{z}{\normalcolor \type{\Gamma} , \type{\Delta} \vdashM \term{M \tSub{ N/x }} : \type{U}}
        && :: \enct*{\Gamma} , \enct*{\Delta} , z:\enct{U}
        \\ \displaybreak[1]
        & \tinyRL{typ-main}
        && \vdash* \encc{z}{\normalcolor \type{\Gamma} \vdashC{\tMain} \term{\tMain \, M} : \type{\hat{T}}}
        && :: \enct*{\Gamma} , z:\enct{\hat{T}}
        \\ \displaybreak[1]
        & \tinyRL{typ-child}
        && \vdash* \encc{z}{\normalcolor \type{\Gamma} \vdashC{\tChild} \term{\tChild \, M} : \type{\1}}
        && :: \enct*{\Gamma} , z:\bullet = \enct{\1}
        \\ \displaybreak[1]
        & \begin{array}[t]{@{}l@{}}
            \tinyRL{typ-par}
            \\
            \mathrlap{
                (\type{T_1} = \type{\1})
            }
        \end{array}
        && \vdash* \encc{z}{\normalcolor \type{\Gamma} , \type{\Delta} \vdashC{\phi_1+\phi_2} \term{C \prl D} : \type{T_2}}
        && :: \enct*{\Gamma} , \enct*{\Delta} , z:\enct{T_2}
        \\ \displaybreak[1]
        & \begin{array}[t]{@{}l@{}}
            \tinyRL{typ-par}
            \\
            \mathrlap{
                (\type{T_2} = \type{\1})
            }
        \end{array}
        && \vdash* \encc{z}{\normalcolor \type{\Gamma} , \type{\Delta} \vdashC{\phi_1+\phi_2} \term{C \prl D} : \type{T_1}}
        && :: \enct*{\Gamma} , \enct*{\Delta} , z:\enct{T_1}
        \\ \displaybreak[1]
        & \tinyRL{typ-res}
        && \vdash* \encc{z}{\normalcolor \type{\Gamma'} \vdashC{\phi} \term{\pRes{x\tBfr{\vec{m}}y} C} : \type{T}}
        && :: \enct*{\Gamma'} , z:\enct{T}
        \\ \displaybreak[1]
        & \tinyRL{typ-conf-sub}
        && \vdash* \encc{z}{\normalcolor \type{\Gamma} , \type{\Delta} \vdashC{\phi} \term{C \tSub{ M/x }} : \type{U}}
        && :: \enct*{\Gamma} , \enct*{\Delta} , z:\enct{U}
        \\ \displaybreak[1]
        & \begin{array}[t]{@{}l@{}}
            \tinyRL{typ-buf}
            \\
            (\type{S'} = \type{{!}T.S})
        \end{array}
        && \vdash* \encc{a\>b}{\normalcolor \type{\emptyset} \vdashB \term{\tBfr{\epsi}} : \type{S'}\>\type{S'}}
        && :: \begin{array}[t]{@{}l@{}}
            a:\begin{array}[t]{@{}l@{}}
                \bullet \parr \bullet \tensor \enct*{T} \parr \ol{\enct*{S}} = \bullet \parr \enct{S'}
                \\
                = \ol{\enct*{S'}} ,
            \end{array}
            \\
            b:\begin{array}[t]{@{}l@{}}
                \bullet \parr \ol{\enct*{T}} \tensor \ol{\enct*{\ol{S}}}
                = \bullet \parr \enct{{?}T.\ol{S}}
                \\
                = \bullet \parr \enct{\ol{S'}} = \ol{\enct*{\ol{S'}}}
            \end{array}
        \end{array}
        \\ \displaybreak[1]
        & \begin{array}[t]{@{}l@{}}
            \tinyRL{typ-buf}
            \\
            \mathrlap{
                (\type{S'} = \type{\oplus \{ i : S_i \}_{i \in I}})
            }
        \end{array}
        && \vdash* \encc{a\>b}{\normalcolor \type{\emptyset} \vdashB \term{\tBfr{\epsi}} : \type{S'}\>\type{S'}}
        && :: \begin{array}[t]{@{}l@{}}
            a:\begin{array}[t]{@{}l@{}}
                \bullet \parr \bullet \tensor \& \{ i : \ol{\enct*{S_i}} \}_{i \in I}
                \\
                = \bullet \parr \enct{S'} = \ol{\enct*{S'}} ,
            \end{array}
            \\
            b:\begin{array}[t]{@{}l@{}}
                \bullet \parr \oplus \{ i : \ol{\enct*{\ol{S_i}}} \}_{i \in I}
                \\
                = \bullet \parr \enct{\& \{ i : \ol{S_i} \}_{i \in I}}
                \\
                = \bullet \parr \enct{\ol{S'}} = \ol{\enct*{\ol{S'}}}
            \end{array}
        \end{array}
        \\ \displaybreak[1]
        & \begin{array}[t]{@{}l@{}}
            \tinyRL{typ-buf}
            \\
            \mathrlap{
                (\type{S'} \in \{\type{{?}T.S},\type{\& \{ i : S_i \}_{i \in I}}\})
            }
        \end{array}
        && \vdash* \encc{a\>b}{\normalcolor \type{\emptyset} \vdashB \term{\tBfr{\epsi}} : \type{S'}\>\type{S'}}
        && :: a:\ol{\enct*{\ol{\ol{S'}}}} = \ol{\enct*{S'}} , b:\ol{\enct*{\ol{S'}}}
        \\ \displaybreak[1]
        & \begin{array}[t]{@{}l@{}}
            \tinyRL{typ-buf}
            \\
            (\type{S'} = \type{\tEnd})
        \end{array}
        && \vdash* \encc{a\>b}{\normalcolor \type{\emptyset} \vdashB \term{\tBfr{\epsi}} : \type{S'}\>\type{S'}}
        && :: \begin{array}[t]{@{}l@{}}
            a:\bullet \parr \bullet \tensor \bullet = \ol{\enct*{S'}} ,
            \\
            b:\bullet \parr \bullet \tensor \bullet = \ol{\enct*{S'}} = \ol{\enct*{\ol{S'}}}
        \end{array}
        \\ \displaybreak[1]
        & \begin{array}[t]{@{}l@{}}
            \tinyRL{typ-buf}
            \\
            (\type{S'} = \type{\tNil})
        \end{array}
        && \vdash* \encc{a\>b}{\normalcolor \type{\emptyset} \vdashB \term{\tBfr{\epsi}} : \type{S'}\>\type{S'}}
        && :: a:\bullet = \ol{\enct*{S'}} , b:\bullet = \ol{\enct*{S'}} = \ol{\enct*{\ol{S'}}}
        \\ \displaybreak[1]
        & \tinyRL{typ-buf-send}
        && \vdash* \encc{a\>b}{\normalcolor \type{\Gamma} , \type{\Delta} \vdashB \term{\tBfr{\vec{m},M}} : \type{S'}\>\type{{!}T.S}}
        && :: \begin{array}[t]{@{}l@{}}
            \enct*{\Gamma} , \enct*{\Delta} , a:\ol{\enct*{S'}} ,
            \\
            b:\begin{array}[t]{@{}l@{}}
                \bullet \parr \ol{\enct*{T}} \tensor \ol{\enct*{\ol{S}}} = \bullet \parr \enct{{?}T.\ol{S}}
                \\
                = \bullet \parr \enct{\ol{{!}T.S}} = \ol{\enct*{\ol{{!}T.S}}}
            \end{array}
        \end{array}
        \\ \displaybreak[1]
        & \tinyRL{typ-buf-sel}
        && \vdash* \encc{a\>b}{\normalcolor \type{\Gamma} \vdashB \term{\tBfr{\vec{m},j}} : \type{S'}\>\type{\oplus \{ i : S_i \}_{i \in I}}}
        && :: \begin{array}[t]{@{}l@{}}
            \enct*{\Gamma} , a:\ol{\enct*{S'}} ,
            \\
            b:\begin{array}[t]{@{}l@{}}
                \bullet \parr \oplus \{ i : \ol{\enct*{\ol{S_i}}} \}_{i \in I}
                \\
                = \bullet \parr \enct{\& \{ i : \ol{S_i} \}_{i \in I}}
                \\
                = \bullet \parr \enct{\ol{\oplus \{ i : S_i \}_{i \in I}}}
                \\
                = \ol{\enct*{\ol{\oplus \{ i : S_i \}_{i \in I}}}}
            \end{array}
        \end{array}
        \\ \displaybreak[1]
        & \tinyRL{typ-buf-end-L}
        && \vdash* \encc{a\>b}{\normalcolor \type{\emptyset} \vdashB \term{\tBfr{\epsi}} : \type{\tEnd}\>\type{\tNil}}
        && :: \begin{array}[t]{@{}l@{}}
            a:\bullet \parr \bullet \tensor \bullet = \ol{\enct*{\tEnd}} ,
            \\
            b:\bullet = \ol{\enct*{\tNil}} = \ol{\enct*{\ol{\tNil}}}
        \end{array}
        \\ \displaybreak[1]
        & \tinyRL{typ-buf-end-R}
        && \vdash* \encc{a\>b}{\normalcolor \type{\emptyset} \vdashB \term{\tBfr{\epsi}} : \type{\tNil}\>\type{\tEnd}}
        && :: \begin{array}[t]{@{}l@{}}
            a:\bullet = \ol{\enct*{\tNil}} ,
            \\
            b:\bullet \parr \bullet \tensor \bullet = \ol{\enct*{\tEnd}} = \ol{\enct*{\ol{\tEnd}}}
            \qedhere
        \end{array}
    \end{align*}
\end{proof}

\subsection{Operational Correspondence}
\label{as:FIRST:oc}

\Cref{as:FIRST:oc:compl,as:FIRST:oc:sound} prove completeness and soundness, respectively.
Both results rely on the following lemma, which ensures that \FIRST contexts translate to evaluation contexts in \APCP.

\lFIRSTTransCtxs*

\begin{sketch}
    By induction on the structure of the contexts.
\end{sketch}

\subsubsection{Completeness}
\label{as:FIRST:oc:compl}

Here we prove the completeness of the translation.
The proof relies on the following intermediate results:
\begin{itemize}

    \item
        \Cref{t:FIRST:transTermSc,t:FIRST:transConfSc} prove that the translation preserves structural congruence for terms and configurations, respectively.

    \item
        \Cref{t:FIRST:transTermRed} then shows that the translation is complete with respect to term reduction.

    \item
        Finally, we prove completeness (\Cref{t:FIRST:transCompl}).

\end{itemize}
\noindent 
\begin{restatable}[Preservation of Structural Congruence for Terms]{theorem}{tFIRSTTransTermSc}
\label{t:FIRST:transTermSc}
    Given $\type{\Gamma} \vdashM \term{M}: \type{T}$, if $\term{M} \equivM \term{N}$, then $\encc{z}{M} \equiv \encc{z}{N}$.
\end{restatable}

\begin{proof}
    By induction on the derivation of $\term{M} \equivM \term{N}$ (\ih1).
    \proofFIRSTTransTermSc12{\qedhere}
\end{proof}
\noindent
\begin{restatable}[Preservation of Structural Congruence for Configurations]{theorem}{tFIRSTTransConfSc}
    \label{t:FIRST:transConfSc}
    Given $\type{\Gamma} \vdashC{\phi} \term{C}: \type{T}$, if $\term{C} \equivC \term{D}$, then $\encc{z}{C} \equiv \encc{z}{D}$.
\end{restatable}

\begin{proof}
    By induction on the derivation of $\term{C} \equivC \term{D}$ (\ih1).
    The inductive cases follow from \ih1 and \Cref{l:FIRST:transCtxs} straightforwardly.
    We detail the base cases, induced by the ten rules in \Cref{f:FIRST:confs}:
    \begin{itemize}

        \item
            Rule~\ruleLabel{sc-term-sc}:
            $\term{M} \equivM \term{M'}$ implies $\term{\phi \, M} \equivC \term{\phi \, M'}$.
            We have $\encc{z}{\phi \, M} = \encc{z}{M}$ and $\encc{z}{\phi \, M'} = \encc{z}{M'}$.
            By the assumption that $\term{M} \equivM \term{M'}$ and \Cref{t:FIRST:transTermSc}, $\encc{z}{M} \equiv \encc{z}{M'}$.
            The thesis follows immediately.

        \item
            \proofFIRSTTransConfScResSwap{}{2}

        \item
            Rule~\ruleLabel{sc-res-comm}:
            $\term{\pRes{x\tBfr{\vec{m}}y} \pRes{z\tBfr{\vec{n}}w} C} \equivC \term{\pRes{z\tBfr{\vec{n}}w} \pRes{x\tBfr{\vec{m}}y} C}$.
            The thesis holds as follows:
            \begin{align*}
                \encc{z}{\pRes{x\tBfr{\vec{m}}y} \pRes{z\tBfr{\vec{n}}w} C} &= \pRes{a_1x} \pRes{b_1y} \big( \encc{a_1\>b_1}{\tBfr{\vec{m}}} \| \pRes{a_2z} \pRes{b_2w} ( \encc{a_2\>b_2}{\tBfr{\vec{n}}} \| \encc{z}{C} ) \big)
                \\
                &\equiv \pRes{a_2z} \pRes{b_2w} \big( \encc{a_2\>b_2}{\tBfr{\vec{n}}} \| \pRes{a_1x} \pRes{b_1y} ( \encc{a_1\>b_1}{\tBfr{\vec{m}}} \| \encc{z}{C} ) \big)
                \\
                &= \encc{z}{\pRes{z\tBfr{\vec{n}}w} \pRes{x\tBfr{\vec{m}}y} C}
            \end{align*}

        \item
            Rule~\ruleLabel{sc-res-ext}:
            $\term{x},\term{y} \notin \fv(\term{C})$ implies $\term{\pRes{x\tBfr{\vec{m}}y} ( C \prl D )} \equivC \term{C \prl \pRes{x\tBfr{\vec{m}}y} D}$.
            The analysis depends on which of $\term{C},\term{D}$ is a child thread; w.l.o.g., assume that $\term{C}$ is.
            Assume the condition; by \Cref{l:FIRST:transCtxs}, then $x,y \notin \fn(\encc{\_}{C})$ ($\ast$).
            The thesis holds as follows:
            \begin{align*}
                \encc{z}{\term{\pRes{x\tBfr{\vec{m}}y} ( C \prl D )}} &= \pRes{ax} \pRes{by} ( \encc{a\>b}{\tBfr{\vec{m}}} \| \pRes{\_\_} \encc{\_}{C} \| \encc{z}{D} )
                \\
                &\equiv \pRes{\_\_} \encc{\_}{C} \| \pRes{ax} \pRes{by} ( \encc{a\>b}{\tBfr{\vec{m}}} \| \encc{z}{D} )
                \tag{$\ast$}
                \\
                &= \encc{z}{C \prl \pRes{x\tBfr{\vec{m}}y} D}
            \end{align*}

        \item
            Rule~\ruleLabel{sc-par-comm}:
            $\term{C \prl D} \equivC \term{D \prl C}$.
            The analysis depends on which of $\term{C},\term{D}$ is a child thread; w.l.o.g., assume that $\term{C}$ is.
            The thesis holds as follows:
            \begin{align*}
                \encc{z}{C \prl D} &= \pRes{\_\_} \encc{\_}{C} \| \encc{z}{D}
                \\
                &\equiv \encc{z}{D} \| \pRes{\_\_} \encc{\_}{C}
                \\
                &= \encc{z}{D \prl C}
            \end{align*}

        \item
            Rule~\ruleLabel{sc-par-assoc}:
            $\term{C \prl (D \prl E)} \equivC \term{(C \prl D) \prl E}$.
            The analysis depends on which of $\term{C},\term{D},\term{E}$ are child threads; w.l.o.g., assume that $\term{C},\term{D}$ are.
            The thesis holds as follows:
            \begin{align*}
                \encc{z}{C \prl (D \prl E)} &= \pRes{\_\_} \encc{\_}{C} \| ( \pRes{\_\_} \encc{\_\_}{D} \| \encc{z}{E} )
                \\
                &\equiv \pRes{\_\_} ( \pRes{\_\_} \encc{\_}{C} \| \encc{\_\_}{D} ) \| \encc{z}{E}
                \\
                &= \encc{z}{(C \prl D) \prl E}
            \end{align*}

        \item
            Rule~\ruleLabel{sc-conf-sub}:
            $\term{\phi \, (M \tSub{ N/x })} \equivC \term{(\phi \, M) \tSub{ N/x }}$.
            The thesis holds as follows:
            \begin{align*}
                \encc{z}{\phi \, (M \tSub{ N/x })} &= \pRes{xa} ( \encc{z}{M} \| \pIn a(\_,b) ; \encc{b}{N} )
                \\
                &= \pRes{xa} ( \encc{z}{\phi \, M} \| \pIn a(\_,b) ; \encc{b}{N} )
                \\
                &= \encc{z}{(\phi \, M) \tSub{ N/x }}
            \end{align*}

        \item
            Rule~\ruleLabel{sc-conf-sub-ext}:
            $\term{x} \notin \fv(\tCtx{G})$ implies $\term{(\tCtx{G}[C]) \tSub{ M/x }} \equivC \tCtx{G}[C \tSub{ M/x }]$.
            By \Cref{l:FIRST:transCtxs}, for any $\term{D}$, $\encc{z}{\tCtx{G}[D]} = \evalCtx{E}[\encc{z'}{D}]$ for some $\evalCtx{E},z'$ ($\ast_1$).
            Assume the condition; by \Cref{l:FIRST:transCtxs}, then $x \notin \fn(\evalCtx{E})$ ($\ast_2$).
            The thesis holds as follows:
            \begin{align*}
                \encc{z}{(\tCtx{G}[C]) \tSub{ M/x }} &= \pRes{xa} ( \encc{z}{\tCtx{G}[C]} \| \pIn a(\_,b) ; \encc{b}{M} )
                \\
                &= \pRes{xa} ( \evalCtx{E}[\encc{z'}{C}] \| \pIn a(\_,b) ; \encc{b}{M} )
                \tag{$\ast_1$}
                \\
                &\equiv \evalCtx{E}[\pRes{xa} ( \encc{z'}{C} \| \pIn a(\_,b) ; \encc{b}{M} )]
                \tag{$\ast_2$}
                \\
                &= \evalCtx{E}[\encc{z'}{C \tSub{ M/x }}]
                \\
                &= \encc{z}{\tCtx{G}[C \tSub{ M/x }]}
                \tag{$\ast_1$}
                \\
                \tag*{\qedhere}
            \end{align*}

    \end{itemize}
\end{proof}
\noindent 
\begin{restatable}[Completeness of Reduction for Terms]{theorem}{tFIRSTTransTermRed}
    \label{t:FIRST:transTermRed}
    Given $\type{\Gamma} \vdashM \term{M}: \type{T}$, if $\term{M} \reddM \term{N}$, then $\encc{z}{M} \redd* \encc{z}{N}$.
\end{restatable}

\begin{proof}
    By induction on the derivation of $\term{M} \reddM \term{N}$.
    We detail each rule:
    \begin{itemize}

        \item
            \proofTransTermRedLam

        \item
            Rule~\ruleLabel{red-pair}:
            $\term{\tLet (x,y) = (M_1,M_2) \tIn N} \reddM \term{N \tSub{ M_1/x,M_2/y }}$.
            The thesis holds as follows:
            \begin{align*}
                & \encc{z}{\tLet (x,y) = (M_1,M_2) \tIn N}
                \\
                &= \pRes{a_1b_1} \big(
                    \begin{array}[t]{@{}l@{}}
                        \pIn a_1(x,y) ; \encc{z}{N}
                        \\
                        {} \| \pRes{a_2b_2} \pRes{c_2d_2} ( \begin{array}[t]{@{}l@{}}
                            \pOut b_1[a_2,c_2]
                            \\
                            {} \| \pIn b_2(\_,e_2) ; \encc{e_2}{M_1}
                            \\
                            {} \| \pIn d_2(\_,f_2) ; \encc{f_2}{M_2} )
                        \big)
                    \end{array}
                \end{array}
                \\
                &\redd \pRes{a_2b_2}\pRes{c_2d_2} ( \begin{array}[t]{@{}l@{}}
                    \encc{z}{N} \{ a_2/x,c_2/y \}
                    \\
                    {} \| \pIn b_2(\_,e_2) ; \encc{e_2}{M_1} \| \pIn d_2(\_,f_2) ; \encc{f_2}{M_2} )
                \end{array}
                \\
                &\equiv \pRes{yd_2} \big( \pRes{xb_2} ( \encc{z}{N} \| \pIn b_2(\_,e_2) ; \encc{e_2}{M_1} ) \| \pIn d_2(\_,f_2) ; \encc{f_2}{M_2} \big)
                \\
                &= \encc{z}{N \tSub{ M_1/x,M_2/y }}
            \end{align*}

        \item
            Rule~\ruleLabel{red-name-sub}:
            $\term{x \tSub{ M/x }} \reddM \term{M}$.
            The thesis holds as follows:
            \begin{align*}
                \encc{z}{x \tSub{ M/x }} &= \pRes{xa} ( \pOut x[\_,z] \| \pIn a(\_,b) ; \encc{b}{M} )
                \\
                &\redd \encc{z}{M}
            \end{align*}

        \item
            Rule~\ruleLabel{red-lift}:
            $\term{M} \reddM \term{N}$ implies $\tCtx{R}[M] \reddM \tCtx{R}[N]$.
            By \Cref{l:FIRST:transCtxs}, for any $\term{L}$, $\encc{z}{\tCtx{R}[L]} = \evalCtx{E}[\encc{z'}{L}]$ for some $\evalCtx{E},z'$ ($\ast_1$).
            Assume the condition; by the IH, $\encc{z'}{M} \redd* \encc{z'}{N}$ ($\ast_2$).
            The thesis holds as follows:
            \begin{align*}
                \encc{z}{\tCtx{R}[M]} &= \evalCtx{E}[\encc{z'}{M}]
                \tag{$\ast_1$}
                \\
                &\redd* \evalCtx{E}[\encc{z'}{N}]
                \tag{$\ast_2$}
                \\
                &= \encc{z}{\tCtx{R}[N]}
                \tag{$\ast_1$}
            \end{align*}

        \item
            Rule~\ruleLabel{red-lift-sc}:
            $\term{M} \equivM \term{M'}$, $\term{M'} \reddM \term{N'}$, and $\term{N'} \equivM \term{N}$ imply $\term{M} \reddM \term{N}$.
            Assume the conditions.
            By \Cref{t:FIRST:transTermSc}, $\encc{z}{M} \equiv \encc{z}{M'}$ ($\ast_1$) and $\encc{z}{N'} \equiv \encc{z}{N}$ ($\ast_2$).
            By the IH, $\encc{z}{M'} \redd* \encc{z}{N'}$ ($\ast_3$).
            The thesis holds as follows:
            \begin{align*}
                \encc{z}{M} &\equiv \encc{z}{M'}
                \tag{$\ast_1$}
                \\
                &\redd* \encc{z}{N'}
                \tag{$\ast_3$}
                \\
                &\equiv \encc{z}{N}
                \tag{$\ast_2$}
                \\
                \tag*{\qedhere}
            \end{align*}

    \end{itemize}
\end{proof}
\noindent 
\tFIRSTTransCompl*

\begin{proof}
    By induction on the derivation of $\term{C} \reddC \term{D}$.
    We detail every rule:
    \begin{itemize}

        \item
            Rule~\ruleLabel{red-new}:
            $\tCtx{F}[\tNew] \reddC \term{\pRes{x\tBfr{\epsi}y} ( \tCtx{F}[(x,y)] )}$.
            By \Cref{l:FIRST:transCtxs}, for any $\term{M}$, $\encc{z}{\tCtx{F}[M]} = \evalCtx{E}[\encc{z'}{M}]$ for some $\evalCtx{E},z'$ ($\ast$).
            The thesis holds as follows:
            \begin{align*}
                \encc{z}{\tCtx{F}[\tNew]} &= \evalCtx{E}[\encc{z'}{\tNew}]
                \tag{$\ast$}
                \\
                &= \evalCtx{E}[\pRes{ab} \big( \pOut a[\_,z'] \| \pIn b(\_,c) ; \pRes{dx} \pRes{ey} ( \encc{d\>e}{\tBfr{\epsi}} \| \encc{c}{(x,y)} ) \big)]
                \\
                &\redd \evalCtx{E}[\pRes{dx} \pRes{ey} ( \encc{d\>e}{\tBfr{\epsi}} \| \encc{z'}{(x,y)} )]
                \\
                &\equiv \pRes{dx} \pRes{ey} ( \encc{d\>e}{\tBfr{\epsi}} \| \evalCtx{E}[\encc{z'}{(x,y)}] )
                \\
                &= \pRes{dx} \pRes{ey} ( \encc{d\>e}{\tBfr{\epsi}} \| \encc{z}{\tCtx{F}[(x,y)]} )
                \tag{$\ast$}
                \\
                &= \encc{z}{\pRes{x\tBfr{\epsi}y} (\tCtx{F}[(x,y)])}
            \end{align*}

        \item
            Rule~\ruleLabel{red-fork}:
            $\tCtx{\hat{F}}[\tFork M ; N] \reddC \term{\tCtx{\hat{F}}[N] \prl \tChild \, M}$.
            By \Cref{l:FIRST:transCtxs}, for any $\term{L}$, $\encc{z}{\tCtx{F}[L]} = \evalCtx{E}[\encc{z'}{L}]$ for some $\evalCtx{E},z'$ ($\ast_1$).
            Moreover, since $\tCtx{\hat{F}}$ does not have its hole under an explicit substitution, it does not capture any free variables of $M$; hence, by \Cref{l:FIRST:transCtxs}, $\evalCtx{E}$ does not capture any free names of $\encc{u}{M}$ for any $u$ ($\ast_2$).
            The thesis holds as follows:
            \begin{align*}
                \encc{z}{\tCtx{\hat{F}}[\tFork M ; N]} &= \evalCtx{E}[\encc{z'}{\tFork M ; N}]
                \tag{$\ast_1$}
                \\
                &= \evalCtx{E}[\pRes{ab} \big( \pOut a[\_,z'] \| \pIn b(\_,c); ( \pRes{\_\_} \encc{\_}{M} \| \encc{c}{N} ) \big) ]
                \\
                &\redd \evalCtx{E}[\pRes{\_\_} \encc{\_}{M} \| \encc{z'}{N}]
                \\
                &\equiv \evalCtx{E}[\encc{z'}{N}] \| \pRes{\_\_} \encc{\_}{M}
                \tag{$\ast_2$}
                \\
                &= \encc{z}{\tCtx{\hat{F}}[N]} \| \pRes{\_\_} \encc{\_}{M}
                \tag{$\ast_1$}
                \\
                &= \encc{z}{\tCtx{\hat{F}}[N] \prl \tChild \, M}
            \end{align*}

        \item
            \proofTransComplSend

        \item
            Rule~\ruleLabel{red-recv}:
            $\term{\pRes{x\tBfr{\vec{m},M}y} ( \tCtx{\hat{F}}[\tRecv y] \prl C )} \reddC \term{\pRes{x\tBfr{\vec{m}}y} ( \tCtx{\hat{F}}[(M,y)] \prl C )}$.
            W.l.o.g., assume $\term{C}$ is a child thread.
            By \Cref{l:FIRST:transCtxs}, for any $\term{L}$, $\encc{z}{\tCtx{F}[L]} = \evalCtx[\big]{E}[\encc{z'}{L}]$ for some $\evalCtx{E},z'$ ($\ast_1$).
            Moreover, since $\tCtx{\hat{F}}$ does not have its hole under an explicit substitution, it does not capture any free variables of $M$; hence, by \Cref{l:FIRST:transCtxs}, $\evalCtx{E}$ does not capture any free names of $\encc{u}{M}$ for any $u$ ($\ast_2$).
            The thesis holds as follows:
            \begin{align*}
                & \encc{z}{\pRes{x\tBfr{\vec{m},M}y} ( \tCtx{\hat{F}}[\tRecv y] \prl C )}
                \\ \displaybreak[1]
                &= \pRes{ax} \pRes{by} ( \encc{a\>b}{\tBfr{\vec{m},M}} \| \encc{z}{\tCtx{\hat{F}}[\tRecv y]} \| \pRes{\_\_} \encc{\_}{C} )
                \\ \displaybreak[1]
                &= \pRes{ax} \pRes{by} ( \encc{a\>b}{\tBfr{\vec{m},M}} \| \evalCtx[\big]{E}[\encc{z'}{\tRecv y}] \| \pRes{\_\_} \encc{\_}{C} )
                \tag{$\ast_1$}
                \\ \displaybreak[1]
                &= \begin{array}[t]{@{}l@{}}
                    \pRes{ax} \pRes{by} (
                    \\ ~~
                    \pRes{c_1d_1} \pRes{e_1f_1} ( \pIn c_1(\_,g_1) ; \encc{g_1}{M} \| \pIn b(\_,h_1) ; \pOut h_1[d_1,e_1] \| \encc{a\>f_1}{\tBfr{\vec{m}}} )
                    \\ ~~
                    {} \| \evalCtx[\big]{E}[\pRes{a_2b_2} \big( \pOut y[\_,a_2] \| \pIn b_2(c_2,d_2) ; \pRes{e_2,f_2} ( \pOut z'[c_2,e_2] \| \pIn f_2(\_,g_2) ; \pOut d_2[\_,g_2] ) \big)]
                    \\ ~~
                    {} \| \pRes{\_\_} \encc{\_}{C} )
                \end{array}
                \\ \displaybreak[1]
                &\redd \begin{array}[t]{@{}l@{}}
                    \pRes{ax} \pRes{a_2b_2} (
                    \\ ~~
                    \pRes{c_1d_1} \pRes{e_1f_1} ( \pIn c_1(\_,g_1) ; \encc{g_1}{M} \| \pOut a_2[d_1,e_1] \| \encc{a\>f_1}{\tBfr{\vec{m}}} )
                    \\ ~~
                    {} \| \evalCtx[\big]{E}[\pIn b_2(c_2,d_2) ; \pRes{e_2,f_2} ( \pOut z'[c_2,e_2] \| \pIn f_2(\_,g_2) ; \pOut d_2[\_,g_2] )]
                    \\ ~~
                    {} \| \pRes{\_\_} \encc{\_}{C} )
                \end{array}
                \\ \displaybreak[1]
                &\redd \begin{array}[t]{@{}l@{}}
                    \pRes{ax} \pRes{d_1c_1} \pRes{f_1e_1} (
                    \\ ~~
                    \pIn c_1(\_,g_1) ; \encc{g_1}{M} \| \encc{a\>f_1}{\tBfr{\vec{m}}}
                    \\ ~~
                    {} \| \evalCtx[\big]{E}[\pRes{e_2,f_2} ( \pOut z'[d_1,e_2] \| \pIn f_2(\_,g_2) ; \pOut e_1[\_,g_2] )]
                    \\ ~~
                    {} \| \pRes{\_\_} \encc{\_}{C} )
                \end{array}
                \\ \displaybreak[1]
                &\equiv \begin{array}[t]{@{}l@{}}
                    \pRes{ax} \pRes{f_1e_1} (
                    \\ ~~
                    \encc{a\>f_1}{\tBfr{\vec{m}}}
                    \\ ~~
                    {} \| \evalCtx[\big]{E}[ \pRes{d_1c_1} \pRes{e_2,f_2} ( \pOut z'[d_1,e_2] \| \pIn c_1(\_,g_1) ; \encc{g_1}{M} \| \pIn f_2(\_,g_2) ; \pOut e_1[\_,g_2] )]
                    \\ ~~
                    {} \| \pRes{\_\_} \encc{\_}{C} )
                \end{array}
                \tag{$\ast_2$}
                \\ \displaybreak[1]
                &\equiv \begin{array}[t]{@{}l@{}}
                    \pRes{ax} \pRes{by} (
                    \\ ~~
                    \encc{a\>b}{\tBfr{\vec{m}}}
                    \\ ~~
                    {} \| \evalCtx[\big]{E}[ \pRes{d_1c_1} \pRes{e_2,f_2} ( \pOut z'[d_1,e_2] \| \pIn c_1(\_,g_1) ; \encc{g_1}{M} \| \pIn f_2(\_,g_2) ; \pOut y[\_,g_2] )]
                    \\ ~~
                    {} \| \pRes{\_\_} \encc{\_}{C} )
                \end{array}
                \\ \displaybreak[1]
                &= \pRes{ax} \pRes{by} ( \encc{a\>b}{\tBfr{\vec{m}}} \| \evalCtx[\big]{E}[\encc{z'}{(M,y)}] \| \pRes{\_\_} \encc{\_}{C} )
                \\ \displaybreak[1]
                &= \pRes{ax} \pRes{by} ( \encc{a\>b}{\tBfr{\vec{m}}} \| \encc{z}{\tCtx{\hat{F}}[(M,y)]} \| \pRes{\_\_} \encc{\_}{C} )
                \tag{$\ast_1$}
                \\
                &= \encc{z}{\pRes{x\tBfr{\vec{m}}y} ( \tCtx{\hat{F}}[(M,y)] \prl C )}
            \end{align*}

        \item
            Rule~\ruleLabel{red-select}:
            $\term{\pRes{x\tBfr{\vec{m}}y} ( \tCtx{F}[\tSel \ell \, x] \prl C )} \reddC \term{\pRes{x\tBfr{\ell,\vec{m}}y} ( \tCtx{F}[x] \prl C )}$.
            W.l.o.g., assume $\term{C}$ is a child thread.
            By \Cref{l:FIRST:transCtxs}, for any $\term{L}$, $\encc{z}{\tCtx{F}[L]} = \evalCtx[\big]{E_1}[\encc{z'}{L}]$ for some $\evalCtx{E_1},z'$ ($\ast_1$).
            By inversion of typing, $\type{\Gamma} \vdashB \term{\tBfr{\vec{m}}} : \type{S_1}\>\type{S}$ where $\type{S_1} = \type{\oplus \{ i : S_i \} \cup \{ \ell : S_2 \}}$ ($\ast_2$).
            By \Cref{l:FIRST:transBuf},
            \[
                \encc{a\>b}{\type{\Gamma} \vdashB \term{\tBfr{\vec{m}}} : \type{S_1}\>\type{S}} = \evalCtx[\big]{E_2}[\encc{a\>c}{\type{\emptyset} \vdashB \term{\tBfr{\epsi}} : \type{S_1}\>\type{S_1}}] \tag*{($\ast_3$)}
            \]
            and
            \[
                \encc{a\>b}{\type{\Gamma} \vdashB \term{\tBfr{\ell,\vec{m}}} : \type{S_0}\>\type{S}} = \evalCtx[\big]{E_2}[\encc{a\>c}{\type{\emptyset} \vdashB \term{\tBfr{\ell}} : \type{S_1}\>\type{S_1}}] \tag*{($\ast_4$)}
            \]
            for some $\evalCtx{E_2},c$.
            Below, we omit types from translations of buffers.
            The thesis holds as follows:
            \begin{align*}
                & \encc{z}{\pRes{x\tBfr{\vec{m}}y} ( \tCtx{F}[\tSel \ell \, x] \prl C )}
                \\ \displaybreak[1]
                &= \pRes{ax} \pRes{by} ( \encc{a\>b}{\tBfr{\vec{m}}} \| \encc{z}{\tCtx{F}[\tSel \ell \, x]} \| \pRes{\_\_} \encc{\_}{C} )
                \\ \displaybreak[1]
                &= \pRes{ax} \pRes{by} ( \evalCtx[\big]{E_2}[\encc{a\>c}{\tBfr{\epsi}}] \| \evalCtx[\big]{E_1}[\encc{z'}{\tSel \ell \, x}] \| \pRes{\_\_} \encc{\_}{C} )
                \tag{$\ast_1$,$\ast_3$}
                \\ \displaybreak[1]
                &= \begin{array}[t]{@{}l@{}}
                    \pRes{ax} \pRes{by} (
                    \\ ~~
                    \evalCtx[\big]{E_2}[\pIn a(\_,c_1) ; \pRes{d_1e_1} \big( \pOut c_1[\_,d_1] \| \pBra e_1(f_1) > \{ i : \ldots \}_{i \in I} \cup \{ \ell : \pRes{g_1h_1} \left( \begin{array}{@{}l@{}}
                        \pIn c(\_,k_1) ; \pSel k_1[g_1] < \ell
                        \\
                        {}\| \encc{f_1\>h_1}{\tBfr{\epsi}}
                    \end{array} \right) \!\} \!\big)\!] 
                    \\ ~~
                    {} \| \evalCtx[\big]{E_1}[\pRes{a_2b_2} \big( \pOut x[\_,a_2] \| \pIn b_2(\_,c_2) ; \pRes{d_2e_2} ( \pSel c_2[d_2] < \ell \| \pOut e_2[\_,z'] ) \big)]
                    \\ ~~
                    {} \| \pRes{\_\_} \encc{\_}{C} )
                \end{array}
                \tag{$\ast_2$}
                \\ \displaybreak[1]
                &\redd \begin{array}[t]{@{}l@{}}
                    \pRes{a_2b_2} \pRes{by} (
                    \\ ~~
                    \evalCtx[\big]{E_2}[\pRes{d_1e_1} \big( \pOut a_2[\_,d_1] \| \pBra e_1(f_1) > \{ i : \ldots \}_{i \in I} \cup \{ \ell : \pRes{g_1h_1} \left( \begin{array}{@{}l@{}}
                        \pIn c(\_,k_1) ; \pSel k_1[g_1] < \ell
                        \\
                        {}\| \encc{f_1\>h_1}{\tBfr{\epsi}}
                    \end{array} \right) \} \big)]
                    \\ ~~
                    {} \| \evalCtx[\big]{E_1}[\pIn b_2(\_,c_2) ; \pRes{d_2e_2} ( \pSel c_2[d_2] < \ell \| \pOut e_2[\_,z'] )]
                    \\ ~~
                    {} \| \pRes{\_\_} \encc{\_}{C} )
                \end{array}
                \\ \displaybreak[1]
                &\redd \begin{array}[t]{@{}l@{}}
                    \pRes{a_2b_2} \pRes{by} (
                    \\ ~~
                    \evalCtx[\big]{E_2}[\pRes{d_1e_1} \big( \pOut a_2[\_,d_1] \| \pBra e_1(f_1) > \{ i : \ldots \}_{i \in I} \cup \{ \ell : \pRes{g_1h_1} \left( \begin{array}{@{}l@{}}
                        \pIn c(\_,k_1) ; \pSel k_1[g_1] < \ell
                        \\
                        {}\| \encc{f_1\>h_1}{\tBfr{\epsi}}
                    \end{array} \right) \} \big)]
                    \\ ~~
                    {} \| \evalCtx[\big]{E_1}[\pIn b_2(\_,c_2) ; \pRes{d_2e_2} ( \pSel c_2[d_2] < \ell \| \pOut e_2[\_,z'] )]
                    \\ ~~
                    {} \| \pRes{\_\_} \encc{\_}{C} )
                \end{array}
                \\ \displaybreak[1]
                &\redd \begin{array}[t]{@{}l@{}}
                    \pRes{d_2e_2} \pRes{by} (
                    \\ ~~
                    \evalCtx[\big]{E_2}[\pRes{g_1h_1} ( \pIn c(\_,k_1) ; \pSel k_1[g_1] < \ell \| \encc{d_2\>h_1}{\tBfr{\epsi}} )]
                    \\ ~~
                    {} \| \evalCtx[\big]{E_1}[\pOut e_2[\_,z']]
                    \\ ~~
                    {} \| \pRes{\_\_} \encc{\_}{C} )
                \end{array}
                \\ \displaybreak[1]
                &\equiv \begin{array}[t]{@{}l@{}}
                    \pRes{ax} \pRes{by} (
                    \\ ~~
                    \evalCtx[\big]{E_2}[\pRes{g_1h_1} ( \pIn c(\_,k_1) ; \pSel k_1[g_1] < \ell \| \encc{a\>h_1}{\tBfr{\epsi}} )]
                    \\ ~~
                    {} \| \evalCtx[\big]{E_1}[\pOut x[\_,z']]
                    \\ ~~
                    {} \| \pRes{\_\_} \encc{\_}{C} )
                \end{array}
                \\ \displaybreak[1]
                &= \pRes{ax} \pRes{by} ( \evalCtx[\big]{E_2}[\encc{a\>c}{\tBfr{\ell}} )] \| \evalCtx[\big]{E_1}[\encc{z'}{x} )] \| \pRes{\_\_} \encc{\_}{C} )
                \\ \displaybreak[1]
                &= \pRes{ax} \pRes{by} ( \encc{a\>b}{\tBfr{\ell,\vec{m}}} )] \| \encc{z}{\tCtx{F}[x]} \| \pRes{\_\_} \encc{\_}{C} )
                \tag{$\ast_1$,$\ast_4$}
                \\ \displaybreak[1]
                &= \encc{z}{\pRes{x\tBfr{\ell,\vec{m}}y} ( \tCtx{F}[x] \prl C )}
            \end{align*}

        \item
            Rule~\ruleLabel{red-case}:
            $j \in I$ implies
            \[
                \term{\pRes{x\tBfr{\vec{m},j}y} ( \tCtx{F}[\tCase y \tOf \{ i : M_i \}_{i \in I}] \prl C )} \reddC \term{\pRes{x\tBfr{\vec{m}}y} ( \tCtx{F}[M_j\ y] \prl C )}.
            \]
            W.l.o.g., assume $\term{C}$ is a child thread.
            By \Cref{l:FIRST:transCtxs}, for any $\term{L}$, $\encc{z}{\tCtx{F}[L]} = \evalCtx[\big]{E}[\encc{z'}{L}]$ for some $\evalCtx{E},z'$ ($\ast$).
            Assume the condition.
            The thesis holds as follows:
            \begin{align*}
                & \encc{z}{\pRes{x\tBfr{\vec{m},j}y} ( \tCtx{F}[\tCase y \tOf \{ i : M_i \}_{i \in I}] \prl C )}
                \\ \displaybreak[1]
                &= \pRes{ax} \pRes{by} ( \encc{a\>b}{\tBfr{\vec{m},j}} \| \encc{z}{\tCtx{F}[\tCase y \tOf \{ i : M_i \}_{i \in I}]} \| \pRes{\_\_} \encc{\_}{C} )
                \\ \displaybreak[1]
                &= \pRes{ax} \pRes{by} ( \encc{a\>b}{\tBfr{\vec{m},j}} \| \evalCtx[\big]{E}[\encc{z'}{\tCase y \tOf \{ i : M_i \}_{i \in I}}] \| \pRes{\_\_} \encc{\_}{C} )
                \tag{$\ast$}
                \\ \displaybreak[1]
                &= \begin{array}[t]{@{}l@{}}
                    \pRes{ax} \pRes{by} (
                    \\ ~~
                    \pRes{c_1d_1} ( \pIn b(\_,e_1) ; \pSel e_1[c_1] < j \| \encc{a\>d_1}{\tBfr{\vec{m}}} )
                    \\ ~~
                    {} \| \evalCtx[\big]{E}[\pRes{a_2b_2} ( \pOut y[\_,a_2] \| \pBra b_2(c_2) > \{ i : \encc{z'}{M_i\ c_2} \}_{i \in I} )]
                    \\ ~~
                    {} \| \pRes{\_\_} \encc{\_}{C} )
                \end{array}
                \\ \displaybreak[1]
                &\redd \begin{array}[t]{@{}l@{}}
                    \pRes{ax} \pRes{a_2b_2} (
                    \\ ~~
                    \pRes{c_1d_1} (\pSel a_2[c_1] < j \| \encc{a\>d_1}{\tBfr{\vec{m}}} )
                    \\ ~~
                    {} \| \evalCtx[\big]{E}[\pBra b_2(c_2) > \{ i : \encc{z'}{M_i\ c_2} \}_{i \in I}]
                    \\ ~~
                    {} \| \pRes{\_\_} \encc{\_}{C} )
                \end{array}
                \\ \displaybreak[1]
                &\redd \pRes{ax} \pRes{d_1c_1} ( \encc{a\>d_1}{\tBfr{\vec{m}}} \| \evalCtx[\big]{E}[\encc{z'}{M_j\ c_2} \{ c_1/c_2 \}] \| \pRes{\_\_} \encc{\_}{C} )
                \\ \displaybreak[1]
                &= \pRes{ax} \pRes{d_1c_1} ( \encc{a\>d_1}{\tBfr{\vec{m}}} \| \evalCtx[\big]{E}[\encc{z'}{M_j\ c_1}] \| \pRes{\_\_} \encc{\_}{C} )
                \tag{\Cref{l:FIRST:transCtxs}}
                \\ \displaybreak[1]
                &\equiv \pRes{ax} \pRes{by} ( \encc{a\>b}{\tBfr{\vec{m}}} \| \evalCtx[\big]{E}[\encc{z'}{M_j\ y}] \| \pRes{\_\_} \encc{\_}{C} )
                \tag{\Cref{l:FIRST:transCtxs}}
                \\ \displaybreak[1]
                &= \pRes{ax} \pRes{by} ( \encc{a\>b}{\tBfr{\vec{m}}} \| \encc{z}{\tCtx{F}[M_j\ y]} \| \pRes{\_\_} \encc{\_}{C} )
                \tag{$\ast$}
                \\ \displaybreak[1]
                &= \encc{z}{\pRes{x\tBfr{\vec{m}}y} ( \tCtx{F}[M_j\ y] \prl C )}
            \end{align*}

        \item
            Rule~\ruleLabel{red-close}:
            $\term{\pRes{x\tBfr{\vec{m}}y} ( \tCtx{F}[\tClose x ; M] \prl C )} \reddC \term{\pRes{\tNil\tBfr{\vec{m}}y} ( \tCtx{F}[M] \prl C )}$.
            W.l.o.g., assume $\term{C}$ is a child thread.
            By \Cref{l:FIRST:transCtxs}, for any $\term{L}$, $\encc{z}{\tCtx{F}[L]} = \evalCtx[\big]{E_1}[\encc{z'}{L}]$ for some $\evalCtx{E_1},z'$ ($\ast_1$).
            The analysis depends on whether $\term{y} = \term{\tNil}$; w.l.o.g., assume not.
            By inversion of typing, $\type{\Gamma} \vdashB \term{\tBfr{\vec{m}}} : \type{\tEnd}\>\type{S}$ where $\type{S} \neq \type{\tNil}$ ($\ast_2$).
            By \Cref{l:FIRST:transBuf}, $\encc{a\>b}{\type{\Gamma} \vdashB \term{\tBfr{\vec{m}}} : \type{\tEnd}\>\type{S}} = \evalCtx[\big]{E_2}[\encc{a\>c}{\type{\emptyset} \vdashB \term{\tBfr{\epsi}} : \type{\tEnd}\>\type{\tEnd}}]$ ($\ast_3$) and $\encc{a\>b}{\type{\Gamma} \vdashB \term{\tBfr{\vec{m}}} : \type{\tNil}\>\type{S}} = \evalCtx[\big]{E_2}[\encc{a\>c}{\type{\emptyset} \vdashB \term{\tBfr{\epsi}} : \type{\tNil}\>\type{\tEnd}}]$ ($\ast_4$) for some $\evalCtx{E_2},c$.
            Below, we omit types from translations of buffers.
            The thesis holds as follows:
            \begin{align*}
                & \encc{z}{\pRes{x\tBfr{\vec{m}}y} ( \tCtx{F}[\tClose x ; M] \prl C )}
                \\ \displaybreak[1]
                &= \pRes{ax} \pRes{by} ( \encc{a\>b}{\tBfr{\vec{m}}} \| \encc{z}{\tCtx{F}[\tClose x ; M]} \| \pRes{\_\_} \encc{\_}{C} )
                \\ \displaybreak[1]
                &= \pRes{ax} \pRes{by} ( \evalCtx[\big]{E_2}[\encc{a\>c}{\tBfr{\epsi}}] \| \evalCtx[\big]{E_1}[\encc{z'}{\tClose x ; M}] \| \pRes{\_\_} \encc{\_}{C} )
                \tag{$\ast_1$,$\ast_3$}
                \\ \displaybreak[1]
                &= \begin{array}[t]{@{}l@{}}
                    \pRes{ax} \pRes{by} (
                    \\ ~~
                    \evalCtx[\big]{E_2}[\pIn a(\_,c_1) ; \pOut c_1[\_,\_] \| \pIn b(\_,d_1) ; \pOut d_1[\_,\_]]
                    \\ ~~
                    {} \| \evalCtx[\big]{E_1}[\pRes{a_2b_2} ( \pOut x[\_,a_2] \| \pIn b_2(\_,\_) ; \encc{z'}{M} )]
                    \\ ~~
                    {} \| \pRes{\_\_} \encc{\_}{C} )
                \end{array}
                \tag{$\ast_2$}
                \\ \displaybreak[1]
                &\redd \begin{array}[t]{@{}l@{}}
                    \pRes{a_2b_2} \pRes{by} (
                    \\ ~~
                    \evalCtx[\big]{E_2}[\pOut a_2[\_,\_] \| \pIn b(\_,d_1) ; \pOut d_1[\_,\_]]
                    \\ ~~
                    {} \| \evalCtx[\big]{E_1}[\pIn b_2(\_,\_) ; \encc{z'}{M}]
                    \\ ~~
                    {} \| \pRes{\_\_} \encc{\_}{C} )
                \end{array}
                \\ \displaybreak[1]
                &\redd \pRes{by} ( \evalCtx[\big]{E_2}[\pIn b(\_,d_1) ; \pOut d_1[\_,\_]] \| \evalCtx[\big]{E_1}[\encc{z'}{M}] \| \pRes{\_\_} \encc{\_}{C} )
                \\ \displaybreak[1]
                &\equiv \pRes{a\_} \pRes{by} ( \evalCtx[\big]{E_2}[\pIn b(\_,d_1) ; \pOut d_1[\_,\_]] \| \evalCtx[\big]{E_1}[\encc{z'}{M}] \| \pRes{\_\_} \encc{\_}{C} )
                \\ \displaybreak[1]
                &= \pRes{a\_} \pRes{by} ( \evalCtx[\big]{E_2}[\encc{a\>c}{\tBfr{\epsi}}] \| \evalCtx[\big]{E_1}[\encc{z'}{M}] \| \pRes{\_\_} \encc{\_}{C} )
                \\ \displaybreak[1]
                &= \pRes{a\_} \pRes{by} ( \encc{a\>b}{\tBfr{\vec{m}}} \| \encc{z}{\tCtx{F}[M]} \| \pRes{\_\_} \encc{\_}{C} )
                \tag{$\ast_1$,$\ast_4$}
                \\ \displaybreak[1]
                &= \encc{z}{\pRes{\tNil\tBfr{\vec{m}}y} ( \tCtx{F}[M] \| C )}
            \end{align*}

        \item
            Rule~\ruleLabel{red-par-nil}:
            $\term{C \prl \tChild \, ()} \reddC \term{C}$.
            The thesis holds as follows:
            \begin{align*}
                \encc{z}{C \prl \tChild \, ()} &= \encc{z}{C} \| \pRes{\_\_} \0
                \\
                &\equiv \encc{z}{C}
            \end{align*}

        \item
            Rule~\ruleLabel{red-res-nil}:
            $\term{\pRes{\tNil\tBfr{\epsi}\tNil} C} \reddC \term{C}$.
            The thesis holds as follows:
            \begin{align*}
                \encc{z}{\pRes{\tNil\tBfr{\epsi}\tNil} C} &= \pRes{a\_} \pRes{b\_} ( \0 \| \encc{z}{C} )
                \\
                &\equiv \encc{z}{C}
            \end{align*}

        \item
            Rule~\ruleLabel{red-lift-C}:
            $\term{C} \reddC \term{C'}$ implies $\term{\tCtx{G}[C]} \reddC \term{\tCtx{G}[C']}$.
            By \Cref{l:FIRST:transCtxs}, for any $\term{D}$, $\encc{z}{\tCtx{G}[D]} = \evalCtx[\big]{E}[\encc{z'}{D}]$ for some $\evalCtx{E},z'$ ($\ast_1$).
            Assume the condition.
            By the IH, $\encc{z'}{C} \redd* \encc{z'}{C'}$ ($\ast_2$).
            The thesis holds as follows:
            \begin{align*}
                \encc{z}{\tCtx{G}[C]} &= \evalCtx[\big]{E}[\encc{z'}{C}]
                \tag{$\ast_1$}
                \\ \displaybreak[1]
                &\redd* \evalCtx[\big]{E}[\encc{z'}{C'}]
                \tag{$\ast_2$}
                \\ \displaybreak[1]
                &= \encc{z}{\tCtx{G}[C']}
                \tag{$\ast_1$}
            \end{align*}

        \item
            Rule~\ruleLabel{red-lift-M}:
            $\term{M} \reddM \term{M'}$ implies $\term{\tCtx{F}[M]} \reddC \term{\tCtx{F}[M']}$.
            By \Cref{l:FIRST:transCtxs}, for any $\term{N}$, $\encc{z}{\tCtx{F}[N]} = \evalCtx[\big]{E}[\encc{z'}{N}]$ for some $\evalCtx{E},z'$ ($\ast_1$).
            Assume the condition.
            By \Cref{t:FIRST:transTermRed}, $\encc{z'}{M} \redd* \encc{z'}{M'}$ ($\ast_2$).
            The thesis holds as follows:
            \begin{align*}
                \encc{z}{\tCtx{F}[M]} &= \evalCtx[\big]{E}[\encc{z'}{M}]
                \tag{$\ast_1$}
                \\ \displaybreak[1]
                &\redd* \evalCtx[\big]{E}[\encc{z'}{M'}]
                \tag{$\ast_2$}
                \\ \displaybreak[1]
                &= \encc{z}{\tCtx{F}[M']}
                \tag{$\ast_1$}
            \end{align*}

        \item
            Rule~\ruleLabel{red-conf-lift-sc}:
            $\term{C} \equivC \term{C'}$, $\term{C'} \reddC \term{D'}$, and $\term{D'} \equivC \term{D}$ imply $\term{C} \reddC \term{D}$.
            Assume the conditions.
            By \Cref{t:FIRST:transConfSc}, $\encc{z}{C} \equiv \encc{z}{C'}$ ($\ast_1$) and $\encc{z}{D'} \equiv \encc{z}{D}$ ($\ast_2$).
            By the IH, $\encc{z}{C'} \redd* \encc{z}{D'}$ ($\ast_3$).
            The thesis holds as follows:
            \begin{align*}
                \encc{z}{C} &\equiv \encc{z}{C'}
                \tag{$\ast_1$}
                \\
                &\redd* \encc{z}{D'}
                \tag{$\ast_3$}
                \\
                &\equiv \encc{z}{D}
                \tag{$\ast_2$}
                \\
                \tag*{\qedhere}
            \end{align*}

    \end{itemize}
\end{proof}

\subsubsection{Soundness}
\label{as:FIRST:oc:sound}

\tFIRSTTransSound*

\begin{proof}
    \proofFIRSTTransSoundIntro
    \begin{itemize}

        \item
            Case $\term{C} = \term{\phi \, M}$.
            By construction, we can identify a maximal context $\tCtx{R}$ and a term $\term{M_0}$ such that $\term{M} = \tCtx{R}[M_0]$ and the observed reduction $\encc{z}{\phi \, (\tCtx{R}[M_0])} \redd Q_0$ originates from the translation of $M_0$ directly (i.e., not from inside an evaluation context in the translation of $\term{M_0}$ or from interaction with the translation of $\tCtx{R}$).

            We detail every case for $\term{M_0}$, though not all cases may be applicable to show a reduction.
            In each case, we rely on \Cref{l:FIRST:transCtxs} to work with $\evalCtx{E},z'$ such that $\encc{z}{\phi \, (\tCtx{R}[M_0])} = \encc{z}{\tCtx{R}[M_0]} = \evalCtx[\big]{E}[\encc{z'}{M_0}]$.
            Also, in many cases, subterms that partake in the reduction may appear in sequences of explicit substitutions; since structural congruence can always extrude the scope of explicit substitutions, they can, w.l.o.g., be factored out of the proofs below.
            \begin{itemize}

                \item
                    Case $\term{M_0} = \term{x}$.
                    We have $\encc{z'}{x} = \pOut x[\_,z']$, so no reduction is possible.

                \item
                    Case $\term{M_0} = \term{()}$.
                    We have $\encc{z'}{()} = \0$, so no reduction is possible.

                \item
                    Case $\term{M_0} = \term{\lam x . M_1}$.
                    We have $\encc{z'}{\lam x . M_1} = \pIn z'(x,a) ; \encc{a}{M_1}$, so no reduction is possible.

                \item
                    Case $\term{M_0} = \term{M_1\ M_2}$.
                    We have
                    \[
                        \encc{z'}{M_1\ M_2} = \pRes{a_1b_1} \pRes{c_1d_1} ( \encc{a_1}{M_1} \| \pOut b_1[c_1,z'] \| \pIn d_1(\_,e_1) ; \encc{e_1}{M_2} ).
                    \]
                    The reduction can only originate from a synchronization between the send on $b_1$ and a receive on $a_1$ in $\encc{a_1}{M_1}$.
                    By well typedness, $\term{M_1}$ must be of type $\type{T_2 \lolli T_1}$ and $\term{M_2}$ of type $\type{T_2}$.
                    It must then be the case that $\term{M_1} = \term{\lam x . M_{1.1}}$: this is the only possibility for a receive on $a_1$ in $\encc{a_1}{M_1}$.

                    Let $\term{D_0} \deq \term{\phi \, (\tCtx{R}[M_{1.1} \tSub{ M_2/x })]}$.
                    We have $\term{C} \reddC \term{D_0}$.
                    Moreover:
                    \begin{align*}
                        \encc{z}{C}
                        &= \evalCtx[\big]{E}[\pRes{a_1b_1} \pRes{c_1d_1} ( \pIn a_1(x,a_2) ; \encc{a_2}{M_{1.1}} \| \pOut b_1[c_1,z'] \| \pIn d_1(\_,e_1) ; \encc{e_1}{M_2} )]
                        \\
                        &\redd \evalCtx[\big]{E}[\pRes{xd_1} ( \encc{z'}{M_{1.1}} \| \pIn d_1(\_,e_1) ; \encc{e_1}{M_2} )]
                        \\
                        &= \encc{z}{D_0}
                    \end{align*}

                \item
                    Case $\term{M_0} = \term{\tNew}$.
                    We have
                    \[
                        \encc{z'}{\tNew} = \pRes{a_1b_1} \big( \pOut a_1[\_,z'] \| \pIn b_1(\_,c_1) ; \pRes{d_1x} \pRes{e_1y} ( \encc{d_1\>e_1}{\tBfr{\epsi}} \| \encc{c_1}{(x,y)} ) \big).
                    \]
                    The reduction can only originate from a synchronization between the send on $a_1$ and the receive on $b_1$.
                    Let $\term{D_0} \deq \term{\pRes{x\tBfr{\epsi}y} ( \phi \, (\tCtx{R}[(x,y)]) )}$.
                    We have $\term{C} \reddC \term{D_0}$.
                    Moreover:
                    \begin{align*}
                        \encc{z}{C}
                        &= \evalCtx[\big]{E}[\pRes{a_1b_1} \big( \pOut a_1[\_,z'] \| \pIn b_1(\_,c_1) ; \pRes{d_1x} \pRes{e_1y} ( \encc{d_1\>e_1}{\tBfr{\epsi}} \| \encc{c_1}{(x,y)} ) \big)]
                        \\
                        &\redd \pRes{d_1x} \pRes{e_1y} ( \encc{d_1\>e_1}{\tBfr{\epsi}} \| \evalCtx[\big]{E}[\encc{z'}{(x,y)}] )
                        \\
                        &= \encc{z}{D_0}
                    \end{align*}

                \item
                    Case $\term{M_0} = \term{\tFork M_1 ; M_2}$.
                    We have
                    \[
                        \encc{z'}{\tFork M_1 ; M_2} = \pRes{a_1} \big( \pOut a_1[\_,z'] \| \pIn b_1(\_,c_1) ; ( \pRes{\_\_} \encc{\_}{M_1} \| \encc{c_1}{M_2} ) \big).
                    \]
                    The reduction can only originate from a synchronization between the send on $a_1$ and the receive on $b_1$.
                    Let $\term{D_0} \deq \term{\phi \, (\tCtx{R}[M_2]) \prl \tChild \, M_1}$.
                    We have $\term{C} \reddC \term{D_0}$.
                    Moreover:
                    \begin{align*}
                        \encc{z}{C}
                        &= \evalCtx[\big]{E}[\pRes{a_1} \big( \pOut a_1[\_,z'] \| \pIn b_1(\_,c_1) ; ( \pRes{\_\_} \encc{\_}{M_1} \| \encc{c_1}{M_2} ) \big)]
                        \\
                        &\redd \evalCtx[\big]{E}[\encc{z'}{M_2}] \| \pRes{\_\_} \encc{\_}{M_1}
                        \\
                        &= \encc{z}{D_0}
                    \end{align*}

                \item
                    Case $\term{M_0} = \term{(M_1,M_2)}$.
                    We have
                    \[
                        \encc{z'}{(M_1,M_2)} = \pRes{a_1b_1} \pRes{c_1d_1} ( \pOut z'[a_1,c_1] \| \pIn b_1(\_,e_1) ; \encc{e_1}{M_1} \| \pIn d_1(\_,f_1) ; \encc{f_1}{M_2} ),
                    \]
                    so no reduction is possible.

                \item
                    Case $\term{M_0} = \term{\tLet (x,y) = M_1 \tIn M_2}$.
                    We have
                    \[
                        \encc{z'}{\tLet (x,y) = M_1 \tIn M_2} = \pRes{a_1b_1} ( \pIn a_1(x,y) ; \encc{z'}{M_2} \| \encc{b_1}{M_1} ).
                    \]
                    The reduction can only originate from a synchronization between the receive on $a_1$ and a send on $b_1$ in $\encc{b_1}{M_1}$.
                    By well typedness, $\term{M_1}$ must be of type $\type{T_{1.1} \times T_{1.2}}$.
                    It must then be the case that $\term{M_1} = \term{(M_{1.1},M_{1.2})}$: this is the only possibility for a send on $b_1$ in $\encc{b_1}{M_1}$.

                    Let $\term{D_0} \deq \term{\phi \, (\tCtx{R}[M_1 \tSub{ M_{1.1}/x,M_{1.2}/y }])}$.
                    We have $\term{C} \reddC \term{D_0}$.
                    Moreover:
                    \begin{align*}
                        \encc{z}{C}
                        &= \evalCtx[\big]{E}[\pRes{a_1b_1} ( \pIn a_1(x,y) ; \encc{z'}{M_2} \| \pRes{a_2b_2} \pRes{c_2d_2} \left(
                            \begin{array}{@{}l@{}}
                                \pOut b_1[a_2,c_2] \| \pIn b_2(\_,e_2) ; \encc{e_2}{M_{1.1}}
                                \\
                                {} \| \pIn d_2(\_,f_2) ; \encc{f_2}{M_{1.2}}
                            \end{array}
                        \right) )]
                        \\
                        &\redd \evalCtx[\big]{E}[\pRes{yd_2} \big( \pRes{xb_2} ( \encc{z'}{M_2} \| \pIn b_2(\_,e_2) ; \encc{e_2}{M_{1.1}} ) \| \pIn d_2(\_,f_2) ; \encc{f_2}{M_{1.2}} \big)]
                        \\
                        &= \encc{z}{D_0}
                    \end{align*}

                \item
                    Case $\term{M_0} = \term{\tSend M_1 \, M_2}$.
                    We have
                    \[
                        \encc{z'}{\tSend M_1 \, M_2} = \pRes{a_1b_1} \pRes{c_1d_1} \left(
                            \begin{array}{@{}l@{}}
                                \pIn a_1(\_,e_1) ; \encc{e_1}{M_1} \| \encc{c_1}{M_2}
                                \\
                                {} \| \pIn d_1(\_,f_1) ; \pRes{g_1h_1} ( \pOut f_1[b_1,g_1] \| \pOut h_1[\_,z'] )
                            \end{array}
                        \right).
                    \]
                    The reduction can only originate from a synchronization between the receive on $d_1$ and a send on $c_1$ in $\encc{c_1}{M_2}$.
                    By well typedness, $\term{M_2}$ must be of type $\type{{!}T_1 . S_2}$.
                    No reduction is possible: no $\term{M_2}$ can satisy these conditions.

                \item
                    Case $\term{M_0} = \term{\tRecv M_1}$.
                    We have
                    \[
                        \encc{z'}{\tRecv M_1} = \pRes{a_1b_1} \big( \encc{a_1}{M_1} \| \pIn b_1(c_1,d_1) ; \pRes{e_1f_1} ( \pOut z'[c_1,e_1] \| \pIn f_1(\_,g_1) ; \pOut d_1[\_,g_1] ) \big).
                    \]
                    The reduction can only originate from a synchronization between the receive on $b_1$ and a send on $a_1$ in $\encc{a_1}{M_1}$.
                    By well typedness, $\term{M_1}$ must be of type $\type{{?}T_1.S_1}$.
                    No reduction is possible: no $\term{M_1}$ can satisfy these conditions.

                \item
                    Case $\term{M_0} = \term{\tSel j \, M_1}$.
                    We have
                    \[
                        \encc{z'}{\tSel j \, M_1} = \pRes{a_1b_1} \big( \encc{a_1}{M_1} \| \pIn b_1(\_,c_1) ; \pRes{d_1e_1} ( \pSel c_1[d_1] < j \| \pOut e_1[\_,z'] ) \big).
                    \]
                    The reduction can only originate from a synchronization between the receive on $b_1$ and a send on $a_1$ in $\encc{a_1}{M_1}$.
                    By well typedness, $\term{M_1}$ must be of type $\type{\oplus \{ i : S_1^i \}_{i \in I}}$ with $j \in I$.
                    No reduction is possible: no $\term{M_1}$ can satisfy these conditions.

                \item
                    Case $\term{M_0} = \term{\tCase M_1 \tOf \{ i : M_2^i \}_{i \in I}}$.
                    We have
                    \[
                        \encc{z'}{\tCase M_1 \tOf \{ i : M_2^i \}_{i \in I}} = \pRes{a_1b_1} ( \encc{a_1}{M_1} \| \pBra b_1(c_2) > \{ i : \encc{z'}{M_2^i\ c_2} \}_{i \in I} ).
                    \]
                    The reduction can only originate from a synchronization between the branch on $b_1$ and a selection on $a_1$ in $\encc{a_1}{M_1}$.
                    By well typedness, $M_1$ must be of type $\type{\& \{ i : S_2^i \}_{i \in I}}$.
                    No reduction is possible: no $\term{M_1}$ can satisfy these conditions.

                \item
                    Case $\term{M_0} = \term{\tClose M_1 ; M_2}$.
                    We have
                    \[
                        \encc{z'}{\tClose M_1 ; M_2} = \pRes{a_1b_1} ( \encc{a_1}{M_1} \| \pIn b_1(\_,\_) ; \encc{z'}{M_2} ).
                    \]
                    The reduction can only originate from a synchronization between the receive on $b_1$ and a send on $a_1$ in $\encc{a_1}{M_1}$.
                    By well typedness, $\term{M_1}$ must be of type $\type{\tEnd}$.
                    No reduction is possible: no $\term{M_1}$ can satisfy these conditions.

                \item
                    Case $\term{M_0} = \term{M_1 \tSub{ M_2/x }}$.
                    We have
                    \[
                        \encc{z'}{M_1 \tSub{ M_2/x }} = \pRes{xa_1} ( \encc{z'}{M_1} \| \pIn a_1(\_,b_1) ; \encc{b_1}{M_2} ).
                    \]
                    The reduction can only originate from a synchronization between the receive on $a_1$ and a send on $x$ in $\encc{z'}{M_1}$.
                    It must then be the case that $\term{M_1} = \tCtx{R_1}[x]$.
                    By \Cref{l:FIRST:transCtxs}, $\encc{z'}{\tCtx{R_1}[x]} = \evalCtx{E_1}[\encc{z_1}{x}] = \evalCtx{E_1}[\pOut x[\_,z_1]]$.

                    Let $\term{D_0} \deq \term{\phi \, (\tCtx[\big]{R}[\tCtx{R_1}[M_2]])}$.
                    We have $\term{C} \equivC \term{\phi \, (\tCtx[\big]{R}[\tCtx{R_1}[x \tSub{ M_1/x }]])} \reddC \term{D_0}$.
                    Moreover:
                    \begin{align*}
                        \encc{z}{C}
                        &= \evalCtx[\big]{E}[\pRes{xa_1} ( \evalCtx{E_1}[\pOut x[\_,z_1]] \| \pIn a_1(\_,b_1) ; \encc{b_1}{M_2} )]
                        \\
                        &\redd \evalCtx[\big]{E}[\evalCtx{E_1}[\encc{z_1}{M_2}]]
                        \\
                        &= \encc{z}{D_0}
                    \end{align*}

            \end{itemize}

        \item
            Case $\term{C} = \term{C_1 \prl C_2}$.
            Assume, w.l.o.g., that $\term{C_2}$ is a child thread.
            We have
            \[
                \encc{z}{C} = \encc{z}{C_1} \| \pRes{\_\_} \encc{\_}{C_2}.
            \]
            The reduction may originate from $\encc{z}{C_1}$ or from $\encc{\_}{C_2}$; w.l.o.g., assume the former.

            We thus have $\encc{z}{C_1} \redd Q_1$.
            By \ih2, there are $\term{D_1},k_1 \geq 0$ such that $\term{C_1} \reddC \term{D_1}$ and $\encc{z}{C_1} \redd Q_1 \redd^{k_1} \encc{z}{D_1}$.
            Let $\term{D_0} \deq \term{D_1 \prl C_2}$.
            We have $\term{C} \reddC \term{D_0}$.
            Moreover:
            \begin{align*}
                \encc{z}{C}
                &= \encc{z}{C_1} \| \pRes{\_\_} \encc{\_}{C_2}
                \\
                &\redd^{k_1+1} \encc{z}{D_1} \| \pRes{\_\_} \encc{\_}{C_2}
                \\
                &= \encc{z}{D_0}
            \end{align*}

        \item
            Case \proofFIRSTTransSoundBuf{}

        \item
            Case $\term{C} = \term{C_1 \tSub{ M/x }}$.
            We have
            \[
                \encc{z}{C_1 \tSub{ M/x }} = \pRes{xa_1} ( \encc{z}{C_1} \| \pIn a_1(\_,b_1) ; \encc{b_1}{M} ).
            \]
            The reduction may originate from (i)~$\encc{z}{C_1}$ or (ii)~a synchronization between the receive on $a_1$ and a send on $x$ in $\encc{z}{C_1}$.
            We detail both cases:
            \begin{enumerate}[label=(\roman*)] 

                \item
                    The reduction originates from $\encc{z}{C_1}$.
                    We thus have $\encc{z}{C_1} \redd Q_1$.
                    By \ih2, there are $\term{D_1},k_1 \geq 0$ such that $\term{C_1} \reddC \term{D_1}$ and $\encc{z}{C_1} \redd Q_1 \redd^{k_1} \encc{z}{D_1}$.
                    Let $\term{D_0} \deq \term{D_1 \tSub{ M/x }}$.
                    We have $\term{C} \reddC \term{D_0}$.
                    Moreover:
                    \begin{align*}
                        \encc{z}{C}
                        &= \pRes{xa_1} ( \encc{z}{C_1} \| \pIn a_1(\_,b_1) ; \encc{b_1}{M} )
                        \\
                        &\redd^{k_1+1} \pRes{xa_1} ( \encc{z}{D_1} \| \pIn a_1(\_,b_1) ; \encc{b_1}{M} )
                        \\
                        &= \encc{z}{D_0}
                    \end{align*}

                \item
                    The reduction originates from a synchronization between the receive on $a_1$ and a send on $x$ in $\encc{z}{C_1}$.
                    It must then be the case that $\term{C_1} = \tCtx[\big]{G}[\tCtx{F}[x]]$.
                    By \Cref{l:FIRST:transCtxs}, $\encc{z}{\tCtx[\big]{G}[\tCtx{F}[x]]} = \evalCtx{E}[\encc{z'}{x}] = \evalCtx{E}[\pOut x[\_,z']]$.

                    Let $\term{D_0} \deq \tCtx[\big]{G}[\tCtx{F}[M]]$.
                    We have $\term{C} \equivC \tCtx[\big]{G}[\tCtx{F}[x \tSub{ M/x }]] \reddC \term{D_0}$.
                    Moreover:
                    \begin{align*}
                        \encc{z}{C}
                        &= \pRes{xa_1} ( \evalCtx{E}[\pOut x[\_,z']] \| \pIn a_1(\_,b_1) ; \encc{b_1}{M} )
                        \\
                        &\redd \evalCtx{E}[\encc{z'}{M}]
                        \\
                        &= \encc{z}{D_0}
                        \tag*{\qedhere}
                    \end{align*}

            \end{enumerate}

    \end{itemize}
\end{proof}

\end{document}